%% file: main.tex
\documentclass[a4paper,fleqn]{cas-sc}

\usepackage{include} 

\def\tsc#1{\csdef{#1}{\textsc{\lowercase{#1}}\xspace}}
\tsc{WGM}
\tsc{QE}
\tsc{EP}
\tsc{PMS}
\tsc{BEC}
\tsc{DE}

\newcolumntype{P}{l*{4}{>{\raggedright\arraybackslash}p{0.12\linewidth}}}

\begin{document}

\let\WriteBookmarks\relax
\def\floatpagepagefraction{1}
\def\textpagefraction{.001}

\shorttitle{Energy and Flow Effects of Optimal Automated Driving in Mixed Traffic: Vehicle-in-the-Loop Experimental Results}
\shortauthors{T. Ard et~al.}
\title[mode = title]{Energy and Flow Effects of Optimal Automated Driving in Mixed Traffic: Vehicle-in-the-Loop Experimental Results}

\author[1]{Tyler Ard}[type=editor,orcid=0000-0001-9816-7578]
\ead{trard@g.clemson.edu}
\credit{Methodology, Software, Validation, Formal Analysis, Investigation, Data Curation, Writing - original draft}
\author[2]{Longxiang Guo}[]
\credit{Methodology, Software, Investigation, Writing - Review \& Editing}
\author[1]{Robert Austin Dollar}[]
\credit{Methodology, Data Curation, Formal Analysis, Writing - Review \& Editing}
\author[1]{Alireza Fayazi}[]
\credit{Software, Data Curation, Validation, Formal Analysis, Writing - Review \& Editing, Supervision}
\author[2]{Nathan Goulet}[]
\credit{Investigation, Writing - Review \& Editing}
\author[2]{Yunyi Jia}[]
\credit{Conceptualization, Supervision, Funding Acquisition, Writing - Review \& Editing}
\author[2]{Beshah Ayalew}[]
\credit{Conceptualization, Supervision, Funding Acquisition, Writing - Review \& Editing}
\author[1]{Ardalan Vahidi}[type=editor,orcid=0000-0002-1669-3345]
\cormark[1]
\ead{avahidi@g.clemson.edu}
\credit{Conceptualization, Supervision, Project Administration, Funding Acquisition, Writing - review \& editing}
\cortext[cor1]{Corresponding author}
\address[1]{Department of Mechanical Engineering, Clemson University, Clemson, SC}
\address[2]{Department of Automotive Engineering, Clemson University, Greenville, SC}

\begin{abstract}
This paper experimentally demonstrates the effectiveness of an anticipative car-following algorithm in reducing energy use of gasoline engine and electric Connected and Automated Vehicles (CAV), without sacrificing safety and traffic flow. We propose a Vehicle-in-the-Loop (VIL) testing environment in which experimental CAVs driven on a track interact with surrounding virtual traffic in real-time. We explore the energy savings when following city and highway drive cycles, as well as in emergent highway traffic created from microsimulations. Model predictive control handles high level velocity planning and benefits from communicated intentions of a preceding CAV or estimated probable motion of a preceding human driven vehicle. A combination of classical feedback control and data-driven nonlinear feedforward control of pedals achieve acceleration tracking at the low level. The controllers are implemented in ROS and energy is measured via calibrated OBD-II readings. We report up to $30\%$ improved energy economy compared to realistically calibrated human driver car-following without sacrificing following headway. 
\end{abstract}

\begin{highlights}
    \item Mixes physical vehicles with virtual traffic environments for energy and flow evaluation from integrated vehicle control algorithms.
    \item Measures conventional and electric powertrain types for energy effects.
    \item Introduces probabilistic constraints to balance safety and traffic flow considerations.
    \item Combines data-driven techniques with classical techniques for automated vehicle actuation.
\end{highlights}


\maketitle


\section{Introduction} \label{sec:introduction}

Modern camera-based automated vehicles were popularized by efforts such as \citep{Pomerleau1990} or \citep{stanleyrobot}, conceptually demonstrating that challenges of vehicle localization and control in off-road and low-traffic environments could be overcome. Since then, community perspective has become commercially motivated to automate vehicle control on roadways mixed with human-driven traffic. Promising advances in automation could realize unprecedented improvements in network flow efficiency, energy utilization, and safety by leveraging data as available from traffic infrastructure and other drivers \citep{vahidi2018energy}. 

Traffic throughput issues have been experimentally shown in commercially produced adaptive cruise controllers, in which they can be string unstable - causing disturbance propagation to traffic upstream and reducing overall network efficiency \citep{Gunter2019}. However, experiments in automated vehicles have shown string stabilizing effects by leveraging vehicle-to-vehicle connectivity. \citep{Ploeg2011} give a classically-based connected and adaptive cruise controller for improving string stability, whereas \citep{Hajdu2020} additionally account for uncertainty in human driving to show robust string stabilizing cruise control. More specifically, improved traffic throughput has been quantified during simulation of highway scenarios with partially mixed strings of automated vehicles. Such topics have actuated vehicle motion using reactive car-following controllers \citep{Talebpour2016,Liu2020}, and could additionally utilize centralized prioritization schemes when in highway merging zones \citep{Rios-Torres2017,Rios-Torres2018}.

We focus on additionally quantifying improved energy efficiency in automated driving from flow experiments via anticipative car-following. Motion planning is done by casting an optimal control problem (OCP) through model predictive control (MPC). Such approaches form a look-ahead control scheme, and as such can utilize preview through connectivity or prediction to better plan vehicle maneuvers - similar to those presented in \citep{Zheng2017} for traffic smoothing. Energy-motivated nonlinear approaches such as \citep{Turri2017} are experimentally shown to improve energy economy by up to 50\% over the baseline control for a high-disturbance scenario when utilizing connected preview, whereas linearizations of similar OCPs would not rely on nonlinear program solvers while still showing notable energy improvements \citep{ard2019platooning}. Further experiments utilizing hardware-in-the-loop setups have quantified fuel and emission reductions from a combustion vehicle \citep{Schmied2015} using preview-based control. Specifically addressing energy economy and traffic compactness, \citep{Kamal2011} introduce a multi-objective approach that minimizes error from desired time gap behind the preceding vehicle and vehicle acceleration. This work employs the MPC approach proposed in \citep{dollar2018efficient,ard2019microsimulation}.

When connected preview is not available, challenges of safety can arise due to improper anticipation of the behavior of other vehicles, as well as sensing and recognition error of obstacles on the road \citep{TeslaCrash}. To address issues of safety during motion planning, barrier methods \citep{Schmied2015} or invariant set constraints \citep{Gao2014} can be introduced with MPC for control robustness. To further support the multi-objective of promoting both traffic compactness and energy economy, we instead introduce a probabilistic approach, similar to that of \citep{Zhang2011}, for robustness by quantifying uncertainty in prediction of the preceding vehicle \citep{ard2019microsimulation}.

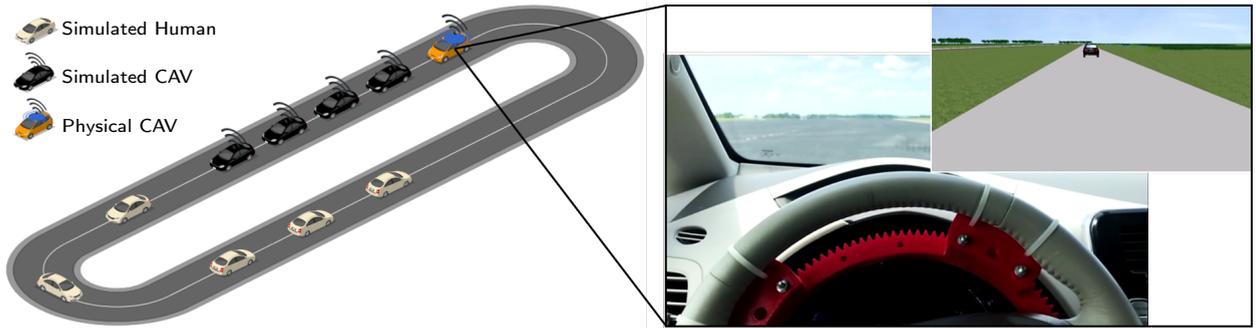
\begin{figure}
    \input{Images/pnp}
    \caption{Visualization of the Vehicle-in-the-Loop environment from the real vehicle perspective in both simulation and reality. The physical vehicle is embedded into a virtual environment and interacts with virtual drivers. The diagram on the left was generated at \texttt{icograms.com}}
    \label{fig:vilconcept}
\end{figure}

Ring experiments such as \citep{Stern2018} leverage a small string of vehicles to create macroscopic traffic effects from vehicles interacting with each other because no vehicle is in free flow. As a complement to experiments of this nature, a Vehicle-in-the-Loop (VIL) approach is presented here to additionally co-simulate virtual drivers and create a larger string of vehicles for more realistic traffic scenarios. This is conducted in a safe virtual reality mixed with a traffic microsimulation \citep{ard2019microsimulation,fayazi2019,Liu2020}, and has potentially additional applications for studying urban environments with intersection control, highway and merging scenarios, parking maneuvers, etc. 

We report on experimental details and analysis of the implementation of anticipative car-following algorithms proposed in \citep{dollar2018efficient,ard2019microsimulation}. Figure \ref{fig:vilconcept} demonstrates the virtual reality viewpoint of the VIL environment, where the physical vehicle is embedded in simulation. This work is focused on the impact of connectivity for predicting traffic events and improving subsequent motion planning - with an emphasis on increasing energy efficiency.

The contributions of this paper are the following:
\begin{enumerate}
    \item A VIL framework is proposed for real time co-simulation of cyber-physical systems of traffic simulation environments and physical automated vehicles.
    \item An anticipative and connected vehicle guided with MPC is experimentally studied for traffic flow and energy benefits. Safety and compactness robustification is handled through the use of probabilistic constraints in the absence of connectivity: explicitly considering the uncertainty in predicted position of the preceding vehicle.
    \item Instrumentation of combustion engine and electric vehicles is shown through pedal and steering actuators, and vehicle localization is handled through positioning and inertial measurements. Nonlinearities in pedal-to-engine transitions are approximated through data-driven maps for command of vehicle acceleration.
    \item Quantitative analysis is done to verify measured energy improvements of the automated control scheme for both a combustion engine and an electric vehicle in several traffic environments. Additionally, traffic flow impacts of the control scheme are shown. 
\end{enumerate}

The remainder of this paper is organized as follows. Section \ref{sec:VILSetup} describes the VIL architecture and how major system components relate to each other - motivating later discussion. Section \ref{sec:chanceMPC} then describes the MPC designed for high-level longitudinal control, and presents both the connected and unconnected variants of the controller. Baseline high-level control modeled to capture human drivers are given in Section \ref{sec:baselineCarFollowing}. The vehicle instrumentation and low-level control design are presented in Section \ref{sec:expVehicles}. To study energy for both combustion engine and electric vehicles, Section \ref{sec:OBDEnergy} presents the modeling techniques used to analyze on-board diagnostics (OBD) signals as collected during experiments. Then, experimental scenario descriptions are given in Section \ref{sec:expDescription}. Finally, Section \ref{sec:energyresults} discusses the results for the experiments, and Section \ref{sec:conclusion} concludes the work as presented here.


\section{Vehicle-in-the-Loop simulator setup}\label{sec:VILSetup}

Vehicle-in-the-Loop is an automated driving virtual simulation in which a physical vehicle (ego) is embedded into a traffic scenario with the goal of evaluating performance of such cyber-physical systems in a realistic manner without compromising safety. VIL is used here to evaluate energy performance of both a Mazda CX-7 combustion engine vehicle and a Nissan Leaf electric vehicle. 

VIL is developed in three component layers: 
\begin{enumerate}
    \item the server layer, responsible for conducting remote simulation of a traffic environment and communicating simulation status to the ego vehicle at fixed time intervals,
    \item the autonomous client layer, responsible for control decisions for the ego vehicle at a high-level trajectory planning and a low-level execution, as well as vehicle localization in the traffic environment and communication of vehicle status at fixed intervals in time, and
    \item the vehicle hardware layer, which features pedal and steering actuators fit to command the vehicle autonomously, as well as an OBD-II port which publishes select signals from the vehicle electronic-control-unit (ECU).
\end{enumerate} 

\begin{figure}
    \centering
    \includegraphics[page=1]{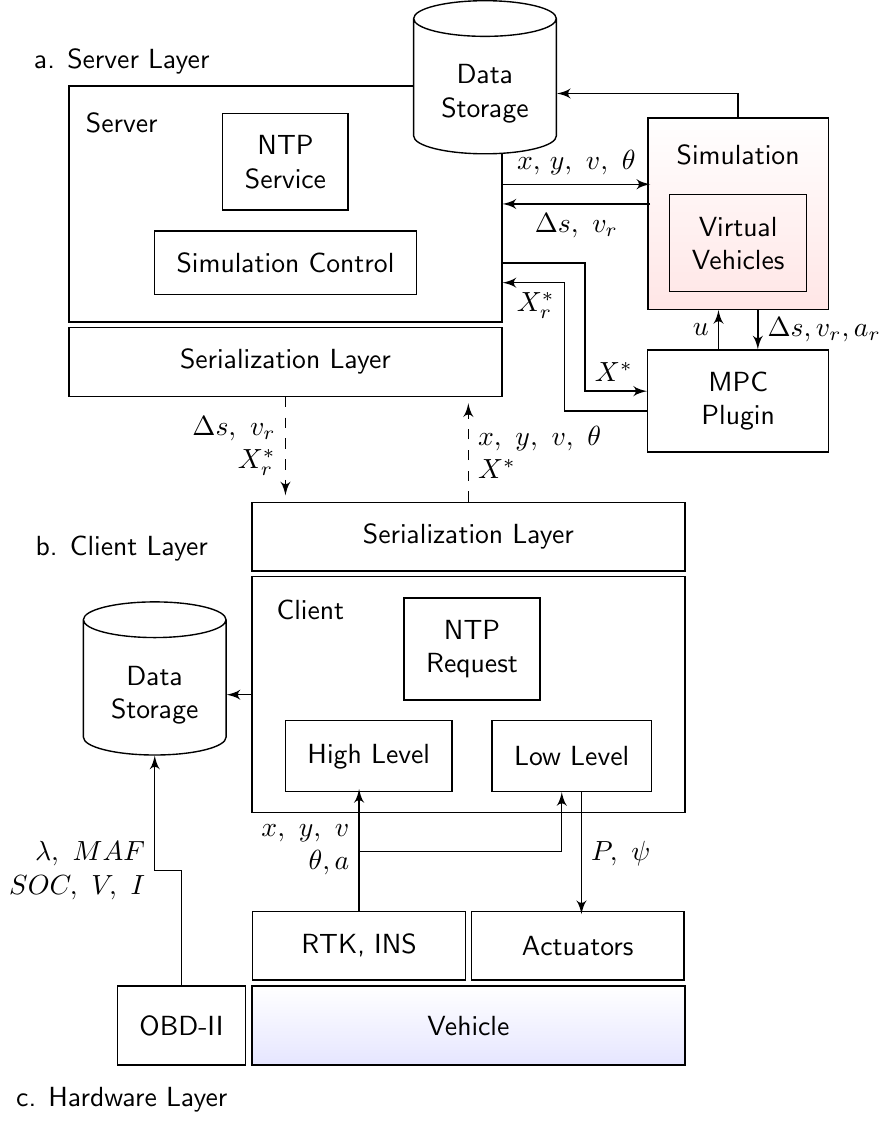}
    \caption{Local network of the VIL software and hardware}
    \label{fig:vilarchitecture}
\end{figure}

Figure \ref{fig:vilarchitecture} depicts the major nodes and signals of the VIL architecture. Software specifics are detailed for the remainder of this section.

\subsection{Simulation-reality synchronization}
To restrict both the ego and virtual vehicles to drive on the same road map and allow them to directly interact, a key design is made in how the roads of the simulation network are defined. First, a desired geometry is driven on a test track and positional data from the RTK-GPS servicing the ego vehicle is recorded. Then, the RTK signals are mapped directly to the simulation environment to construct the roadways $(x, y)_{\text{RTK}} \mapsto (x, y)_{\text{Sim}}$. It then follows that the ego vehicle can be embedded in simulation in real-time by communicating its position, velocity, and heading $(x, y, v, \theta)$ tuple.

\subsection{Simulation data exchange}
A client-server architecture was designated between the ego vehicle and simulation computer so that computational load could be split between multiple computers. Such a setup also has the advantage of allowing co-simulation of multiple clients at once - suitable for future experiments.

In this case, exchange of key data between simulation and the ego vehicle was defined using a Google Protocol Buffer (Protobuf) serialization to byte arrays and then broadcast through the User-Datagram Protocol (UDP) socket communication. UDP was chosen because of its low-latency data exchange and is suitable in systems with lower chances of packet loss. Protobuf is a class-based API that allows the user to define the data structures to be included in a message, then handles packing and unpacking the messages when sending or receiving \citep{protobuf}. 

Appendix \ref{app:CAVgpbcommunication} details the structure of the Protobuf message used. The data exchanged between the physical vehicle and the simulation server lies in four categories: 1) Subscription/Unsubscription Message: A physical vehicle subscribes to the simulation server at the beginning of VIL simulations, or unsubscribes to end the simulation, 2) V2Sim Message: A physical vehicle transmits its updates ($x$, $y$, $v$, $\theta$) to the simulation server, 3) Sim2V Message: the simulation server transmits information of the simulated vehicles surrounding the physical vehicle, and 4) V2V Message: the simulated and physical vehicles exchange planned trajectories for connected vehicle guidance (MPC-C). The simulation server was connected to the vehicle-control computer via ethernet, and V2V communication delay in MPC-C messages was emulated by delaying their transmission by \unit[100]{ms}. This was done to account for typical latencies of cellular or DSRC communication technologies that would be used in V2V experiments - as opposed to the VIL approach espoused here. 

When conducting experimental microsimulations - using commercial software PTV VISSIM - a plugin was created to define automated vehicle behavior using model predictive control \citep{ard2019microsimulation}. A shared memory interface was then established to handle interprocess communication between the server process and the plugin - so that communication between physical and virtual CAVs could occur. By this, the state preview generated from the MPC preceding vehicle, $X_r^*$, is extracted and communicated to the ego, and the ego shares its state preview, $X^*$, with simulation. 

\subsection{Simulation and control timing}
Timing mechanisms are defined in the server layer for precisely controlling the simulation environment to run in real-time and broadcast updates on its status in regular intervals of \unit[10]{hz}, and are defined in the client layer for broadcasting the ego vehicle's status in regular intervals of \unit[10]{hz} - so that the vehicle can provide feedback to the simulation for surrounding virtual traffic to react to. Because multiple computers are involved, a Network Time Protocol (NTP) was introduced to synchronize the clocks and regularly measure communication delay \citep{ntpmills}. In this case, the client polls the server for its clock time, and makes an adjustment to its own clock by measuring a time-offset $t'$, and round-trip delay $\Delta t$,
\begin{subequations}
\begin{align}
    &t' = \frac{(t_1 - t_0) - (t_3 - t_2)}{2} \\
    &\Delta t = (t_3 - t_0) - (t_2 - t_1)
\end{align}
\end{subequations}
where $t_0, \ t_3$ are the client's request and reception message timestamps, and $t_1, \ t_2$ are the server's reception and response message timestamps. Most recent round-trip delays are used by the client and server to interpolate positional data from the ego vehicle and its preceding vehicle using their respective current velocity $s^+ = s + v\Delta t/2$.

Further design elements of the VIL architecture are described in the following sections.


\section{Chance-constrained MPC for car-following}\label{sec:chanceMPC}

The proposed energy-efficient CAV controller uses model predictive control based on \citep{dollar2018efficient,ard2019microsimulation} to consider preview information when computing the ego vehicle's control move. It requires a longitudinal vehicle dynamics model that predicts how the ego vehicle will respond to a given sequence of control inputs over a finite time called the prediction horizon. Armed with such a model, the controller solves a mathematical program to minimize an objective subject to constraints. Since the preceding vehicle (PV) may be either connected or not, different algorithm variants are used to handle the connected (MPC-C) and unconnected (MPC-U) cases. The unconnected case involves particularly significant uncertainty that chance constraints help to manage.

This section will first introduce the model before developing the optimal control problem (OCP). Then, the connected and unconnected variants are described including the chance constraint formulation in the latter case.

\subsection{Control modeling}
Since the optimal control problem is solved numerically, computation time is an important consideration in MPC design. Problems using linear models, quadratic objectives, and affine constraints become quadratic programs (QPs), which can be solved quickly online\footnote{These experiments used the Gurobi optimizer \citep{gurobiref} to solve the QP - solving in \unit[2]{ms} on average.}. Hence, the following linear double-integrator with low-pass filter is chosen to approximate the vehicle dynamics.
\begin{equation} \label{eqn:linmodel}
    \frac{d}{dt}\begin{bmatrix} s \\ v \\ a
    \end{bmatrix} = \begin{bmatrix} 0 & 1 & 0 \\
    0 & 0 & 1 \\
    0 & 0 & -\tau_a^{-1}
    \end{bmatrix} \begin{bmatrix} s \\ v \\ a
    \end{bmatrix} + 
    \begin{bmatrix} 0 \\ 0 \\ \tau_a^{-1}
    \end{bmatrix} u 
\end{equation}
In Eqn. \eqref{eqn:linmodel}, the position $s$, velocity $v$, and actual acceleration $a$ form the state vector. The acceleration state lags the commanded acceleration $u$ with time constant $\tau_a$. In this experimental study, $\tau_a = \unit[0.275]{s}$ is used to approximate the lumped response of the acceleration controller, actuator, and vehicle powertrain or brake system.  

\subsection{Optimal control problem}
The optimal control problem (OCP) should fulfill several purposes: improving the ego vehicle's energy efficiency, smoothing the overall traffic stream, maintaining road throughput, guaranteeing safe and feasible operation, and limiting computation time.  In the proposed formulation, a weighted objective minimizes the squares of acceleration and headway tracking error, which can both be expressed as quadratic functions of state variables. Such an objective reduces unneeded braking, and reduces electric resistance losses in electric vehicles (EVs). While higher acceleration can sometimes improve efficiency in internal combustion engine vehicles (ICEVs) \citep{han2019fundamentals}, in some applications minimizing acceleration helps to avoid inefficient high-power enrichment as demonstrated in Section \ref{sec:energyresults}.

The following OCP yields the vector $U$ of control variables $u$ at each stage $i$ in the $N$-step prediction horizon and utilizes additional decision variables $\epsilon$ to soften pure state constraints
\begin{mini!} 
  {U,\epsilon}{q_g \left( s\left(N\right) - s_t\left(N\right) \right)^2 + q_a a^{2}\left(N\right) + \sum_{i=0}^{N-1} \left[ q_g \left( s\left(i\right) - s_t\left(i\right) \right)^2 + q_a \left(u^2\left(i\right) + a^2\left(i\right) \right) \right] + \sum_{j = 1}^{4}\rho_j \epsilon_j}{}{}
  \addConstraint{0 - \epsilon_3 \leq v(i) \leq \overline{v}(i) + \epsilon_2}{}{\label{eqn:vlim}}
  \addConstraint{\underline{u} \leq u(i) \leq \min\left\{m_1v(i) + b_1 , \ m_2v(i) + b_2 \right\}}{}{\label{eqn:ulim}}
  \addConstraint{a(i) \leq \min\left\{m_1v(i) + b_1 , \ m_2v(i) + b_2 \right\} + \epsilon_4}{}{\label{eqn:alim}}
  \addConstraint{s(i) \leq s_\alpha(i) - \underline{d} + \epsilon_1}{}{\label{eqn:slim}}
\end{mini!}
where $s_t(i) \triangleq s_r(i) - T v(i) - d_r $, $T$ is the target time headway, $s_r$ is the preceding vehicle's (PV's) position, and $d_r$ is the minimum target distance.  Combining $U$ with the ego's current state and linear model yields the ego's future state trajectory. Table \ref{tab:mpc} defines the MPC parameters and lists their values.

Equation \eqref{eqn:vlim} prevents the controller from driving in reverse or exceeding the speed limit $\overline{v}$ and is softened with slack variables $\epsilon_3$ and $\epsilon_2$, respectively.  The vehicle hardware has limited braking capacity $\underline{u}$ and acceleration capacities that Eqns. (\ref{eqn:ulim}, \ref{eqn:alim}) approximate.  The slack variable $\epsilon_4$ softens the acceleration state constraints; however, the input constraints are hard.  Table \ref{tab:mpc} provides the acceleration constraint coefficients $m$ and $b$. Collision avoidance is accomplished using Eqn. \eqref{eqn:slim}, where $s_\alpha$ denotes the position ahead of which the PV will lie with probability $\alpha$ and $\underline{d}$ is the minimum safe distance.  The collision avoidance constraint is softened with slack $\epsilon_1$ to guarantee feasibility even under unexpectedly harsh PV braking. The rest of this section will discuss how $s_\alpha$ is obtained in MPC-C and MPC-U.

\begin{table}[]
    \centering
    \caption{MPC parameters for the unconnected (MPC-U) and connected (MPC-C) cases.}
    \begin{tabular}{llll}
    \toprule
        Symbol & Description & MPC-U & MPC-C \\
    \midrule
        $N$ & horizon length & 16 [s] & 17 [s] \\
        $q_a$ & acceleration penalty & 2050 & 4000 \\
        $q_g$ & gap penalty & 1 & 1 \\
        $T$ & reference headway & 1.3 [s] & 0.0 [s] \\
        $d_r$ & reference gap & 2.0 [m] & 6.0 [m] \\
        $\underline{d}$ & minimum gap & 2.0 [m] & 2.0 [m] \\
    \midrule
        $\boldsymbol{\rho}$ & soft constraint penalty & \multicolumn{2}{l}{$\{1e6, \ 5e5, \ 5e5, \ 1e6\}$} \\
        $\boldsymbol{m}$ & acceleration constraint slope & \multicolumn{2}{l}{$\{0.285, \ -0.121\}$} \\
        $\boldsymbol{b}$ & acceleration constraint intercept & \multicolumn{2}{l}{$\{2.00, \ 4.83\}$} \\
        $\underline{u}$ & minimum acceleration constraint & \multicolumn{2}{l}{-5.5 [m/s$^2$]} \\
        $\tau$ & acceleration lag & \multicolumn{2}{l}{0.275 [s]} \\
        $\Delta t_h$ & discretization stepsize & \multicolumn{2}{l}{1.0 [s]} \\
    \bottomrule
    \end{tabular}
    \label{tab:mpc}
\end{table}

\subsection{Connected case (MPC-C)}
When a predictive CAV leads the ego vehicle in the string, the PV's planned trajectory is available through connectivity for all stages $i$. In this case, $s_r$ and $s_\alpha$ are both equal to the planned positions of the PV. This trajectory is accurate enough that no probabilistic safety margin is needed.

\subsection{Unconnected case (MPC-U)}
If a CAV follows a conventional vehicle that does not communicate its future plans, that CAV must predict $s_r$ and $s_\alpha$. To obtain the nominal PV position $s_r$, prediction assumes a constant acceleration $a_r$ with saturation of the predicted PV speed $v_r$ to its minimum or maximum.
\begin{subequations}
\begin{align}
    &a_r\left(i\right) = \begin{cases} a_r\left(0\right) & 0 < v\left(i\right) < \overline{v} \\
    0 & \mathrm{otherwise} \end{cases} \\
    &v_r\left(i+1\right) = \min\left\{ \max \left\{v_r\left(i\right) + \Delta t_h a_r\left(i\right), \ 0 \right\}, \ \overline{v} \right\} \\
    &s_r\left(i+1\right) = s_r\left(i\right) +  v_r \left(i\right) \Delta t_h + \frac{1}{2} a_r\left(i\right) \Delta t_h^2
\end{align}
\end{subequations}
In reality, the PV's acceleration can change significantly over the prediction horizon.  This mismatch between preview and reality can cause collisions if not addressed. A worst-case constraint was used in \citep{dollar2018efficient} to prevent collisions, but the resulting conservatism led to reduced traffic flow in the presence of unconnected vehicles.  To improve throughput, \citep{ard2019microsimulation} replaced the worst-case constraint with a less-conservative chance constraint.  

In the less-certain unconnected case, $s_\alpha \left(i\right)$ in Eqn. \eqref{eqn:slim} approximates the position that the PV will have reached by prediction step $i$ with probability $\alpha$. Therefore, the resulting open-loop ego vehicle trajectory avoids the realized PV trajectory at stage $i$ with probability $\alpha$.  The chance constraints consider a stochastic PV state vector $Z_r = \begin{bmatrix} S_r & V_r & A_r\end{bmatrix}^\mathrm{T}$. 
The uncertainty at the current time comes from the acceleration, whose deviation from $a_r$ is assumed normally distributed i.e. $\tilde{A}_r \sim \mathcal{N}\left(0, \ \sigma^2_A\right)$. Thus
\begin{equation}
    Z_r\left(0\right) = \begin{bmatrix} s_r & v_r & a_r \end{bmatrix}^\mathrm{T} + \begin{bmatrix} 0 & 0 & \tilde{A}_r \end{bmatrix}^\mathrm{T}.
\end{equation}
This initial distribution is used in \citep{ard2019microsimulation} to derive the future random PV position variable $S_r\left(i\right) = s_r + \tilde{S}_r\left(i\right)$, where $\tilde{S}_r\left(i\right) \sim \mathcal{N}\left( 0 , \ \Lambda_{11}\left(i\right)\right)$ with the covariance of the PV's future state as
\begin{equation}
    \Lambda\left(i\right) = A^i \begin{bmatrix} 0 & 0 & 0 \\ 0 & 0 & 0 \\ 0 & 0 & \sigma_A^2\end{bmatrix} \left(A^i\right)^\mathrm{T},
\end{equation}
which is obtained by propagating a Gaussian signal through the linear state-space model in Eqn. \eqref{eqn:linmodel}. The probability $\alpha$ can then be expressed in terms of the cumulative distribution function $F_{\tilde{S}_r}\left(s_\alpha - s_r\right)$ of $\tilde{S}_r$, where the argument is the position safety adjustment from the nominal estimate $s_r$.
\begin{equation}
    \alpha = 1 - F_{\tilde{S}_r}\left(s_\alpha - s_r\right) 
\end{equation}
Inverting the cumulative density function and exploiting its symmetry yields $s_\alpha$ for a chosen $\alpha$.
\begin{equation}
    s_\alpha\left(\alpha\right) = s_r - F_{\tilde{S}_r}^{-1}\left(\alpha\right)
\end{equation}
While \citep{wan2017probabilistic} used a constant value of $\alpha$, the linear function proposed in \citep{dollar2019automated} is used here,
%
%
\begin{equation}\label{eq:alpha}
    \alpha(t) = \left( \underline{\alpha} - \overline{\alpha} \right) \cdot \frac{t}{t_f} + \overline{\alpha}; \ \ t \in [1, 10]
\end{equation}
where $\underline{\alpha} = 0.5$, $\overline{\alpha} = 0.99999$, and $t_f = \unit[10]{s}$. Equation \eqref{eq:alpha} thus sets $\alpha$ to a conservative value early in prediction when the controller has little opportunity for closed-loop correction but the variance in PV position is small - and gradually relaxes $\alpha$ later in prediction when the variance in PV position is high but more risk is acceptable. This results in a maximum buffer distance of approximately $s_r - s_\alpha = \unit[9.5]{m}$ when $t=\unit[6]{s}$.


\section{Baseline car-following control}\label{sec:baselineCarFollowing}

To evaluate the energy and traffic flow impacts of the MPC automated vehicles, both the Wiedemann model (WIE) and Intelligent driver model (IDM) were designated as baseline longitudinal controllers. Both models were tuned to replicate human driving, and have been shown in various literature to well-realize macroscopic traffic effects \citep{WashingtonStateDepartmentofTransportationWSDOT2014ProtocolSimulation,FellendorfVISSIM,treiber2000congested}.

Real human drivers were not introduced in this study due to safety concerns of interfacing humans with virtual environments - particularly when driving cars. Human drivers are known to perceive upcoming traffic events by observing behavior of multiple vehicles downstream \citep{alexanderExpectancy}, so realistic human-trial experiments would likely either: 1) involve instrumenting a string of several automated vehicles to drive in front of a test subject - all of which interact with the VIL environment, or 2) introduce a mixed reality headset for the driver to wear. Both cases are interesting and the authors leave them for future studies.

\subsection{Wiedemann human driver model}

The psycho-physical Wiedemann 74 model, originally published in \citep{Wiedemann1974}, features distinct modes of: car-following headway tracking, free-driving velocity tracking, emergency braking, and catch-up acceleration. The model additionally limits acceleration capabilities of the vehicle as a function of velocity to model the powertrain. Additionally, by modeling imperfections in human perception of the velocity and acceleration of the preceding vehicle, it features oscillation in the commanded acceleration of the driver at steady tracking.

To build on the model, an updated version was then created specifically for microsimulation software PTV VISSIM - denoted the Wiedemann 99 model (WIE 99). This version expands on the number of parameters available to tune the longitudinal behavior, and because of refined thresholds for the different modes of driving, is preferred for longitudinal modeling \citep{FellendorfVISSIM}. The virtual vehicles in VISSIM are controlled by the WIE 99 model and are tuned by the procedure defined in \citep{ard2019microsimulation}. Several studies note that the default parameters in VISSIM are sufficient for general simulations, and only careful changes to driver headway are typically needed \citep{Woody2006CALIBRATINGVISSIM}. Here, it was observed that WIE 99 vehicles on average drive with more aggressive headways compared to empirical data. As such, the headways of the VISSIM drivers were adjusted to replicate the data as presented in \citep{Chen2010ATraffic}.

\begin{table}[]
    \centering
    \caption{Wiedemann 99 model parameters}
    \begin{tabular}{lll}
    \toprule
        Symbol & Description & Value \\
    \midrule
        $CC_0$ & standstill gap & 3.00 [m] \\
        $CC_1$ & desired headway (mean) & 1.35 [s] \\
        $CC_2$ & following oscillation amplitude & 8.00 [m] \\
        $CC_3$ & distance threshold & -8.00 [m] \\
        $CC_4$ & velocity threshold & -0.35 [m/s] \\
        $CC_5$ & velocity threshold & 0.35 [m/s] \\
        $CC_6$ & speed perception threshold & 11.4 [1/(m$\cdot$s)] \\
        $CC_7$ & acceleration oscillation amplitude & 0.25 [m/s$^2$] \\
        $CC_8$ & maximum acceleration (0 km/h) & 3.50 [m/s$^2$] \\
        $CC_9$ & maximum acceleration (80 km/h) & 1.50 [m/s$^2$] \\
        $\mathcal{X}$ & random driver variable in experiments & 0.35 [m/s$^2$] \\
    \bottomrule
    \end{tabular}
    \label{tab:wie}
\end{table}

In addition to using the WIE 99 model for modeling the human drivers in VISSIM, the WIE 99 model was implemented for control of the experimental vehicles as a baseline. The model is reproduced as given in \citep{vortischpdf} and shown in Appendix \ref{app:wiestate}. Although VISSIM has since expanded the model and uses additional features and considerations in car-following behavior, it was found that the implementation for the physical cars well-matched VISSIM behavior.

Table \ref{tab:wie} gives the parameters for the WIE 99 model used in simulation and in the physical car for experiments.

\subsection{Intelligent driver model}

The Intelligent driver model (IDM), as originally introduced in \citep{treiber2000congested}, is a popular longitudinal controller to model both human behavior and adaptive cruise controllers. For this study, the IDM was tuned to replicate human driving as introduced in \citep{dollar2018efficient} and based on the study by \citep{pourabdollahcalibration}. The IDM differs from the WIE model by precisely tracking desired headway in steady conditions and features less aggressive acceleration behavior.

The model computes an acceleration command given instantaneous ego velocity, $v$, and the differences in the relative position and velocity with respect to the PV, $\Delta s$ and $\Delta v$,
\begin{subequations}\label{eq:idm}
\begin{align}
    \dot{s} &= v \\
    \dot{v} &= a_0\left(1 - \left(\frac{v}{v_0}\right)^\delta - \frac{s^*(v, \Delta v)}{\Delta s} \right)
\end{align}
\end{subequations}
with $$s^*(v, \Delta v) = s_0 + \max\left\{0, \ Tv + \frac{v\Delta v}{2\sqrt{a_0 b_0}}\right\} .$$ 
Here, $a_0$ is the maximum acceleration, $b_0$ is the comfortable deceleration, $T$ is the desired time headway, $\delta$ is an acceleration exponent, $s_0$ is the desired vehicle-to-vehicle gap, and $v_0$ is the desired velocity. 

Table \ref{tab:idm} lists the parameters for the IDM used in the experiments.

\begin{table}[]
    \centering
    \caption{Intelligent driver model parameters}
    \begin{tabular}{lll}
    \toprule
        Symbol & Description & Value \\
    \midrule
        $a_0$ & maximum acceleration & 1.52 [m/s$^2]$ \\
        $b_0$ & comfortable deceleration & 3.24 [m/s$^2]$ \\
        $T$ & desired headway & 1.02 [s] \\
        $s_0$ & desired standstill distance & 10.0 [m] \\
        $\delta$ & acceleration exponent & 4 [-] \\
    \bottomrule
    \end{tabular}
    \label{tab:idm}
\end{table}


\section{Experimental vehicle adaptations}\label{sec:expVehicles}


As mentioned in Section \ref{sec:VILSetup}, the vehicle required pedal and steering actuators to command the vehicle autonomously. This section introduces the instrumentation of the experimental vehicles and the design and performance of the low-level controllers.

\subsection{Vehicle instrumentation}\label{sec:vehicleinstrumentation}
Testing consisted of one electric vehicle (Nissan Leaf) and one gasoline engine vehicle (Mazda CX-7). Both vehicles were not equipped with automated driving capabilities from the manufacturer, so they were modified to execute the commands from the high-level controller autonomously. The modifications include adding necessary sensors, actuators, and designing control algorithms.

\subsubsection{Structure of low-level controller}
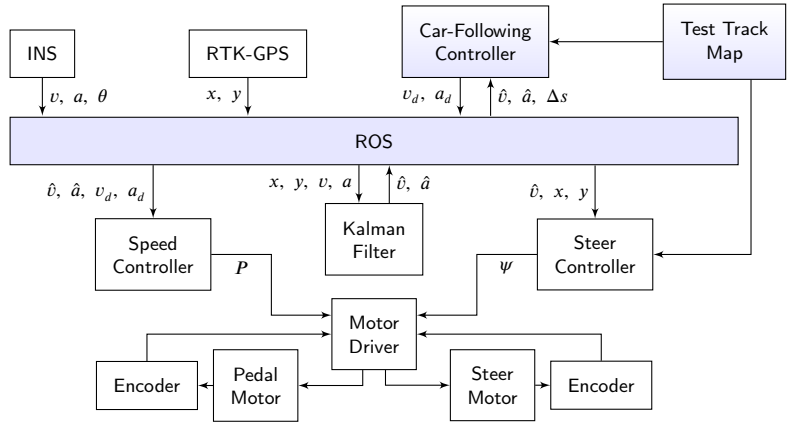
\begin{figure}
    \centering
    \input{Images/LLArchitecture}
    \caption{Structure of the low-level vehicle control system}
    \label{fig:lowlevelStruct}
\end{figure}

The relationship between the sensors, actuators and controls in the low-level controller is shown in Figure \ref{fig:lowlevelStruct}. An inertial navigation system (INS) and RTK-GPS were added to each experimental vehicle. The INS measures vehicle speed $v$, acceleration $a$, and heading direction $\theta$, while the RTK-GPS measures vehicle position $(x$, $y)$ and publishes through ROS. A Kalman Filter (KF) fuses the speed, acceleration, and location information to obtain a more accurate estimate of the true vehicle speed $\hat{v}$ and acceleration $\hat{a}$. 

The low-level speed controller computes the desired position of the motor in the pedal actuator from the localization information, the estimate from the KF, and the desired speed $v_d$ and acceleration $a_d$ issued by the high-level car-following controller. The low-level steering controller takes input from a pre-configured waypoint map of the test track and computes the desired position of the motor in the steering wheel actuator from the localization information and the estimated speed. The motors in both the pedal actuator and the steering wheel actuator are driven by a two-channel motor driver. They both work under position control - and the position feedback is provided by high precision optical encoders. 

\subsubsection{Design of pedal and steering actuators}
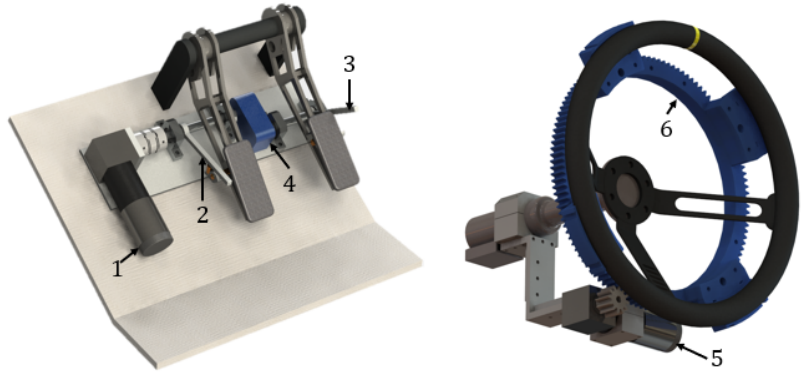
\begin{figure}
    \centering
    \input{Images/MechStruct}
    \caption{Specially designed actuators for pedal and steering wheel control}
    \label{fig:MechStruct}
\end{figure}

The pedal and steering wheel actuators were specially designed. They share a similar design for both vehicles, which is shown in Figure \ref{fig:MechStruct}, despite minor adjustments made to accommodate different vehicle geometries. Both actuators are additionally powered with an emergency stop if intervention is required.

Figure \ref{fig:MechStruct} left shows the structure of the pedal actuator. The actuator was designed to only push either the accelerator or brake pedal, and so features a release mechanism which releases the engaged pedal before actuation switches to the other. This is accomplished through levers 2 and 3, which have different rotation directions: lever 2 is attached to the output shaft of electric motor 1, whereas lever 3 is attached to another shaft that is parallel to and connected to the motor shaft via gear drive 4. A lever can push down the corresponding pedal by applying force on a small cylinder that is protruding from the side of the pedal, but the lever itself is not rigidly fixed to the pedal. Thus when lever 2 rotates down, the brake pedal is pressed and the accelerator pedal is released - and vice versa. 

Figure \ref{fig:MechStruct} right shows the structure of the steering wheel actuator. The electric motor 5 drives the gear ring 6 that is mounted behind the steering wheel so that the steering wheel can be rotated. 

\subsection{Low-level pedal controller design}

\begin{figure}
    \centering
    \input{Images/SpeedControl}
    \caption{Pedal control system}
    \label{fig:LowLevelSpeed}
\end{figure}
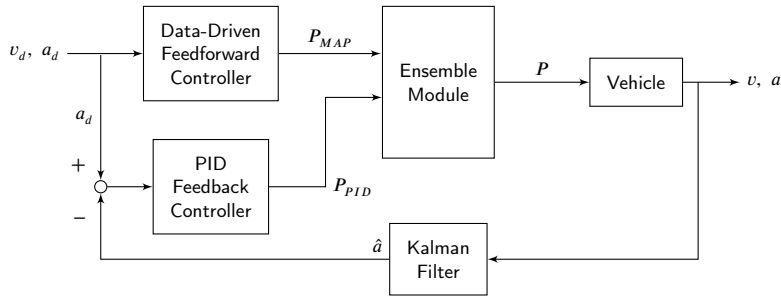

Section \ref{sec:vehicleinstrumentation} shows that longitudinal control of the vehicle is sought to be achieved by directly controlling the brake and accelerator pedals. However, the dynamics from pedals to vehicle motion is highly nonlinear due to the internal combustion engine, transmission \citep{li2010model,mcmahon1990vehicle}, or batteries \citep{szumanowski2008battery,bhangu2005nonlinear}. Moreover, the calibration map of the engine, transmission, or the battery was not available to us. 


These factors make the implementation of solely a classical controller or solely a data-driven controller difficult here. Thus, the two approaches are fused to combine a data-driven feedforward controller with a classical PID feedback controller to solve the speed and acceleration tracking problem. The structure of the pedal controller is shown in Figure \ref{fig:LowLevelSpeed}, which commands pedal position $P$ given desired velocity $v_d$ and acceleration $a_d$ from the car-following control.


\subsubsection{Data-driven feedforward control}


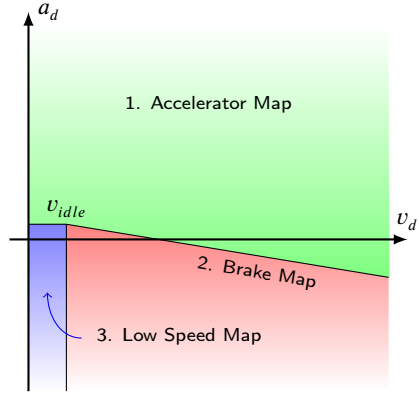
\begin{figure}
    \centering
    \input{Images/SSLongiCtrl}
    \caption{State space of the three different cases in longitudinal control}
    \label{fig:SSLongiCtrl}
\end{figure}

With longitudinal control actuated through the pedals, we classify three different cases. Case 1 is control of the vehicle speed using the accelerator pedal - shown by the green area in Figure \ref{fig:SSLongiCtrl}. This case is characterized by vehicle acceleration due to the accelerator pedal and deceleration due to coasting. Case 2 is high speed control of the vehicle deceleration using the brake pedal - shown by the red area in Figure \ref{fig:SSLongiCtrl}. This occurs when vehicle deceleration needs to be achieved by the brake pedal. Case 3 occurs when attempting to control vehicle speed below idling, and is shown by the blue area in Figure \ref{fig:SSLongiCtrl}.


The candidate calibration map is required to map the target speed and acceleration to proper pedal position as accurately as possible. However, because the response of the vehicle is significantly different in each case, three separate maps (1. accelerator map, 2. brake map, and 3. low speed map) were identified and fitted. Thresholds between the maps are defined from coasting and idling data to determine the active region of the state space. Switching between maps was then introduced, with hysteresis bands included so that high-frequency toggling between modes does not occur.


The collection and classification of the calibration data for the maps was first done through calibration tests on a chassis dynamometer. Collecting data from such a controlled environment can help ensure minimum influence from environmental disturbances such as varying wind speed, road roughness, and grade. Data was gathered for the various configurations of accelerator and brake pedal positions to map the state space. When collecting data for the accelerator map, the vehicle was accelerated from standing to maximum speed through a range of accelerator positions. Likewise, data was collected for the brake map when decelerating the vehicle from maximum speed to standing through a range of brake positions. For the low speed map, the vehicle was brought to idle speed and then stopped through a range of brake positions.



In this paper, the three maps were fit with three-dimensional polynomial surfaces. They were calibrated from labeled data with vehicle speed and acceleration as input and pedal position $P_{MAP}$ as target output. The order of the polynomial in each dimension was selected such that the surface fitting error is the minimum.  



\subsubsection{PID feedback control}


While the three maps can describe the inverse vehicle response in a controlled environment, there remains inaccuracy in the calibration. Moreover, when the vehicle is driving on open roads, the air drag, road resistance, and grade will vary. An open-loop feedforward controller that only employs the calibration maps will not overcome these disturbances, particularly without a significant increase in the required calibration data. Thus, combining with a feedback controller can improve the low-level tracking performance. Here, a PID controller is introduced in feedback
\begin{equation}
    P_{PID} = K_pe + K_i \int_0^t e \, dt + K_d \frac{de}{dt}
\end{equation}
where $e=a_d-\hat{a}$ quantifies the tracking error from the car-following command $a_d$. The desired speed is then calculated open-loop by
\begin{equation}
    v_d = v + a_d \Delta t
\end{equation}
where $\Delta t$ is the runtime of a control loop. 

\subsubsection{Fusion of data-driven feedforward and PID feedback controllers}

The final pedal position output, $P$, is calculated by the ensemble module. In this paper, a complementary ensemble fuses the outputs of the data-driven feedforward and the PID controllers by
%
\begin{equation}
    P = wP_{MAP} + (1 - w)P_{PID}
\end{equation}
where $w$ is the weight coefficient of the fused controller. The weight coefficient $w$ was tuned manually when driving on the test track so that the acceleration response of the vehicle did not show significant overshoot or overdamping. 

\subsection{Low-level steering controller design}

A Pure Pursuit controller was implemented for the experimental vehicles to track the designated path of the test track. The steering input $\psi$ to the vehicle is computed with the location of the target point $l_d$ and the angle $\alpha$ between the vehicle’s heading direction and the look-ahead direction
\begin{equation}
    \psi = k \, \arctan{\left( \frac{2L\sin\alpha}{l_d} \right)}
\end{equation}
where $l_d$ is the look ahead distance given by
\begin{equation}
    l_d = l_d^{min} + k_l\,v.
\end{equation}
$k_l$ and $l_d^{min}$ are tunable gain and minimum look-ahead distance parameters. More complete details can be found in the literature, such as in \citep{Amidi1991}. 

\subsection{Low-level speed profile tracking performance}
Both experimental vehicles were calibrated to follow a short square-wave speed profile while driving on a rough road with notable grade. A dynamic target generator was created based on the IDM from Eqn. \eqref{eq:idm}, which calculates desired speed and desired acceleration from the speed profile.
\begin{subequations}
\begin{align}
    &a_d = a_0\left( 1 - \left(\frac{v}{v_0}\right)^\delta\right) \\
    &v_d = v + a_d\Delta t
\end{align}
\end{subequations}
Here, $a_{0}=\pm \unit[2.0]{m/s^2}$ is the maximum acceleration, and $v_{0}$ is the target speed from the square-wave-shaped profile. The PID controllers of both vehicles were tuned to reject the road disturbance without notable overshoot.

\begin{figure}
    \centering
    \input{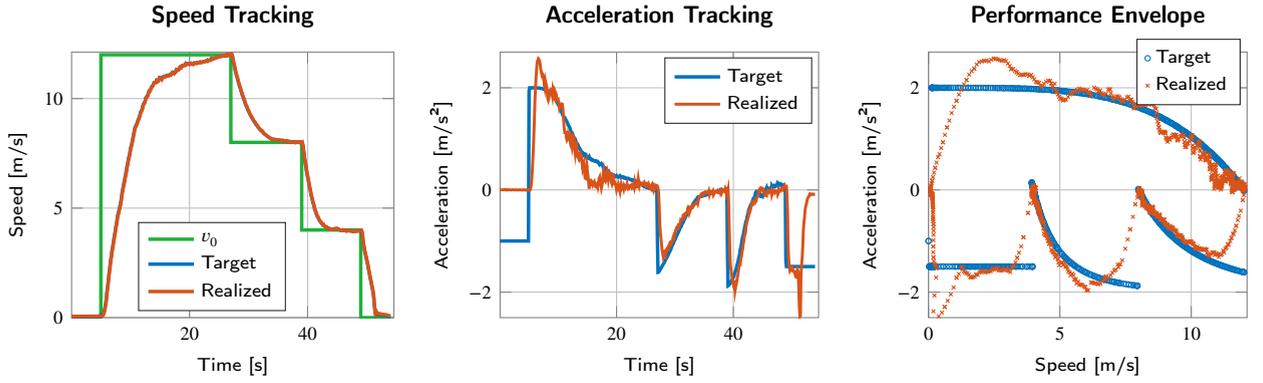}
    \caption{Mazda CX-7 speed profile tracking performance}
    \label{fig:SpeedTrackingMazda}
\end{figure}

\begin{figure}
    \centering
    \input{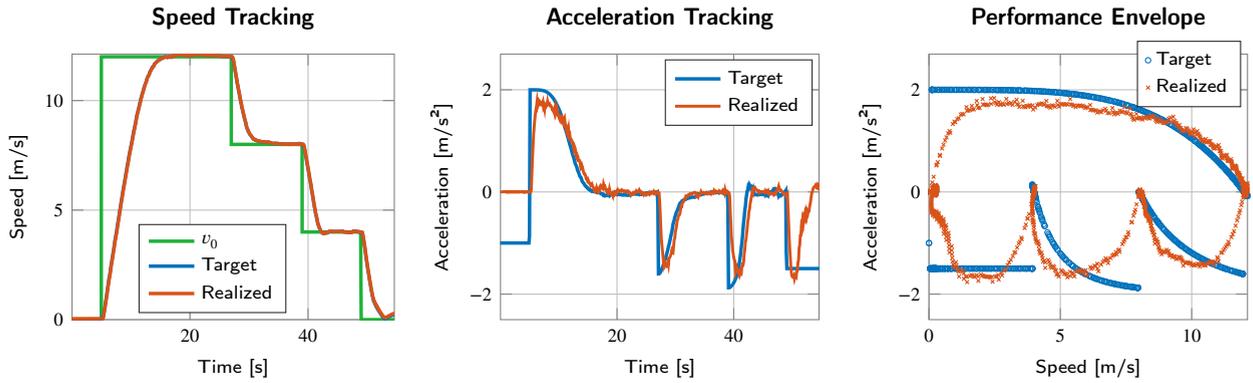}
    \caption{Nissan Leaf speed profile tracking performance}
    \label{fig:SpeedTrackingNissan}
\end{figure}

The result is shown in Figure \ref{fig:SpeedTrackingMazda} and \ref{fig:SpeedTrackingNissan}. The combined controller showed acceptable tracking response and achieved a steady state speed tracking error of $\unit[\pm0.06]{m/s}$ for both vehicles. The acceleration tracking was also accurate, where the performance envelope shows that the acceleration tracking was acceptable over the entire domain of velocity.

\section{OBD-II energy measurement methods}\label{sec:OBDEnergy}


An iOS app was created to pair with the On-Board Diagnostics (OBD-II) port of the vehicles with 29-bit ISO 15765-4 CAN protocol as in \citep{fayazi2019,fayazi2017vehicle}. The implemented iOS app connects to commercial WiFi OBD-II dongles supporting the ELM327 chip \citep{elm327}, as depicted in Figure \ref{fig:avlfuelmeter}. The app was extended to read the combustion engine Mazda CX7's 11-bit CAN protocol and the electric motor Nissan Leaf's unstandardized protocol. 
A custom procedure was programmed to increase the default data sampling frequency of the ELM chip. In total, data was collected from the Mazda at a rate of \unit[8]{hz}, and data was collected from the Nissan at a rate of \unit[4]{hz}. Finally, the app collected iOS GPS and timestamp data so that OBD-II readings could be correlated with simulation. 

\subsection{Internal combustion engine vehicle}

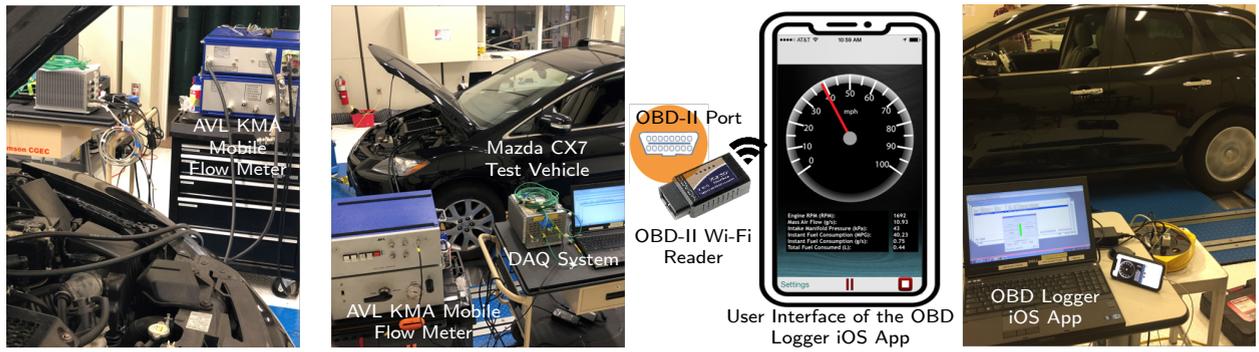
\begin{figure}
    \input{Images/AVL}
    \caption{The AVL KMA Mobile fuel measurement system and the OBD-based fuel rate estimator connected to the combustion test vehicle}
    \label{fig:avlfuelmeter}
\end{figure}

First, fuel flow rate was modeled as a function of available OBD-II signals and one data-driven parameter. The fuel estimation model was then validated using fuel volume and flow data available from a flow meter during a chassis dynamometer calibration test. Finally, the calibrated model was applied to OBD-II data from the test track to record fuel consumption.

Figure \ref{fig:avlfuelmeter} shows the flow meter calibration experiment, with an AVL KMA Mobile flow meter connected between the fuel tank and the high pressure fuel pump of the engine \citep{avlkma}. The SoMat eDAQ system \citep{somateDaq} was used with the SoMat Test Control Environment software to collect fuel data.

Fuel injector flow rate, pulse width, or a similar quantity was not available on the test vehicle, so fuel rate was modeled using ECU-estimated mass airflow (MAF), commanded $\lambda$, and a derived MAF correction curve. 
As background, the air-fuel ratio of a conventional gasoline engine is typically near stoichiometric to promote stable combustion, high efficiency, and low emissions. However, a mildly rich $\lambda$ of 0.85 to 0.91 can increase maximum torque. Even richer mixtures are sometimes used to limit exhaust gas temperatures, particularly in turbocharged applications. The test vehicle's OBD-II data includes the ECU-commanded air-fuel ratio in the form of $\lambda = \dot{m}_{fs} / \dot{m}_f$ where $\dot{m}_{fs}$ and $\dot{m}_f$ denote the stoichiometric and actual mass fuel flow, respectively.  The fuel flow that the ECU commands is modeled as follows.
\begin{equation} \label{eqn:fuelest}
    \dot{m}_f = \frac{\dot{m}_a}{AFR_s \lambda_{c}} \overline{E}_A \left(\dot{m}_a\right)
\end{equation}
A stoichiometric air-fuel ratio $AFR_s = 14.1$ was used for the \unit[10]{\%} ethanol pump fuel that is commonly available in the United States.

In Eqn. \eqref{eqn:fuelest}, $\overline{E}_A$ is a correction factor to the ECU-estimated mass airflow. $\lambda$ is normally closed-loop controlled to stoichiometric using an exhaust oxygen sensor. Long-term and short-term correction factors called \textit{trims} are applied to the fuel pulse width such that the desired $\lambda$ is delivered. These trims are denoted $LTFT$ and $STFT$, respectively. Errors in both airflow measurement and fuel system modeling contribute to these fuel trims. Assuming that fuel system model deviation results from a change in effective orifice size, Eqn. \eqref{eqn:mafcorrdef} assumes that the trim due to the fuel system is a constant $e_F$ with respect to mass airflow.
\begin{equation}  \label{eqn:mafcorrdef}
    E_A = 1 + \frac{LTFT + STFT}{100} - e_F
\end{equation}
Transient gas exchange dynamics and measurement delays generally affect fuel trims and their time-alignment with airflow estimates.  Fuel trims are therefore scattered as a function of airflow, although trends do emerge over large datasets. So, $E_A$ is not used directly, but rather averaged into bins to calibrate $\overline{E}_A$ - where $M$ is the number of samples in the bin.
\begin{equation}
\overline{E}_A\left(\dot{m}_{aj}\right) = \frac{1}{M} \sum_{j=1}^M E_A\dot{m}_{aj}
\end{equation}

\subsubsection{Calibration and Validation}
\begin{figure}
    \centering
    \input{Images/mafcorr}
    \vspace{-0.15in}
    \caption{MAF correction factor and source fuel trim data}
    \label{fig:mafcorr}
\end{figure}
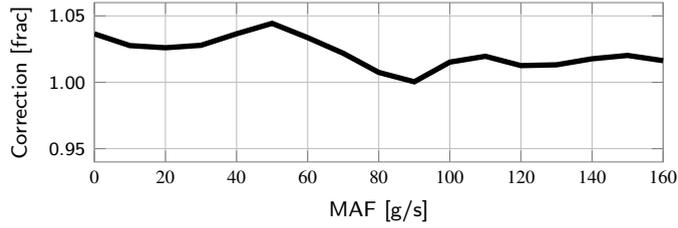

Closed-loop data from track tests of various algorithms was combined into a calibration dataset for $\overline{E}_A$. Figure \ref{fig:mafcorr} shows the resulting correction, which is within \unit[5]{\%} for all $\dot{m}_a$.

\begin{figure}
    \includegraphics[page=5]{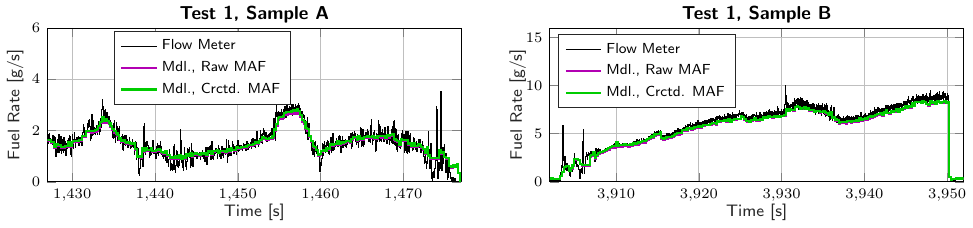}
    \vspace{-0.15in}
    \caption{Comparison of OBD model-based fuel rate with and without MAF correction against fuel flow measurements}
    \label{fig:fuelval}
\end{figure}

The OBD-based fuel flow model was validated in two ways: dynamically by comparison against measurements from a volume flow meter, and cumulatively by comparison against a measured volume of fuel placed in the empty tank.  This fuel was also weighed to ensure accurate density. Three tests were performed, all of which began with a certain amount of fuel in the tank. For each test, the vehicle was run on a chassis dynamometer until it ran out of fuel. Test 3, during which \unit[3]{US gallons} of fuel was consumed, was used to calibrate $e_F$. Tests 1 and 2 were reserved for validation. Figure \ref{fig:fuelval} demonstrates qualitative model performance in lower and higher power samples from Test 1.

Table \ref{tab:fuelval} lists the model's accuracy in the three chassis dynamometer tests where total fuel volume was directly measured.  Test 2's flow meter and volume-based cumulative fuel measurements differed by \unit[8.0]{\%}, exceeding the differences observed in the other tests and indicating a possible ground truth measurement error in Test 2.  Therefore, the proposed OBD-based technique was adopted by virtue of its close match to the validation data in test 1 and acceptable match to the instantaneous fuel flow measurement in Figure \ref{fig:fuelval}.  

\begin{table}[]
    \centering
    \caption{OBD model accuracy in cumulative fuel consumption}
    \begin{tabular}{lll}
    \toprule
        Test No. & Meas. Fuel Vol. [L] & Est. Fuel Vol. [L]\\
    \midrule
        1 (validation)& 3.79 & 3.79 \\
        2 (validation)& 9.46 & 8.90 \\
        3 (calibration)& 11.36 & 11.38 \\
    \bottomrule
    \end{tabular}
    \label{tab:fuelval}
\end{table}

\subsection{Battery electric vehicle}
Unlike the combustion engine vehicle, the specification of the electric vehicle's OBD data is not published by the vehicle manufacturer - mainly because the electric car manufacturers have not established a standard for messages exchanged from its CAN bus \citep{tseng2016data}. The existing smart phone applications for Nissan Leaf's OBD data collection, such as Leaf Spy \citep{leafspy}, provide insufficient sample rates. Following the guidelines in \citep{tseng2016data} and verifying the results with that of Leaf Spy, OBD-based measurements were collected for the signed battery current, $I$, terminal voltage, $V_T$, state-of-charge, $SOC$, and capacity via the custom iOS app. 

For the Li-ion battery of the Nissan Leaf, a lumped resistance $R_s$ is considered. As shown in \citep{parvini2017heuristic}, the open-circuit voltage, $V_{OC}$, is assumed to have a linear relationship with $SOC$ in the mid-range of $SOC$ levels when considering fixed battery temperature. So, the resulting linear model is fit to the collected OBD data to give an estimated value of $R_s=\unit[0.1]{\Omega}$. 
As the obtained resistive loss is negligible compared to the battery net energy, and each test consisted of similar ambient temperatures, $R_s$ was assumed constant for all tests. Considering the resistive energy loss, the battery net energy is then obtained by integrating over the entire test interval.
\begin{align}\label{eqn:ebattery}
  E_{battery} &= \int_{t_0}^{t_1} \left[ V_T(t) \, I(t) + R_sI^2(t) \right] dt \\
  &\approx \sum_{k=1}^M \left[ V_T(t_{k-1})I(t_{k-1}) + V_T(t_{k})I(t_{k}) + R_sI^2(t_{k-1})+R_sI^2(t_{k}) \right]\frac{\Delta t_k}{2} 
\end{align}
%


\section{Experimental scenario description}\label{sec:expDescription}

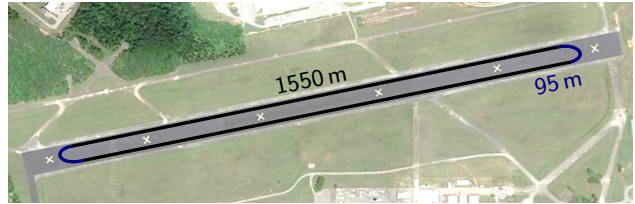
\begin{figure}
    \centering
    \input{Images/TestTrack}
    \caption{International Transportation Innovation Center (ITIC) testing grounds with virtual road markings and speed limit zones denoted. Black straight-aways had a speed limit of \unit[$\overline{v}_0=22.3$]{m/s}, while blue U-turns had a speed limit of \unit[$\overline{v}_1=7.0$]{m/s}. Base image retrieved from \texttt{Google Maps}}
    \label{fig:testtrack}
\end{figure}

As mentioned in Section \ref{sec:VILSetup}, the test track and simulation roads are mapped one-to-one by creating waypoints for the real and simulated vehicles to drive the same road network. 

A picture of the testing grounds with the network overlay is given in Figure \ref{fig:testtrack}. Because the track was long and narrow in geometry, it requires making U-turns to continue a simulation, and velocity is physically limited to a lower speed while making a U-turn. As such, control logic was defined for the high-level controllers to interact with the varied speed limits. 

Three simulation variants of 1) a microsimulation environment, 2) the EPA US06, and 3) the EPA UDDS are taken to model traffic in the network, and are modified to match the test track requirements. 6 laps were run for the microsimulated experiments, and 3 laps were run for each of the drive cycle experiments. 

\subsection{Speed limit control logic}
\label{subsec:spcLogic}
Recall particle kinematic equations of constant acceleration,
\begin{equation}\label{eqn:vdv}
    a_c \, ds = v \, dv \rightarrow \overline{v}_1^2 = \overline{v}_0^2 + 2a_c\left(s-s_0\right)
\end{equation}
where $a_c = \unit[-2.0]{m/s^2}$ is chosen as a comfortable deceleration to reach the U-turn velocity from straight-away velocity. The quantity $\delta s \triangleq s-s_0$ then describes the distance away from the U-turn speed limit $\overline{v}_1$ needed to slow down from straight-away speed limit $\overline{v}_0$.
\begin{equation}\label{eqn:deltas}
    \delta s = \frac{\overline{v}_1^2-\overline{v}_0^2}{2a_c}   
\end{equation}
For the WIE and IDM controllers, switching control can then be applied when within $\delta s$ as
\begin{equation}
    u^+ = \min\left\{u, \ a_c\right\}
\end{equation}
where $u$ is the control of the WIE or IDM continuous model, and $u^+$ is the control command to apply when approaching the U-turns. This logic limits the control to only engage in the most conservative acceleration that occurs due to either car-following or velocity maintenance.

For the MPC, Eqn. \eqref{eqn:vlim} limits maximum velocity for each optimization stage. This poses challenges because the MPC optimization horizon is a function of time, whereas speed limit transitions are a function of distance instead. 

An approximation is made to convert the velocity constraint to a function of time using an estimate of the ego's speed trajectory. In this case, a constant velocity is assumed in the speed estimate $\tilde{v}(i) = v, \ i = 0,...,N$. The position estimate then follows as $\tilde{s}(i+1) = \tilde{s}(i) + \tilde{v}(i)\Delta t_h$. Combining with Eqns. (\ref{eqn:vdv}, \ref{eqn:deltas}), the following can then be used to define the MPC moving velocity constraint with $s_l$ describing the current distance from the speed limit.
\begin{equation}
    \overline{v}(i) = 
    \begin{cases}
    \overline{v}_0 & \tilde{s}(i) < s_l \\
    \sqrt{\,\overline{v}_0^2 + 2a_c\left(\delta s-\left(s_l-\tilde{s}(i)\right)\right)} & \tilde{s}(i) < \delta s + s_l \\
    \overline{v}_1 & \tilde{s}(i) \geq \delta s + s_l 
    \end{cases}
\end{equation}
Similar approaches can follow for the moving velocity constraint when approaching a higher velocity speed zone.

\subsection{Microsimulator experiment}
A PTV VISSIM microsimulation environment was set up which comprised of a single-lane circuit of 74 vehicles as depicted in Figure \ref{fig:vilconcept}. The VISSIM human-driver model (as introduced in Section \ref{sec:baselineCarFollowing}) controlled all simulated vehicles, with the exception of the MPC-C scenario, in which a string of 5 simulated CAVs plus the experimental CAV were controlled by the MPC. Because the vehicles start from rest, lap 1 is discarded to remove unwanted transient effects in simulation, and then the next 5 laps are considered for energy and traffic flow analysis. The scenario presented here can be physically motivated as a general car-following study consisting of a combination of on-ramps and off-ramps into highway driving, or a car-following study with forced slow-downs due to bottlenecks, etc.

\begin{figure}
    \centering
    \input{Images/vissimvel}
    \caption{Microsimulation experimental velocity profiles of the PV for MPC-U, MPC-C, WIE, and IDM ego vehicle cases for lap 3}
    \label{fig:vissimvelocity}
\end{figure}
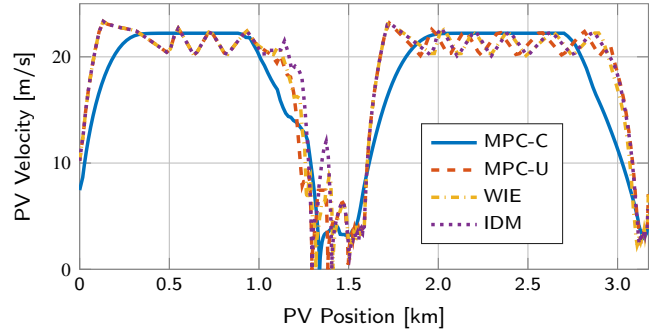

As an improvement over some ring experiments, which can feature tight strings of vehicles which all interact together and thus form unrealistic disturbance propagation/attenuation, the experiments here use a larger road network with significant gap between the first and last vehicle in the string. This largely removed the ego vehicle's influence on traffic downstream. It was found the PV velocity profiles well-matched in behavior, as depicted in Figure \ref{fig:vissimvelocity}, though some differences do exist due to bottle-necking that occurred at the U-turns.

\subsection{Modified EPA drive cycle experiments}
Separate from the VISSIM microsimulation environment, a simulation environment consisting of a single preceding vehicle was created. The EPA US06 and EPA UDDS drive cycles were used to set the velocity profile of the PV \citep{epacycle}. These cycles target more specific acceleration-aggressive and urban-driving car-following scenarios and can serve as an upper-bound on energy benefits found with the MPC.

After setting an initial distance between the simulated PV and the physical ego vehicle of the car-following controller minimum distance $\underline{d}$, the real-time position of the PV at time $t$ and the gap between the PV and ego vehicle are computed by interpolation. The drive cycles were looped until 3 laps were completed.

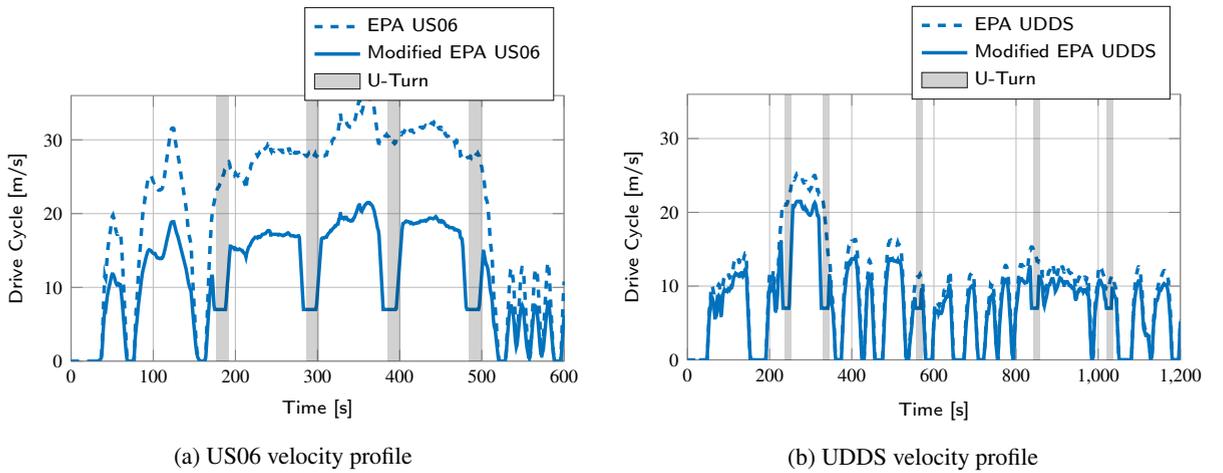
\begin{figure}
\centering
    \begin{subfigure}{0.49\linewidth} 
        \centering
        \input{Images/EPA_US06_modified}
        \caption{US06 velocity profile}\label{fig:VILDriveCyclesA}
  \end{subfigure}
  \begin{subfigure}{0.49\linewidth} 
        \centering
        \input{Images/EPA_UDDS_modified}
        \caption{UDDS velocity profile}\label{fig:VILDriveCyclesB}
  \end{subfigure}
\caption{Modified EPA drive cycle experimental velocity profiles used for the PV} \label{fig:VILDriveCycles}
\end{figure}

The velocity trajectories of US06 and UDDS cycles are scaled down by 40\% and 20\%, respectively, to match the speed limits at the test track. Due to speed limits at the U-turns, the drive cycles were also modified by the similar logic as given in Section \ref{subsec:spcLogic} - the result of which is shown in Figure \ref{fig:VILDriveCycles}. For the MPC-C scenario, the PV's future position over the horizon $S_r$ is also communicated. While this is physically unrealistic with fixed drive cycles, a priori knowledge of the PV position provides an upper bound on the possible energy benefits the anticipative car-following MPC is capable of. Experimental results are given in Section \ref{sec:energyresults}.

\section{VIL experimental results for energy efficiency and traffic flow}\label{sec:energyresults}

The results for the VIL experiments with the 3 simulation variants of microsimulator environment, EPA US06, and EPA UDDS are given for the ICEV Mazda CX7 and the EV Nissan Leaf. Here, all high-level controller types of Wiedemann 99 (WIE), Intelligent driver model (IDM), MPC unconnected (MPC-U), and MPC connected (MPC-C) were examined for both their energy and traffic flow impacts. 

Realized headway for the ego vehicle was calculated as gap divided by ego velocity $T = \Delta s/v$, where data with a velocity below \unit[0.1]{m/s} is discarded to avoid the singularity. 

\subsection{Microsimulated experimental results}
The microsimulated experimental study gives better intuitions and general observations about the performance of the various controllers as compared to the drive cycle studies - largely because the driving patterns of the PV are clearly motivated by naturally emergent traffic stimuli. A summary of the microsimulated experiments is given in Tables \ref{tab:VILcomparisonmazda} and \ref{tab:VILcomparison_nissan} for the ICEV and EV, respectively.

\begin{table}[]
\scriptsize
\begin{minipage}{0.48\linewidth}
    \centering
    \caption{ICEV Mazda microsimulated controller performance}
    \renewcommand\arraystretch{\stretchlength}
    \begin{tabular}{P}
    \toprule
        & WIE & IDM & MPC-U & MPC-C \\
    \midrule
        Travel Time &       24' 01" & \tintg{23' 45"}{-1.1\%} & \tintg{24' 00"}{0\%} & \tintg{23' 34"}{-1.9\%} \\ 
        Avg. Headway [s] &  3.47 & \tintr{5.73}{+65.1\%} & \tintg{3.32}{-4.3\%} & \tintg{2.75}{-20.7\%} \\
        Mean Gap [m] &      28 & \tintr{59}{+111\%} & \tintr{37}{+32\%} & \tintg{28}{0\%} \\ 
        Max. Gap [m] &   83 & \tintr{144}{+73\%} & \tintr{112}{+35\%} & \tintg{74}{-11\%} \\ 
        Net Fuel [L] &      2.556 & \tintg{2.174}{-15\%} & \tintg{2.241}{-12\%} & \tintg{1.978}{-23\%} \\ 
    \bottomrule
    \end{tabular}
    \label{tab:VILcomparisonmazda}
\end{minipage}
\hfill
\begin{minipage}{0.48\linewidth}
    \centering
    \caption{EV Nissan microsimulated controller performance}
    \renewcommand\arraystretch{\stretchlength}
    \begin{tabular}{P}
    \toprule
        & WIE & IDM & MPC-U & MPC-C \\
    \midrule
        Travel Time &       23' 49" & \tintg{23' 49"}{0\%} & \tintg{23' 40"}{-0.9\%} & \tintg{23' 36"}{-1.9\%} \\
        Avg. Headway [s] & 3.96 & \tintr{5.81}{+46.7\%} & \tintg{2.93}{-26.0\%} & \tintg{2.82}{-28.8\%} \\
        Mean Gap [m] &      27 & \tintr{62}{+130\%} & \tintr{31}{+16\%} & \tintr{32}{+21\%} \\
        Max. Gap [m] &      76 & \tintr{151}{+99\%} & \tintr{96}{+26\%} & \tintr{82}{+8\%} \\
        Net Energy [kwh] &      4.090 & \tintg{3.730}{-8.8\%} & \tintg{3.766}{-7.9\%} & \tintg{3.247}{-20.6\%} \\ 
    \bottomrule
    \end{tabular}
    \label{tab:VILcomparison_nissan}
\end{minipage}
\end{table}

\subsubsection{Energy Effects}\label{sec:microsimenergyresults}

\begin{figure}
\centering
\begin{subfigure}{0.49\linewidth}
    \centering
    \input{Images/VIL_mf_Mazda}
    \caption{ICEV microsimulated experiment}
    \label{fig:vilmf}
\end{subfigure}
\begin{subfigure}{0.49\linewidth}
    \centering
    \input{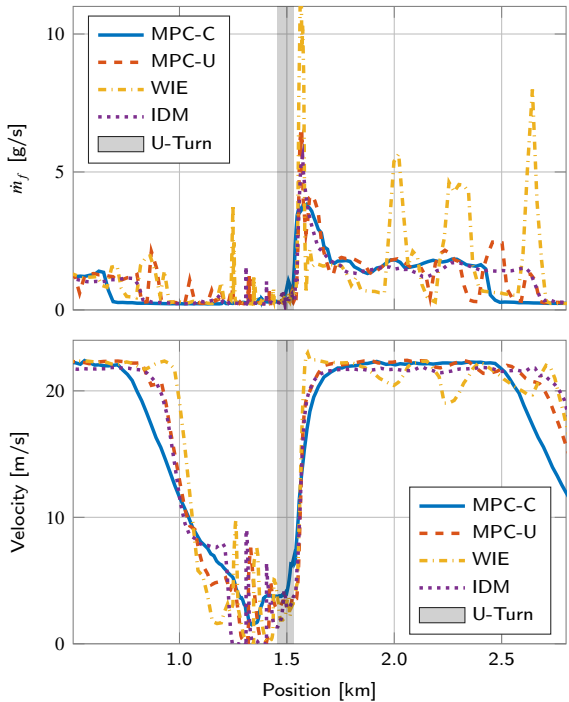}
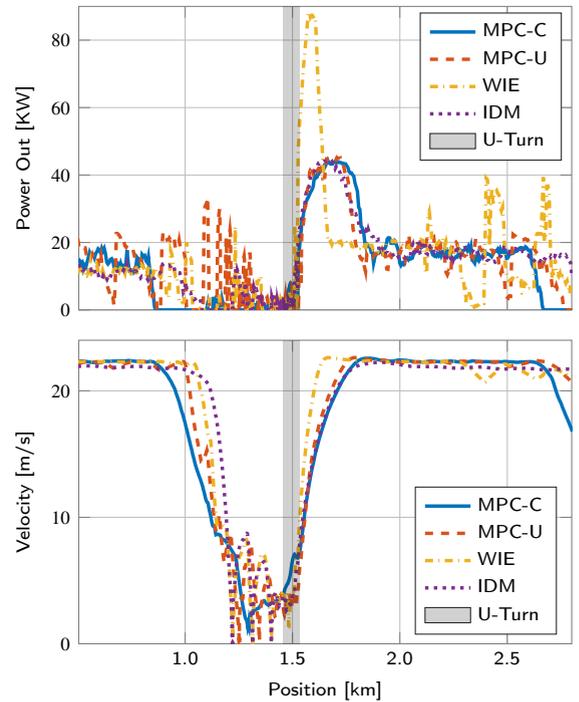
    \caption{EV microsimulated experiment}
    \label{fig:vilpbat}
\end{subfigure}
\caption{Experimental microsimulator comparison of the energy output rates (above) and the velocity profiles (below) for the high-level controllers for lap 3} 
\label{fig:vilenergy}
\end{figure}

The MPC offered energy efficiency improvements over the WIE driver for both the combustion engine and electric motor powertrains - as shown in Tables \ref{tab:VILcomparisonmazda} and \ref{tab:VILcomparison_nissan}. In particular, MPC-U reduced energy usage by 12\% and 8\% for the ICEV and EV, respectively, whereas MPC-C reduced energy usage by 23\% and 21\% for the ICEV and EV, respectively. Additionally, it did not sacrifice realized time headway, and so the OCP defining the MPC was successful at balancing both eco-driving and traffic flow.

The speed limits impose additional disturbance in the network - apart from those arising due to natural longitudinal traffic patterns. While it is not feasible to quantify separate energy effects between the pure car-following behavior and the speed-maintenance behavior, Figure \ref{fig:vilenergy} indicates that the MPC improved the energy efficiency over the WIE in both cases. For the ICEV, WIE's higher fuel rate when accelerating out of the U-turns was caused not only by its higher engine output power, but also by enrichment at high power in this downsized-boosted engine.  The four algorithms IDM, MPC-C, MPC-U, and WIE resulted in richest $\lambda$ commands of 0.90, 0.88, 0.88, and 0.70, respectively - which reduced WIE's tank-to-wheels efficiency during acceleration maneuvers. The EV exhibited similar power consumption trends with the high-level controllers.

The IDM presented here additionally improved energy economies over the WIE, and even slightly out-performed the MPC-U for energy. However, as the IDM significantly increased its commanded gap and headway, it negatively impacted surrounding traffic flow in favor of reducing its own vehicle interactions.

Finally, Figure \ref{fig:VILS} depicts upstream traffic smoothing in the MPC-C case as compared to the baseline WIE case. To this end, the traffic was assumed to be composed of mid-sized combustion vehicles, and energy impact of the virtual upstream traffic smoothing was estimated using the fuel model developed in \citep{dollar2018efficient}. Comparing the fuel usage of the upstream traffic for each high level controller to the WIE case, it was found that the MPC-C scenario improved fuel usage of upstream traffic by 4.5\% on average, the MPC-U scenario improved fuel usage of upstream traffic by 0.4\% on average, and the IDM \emph{worsened} their fuel usage by 0.9\%. By this, automated vehicles and connectivity have the potential to provide secondary benefits for improving the energy usage of neighboring traffic.

\subsubsection{Traffic flow effects}

\begin{figure}
\centering
    \begin{subfigure}{0.49\linewidth}
        \centering\hspace{-2em}
        \includegraphics[page=6]{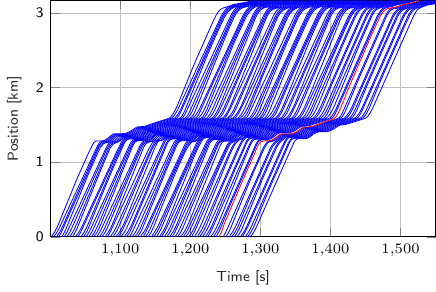}
        \caption{WIE position trajectories}
        \label{fig:VILSWIE}
    \end{subfigure}
    \begin{subfigure}{0.49\linewidth}
        \centering
        \includegraphics[page=7]{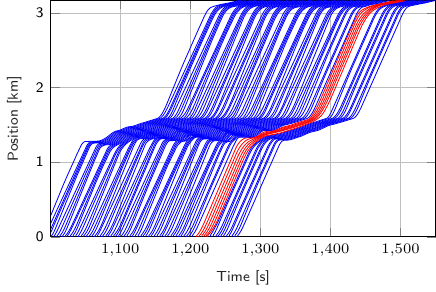}
        \caption{MPC-C position trajectories}
        \label{fig:VIL_S_Connected}
    \end{subfigure}
    \caption{WIE and MPC-C ego (red) position trajectories over time against the virtual human traffic (blue) for lap 5 of the microsimulated experiment. The ego is the rearmost MPC vehicle in (b)}
    \label{fig:VILS}
\end{figure}

The traffic flow behavior in simulation can then be further studied. It can be observed that both MPC cases reduced average time headway as compared to the baseline WIE model. This supports that the MPC can improve traffic flow over human-like drivers, and was accomplished despite less aggressive driving out of the U-turns - which lead to increases in gap.

Figure \ref{fig:VILS} depicts the position trajectories of each vehicle over time. Platoons can be clearly observed as strings of vehicles begin to compact at high velocity, and velocities decrease around the mid and endpoints of the track due to speed limit interactions with the U-turns. One can then observe stop-and-go behavior by the ripples in the position trajectories, which is detrimental to the energy economy of the fleet of vehicles.

Differences in the behavior of the 11 vehicles upstream of the ego highlight the smoothing effects of the CAVs on the fleet: as opposed to the WIE scenario, the CAV string attenuated disturbances of the traffic and subsequently smoothed the driving of the vehicles behind. The MPC-C scenario smoothed traffic the most, and it was found the MPC-U scenario smoothed traffic some - even with only a single automated vehicle present. IDM did not visually affect upstream traffic upon inspection. The energy impact from this traffic attenuation is discussed in Section \ref{sec:microsimenergyresults}.

The position trajectories of the first several vehicles in the string differs between the WIE and MPC-C cases. As discussed in Section \ref{sec:expDescription}, however, this occurred due to bottle-necking at the U-turns, and largely did not influence the ego vehicle results.

The IDM used here was, in short, unacceptable for traffic flow because of its large spacing and headway commanded - increasing them by 130\% and 65\%, respectively. Despite this, IDM was still able to command acceptable travel times because it was able to maintain high velocity for longer periods of time. This largely negates the energy benefits for IDM as seen in Tables \ref{tab:VILcomparisonmazda} and \ref{tab:VILcomparison_nissan}.

\subsection{Modified US06 drive cycle results}


\begin{table}[]
\scriptsize
\begin{minipage}{0.48\linewidth}
    \centering
    \caption{ICEV Mazda US06 controller performance}
    \renewcommand\arraystretch{\stretchlength}
    \begin{tabular}{P}
    \toprule
        & WIE & IDM & MPC-U & MPC-C \\
    \midrule
        Travel Time &       9' 20" & \tintr{9' 33"}{+2.3\%} & \tintg{9' 20"}{0\%} & \tintr{9' 26"}{+1.1\%} \\
        Avg. Headway [s] & 2.45 & \tintr{6.23}{+154.3\%} & \tintr{2.64}{+7.8\%} & \tintr{2.66}{+8.6\%} \\
        Avg. Gap [m] &      23 & \tintr{35}{+52\%} & \tintr{26}{+13\%} & \tintg{23}{0\%} \\
        Max. Gap [m] &   40 & \tintr{93}{+133\%} & \tintr{77}{+93\%} & \tintr{88}{+120\%} \\
        Net Fuel [L] &      1.006 & \tintg{0.840}{-17\%} & \tintg{0.746}{-26\%} & \tintg{0.684}{-32\%} \\ 
    \bottomrule
    \end{tabular}
    \label{tab:us06comparisonmazda}
\end{minipage}
\hfill
\begin{minipage}{0.48\linewidth}
    \centering
    \caption{EV Nissan US06 controller performance\protect\footnotemark}
    \renewcommand\arraystretch{\stretchlength}
    \begin{tabular}{P}
    \toprule
        & WIE & IDM & MPC-U & MPC-C \\
    \midrule
        Travel Time &       8' 39" & \tintg{8' 39"}{0\%} & \tintg{8' 38"}{0\%} & \tintg{8' 39"}{0\%} \\
        Avg. Headway [s] & 2.14 & \tintr{5.80}{+171.0\%} & \tintr{2.76}{+29.0\%} & \tintr{2.58}{+20.6\%} \\
        Avg. Gap [m] &      24 & \tintr{38}{+58\%} & \tintr{28}{+17\%} & \tintr{25}{+4\%} \\
        Max. Gap [m] &      42 & \tintr{92}{+119\%} & \tintr{78}{+86\%} & \tintr{82}{+95\%} \\
        Net Energy [kwh] &      1.286 & \tintg{1.230}{-4\%} & \tintg{1.064}{-17\%} & \tintg{0.963}{-25\%} \\ 
    \bottomrule
    \end{tabular}
    \label{tab:us06comparison_nissan}
\end{minipage}
\end{table}

\begin{figure}
\centering
    \begin{subfigure}{0.49\linewidth}
        \centering
        \includegraphics[page=8]{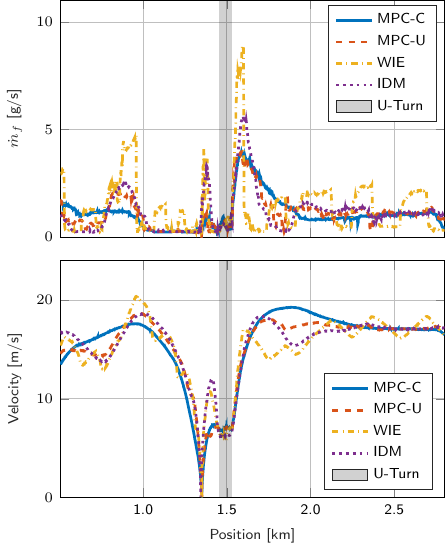}
        \caption{ICEV US06}
        \label{fig:us06mf}
    \end{subfigure}
    \begin{subfigure}{0.49\linewidth}
        \centering
        \includegraphics[page=9]{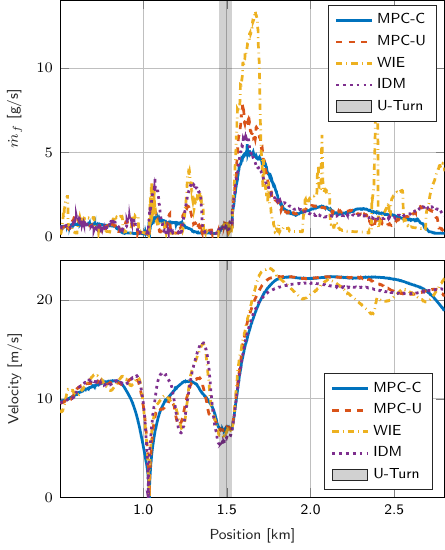}
        \caption{ICEV UDDS}
        \label{fig:uddsmf}
    \end{subfigure}
    \label{fig:us06uddsmf}
    \caption{Drive cycle comparison of the ICEV fuel rates (above) and the velocity profiles (below) for the high-level controllers for lap 1}
\end{figure}
\begin{figure}
\centering
    \begin{subfigure}{0.49\linewidth}
        \centering
        \input{Images/US06_pout_Nissan}
        \caption{EV US06}
        \label{fig:us06bat}
    \end{subfigure}
    \begin{subfigure}{0.49\linewidth}
        \centering
        \input{Images/UDDS_pout_Nissan}
        \caption{EV UDDS}
        \label{fig:uddsbat}
    \end{subfigure}
    \label{fig:us06uddsbat}
    \caption{Drive cycle comparison of the EV battery power (above), regenerative power (middle), and the velocity profiles (below) for the high-level controllers for lap 1}
\end{figure}
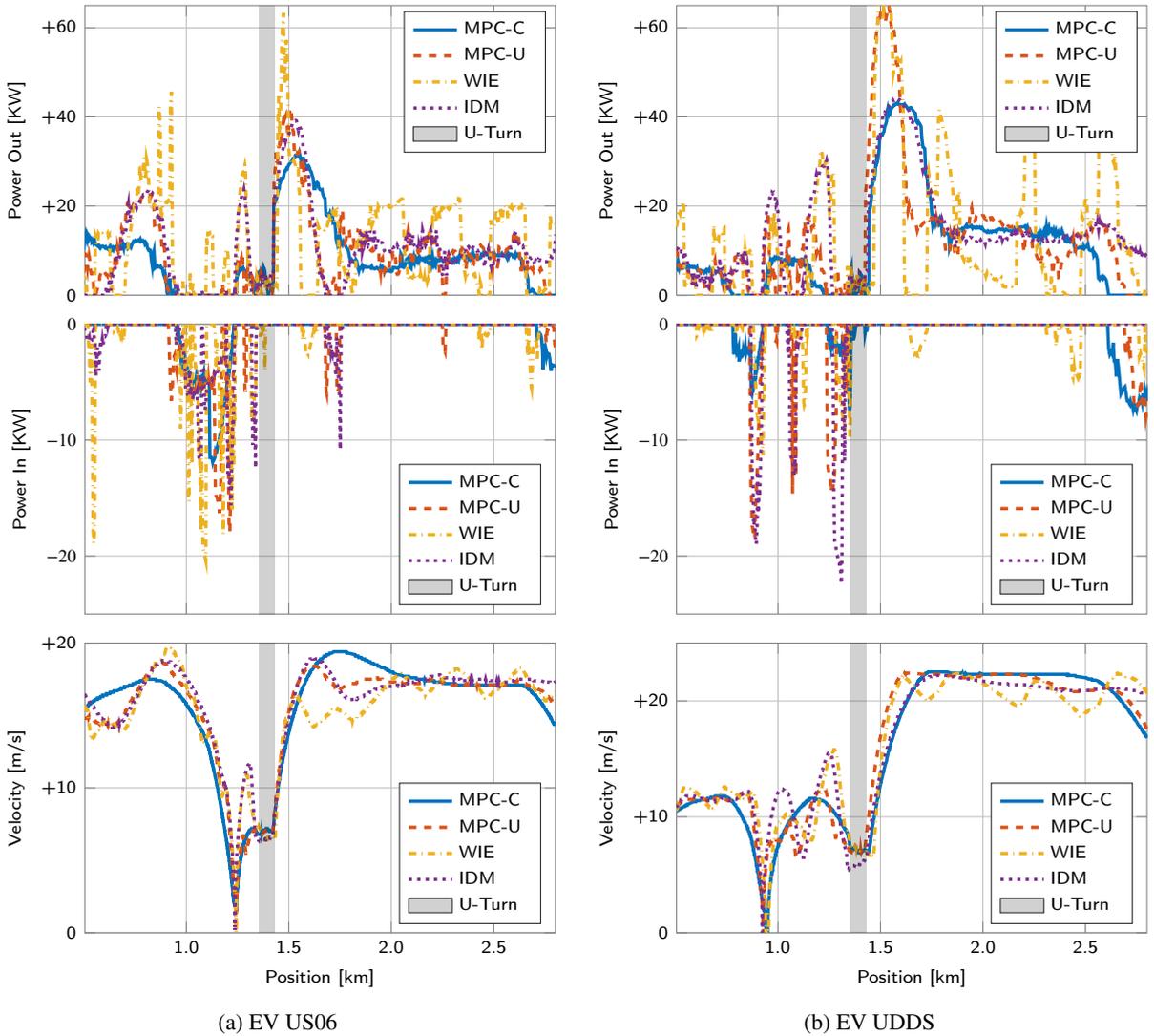

\footnotetext{EV Nissan modified US06 results were shortened to the first 520s due to loss of some OBD-II data.}

As shown in Tables \ref{tab:us06comparisonmazda} and \ref{tab:us06comparison_nissan}, the MPC energy economy was significantly improved over the WIE model - with the modest increase in average headway. Overall, the MPC improved the combustion vehicle energy performance by 26\% and 32\% for the unconnected and connected variants, while the MPC improved the electric vehicle energy performance by 17\% and 25\% for the unconnected and connected variants. Additionally, Figure \ref{fig:us06mf} gives the fuel rate and velocity trajectories of each of the ICEV controllers for a portion of the first lap of the US06. Figure \ref{fig:us06bat} gives the battery output/input power and velocity trajectories of each of the EV controllers for a portion of the first lap of the US06.

In contrast to the microsimulated experiment, it was observed that the MPC cases had worse time headway as compared to the baseline WIE model when simulating the US06. Similar to some humans, the WIE model can violate the speed limit in favor of catching the front vehicle, whereas MPC is soft-constrained to avoid speeding. This can improve the WIE model's headway tracking when considering highly acceleration-aggressive PV cycles such as the US06. In addition, the MPC was heavily penalized against acceleration, which then trades off its ability to track the PV during high acceleration periods. In short, the MPC parameters worked well for the microsimulated experiment, but could be improved for the US06 scenario - and so the MPC performance may benefit from adaptive tunings. This result is interesting and could motivate further studies.

As before, the IDM exhibited poor headway performance. The IDM is constrained with an upper-bound on acceleration capabilities instead of a velocity-dependent acceleration constraint like the MPC and WIE model, which can be too conservative in low-velocity driving and particularly ineffective in scenarios such as this one with large accelerations exhibited by the PV. 

\subsection{Modified UDDS drive cycle results}

\begin{table}[]
\scriptsize
\begin{minipage}{0.48\linewidth}
    \centering
    \caption{ICEV Mazda UDDS controller performance}
    \renewcommand\arraystretch{\stretchlength}
    \begin{tabular}{P}
    \toprule
        & WIE & IDM & MPC-U & MPC-C \\
    \midrule
        Travel Time &       22' 33" & \tintr{22' 36"}{+0.2\%} & \tintr{22' 35"}{+0.1\%} & \tintr{22' 36"}{+0.2\%} \\
        Avg. Headway [s] & 3.39 & \tintr{5.94}{+75.2\%} & \tintg{2.85}{-15.9\%} & \tintr{3.53}{+4.1\%} \\
        Avg. Gap [m] &      16 & \tintr{22}{+38\%} & \tintg{16}{0\%} & \tintr{20}{+25\%} \\
        Max. Gap [m] &   46 & \tintr{99}{+115\%} & \tintr{83}{+80\%} & \tintr{70}{+52\%} \\
        Net Fuel [L] &      1.370 & \tintg{1.329}{-3\%} & \tintg{1.175}{-14\%} & \tintg{1.048}{-24\%} \\ 
    \bottomrule
    \end{tabular}
    \label{tab:uddscomparisonmazda}
\end{minipage}
\hfill
\begin{minipage}{0.48\linewidth}
    \centering
    \caption{EV Nissan UDDS controller performance}
    \renewcommand\arraystretch{\stretchlength}
    \begin{tabular}{P}
    \toprule
        & WIE & IDM & MPC-U & MPC-C \\
    \midrule
        Travel Time &       22' 33" & \tintr{22' 35}{+0.1\%} & \tintr{22' 34"}{+0.1\%} & \tintr{22' 34"}{+0.1\%} \\
        Avg. Headway [s] & 4.22 & \tintr{4.39}{+4.0\%} & \tintg{2.51}{-40.5\%} & \tintg{3.15}{-25.4\%} \\
        Avg. Gap [m] &      16 & \tintr{22}{+38\%} & \tintg{15}{-6\%} & \tintr{19}{+19\%} \\
        Max. Gap [m] &      43 & \tintr{111}{+158\%} & \tintr{85}{+98\%} & \tintr{79}{+84\%} \\
        Net Energy [kwh] &      1.855 & \tintg{1.639}{-12\%} & \tintg{1.389}{-25\%} & \tintg{1.221}{-34\%} \\ 
    \bottomrule
    \end{tabular}
    \label{tab:uddscomparison_nissan}
\end{minipage}
\end{table}

A summary of the modified UDDS experiments is given in Tables \ref{tab:uddscomparisonmazda} and \ref{tab:uddscomparison_nissan} for the ICEV and EV, respectively. Here, the MPC significantly improved energy economy over the baseline human-like driver. The MPC improved the combustion vehicle energy performance by 14\% and 24\% for the unconnected and connected variants, while the MPC improved the electric vehicle energy performance by 25\% and 34\% for the unconnected and connected variants. Additionally, Figure \ref{fig:uddsmf} gives the fuel rate and velocity trajectories of each of the ICEV controllers for a portion of the first lap of the UDDS. Figure \ref{fig:uddsbat} gives the battery output/input power and velocity trajectories of each of the EV controllers for a portion of the first lap of the UDDS. 

Both MPC improved average headway like shown in the microsimulated experiment. Though, of note, the UDDS cycle emerged distinct behavior for the MPC when using connectivity: the MPC-C controller exhibited a noticeably larger average headway than the MPC-U case. The connected case is typically intended to drive closer to the PV to improve traffic compactness but, because it has full preview of the UDDS for its prediction horizon, it sacrifices following distance in favor of improved acceleration minimization. This understanding can be motivated by the frequent stopping of the UDDS, where future stops are in preview, and so the MPC-C cuts wasteful acceleration and subsequent braking.


\section{Conclusion}\label{sec:conclusion}
Experiments were conducted on both a combustion engine vehicle and electric motor vehicle to evaluate their energy and traffic performance using chance-constrained model predictive control for vehicle guidance. Both vehicles were equipped for autonomy by fitting pedal and steering actuators to control their acceleration and pose. In particular, a novel data-driven approach to vehicle pedal control was executed, and fused classical methods with a feedforward map to approximate the powertrain nonlinearities.

A Vehicle-in-the-Loop environment was designed to embed a physical vehicle into virtual reality with simulated traffic. Three separate experimental scenarios were designated to cover a variety of traffic conditions: a microsimulation experiment was conducted to study emergent highway-conditioned car-following, the US06 was simulated for the preceding vehicle to stimulate aggressive acceleration, and the UDDS was simulated for the preceding vehicle to create urban conditions.

The MPC is designed to minimize commanded acceleration and error in headway tracking for each CAV. When following another MPC vehicle it leverages connectivity to better plan control decisions, and in the absence of connectivity it quantifies uncertainty in prediction of the front human driver to balance safety with traffic compactness. 

It was found the MPC outperformed the human-like driver models with regards to both realized headway and energy economy. Overall, the microsimulation experiment showed the MPC improved energy efficiency by 12\% and 8\% for the combustion and electric vehicles when following human drivers. Benefits were even greater in the drive cycle studies, where the US06 showed 26\% and 17\% improved energy efficiency for the combustion and electric vehicles, whereas the UDDS showed 14\% and 25\% improved energy efficiency for the combustion and electric vehicles. 

Finally, once introducing connectivity, between 6-12\% additional improvement in energy efficiency was found for the MPC compared with its unconnected variant. We also estimated secondary impacts of connectivity on traffic upstream of the ego vehicle, and found that 4.5\% energy benefit was achieved for the 11 following drivers on average.

\appendix

\section{Protocol buffer structure}\label{app:CAVgpbcommunication}

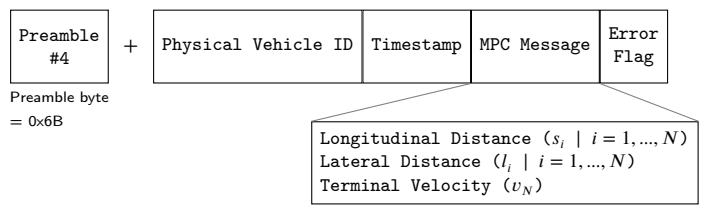
\begin{figure}
    \centering
    \input{Images/Protobuf}
     \caption{Four different message types implemented for the VIL test environment (Protocol buffer structured messages preceded with a preamble)}
    \label{fig:CAVgpbcommunication}
\end{figure}

The data exchanged between the physical vehicles and simulation server lies in four categories with their data structure shown in Figure \ref{fig:CAVgpbcommunication}. Each message is preceded with a predefined preamble after being serialized by Google Protocol Buffers \citep{protobuf}. In order to reduce the bandwidth challenges imposed on the simulation side, we incorporate the vehicle-to-vehicle (V2V) messaging into the Sim2V and V2Sim messages. The resulting buffer is only a few hundred bytes in size.

\section{Wiedemann 99 model}\label{app:wiestate}

\begin{figure}
    \centering
    \includegraphics[page=2]{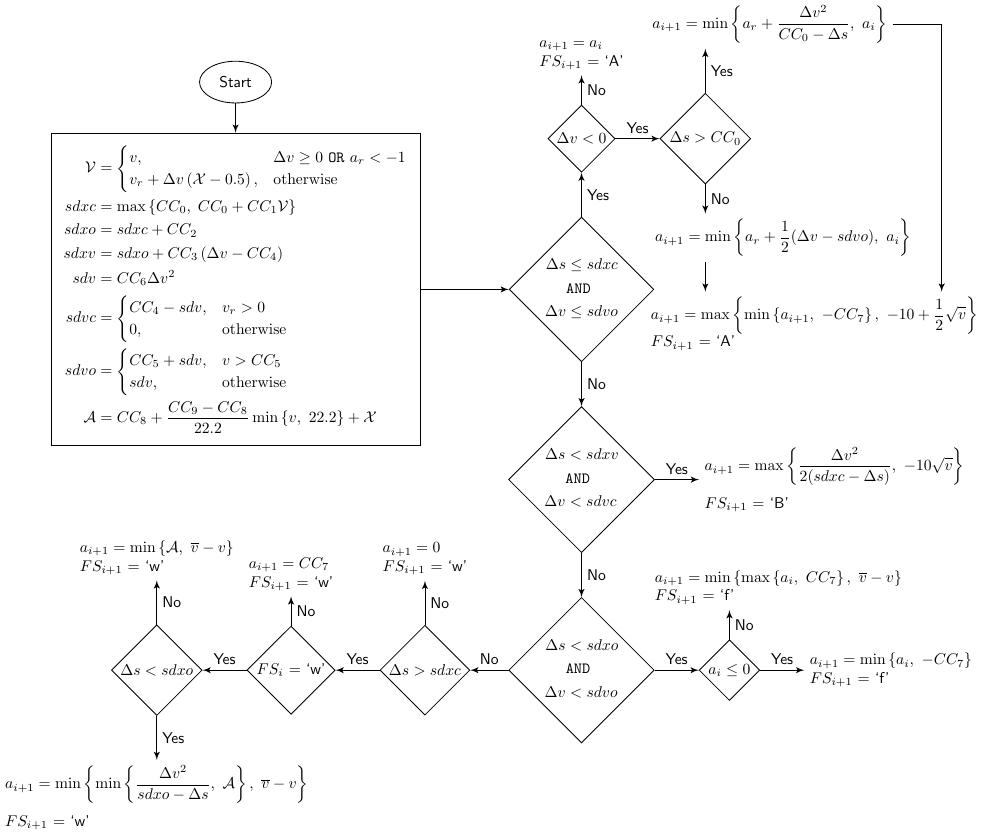}
    \caption{Algorithm of the Wiedemann 99 model used}
    \label{fig:wiestate}
\end{figure}

The Wiedemann 99 model expands on the model proposed in \citep{Wiedemann1974} for use in modeling human drivers.

Figure \ref{fig:wiestate} gives the Wiedemann 99 model used in the experiments. The family of $sd$ variables pertain to following and speed thresholds for the various logic in the model, $\mathcal{X} \in [0, 1]$ is a random variable unique to the driver with expected value 0.5, $\mathcal{V}$ and $\mathcal{A}$ are variables of speed difference perception and max acceleration aggressiveness, $CC_j$ are the tunable Wiedemann 99 parameters, $v$ is the ego velocity, $v_r$ is the PV velocity, $\Delta v \triangleq v_r - v$ is the difference in velocities from the PV, $\overline{v}$ is the max velocity, $\Delta s$ is the gap from the PV, $a_{i}$ is the current acceleration of the ego, $a_{i+1}$ is the acceleration command of the ego, and $a_r$ is the acceleration of the PV. Additionally, $FS$ is the following state of the ego: it is assigned to a string describing the current action of the ego, where `A' is increase distance, `B' is decrease distance, `f' is keep distance, `w' is keep speed \citep{vortischpdf}.

\section*{Acknowledgment}
This work was supported by an award from the U.S. Department of Energy Vehicle Technologies Office (Project DE-EE0008469). The authors would like to thank Dr. Joachim Taiber and the International Transportation Innovation Center (ITIC) for reduced-fee access to the testing grounds throughout the project. They also thank Mr. Dominik Karbowski for contributing to discussion, thank Mr. David Mann for his technical support in dynamometer testing, and thank Professor Zoran Filipi for access to the International Center for Automotive Research (ICAR) facilities. 


\printcredits


\Urlmuskip=0mu plus 1mu\relax
\bibliographystyle{cas-model2-names}
\bibliography{references}

\end{document}

%% file: Images/pnp.tex
\begin{tikzpicture}
\node[anchor=south west,inner sep=0] (pnp) at (0,0) {\includegraphics[width=\textwidth]{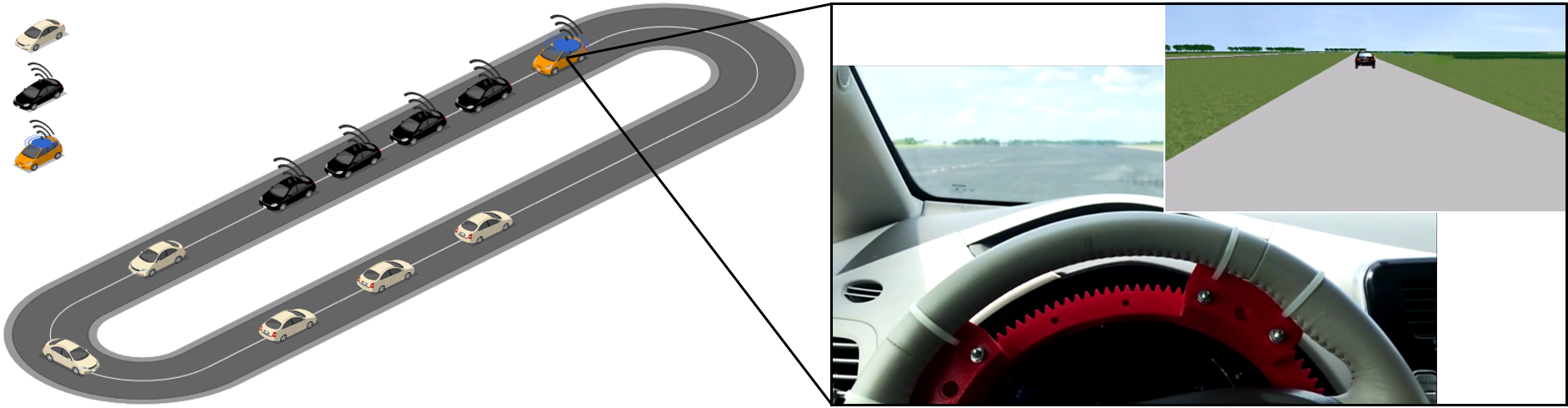}};

\begin{scope}[x={(pnp.south east)},y={(pnp.north west)}]
    \node[anchor=north west]  at (0.04,0.97) {\scriptsize Simulated Human};
    \node[anchor=north west]  at (0.04,0.82) {\scriptsize Simulated CAV};
    \node[anchor=north west]  at (0.04,0.67) {\scriptsize Physical CAV};
\end{scope}

\end{tikzpicture}

%% file: Images/LLArchitecture.tex
\newcommand{\lene}{0.50cm}
\newcommand{\lent}{0.75cm}
\newcommand{\side}{1.50cm}
\newcommand{\xcen}{0.5*\x2-0.5*\x1}
\newcommand{\ycen}{0.5*\y2-0.5*\y1}

\tikzstyle{process} = 
    [
        rectangle
      , draw
      , top color=white, bottom color=white!10,
      , text centered
      , node distance=1.5cm
      , inner sep=8pt
      ]
\tikzstyle{simulation} =
    [
        rectangle
      , draw=red!50
      , top color=white, bottom color=red!10,
      , text centered
      , node distance=1.5cm
      , minimum height=2em
      , inner sep=8pt
    ]
\tikzstyle{database} =
    [
        cylinder
      , draw
      , top color = white, bottom color=white
      , shape border rotate=90
      , minimum height=1.8cm
      , minimum width=0.8cm
      , inner sep=10pt
      , shape aspect=0.25
    ]
\tikzstyle{line} = 
    [
        draw
      , -latex'
    ]
\tikzstyle{dash} =
    [
        draw
      , -latex'
      , dashed
    ]
    
\begin{tikzpicture}[node distance = 1cm, auto, every node/.style={black,scale=0.80}]]
    \node[process, minimum width = 12cm, top color = blue!10, bottom color = blue!10] (ROS) {ROS};
    
    \node[above = 1.15cm of ROS.west, process, anchor = west] (INS) {INS};
    \node[right = \side of INS, process] (RTK) {RTK-GPS};
    \node[right = 2*\side of RTK.south, process, top color=white, bottom color=blue!10, anchor=south, align=center] (High) {Car-Following \\Controller};
    \node[right = \side of High, process, top color = white, bottom color=blue!10, align=center] (Map) {Test Track \\Map};
    
    \path [black, line] let \p1=(INS.south), \p2=($(\p1)-(0,\lene)$) in (\p1) -- +(0, \ycen) -| node[right,align=left]{$v, \ a, \ \theta$} (\p2);
    \path [black, line] let \p1=(RTK.south), \p2=($(\p1)-(0,\lene)$) in (\p1) -- +(0, \ycen) -| node[left,align=right]{$x, \ y$} (\p2);
    
    \path [black, line] let \p1=($(High.south)-(0.20,0.0)$), \p2=($(\p1)-(0,\lene)$) in (\p1) -- +(0, \ycen) -| node[left,align=right]{$v_d, \ a_d$} (\p2);
    \path [black, line] let \p1=($(High.south)+(0.20,-\lene)$), \p2=($(\p1)+(0,\lene)$) in (\p1) -- +(0, \ycen) -| node[right, align=left]{$\hat{v}, \ \hat{a}, \ \Delta s$} (\p2);
    \path [black, line] let \p1=(Map.west), \p2=(High.east) in (\p1) -- +(\xcen, 0) |- (\p2);
    
    \node[below = \lene of ROS, process, align=center] (Kalman) {Kalman\\Filter};
    \node[below left = 0.20cm and \side of Kalman.west, process, align=center, anchor = east] (Low) {Speed\\Controller};
    \node[below right = 0.20cm and \side of Kalman.east, process, align=center, anchor = west] (LowSteer) {Steer\\Controller};
    
    \path [black, line] let \p1=($(Low.north)+(0,\lene+0.20cm)$), \p2=($(\p1)-(0,\lene+0.20cm)$) in (\p1) -- +(0, \ycen) -| node[left,align=right]{$\hat{v}, \ \hat{a}, \ v_d, \ a_d$} (\p2);
    
    \path [black, line] let \p1=($(ROS.south)-(0.20,0.0)$), \p2=($(\p1)-(0,\lene)$) in (\p1) -- +(0, \ycen) -| node[left,align=right]{$x, \ y, \ v, \ a$} (\p2);
    \path [black, line] let \p1=($(Kalman.north)+(0.20,0.0)$), \p2=($(\p1)+(0,\lene)$) in (\p1) -- +(0, \ycen) -| node[right, align=left]{$\hat{v}, \ \hat{a}$} (\p2);
    
    \path [black, line] let \p1=($(LowSteer.north)+(0,\lene+0.20cm)$), \p2=($(\p1)-(0,\lene+0.20cm)$) in (\p1) -- +(0, \ycen) -| node[left, align=right]{$\hat{v}, \ x, \ y$} (\p2);
    
    \path [black, line] let \p1=($(Map.south)+(0.35,0)$), \p2=(LowSteer.east) in (\p1) -- +(0,0) |- (\p2);
    
    \node[below = 0.30cm of Kalman, process, align=center] (Driver) {Motor\\Driver};
    \node[below left = 0.20cm and 2*\lene of Driver.center, process, align=center] (Pedal) {Pedal\\Motor};
    \node[left = 0.20cm of Pedal, process] (Encoder1) {Encoder};
    \node[below right = 0.20cm and 2*\lene of Driver.center, process, align=center] (Steer) {Steer\\Motor};
    \node[right = 0.20cm of Steer, process] (Encoder2) {Encoder};
    
    \path [black, line] let \p1=($(Driver.south)-(0.15, 0)$), \p2=(Pedal.east) in (\p1) -- +(0, 0) |- (\p2);
    \path [black, line] let \p1=($(Driver.south)+(0.15, 0)$), \p2=(Steer.west) in (\p1) -- +(0, 0) |- (\p2);
    \path [black, line] let \p1=(Pedal.west), \p2 = (Encoder1.east) in (\p1) -- +(\xcen, 0) |- (\p2);
    \path [black, line] let \p1=(Steer.east), \p2 = (Encoder2.west) in (\p1) -- +(\xcen, 0) |- (\p2);
    
    \path [black, line] let \p1=(Encoder1.north), \p2 = (Driver.west) in (\p1) -- +(0, \ycen) |- (\p2);
    \path [black, line] let \p1=(Encoder2.north), \p2 = (Driver.east) in (\p1) -- +(0, \ycen) |- (\p2);
    
    \path [black, line] let \p1=(Low.east), \p2=($(Driver.west)+(0,0.25)$) in (\p1) -- node[below, align=right]{$P$} +(\xcen, 0) |- (\p2);
    \path [black, line] let \p1=(LowSteer.west), \p2=($(Driver.east)+(0,0.25)$) in (\p1) -- node[below, align=right]{$\psi$} +(\xcen, 0) |- (\p2);
\end{tikzpicture}

%% file: Images/MechStruct.tex
\begin{tikzpicture}

\node[anchor=south west,inner sep=0] (MechStruct) at (0,0) {\includegraphics[width=0.65\columnwidth]{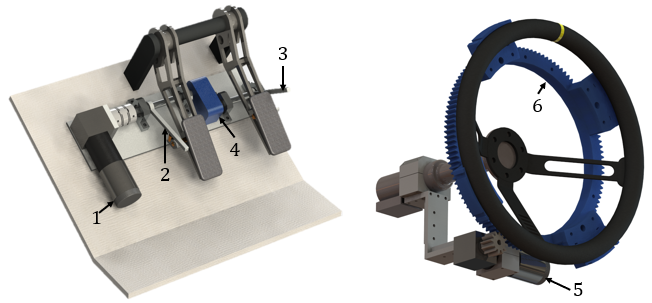}};

\end{tikzpicture}

%% file: Images/SpeedControl.tex
\newcommand{\lene}{0.50cm}
\newcommand{\lent}{0.75cm}
\newcommand{\side}{1.50cm}
\newcommand{\xcen}{0.5*\x2-0.5*\x1}
\newcommand{\ycen}{0.5*\y2-0.5*\y1}

\tikzstyle{process} = 
    [
        rectangle
      , draw
      , top color=white, bottom color=white!10,
      , text centered
      , node distance=1.5cm
      , inner sep=8pt
      ]
\tikzstyle{simulation} =
    [
        rectangle
      , draw=red!50
      , top color=white, bottom color=red!10,
      , text centered
      , node distance=1.5cm
      , minimum height=2em
      , inner sep=8pt
    ]
\tikzstyle{database} =
    [
        cylinder
      , draw
      , top color = white, bottom color=white
      , shape border rotate=90
      , minimum height=1.8cm
      , minimum width=0.8cm
      , inner sep=10pt
      , shape aspect=0.25
    ]
\tikzstyle{line} = 
    [
        draw
      , -latex'
    ]
\tikzstyle{dash} =
    [
        draw
      , -latex'
      , dashed
    ]
    
\begin{tikzpicture}[node distance = 1cm, auto, every node/.style={black,scale=0.80}]]
    \node[] (vd) {$v_d, \ a_d$};
    
    
    
    
    \node[right = 1.0cm of vd, process, align=center] (MAP) {Data-Driven\\ Feedforward\\ Controller};
    \node[right = 2*\side of MAP.north, process, minimum height = 2.5cm, align = center, anchor=north] (Ensemble) {Ensemble\\ Module};
    \node[right=1.25cm of Ensemble, process] (Vehicle) {Vehicle};
    \node[right=\lent of Vehicle] (Output) {$v, \ a$};
    
    \path [black, line] let \p1=(vd.east), \p2 = (MAP.west) in (\p1) -- +(\xcen, 0) |- (\p2);
    \path [black, line] let \p1=(MAP.east), \p2 = ($(MAP.east)+(1.35,0.0)$) in (\p1) -- +(\xcen, 0) |- node[above, align=center]{$P_{MAP}$} (\p2);  
    \path [black, line] let \p1=(Ensemble.east), \p2 = (Vehicle.west) in (\p1) -- +(\xcen, 0) |- node[above, align=center]{$P$} (\p2);
    \path [black, line] let \p1=(Vehicle.east), \p2 = (Output.west) in (\p1) -- +(\xcen, 0) |- (\p2);
    
    \node[below = \lene of MAP, process, align=center] (PID) {PID\\ Feedback\\ Controller};
    \node[left = 0.60cm of PID, draw, circle, inner sep=0mm, minimum size=2mm] (Sum1) {};

    \node[below = 0.85cm of Ensemble, process, align=center] (Kalman) {Kalman\\ Filter};
    
    \path [black, line] let \p1=($(Sum1.north)+(0,1.65)$), \p2=(Sum1.north) in (\p1) -- +(0, \ycen) -| node[left, align=right]{$a_d$} node[below left = 0.45cm and 0.10cm]{+} (\p2);
    \path [black, line] let \p1=(Sum1.east), \p2=(PID.west) in (\p1) -- +(0,0) |- (\p2);
    \path [black, line] let \p1=(PID.east), \p2=($(Ensemble.west)-(0,0.20)$) in (\p1) -- +(\xcen, 0)node[right, align=left]{$P_{PID}$} |- (\p2);
    
    \path [black, line] let \p1=($(Vehicle.east)+(0.20,0)$), \p2=(Kalman.east) in (\p1) |- (\p2);
    \path [black, line] let \p1=(Kalman.west), \p2=(Sum1.south) in (\p1) -- node[above left, align = right]{$\hat{a}$} +(0,0) -| node[above left = 0.38cm and 0.10cm]{$-$} (\p2);
\end{tikzpicture}

%% file: Images/SSLongiCtrl.tex
\begin{tikzpicture}

\shade[top color=red, bottom color=white, opacity=0.5]
    (0.5,0.2) -- (4.75,-0.5) -- (4.75,-2) -- (0.5,-2) -- cycle;
\path (0.5,0.2) -- node[pos=0.60, sloped, below] {\scriptsize 2. Brake Map} (4.75,-0.5);

\shade[left color = green, bottom color = green, top color = white, opacity=0.5]
    (0, 0.2) -- (0, 2.8) -- (4.75, 2.8) -- (4.76, -0.5) -- (0.5, 0.2) -- cycle;
\path (0, 2.0) -- node[pos = 0.50, below] {\scriptsize 1. Accelerator Map} (4.75, 2.0);

\shade[right color = blue, top color = blue, bottom color = white, opacity = 0.5]
    (0.5, 0.2) -- (0.5, -2) -- (0, -2) -- (0, 0.2) -- cycle;
\path (0.5, -1.5) -- node[pos = 0.35, above] {\scriptsize 3. Low Speed Map} (4.75, -1.5);

\draw[->,black,thick,-latex] (-0.25,0)--(5,0) node[above]{$v_d$};
\draw[->,black,thick,-latex] (0,-2)--(0,3) node[right]{$a_d$};

\draw[black] (0.5, 0.2) -- (0.5, -2);
\draw[black] (0.5, 0.2) -- (4.75, -0.5);
\draw[black] (0, 0.2) -- (0.5, 0.2);
\draw[->,blue] (0.70, -1.30) to [out=180,in=270] (0.25, -0.70);

\node[] at (0.5, 0.40) {$v_{idle}$};

\end{tikzpicture}

%% file: Images/AVL.tex
\begin{tikzpicture}

\node[anchor=south west,inner sep=0] (ANNSwitch) at (0,0) {\includegraphics[width=\columnwidth, keepaspectratio]{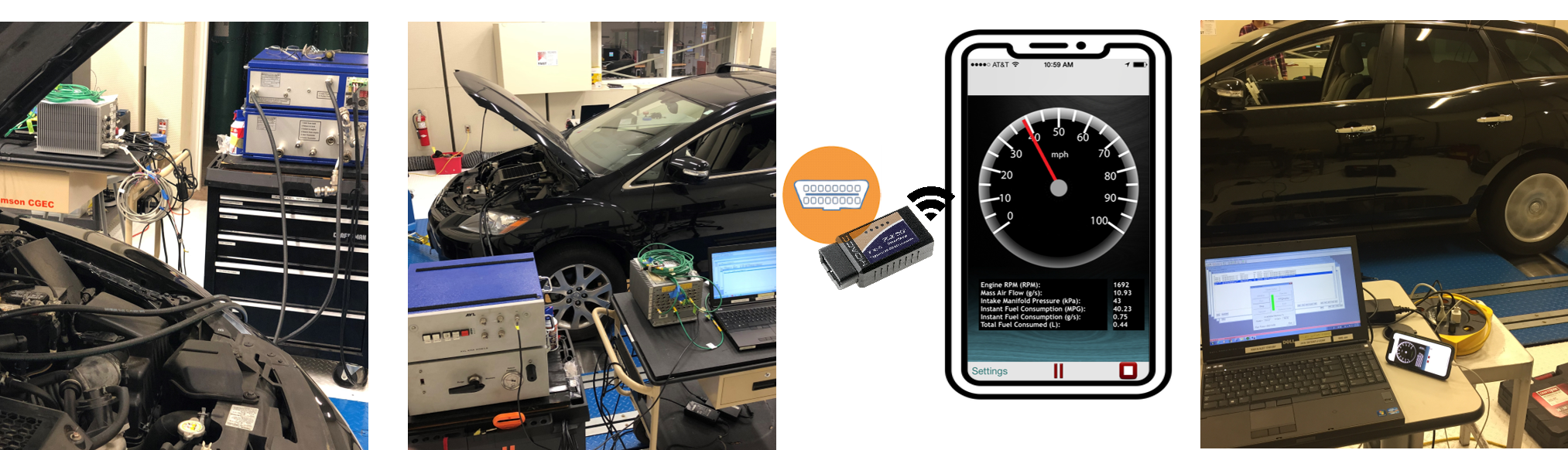}};

\node[color=white,align=center,font=\scriptsize] at (3.05, 2.65) {AVL KMA\\ Mobile\\ Flow Meter};

\node[color=white,align=center,font=\scriptsize] at (5.5, 0.35) {AVL KMA Mobile\\ Flow Meter};
\node[color=white,align=center,font=\scriptsize] at (7.35, 1.15) {DAQ System};
\node[color=white,align=center,font=\scriptsize] at (7.0, 2.5) {Mazda CX7\\ Test Vehicle};

\node[color=black,font=\scriptsize] at (9.00, 3.04) {OBD-II Port};
\node[color=black,align=center,font=\scriptsize] at (9.06, 1.34) {OBD-II Wi-Fi\\ Reader};

\node[color=black,align=center,font=\scriptsize] at (11.0, 0.265) {User Interface of the OBD\\ Logger iOS App};

\node[color=white,align=center,font=\scriptsize] at (13.7, 0.52) {OBD Logger\\ iOS App};

\end{tikzpicture}

%% file: Images/mafcorr.tex
%
%
\begin{tikzpicture}

\begin{axis}[%
width=7.5cm,
height=2.1cm,
at={(0cm,0cm)},
scale only axis,
xmin=0,
xmax=160,
xlabel style={font=\color{white!15!black} \footnotesize},
xlabel={MAF [g/s]},
ymin=0.94,
ymax=1.06,
ylabel style={font=\color{white!15!black} \footnotesize},
ylabel={Correction [frac]},
ytick={0.95, 1.00, 1.05},
yticklabels={0.95, 1.00, 1.05},
ticklabel style = {font = \scriptsize},
axis background/.style={fill=white},
xmajorgrids,
ymajorgrids
]
\addplot [color=black, line width=2.0pt, forget plot]
  table[row sep=crcr]{%
0	1.03647242175779\\
10	1.02761431043024\\
20	1.02599407125419\\
30	1.02783440237786\\
40	1.036502062127\\
50	1.04436243065189\\
60	1.03361117674859\\
70	1.02186165480427\\
80	1.00743693524097\\
90	1.00036479083667\\
100	1.01518203883495\\
110	1.0195649509804\\
120	1.01255969101123\\
130	1.01313701923078\\
140	1.01768342391304\\
150	1.02020833333333\\
160	1.01627840909092\\
};
\end{axis}
\end{tikzpicture}%

%% file: Images/TestTrack.tex
\begin{tikzpicture}

\node[anchor=south west,inner sep=0] (TestTrack) at (0,0) {\includegraphics[width=0.50\columnwidth]{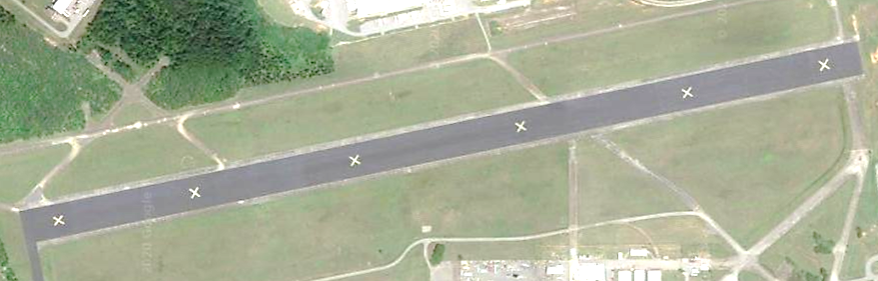}};

\begin{scope}[x={(TestTrack.south east)},y={(TestTrack.north west)}]
    
    \coordinate (A) at (0.10, 0.28);
    \coordinate (B) at (0.88, 0.7750);
    \coordinate (C) at (0.12, 0.21); 
    \coordinate (D) at (0.90, 0.70);
    
    \path[-] (A) edge[line width=1.25pt, color=black] node[sloped, anchor=center, above] {\unit[1550]{m}} (B);
    \draw[line width=1.25pt, color=black] (C) -- (D);
    \draw[line width=1.25pt, color=NavyBlue] (A) .. controls (0.07, 0.26) and (0.08, 0.18) .. (C); %
    \draw[line width=1.25pt, color=NavyBlue] (B) .. controls (0.91, 0.7950) and (0.93, 0.73) .. (D);
    
    \node[color=NavyBlue, rotate=12] (dd) at (0.88, 0.60) {\unit[95]{m}};
\end{scope}

\end{tikzpicture}

%% file: Images/vissimvel.tex
%
%
\definecolor{mycolor1}{rgb}{0.00000,0.44700,0.74100}%
\definecolor{mycolor2}{rgb}{0.85000,0.32500,0.09800}%
\definecolor{mycolor3}{rgb}{0.92900,0.69400,0.12500}%
\definecolor{mycolor4}{rgb}{0.49400,0.18400,0.55600}%
\begin{tikzpicture}

\begin{axis}[%
width=7.5cm,
height=3.5cm,
at={(0cm,0cm)},
scale only axis,
xmin=0,
xmax=3172,
xlabel style={font=\color{white!15!black} \footnotesize},
xlabel={PV Position [km]},
xtick={0, 500, 1000, 1500, 2000, 2500, 3000},
xticklabels={0, 0.5, 1.0, 1.5, 2.0, 2.5, 3.0},
ymin=0,
ymax=25,
ylabel style={font=\color{white!15!black} \footnotesize},
ylabel={PV Velocity [m/s]},
ticklabel style = {font = \scriptsize},
axis background/.style={fill=white},
xmajorgrids,
ymajorgrids,
legend style={at={(0.85,0.55)}, legend cell align=left, align=left, draw=white!15!black}
]
\addplot [color=mycolor1, line width=1.3pt]
  table[row sep=crcr]{%
-0.241984902698277	7.45000596\\
16.8372631720058	8.63889580000023\\
26.0753723631046	9.8388967599999\\
36.5176399885308	11.0416755000001\\
48.1489697301085	12.2166764399999\\
60.939747479179	13.35834402\\
74.8521204948556	14.4583449000002\\
89.8411778149803	15.5111235200002\\
105.85063013775	16.4972354199999\\
122.812432236969	17.4139028200002\\
140.651858348751	18.2527923799998\\
159.291444200659	19.0139041000002\\
178.649405865775	19.6889046400001\\
198.64338072772	20.2833495599998\\
219.188244590661	20.7916832999999\\
240.19751555751	21.2139058600001\\
261.585490879319	21.5500172400002\\
283.26940683799	21.8055730000001\\
305.169639356104	21.9833509199998\\
327.215782527237	22.0972399000002\\
349.347628458114	22.1583510599999\\
371.52360633786	22.1889066399999\\
393.722718504955	22.20557332\\
438.150958158945	22.21946222\\
482.592334769794	22.2222400000001\\
882.589640506862	22.2222400000001\\
904.78581364415	22.1305732599999\\
926.780967174867	21.8277952399999\\
948.525091066496	21.80279522\\
990.807020476866	20.5111275200002\\
1010.98043886019	19.7611269200002\\
1030.50686700343	19.1750153399998\\
1049.25471146702	18.48890368\\
1067.572861865	17.8750143000002\\
1085.12850459065	17.3527916600001\\
1102.29426530987	16.9222357600001\\
1118.69575961415	15.90834606\\
1134.32768784331	15.2222344000002\\
1149.19632931769	14.6889006400002\\
1163.69646432756	14.3889003999998\\
1178.04656938492	14.30834478\\
1192.24154552087	14.0111223200001\\
1206.16823578168	13.86389998\\
1219.9274392186	13.6111219999998\\
1233.38364370618	13.3111217599999\\
1259.37527645307	12.6833434800001\\
1271.86795360586	12.08889856\\
1283.23511676265	10.6861196599998\\
1293.27828416786	9.40834085999995\\
1302.0898389304	8.2250065799999\\
1309.75136454927	7.10833902000013\\
1316.32462711101	6.05000483999993\\
1321.8723956579	5.05278181999984\\
1326.45271815307	4.11666995999985\\
1330.12678223768	3.24166925999998\\
1332.95577546244	2.43055749999985\\
1335.01324331313	1.6972235799999\\
1336.37567029532	1.03888972000004\\
1337.13189877035	0.50277817999995\\
1337.44019728297	0.202777940000033\\
1337.46019729897	0.200000159999945\\
1337.47603064497	0.158333459999994\\
1337.47769731297	0\\
1337.48186398297	0.038888919999863\\
1337.75143482412	0.105555640000148\\
1337.77032372812	0.188889040000049\\
1338.13719751045	0.291666900000109\\
1338.48530808026	0.408333660000153\\
1338.9618724865	0.547222659999989\\
1339.58532724391	0.702778339999895\\
1340.37528913139	0.880556260000048\\
1341.35097375029	1.07500085999982\\
1342.5317972866	1.2888899200002\\
1343.93815561093	1.52500121999992\\
1345.58730497779	1.77777920000017\\
1347.4988621597	2.04722386000003\\
1349.68380416967	2.31944629999998\\
1352.13939259064	2.60000208000019\\
1354.82031519969	2.77500221999981\\
1357.72187421654	2.98889128000019\\
1363.95546538128	3.24166925999998\\
1367.23964747478	3.32222487999979\\
1370.59109580874	3.37778048000018\\
1373.98843484134	3.42222496000022\\
1377.42480683275	3.4472249800001\\
1380.88235379437	3.46389166000017\\
1384.35853784457	3.49166946000014\\
1387.87433336614	3.54444727999999\\
1391.45759571143	3.61944734000008\\
1395.11518262076	3.69166962000008\\
1402.63275691636	3.82222528000011\\
1410.38684301521	3.93055869999989\\
1414.3437107662	3.98055874000011\\
1422.39491776862	4.06666992000009\\
1426.47514039488	4.09166993999997\\
1430.5735989511	4.10000327999978\\
1434.65088253202	4.04166990000022\\
1438.65344039714	3.96666984000012\\
1442.57343512347	3.83333640000001\\
1446.30672743776	3.64166957999987\\
1449.85311676545	3.45000276000019\\
1453.21888231295	3.29166930000019\\
1466.33207519565	3.26944705999995\\
1476.13971045344	3.27222484000004\\
1479.41369461945	3.27500262000012\\
1482.68160012974	3.21944702000019\\
1485.88203881965	3.20278034000012\\
1492.30959129898	3.22500257999991\\
1498.7910724626	3.26111372000014\\
1502.06309727207	3.28333595999993\\
1508.56301495301	3.21944702000019\\
1511.7981667388	3.25278037999988\\
1515.07313056518	3.29444707999983\\
1518.50028430764	3.56111396000006\\
1521.97666393019	3.65000292000013\\
1528.04100598522	3.66389182000012\\
1531.17304863669	3.67222515999993\\
1534.07973605768	3.69444740000017\\
1537.48005689928	3.81666971999994\\
1540.76371459868	3.97778096000002\\
1544.36825703983	4.18333668000014\\
1548.36072332211	4.43055910000021\\
1552.93212371288	4.72500377999995\\
1557.23235775686	5.06667071999982\\
1561.90409442754	5.4583376999999\\
1566.91846811758	5.9000047200002\\
1583.42797198537	7.52778380000018\\
1590.77107778479	8.16389541999979\\
1599.26443194357	8.8333404\\
1608.45310776321	9.55000763999988\\
1618.37505451763	10.3000082399999\\
1629.06440716716	11.08056442\\
1640.54588168896	11.8861206199999\\
1652.84241127083	12.7083435\\
1665.96773512582	13.5388997199998\\
1679.92361168536	14.3722337200002\\
1694.7092620833	15.1972343799998\\
1710.31114798703	16.0027905799998\\
1726.70397534907	16.7777912000001\\
1743.85617071303	17.5194584599999\\
1761.72654084825	18.21390346\\
1780.26841628816	18.8583484199999\\
1799.42786843118	19.4500155599999\\
1819.14938687032	19.9833493199999\\
1839.37610411535	20.4583496999999\\
1860.04879181858	20.8750166999998\\
1881.10724221194	21.2305725400001\\
1902.49006681018	21.5250172199999\\
1924.1372586263	21.75835074\\
1945.98958935751	21.9361286600001\\
1967.99097037681	22.0583509799999\\
1990.09335099491	22.1389066000002\\
2012.25475915176	22.1805733000001\\
2034.44598213544	22.2000177599998\\
2056.65230130301	22.2111288800002\\
2101.08573695131	22.21946222\\
2145.52721053845	22.2222400000001\\
2701.08082078503	22.2222400000001\\
2723.17300347139	21.9277953199999\\
2744.91511630078	21.5416839\\
2766.21949724201	21.0611279599998\\
2786.94729390989	20.3694607399998\\
2806.95867431534	19.6583490600001\\
2826.25559770955	18.90557068\\
2844.70569181884	18.0083477399999\\
2862.28738042909	17.14723594\\
2879.05067350153	16.44167982\\
2895.19247130133	15.8389015600001\\
2910.73219001555	15.2416788599999\\
2925.6729692902	14.6444561600001\\
2940.01754743492	14.0472334599999\\
2953.7643665211	13.4444552\\
2966.9130250839	12.8611213999998\\
2979.48627984571	12.2861209399998\\
2991.49412785406	11.73056494\\
3002.94696784344	11.1750089400002\\
3013.85263698371	10.6388974000001\\
3024.22467361112	10.1055636400001\\
3034.07229594075	9.59167433999983\\
3052.24684535157	8.60000687999991\\
3060.60142723125	8.11667316000012\\
3075.90933184892	7.20278353999993\\
3089.46944861273	6.36667175999992\\
3095.63614998045	5.97500477999984\\
3101.41610909854	5.59167114000002\\
3106.82658212326	5.22500417999981\\
3111.17021886902	4.87778168000023\\
3114.97936605798	4.55000363999989\\
3117.56081555423	4.18611446000023\\
3120.24963641442	4.06389214000001\\
3122.76404802978	3.90278089999993\\
3126.07014768283	3.70833630000016\\
3129.06199027055	3.64722514000005\\
3131.17173862206	3.6138917799999\\
3133.27538276445	3.59166954000011\\
3135.21481264823	3.33055822000006\\
3136.96275384033	3.15000251999982\\
3138.99198585695	2.95278013999996\\
3140.63233001016	2.94166902000006\\
3142.75691889939	2.90000232000011\\
3144.98726538544	2.91944677999982\\
3146.57261165679	2.95555792000005\\
3149.87077875638	3.04444688000012\\
3157.3485026496	3.25833594000005\\
3159.44884746723	3.3111137599999\\
3161.68010800995	3.45555831999991\\
3164.10164609128	3.5944473200002\\
3166.9111632673	4.02222544000006\\
3170.44651070113	4.9972262199999\\
};
\addlegendentry{\scriptsize MPC-C}

\addplot [color=mycolor2, dashed, line width=1.3pt]
  table[row sep=crcr]{%
-0.534154039348323	10.2083415000002\\
19.5611374654272	12.6472323399998\\
33.3487747675972	14.8972341399999\\
49.2921411528109	16.96112468\\
67.2213892629138	18.8805706600001\\
87.0100246384368	20.6805721000001\\
108.544890352245	22.3722401199998\\
131.612809358703	23.3389075599998\\
154.763091031167	23.07224068\\
177.779819867448	22.9750183800002\\
223.514397865198	22.76668488\\
246.249503296133	22.6944626\\
268.902824621597	22.6139069800001\\
291.482801112311	22.53890692\\
313.98825246381	22.43612906\\
358.360202869035	21.9361286600001\\
401.732154105458	21.4361282599998\\
444.103905579359	20.9361278599999\\
485.475457256299	20.4361274600001\\
505.906184482003	20.7333499199999\\
527.25532743491	21.9639064600001\\
549.716847420736	22.6777959199999\\
572.108941242	22.07501766\\
593.868362038757	21.44168382\\
614.99294986931	20.80834998\\
635.483885014853	20.1722383599999\\
655.690057143845	20.4666830400001\\
676.4665344119	21.0861279800001\\
697.86078897359	21.70279514\\
719.874178414441	22.32224008\\
742.228620112289	22.1500177200001\\
764.081002557586	21.5555727999999\\
785.339343786952	20.9611278799998\\
806.003643804138	20.3666829600002\\
826.184120496449	20.2416828599999\\
846.656619131119	20.7027943399999\\
867.590385270817	21.1639058199999\\
888.983459576349	21.6222395200002\\
910.836821714559	22.0833510000002\\
933.137134117121	22.42224016\\
955.306832754157	21.9000175199999\\
976.933479843626	21.35557264\\
998.096900044956	20.9916834599999\\
1018.95928064544	20.7694610600001\\
1039.60220511316	20.51946086\\
1060.00331824111	20.28890512\\
1080.15222146359	19.9194603800001\\
1099.69904548463	19.4333488799998\\
1119.5316121735	20.1722383599999\\
1139.39281332761	19.15001532\\
1158.04063781167	18.2916813000002\\
1176.16070764329	18.1027922600001\\
1193.5102317579	16.6222355200002\\
1209.25335759035	14.4194559799998\\
1222.26771076515	11.78612054\\
1232.91217828921	9.54445208000016\\
1241.52761180699	7.87778407999986\\
1249.48801708049	8.25278437999987\\
1257.75770055438	8.26945105999994\\
1266.0293433861	8.29167330000018\\
1274.68576922909	8.96111827999994\\
1283.51046096396	8.25556215999995\\
1290.29318232344	5.10278186000005\\
1293.85898783046	2.20000175999985\\
1295.13179335692	0.583333800000219\\
1295.41243709889	0.0888889600000766\\
1295.41799265889	0.0555555999999342\\
1295.41799265889	0\\
1295.42299266289	0.0305555800000548\\
1295.42327044089	0.00277778000008766\\
1295.44104823289	0.141666779999923\\
1295.83801184109	0.266666879999775\\
1296.20121872202	0.46111148\\
1296.69210035192	0.533333760000005\\
1297.38910057957	0.891667379999944\\
1298.66582480575	1.79722365999987\\
1300.65288680721	2.02500161999978\\
1302.67682183309	2.02500161999978\\
1304.71487350037	2.14444615999992\\
1307.18555822733	2.58889095999984\\
1310.15752310666	3.65278070000022\\
1314.24480404777	4.23889228000007\\
1318.57409692076	4.65833706000012\\
1324.08205342248	6.35833842000011\\
1331.08814963604	7.43333927999993\\
1338.53021910257	7.44167262000019\\
1368.30201453228	7.44445039999982\\
1375.55306172937	6.61667196000008\\
1380.79665102909	3.7722252399999\\
1383.31656665998	1.4472233800002\\
1384.10848789718	0.336111380000148\\
1384.10848789718	0\\
1384.11209901118	0.0333333600001424\\
1384.35539822616	0.16944457999989\\
1384.66467641592	0.480555940000158\\
1385.36873496191	0.941667420000158\\
1386.56485539683	1.45000115999983\\
1388.26421533963	1.94444599999997\\
1390.43797974949	2.39722414000016\\
1393.04339753929	2.80833557999995\\
1396.03987756854	3.17778031999978\\
1399.39132590279	3.51666948000002\\
1403.06343062462	3.82222528000011\\
1407.0293160449	4.10833662000005\\
1411.26250766018	4.3583368200002\\
1415.74378986213	4.60833701999991\\
1420.35784578013	4.54444808000017\\
1424.77794039745	4.29444788000001\\
1429.04682478117	4.36389237999992\\
1433.56379974925	4.67500374000019\\
1438.40143109795	5.00000399999999\\
1443.56343696172	5.32500425999979\\
1449.04785796271	5.64444895999986\\
1454.84253683025	5.94167142000015\\
1460.92923764593	6.2333383199998\\
1467.23323466553	6.28889392000019\\
1473.43427268987	6.09722710000005\\
1479.37388255595	5.73611570000003\\
1484.77260080033	4.89167058000021\\
1489.02638887937	3.83889196000018\\
1492.61572986725	3.43889163999984\\
1495.37529692227	1.90555708000011\\
1496.74596247755	1.08611198000017\\
1497.66849749434	0.758333939999829\\
1498.26253871635	0.447222580000016\\
1498.7830344705	0.619444939999994\\
1499.54632126062	0.927778520000174\\
1500.67105617026	1.32500105999998\\
1502.19783033964	1.72500137999987\\
1504.11330622321	2.10277945999997\\
1506.39944847532	2.46389085999999\\
1508.9905490491	2.70833549999998\\
1511.85327568748	3.01111351999998\\
1514.97213091501	3.24166925999998\\
1518.3298584764	3.40278050000006\\
1521.60776134146	3.1527802999999\\
1524.10402336196	2.9222245599999\\
1526.66476453168	3.33333600000014\\
1529.75221108094	3.68611405999991\\
1532.78486065631	3.91666979999991\\
1535.95798769597	3.71111407999979\\
1539.15729089506	3.46111388000008\\
1542.02363514554	3.21111367999993\\
1544.99169546906	2.96111348000022\\
1547.71199375277	3.18611366000005\\
1550.92251655966	3.70555852000007\\
1554.69799312434	3.88889199999994\\
1557.95739447557	3.63889179999978\\
1560.9958492348	3.38889160000008\\
1563.84330873398	3.13889139999992\\
1566.645348856	3.24722482000016\\
1569.27643401262	3.49722501999986\\
1572.5144574454	3.74722522000002\\
1575.56677770953	3.91666979999991\\
1579.01582441004	3.88889199999994\\
1582.13610836867	3.91944758\\
1585.32032952566	3.91944758\\
1589.45782023669	4.81389274000003\\
1595.44307665851	7.50278377999985\\
1604.25895008128	10.10278586\\
1615.59843253805	12.55001004\\
1629.29210199931	14.8055674000002\\
1645.1481570919	16.8777912800001\\
1662.99715179121	18.8000150399998\\
1682.70965271975	20.60557204\\
1704.17308107873	22.3027956199999\\
1727.08679902618	23.17224076\\
1772.65790558193	22.3972401400001\\
1794.86422473183	22.0166842799999\\
1816.69008098716	21.6361284200002\\
1838.13625324018	21.25557256\\
1879.88876677983	20.4972386200002\\
1900.19510814548	20.11668276\\
1920.4737946002	20.6972387800001\\
1941.65971898668	21.67223956\\
1963.813089229	22.5722402800002\\
1986.27985693604	22.1889066399999\\
2008.15258355101	21.55835058\\
2029.39361688565	20.92501674\\
2050.00197736206	20.2916829000001\\
2070.16652740041	20.3361273800001\\
2090.81058065153	20.95279454\\
2112.07005057467	21.5666839199998\\
2133.94571604882	22.1833510800002\\
2156.34207796715	22.36390678\\
2178.39836717319	21.7472396200001\\
2199.83944044345	21.1333502399998\\
2220.66529784227	20.51946086\\
2240.91045160113	20.1361272200002\\
2261.30033557312	20.6444609599998\\
2282.19953769751	21.15557248\\
2303.6090375806	21.6639062200002\\
2325.52765478731	22.1722399599998\\
2347.86164875412	22.2694622600002\\
2369.87636737354	21.75835074\\
2391.37960787231	21.2472392200002\\
2412.37059129737	20.7361277\\
2432.85049810358	20.2250161799998\\
2453.01386771298	20.2972384599998\\
2473.50593174712	20.6861276599998\\
2494.3847379927	21.0722390800001\\
2515.65028650303	21.4583505000001\\
2537.30355692755	21.8472397\\
2559.34377037242	22.23335112\\
2581.68482257955	22.2583511399998\\
2603.73523393712	21.8416841399999\\
2625.37006781986	21.42779492\\
2646.58874600482	21.01112792\\
2667.39106784659	20.5944609200001\\
2687.778012928	20.1777939200001\\
2707.9904128694	20.39168298\\
2728.59367441943	20.8139055400002\\
2749.61937097982	21.2361280999999\\
2771.06672359375	21.65835066\\
2792.93631049912	22.0805732200001\\
2815.21323622736	22.3805734600001\\
2837.31757611163	21.8111285599998\\
2858.83415418532	21.2222391999999\\
2879.76395008527	20.63612762\\
2900.11010345899	20.1111271999998\\
2920.40624688213	20.5305719799999\\
2941.1767945174	21.01112792\\
2962.28211516673	20.9166833999998\\
2982.49059655041	19.51112672\\
3001.41509494545	18.3472369000001\\
3019.17740731498	17.1666804000001\\
3035.71830530956	15.86667936\\
3050.83852795521	14.3583448200002\\
3064.41492125734	12.7861213400001\\
3076.39727446855	11.1750089400002\\
3086.75597346658	9.54167430000007\\
3095.47278243797	7.88889520000021\\
3102.53240448018	6.23056054000017\\
3108.03995854399	5.03889291999985\\
3112.00468587812	4.6944481999999\\
3115.07692095967	4.21389226000019\\
3117.90278974173	3.86111419999997\\
3120.30292124969	3.72500298000023\\
3122.52982306272	3.58889176000002\\
3124.61424971086	3.40833605999978\\
3126.4517287018	3.15833586000008\\
3128.00284934838	2.89166897999985\\
3129.26074461191	2.62222431999999\\
3130.23344128592	2.35277966000012\\
3132.55622407688	2.32777963999979\\
3133.40269313059	2.40833526000006\\
3135.23547817452	2.57500205999986\\
3137.02448853697	2.71389106000015\\
3138.34102167963	2.83611337999992\\
3139.86531257176	3.11944694000022\\
3142.24503493801	3.19444699999985\\
3144.68602769189	3.1527802999999\\
3146.61754014298	3.40278050000006\\
3148.79117220472	3.65278070000022\\
3150.61358286507	3.65278070000022\\
3152.82451970271	3.40278050000006\\
3154.19348499107	3.46111388000008\\
3157.19135951694	3.71111407999979\\
3159.11609558269	3.75000300000011\\
3164.31295213069	3.75000300000011\\
3167.13696071537	4.43055910000021\\
3171.73173966029	7.07778344000008\\
};
\addlegendentry{\scriptsize MPC-U}

\addplot [color=mycolor3, dashdotted, line width=1.3pt]
  table[row sep=crcr]{%
-0.534154039348323	10.2083415000002\\
19.5611374654272	12.6472323399998\\
33.3487747675972	14.8972341399999\\
49.2921411528109	16.96112468\\
67.2213892629138	18.8805706600001\\
87.0100246384368	20.6805721000001\\
108.544890352245	22.3722401199998\\
131.612809358703	23.3389075599998\\
154.763091031167	23.07224068\\
177.779819867448	22.9750183800002\\
223.514397865198	22.76668488\\
246.249503296133	22.6944626\\
268.902824621597	22.6139069800001\\
291.482801112311	22.53890692\\
313.98825246381	22.43612906\\
358.360202869035	21.9361286600001\\
401.732154105458	21.4361282599998\\
444.103905579359	20.9361278599999\\
485.475457256299	20.4361274600001\\
505.906184482003	20.7333499199999\\
527.25532743491	21.9639064600001\\
549.716847420736	22.6777959199999\\
572.108941242	22.07501766\\
593.868362038757	21.44168382\\
614.99294986931	20.80834998\\
635.483885014853	20.1722383599999\\
655.690057143845	20.4666830400001\\
676.4665344119	21.0861279800001\\
697.86078897359	21.70279514\\
719.874178414441	22.32224008\\
742.228620112289	22.1500177200001\\
764.081002557586	21.5555727999999\\
785.339343786952	20.9611278799998\\
806.003643804138	20.3666829600002\\
826.184120496449	20.2416828599999\\
846.656619131119	20.7027943399999\\
867.590385270817	21.1639058199999\\
888.983459576349	21.6222395200002\\
910.836821714559	22.0833510000002\\
933.137134117121	22.42224016\\
955.306832754157	21.9000175199999\\
976.933479843626	21.35557264\\
998.063367227995	21.0333501599998\\
1018.97673752989	20.7889055199998\\
1039.66947139185	20.5889053599999\\
1060.13295233795	20.3389051600002\\
1080.36050001677	20.2083495000002\\
1100.90222422454	20.8333499999999\\
1121.77480272957	20.59168314\\
1141.84114286592	19.5416823\\
1160.88786695544	18.5361259400001\\
1178.87830269656	17.4250139400001\\
1195.70027274488	16.2166796400002\\
1211.77425288384	16.2583463400001\\
1228.41441890075	16.0777906399999\\
1242.53486988404	12.9277881200001\\
1254.98793557957	11.4055646800002\\
1264.0260048107	7.1555612799998\\
1270.56475487989	6.59444971999983\\
1277.99740551259	8.31945110000015\\
1286.62539747915	8.70834030000015\\
1294.29692044654	6.30278282000017\\
1299.56914421683	4.40833685999996\\
1303.96158405443	4.41667020000023\\
1308.37853905323	4.41667020000023\\
1312.67115949612	3.87778088000005\\
1315.62978682424	2.05277942000021\\
1316.95750071501	0.725000580000142\\
1317.34228372546	0.158333459999994\\
1317.34228372546	0\\
1317.34922817546	0.0694444999999178\\
1317.37228374946	0.23055574\\
1317.92906604208	0.597222700000202\\
1318.52604629159	0.616667159999906\\
1319.52016742374	1.24166766000008\\
1321.04380197691	1.79444587999978\\
1323.25737843947	2.85278005999999\\
1326.37427431734	3.28055817999984\\
1329.93910014952	3.66944737999984\\
1333.70494562031	4.10833662000005\\
1338.32508019009	4.90833725999983\\
1343.30018304899	5.23611530000017\\
1349.15820945449	6.24722721999979\\
1355.40925743754	6.25278277999996\\
1361.66226482042	6.27222724000012\\
1368.07431978531	6.48056073999987\\
1375.11316975684	7.68333948000009\\
1383.02768433071	7.96389525999984\\
1391.40542580017	8.76667368000017\\
1400.10676265585	8.28611774000001\\
1407.00892018599	5.24722642000006\\
1410.6798446264	2.26944626000022\\
1411.9954012346	0.60555604000001\\
1412.29134187708	0.0888889600000766\\
1412.29134187708	0\\
1412.30717522308	0.133333440000115\\
1412.63827218333	0.419444780000049\\
1413.27310513785	0.872222919999786\\
1414.40489836559	1.39722333999998\\
1416.06326598313	1.9166682\\
1418.22585277953	2.40277969999988\\
1420.85892534939	2.85555784000007\\
1423.92679090692	3.27500262000012\\
1427.39865508387	3.66389182000012\\
1431.2462845933	4.02778099999978\\
1435.44790273547	4.37222572000019\\
1439.98625325779	4.70278154000016\\
1444.84937944722	5.02500401999987\\
1450.03178115786	5.33889316000023\\
1455.52326049855	5.63611562000006\\
1461.30480238922	5.92500474000008\\
1467.35972915103	6.12222711999993\\
1473.44035133184	6.01667147999979\\
1479.35446637121	5.78056018000007\\
1484.913412504	5.28889312000001\\
1489.84694456015	4.55000363999989\\
1493.98325581279	3.73889188000021\\
1497.37157717475	3.01666908000016\\
1499.98405328369	2.20555732000003\\
1501.93130323115	1.71389025999997\\
1503.42748360235	1.27777879999985\\
1504.49103091468	0.900000720000207\\
1506.83836077056	1.50555676000022\\
1508.52928153776	1.87777928000014\\
1510.56733320703	2.19722398000022\\
1512.95859440149	2.57777983999995\\
1515.70600416727	2.90833565999992\\
1518.71346122688	3.10278026000015\\
1521.71797924939	2.88055785999995\\
1524.11958080928	2.83055782000019\\
1526.44788075572	3.2166692400001\\
1529.39201544611	3.51389169999993\\
1532.1228636623	3.45555831999991\\
1535.08226146351	3.20555812000021\\
1537.80737555041	2.95555792000005\\
1540.05648453599	2.93889123999998\\
1543.39041086768	4.09444772000006\\
1546.85999398317	3.58889176000002\\
1550.13915014496	3.05833578000011\\
1553.1190356456	3.38889160000008\\
1556.55797661626	3.83889196000018\\
1559.83403268969	3.80278081999995\\
1563.09594971416	3.55278061999979\\
1565.97730882364	3.30278042000009\\
1568.79693774852	3.07222468000009\\
1571.25116169564	3.40278050000006\\
1574.43614332042	3.75278078000019\\
1577.49044214239	3.8444475199999\\
1580.99894987314	3.91944758\\
1584.13534055013	3.91944758\\
1587.56358912465	4.0861143799998\\
1592.68956457255	6.41667180000013\\
1600.45231356969	9.08056282000007\\
1610.80023651121	11.5916759400002\\
1623.57549045156	13.9250111400002\\
1638.58863488924	16.0694573000001\\
1655.65786570742	18.0472366600002\\
1674.63963307094	19.8972381399999\\
1695.41548044538	21.6361284200002\\
1717.85655264714	23.0916851400002\\
1740.98854624281	23.0250184199999\\
1830.8899288266	21.93335088\\
1874.21392663558	21.3916837800002\\
1895.47025684825	21.1222391199999\\
1937.17276044364	20.5805720200001\\
1957.61795414086	20.3111273600002\\
1977.83231333723	20.3139051399999\\
1998.80408138255	21.6805728999998\\
2021.15023255109	22.8027960200002\\
2043.75879157513	22.2694622600002\\
2065.69976391935	21.6111283999999\\
2086.98178977091	20.95279454\\
2107.60642679148	20.2944606800002\\
2127.75351986314	20.3139051399999\\
2148.4118903601	21.0027945800002\\
2169.75706300979	21.6889062400001\\
2191.78903779505	22.3750178999999\\
2214.130090002	22.10557324\\
2235.93966974608	21.5139061\\
2257.15618792478	20.9194611799999\\
2277.78004604325	20.3277940399998\\
2297.96381152995	20.2972384599998\\
2318.52118241865	20.8194610999999\\
2339.60002861522	21.3389059599999\\
2361.19916970744	21.8583508199999\\
2383.31880651207	22.3805734600001\\
2405.61612810131	22.07501766\\
2427.44002506179	21.57223948\\
2448.75950215031	21.06668352\\
2469.57516157751	20.56390534\\
2489.89170082716	20.1222383200002\\
2510.15704970709	20.4277941199998\\
2530.74815403371	20.7527943800001\\
2572.90584643492	21.4055726800002\\
2594.47341412717	21.73057294\\
2638.58619331519	22.3833512400001\\
2660.83762421806	22.0277953999998\\
2682.62758698512	21.5527950199998\\
2703.94432587554	21.08057242\\
2724.78625901916	20.60557204\\
2745.15378789549	20.13057166\\
2765.3591296101	20.3777940800001\\
2806.85649430593	21.1194613399998\\
2828.16185488969	21.4916838600002\\
2849.83984124354	21.8639063800001\\
2871.88907216696	22.2361289\\
2894.22404580974	22.2444622399998\\
2916.24072367934	21.6972395799999\\
2937.38369640637	20.5805720200001\\
2957.39821642242	19.4500155599999\\
2976.28506278941	18.3222368800002\\
2994.08208821391	17.2944582800001\\
3010.87499539867	16.2750130200002\\
3026.59121855392	15.1416787799999\\
3041.05935213635	13.7527887800002\\
3054.04680321731	12.20556532\\
3065.43943530138	10.56389734\\
3075.13898700444	8.81667371999993\\
3083.05779631323	7.01945006000005\\
3089.20312218606	5.37222651999991\\
3094.2237132537	4.69722597999998\\
3098.52906846898	3.87778088000005\\
3101.96817817919	3.05833578000011\\
3104.85403841924	2.75277998000001\\
3107.48181085133	2.50277977999986\\
3109.56854156629	2.25277958000015\\
3111.24699744573	2.33055741999988\\
3113.16391329737	2.46389085999999\\
3115.01308371303	2.65833546000022\\
3117.03928189704	2.81944669999984\\
3118.78875023296	3.00555795999981\\
3120.94938407736	3.25555815999996\\
3122.83393950048	3.22778036\\
3124.59600936037	3.09722469999997\\
3126.17851126761	2.96666903999994\\
3127.544687326	2.71666883999978\\
3129.09675088291	2.61111320000009\\
3130.70055699517	2.71666883999978\\
3132.60329797057	2.87778007999987\\
3134.09645623328	3.12778028000002\\
3135.84762494315	3.37778048000018\\
3137.78873006586	3.39722493999989\\
3139.64568931598	3.17778031999978\\
3141.97806143133	3.16666919999989\\
3144.4911091644	3.25000259999979\\
3146.51245315375	3.50000279999995\\
3148.78399530284	3.75000300000011\\
3151.03403499862	3.55000284000016\\
3152.67436757648	3.31944710000016\\
3154.06556498308	3.50555836000012\\
3157.10911006121	3.75000300000011\\
3159.09423094685	3.75833633999991\\
3161.59801882921	3.75000300000011\\
3164.21374948451	3.75000300000011\\
3167.03064745829	4.43055910000021\\
3171.61878669764	7.07778344000008\\
};
\addlegendentry{\scriptsize WIE}

\addplot [color=mycolor4, dotted, line width=1.3pt]
  table[row sep=crcr]{%
-0.534154039348323	10.2083415000002\\
19.5611374654272	12.6472323399998\\
33.3487747675972	14.8972341399999\\
49.2921411528109	16.96112468\\
67.2213892629138	18.8805706600001\\
87.0100246384368	20.6805721000001\\
108.544890352245	22.3722401199998\\
131.612809358703	23.3389075599998\\
154.763091031167	23.07224068\\
177.779819867448	22.9750183800002\\
223.514397865198	22.76668488\\
246.249503296133	22.6944626\\
268.902824621597	22.6139069800001\\
291.482801112311	22.53890692\\
313.98825246381	22.43612906\\
358.360202869035	21.9361286600001\\
401.732154105458	21.4361282599998\\
444.103905579359	20.9361278599999\\
485.475457256299	20.4361274600001\\
505.906184482003	20.7333499199999\\
527.25532743491	21.9639064600001\\
549.716847420736	22.6777959199999\\
572.108941242	22.07501766\\
593.868362038757	21.44168382\\
614.99294986931	20.80834998\\
635.483885014853	20.1722383599999\\
655.690057143845	20.4666830400001\\
676.4665344119	21.0861279800001\\
697.86078897359	21.70279514\\
719.874178414441	22.32224008\\
742.228620112289	22.1500177200001\\
764.081002557586	21.5555727999999\\
785.339343786952	20.9611278799998\\
806.003643804138	20.3666829600002\\
826.184120496449	20.2416828599999\\
846.656619131119	20.7027943399999\\
867.590385270817	21.1639058199999\\
888.983459576349	21.6222395200002\\
910.836821714559	22.0833510000002\\
933.137134117121	22.42224016\\
955.306832754157	21.9000175199999\\
976.933479843626	21.35557264\\
998.016296266215	20.8111277600001\\
1018.55410179591	20.2666828800002\\
1038.74085771523	20.2944606800002\\
1059.20315841974	20.6111276000001\\
1100.42663949574	20.6111276000001\\
1121.04720875485	20.6527943000001\\
1142.16159868222	21.4944616399998\\
1163.26677029134	20.3889052\\
1182.70239672456	18.7639039000001\\
1201.50123084259	19.1194597399999\\
1220.60208422703	18.8055706\\
1239.24578939211	18.0916811400002\\
1255.96969887275	15.1805677000002\\
1269.58083169015	12.0555651999998\\
1280.11508128181	9.03334055999994\\
1287.70208174537	6.17500493999978\\
1292.55676877567	3.59722509999983\\
1295.07452446228	1.55555679999998\\
1295.97666360954	0.416666999999961\\
1295.97749694354	0\\
1296.20945730399	0.10277786000006\\
1296.233346212	0.238889079999808\\
1296.77780310626	0.658333859999857\\
1297.58718118163	0.911111840000103\\
1298.55560689257	1.07777863999991\\
1300.21103548996	2.31666851999989\\
1303.22281242343	3.6972251799998\\
1307.35598406066	4.42500354000003\\
1311.83726626365	4.72500377999995\\
1316.84630313338	5.0805596199998\\
1321.99595112062	5.30555980000008\\
1327.39878864974	5.79722686000014\\
1334.1436573246	7.71389506000014\\
1342.57964732798	8.85834041999988\\
1352.08428361419	10.3055638000001\\
1363.0358693764	11.4305647000001\\
1374.90921109151	12.0444540799999\\
1386.50582773913	10.6111196000002\\
1395.52036148628	7.3861170199998\\
1401.37014932265	4.38055906\\
1404.47586759606	1.96111268000004\\
1405.63315564695	0.547222659999989\\
1405.90262178204	0.0750000600000931\\
1405.90262178204	0\\
1405.91984401804	0.144444560000011\\
1406.27190730467	0.447222580000016\\
1406.94361328707	0.916667399999824\\
1408.12835552716	1.45555672\\
1409.85321030607	1.98889048000001\\
1412.09326127336	2.48611309999978\\
1414.81517620897	2.95000235999987\\
1417.98188148837	3.37778048000018\\
1421.56200421295	3.77778080000007\\
1425.53102923924	4.15555888000017\\
1429.86640078243	4.51111472000002\\
1434.54576358838	4.84444832000008\\
1439.55264052828	5.16667079999979\\
1444.8768336283	5.48055994000015\\
1450.51128459525	5.78056018000007\\
1456.43461790558	6.06667152\\
1462.64173455437	6.33056062000014\\
1468.93553363214	6.24167166000007\\
1475.08342205496	6.02778260000014\\
1480.90771504319	5.57222667999986\\
1486.14012676198	4.86111500000015\\
1490.60003342239	4.06666992000009\\
1494.31390954747	3.36666935999983\\
1497.34294273259	2.69444659999999\\
1499.73106430906	2.14166837999983\\
1501.66183710386	1.71944582000015\\
1503.17135501528	1.3000010400001\\
1504.3633561633	1.21111208000002\\
1505.65537728721	1.38333444\\
1508.89464835806	1.88333483999986\\
1510.87249210319	2.03889052000022\\
1513.03703825321	2.28889071999993\\
1515.45085526238	2.53889092000009\\
1518.11452162511	2.78889111999979\\
1520.94135491841	2.75000219999993\\
1523.40023876339	2.7027799399998\\
1525.74657297623	2.88333564000004\\
1528.0356719763	3.09166913999979\\
1530.67249779567	3.34166933999995\\
1533.67919518388	3.3944471599998\\
1536.27992025635	3.20833589999984\\
1539.04094073943	2.95833570000013\\
1541.74564449779	3.0694469\\
1544.65969489061	3.31944710000016\\
1547.81926097781	3.44166941999993\\
1550.90451975641	3.19166922000022\\
1553.86611376365	2.94166902000006\\
1556.51552317551	2.75277998000001\\
1559.03605043709	3.25555815999996\\
1562.04088852988	3.53055838\\
1565.16144304241	3.80833638000013\\
1568.58223669397	3.88611421999985\\
1571.57157793917	3.63611402000015\\
1574.69156573174	3.38611381999999\\
1577.16600730098	3.33611377999978\\
1580.60083561611	4.05278102000011\\
1583.73390240617	3.91944758\\
1587.15228343968	3.98055874000011\\
1592.05412941997	6.13889380000001\\
1599.55017460538	8.8222292800001\\
1609.64608833071	11.3472313000002\\
1622.18756888876	13.7000109599999\\
1636.9843968889	15.8639015799999\\
1653.85574776246	17.85556984\\
1672.65179244737	19.7166824400001\\
1693.25309474446	21.4666838399999\\
1715.54413693248	23.0166850800001\\
1738.67417132583	23.0444628800001\\
1828.54691942178	21.9000175199999\\
1871.78011550571	21.3333504000002\\
1913.88053893146	20.7666832800001\\
1954.84839038777	20.2000161599999\\
1975.17394716631	20.8139055400002\\
1996.69876389396	22.2361289\\
2019.35517288165	22.7027959400002\\
2041.68818724193	22.00279538\\
2063.37009216269	21.3611282000002\\
2104.81450791774	20.10279386\\
2125.06867920487	20.5083497400001\\
2145.84648587047	21.0472390599998\\
2167.16204438725	21.58612838\\
2189.0165351344	22.1222399200001\\
2211.35032834165	22.3472400999999\\
2233.39976006103	21.7500174000002\\
2254.85005161658	21.1527947\\
2275.70218268896	20.5527942200001\\
2295.9718516219	20.0944605200002\\
2316.30780706655	20.5972387000002\\
2337.16503711148	21.1194613399998\\
2358.54492284447	21.6416839799999\\
2380.44726353732	22.16390662\\
2402.77831856199	22.2722400399998\\
2424.8010750634	21.77223964\\
2446.32373175489	21.2722392400001\\
2467.34648935164	20.7722388400002\\
2487.86796675247	20.2722384399999\\
2508.03153712947	20.2639051000001\\
2528.49084732047	20.6555720800002\\
2549.34415878061	21.0472390599998\\
2570.58833181727	21.44168382\\
2592.22532570567	21.8333508000001\\
2614.25416081995	22.2250177800001\\
2636.58913441422	22.2555733600002\\
2658.64052542038	21.8472397\\
2680.28359791882	21.4389060399999\\
2701.51659335987	21.0277946000001\\
2722.340290704	20.61946094\\
2742.75468988162	20.2111272799998\\
2762.94787437788	20.3583496199999\\
2783.53171972186	20.80834998\\
2804.56349485049	21.25557256\\
2826.04340056404	21.70279514\\
2847.97123606433	22.1527955000001\\
2870.28993314555	22.2944622800001\\
2892.30249170996	21.73057294\\
2913.7518036513	21.1666835999999\\
2934.63668849718	20.6027942599999\\
2954.95930637473	20.0389049199998\\
2974.71769795181	19.4777933599999\\
2993.84831482692	18.6639038200001\\
3011.9546184345	17.5111251200001\\
3028.79635521355	16.1444573600002\\
3044.25037125825	14.7639006999998\\
3058.28901155237	13.2888995200001\\
3070.808136984	11.74445384\\
3081.75675788485	10.1361192200002\\
3091.05427042078	8.46111788000007\\
3098.68635741498	6.81389433999993\\
3104.71616567807	5.3638931800001\\
3109.48901618837	4.75278157999992\\
3112.60642193107	3.96944762000021\\
3115.32099445819	3.19444699999985\\
3119.06855236137	2.65555768000013\\
3120.68383525484	2.40555747999997\\
3122.18958643287	2.43889084000011\\
3123.3017025861	2.57222428000023\\
3125.26603431352	2.7027799399998\\
3126.59445492319	2.83611337999992\\
3128.66693251915	3.08611358000007\\
3130.38425016983	3.33611377999978\\
3132.28397635234	3.38889160000008\\
3134.09657461333	3.26111372000014\\
3134.27232850708	3.04444688000012\\
3137.19829852479	2.79444667999996\\
3139.08377782872	2.8500022799999\\
3141.14575439413	2.99166905999982\\
3143.08525079089	3.07500246000018\\
3147.1121387937	3.51666948000002\\
3149.1561770817	3.75000300000011\\
3153.59325900008	3.33333600000014\\
3155.74668380154	3.55278061999979\\
3158.15873649734	3.75000300000011\\
3165.25735786526	3.75000300000011\\
3168.9420830007	5.7000045599998\\
};
\addlegendentry{\scriptsize IDM}

\end{axis}

\begin{axis}[%
width=7.975cm,
height=4cm,
at={(-1.037cm,-0.44cm)},
scale only axis,
xmin=0,
xmax=1,
ymin=0,
ymax=1,
axis line style={draw=none},
ticks=none,
axis x line*=bottom,
axis y line*=left,
legend style={legend cell align=left, align=left, draw=white!15!black}
]
\end{axis}
\end{tikzpicture}%

%% file: Images/EPA_US06_modified.tex
%
%
\definecolor{mycolor1}{rgb}{0.00000,0.44700,0.74100}
\begin{tikzpicture}

\begin{axis}[%
width=6.5cm,
height=3.50cm,
at={(0.0in,0.0in)},
scale only axis,
xmin=0.0,
xmax=600.0,
xlabel style={font=\color{white!10!black} \scriptsize},
xlabel={Time [s]},
ymin=0.0,
ymax=36.0,
ylabel style={font=\color{white!10!black} \scriptsize},
ylabel={Drive Cycle [m/s]},
ticklabel style = {font = \scriptsize},
axis background/.style={fill=white},
xmajorgrids,
ymajorgrids,
legend style={anchor=south east, legend cell align=left, align=left, draw=white!12!black}
]

\addplot [color=mycolor1, dashed, line width=1.3pt]
  table[row sep=crcr]{%
0.0	0.0\\
33.1	0.0\\
34.1	0.1\\
35.3	0.4\\
36.1	0.5\\
37.0	0.8\\
38.0	2.7\\
39.1	6.5\\
40.0	9.2\\
41.0	11.5\\
41.2	11.4\\
42.0	11.2\\
43.2	13.0\\
44.0	14.4\\
45.0	15.5\\
47.0	17.2\\
48.0	17.8\\
49.0	18.9\\
50.0	19.6\\
51.0	19.8\\
53.0	19.0\\
54.0	18.0\\
55.1	17.5\\
56.0	17.2\\
57.0	17.2\\
58.0	17.5\\
59.0	17.3\\
60.0	17.3\\
61.0	16.3\\
62.2	14.0\\
63.1	12.1\\
64.1	9.7\\
65.1	7.4\\
66.9	2.8\\
67.0	2.6\\
68.0	0.5\\
69.0	0.0\\
76.0	0.0\\
76.9	0.3\\
77.0	0.5\\
77.9	3.8\\
78.0	4.2\\
79.0	6.7\\
80.0	8.2\\
81.0	10.0\\
82.0	12.2\\
82.2	12.6\\
83.0	14.1\\
84.0	15.2\\
84.2	15.4\\
85.1	16.8\\
86.0	18.3\\
87.0	19.7\\
87.9	20.6\\
88.0	20.7\\
89.0	21.3\\
90.0	22.2\\
90.1	22.2\\
90.2	22.2\\
91.5	23.2\\
92.1	23.7\\
93.2	24.4\\
94.1	24.9\\
95.0	25.2\\
95.1	25.2\\
96.1	25.1\\
97.1	25.1\\
98.1	24.9\\
98.6	24.8\\
100.1	24.3\\
101.1	24.2\\
102.1	24.3\\
103.1	24.2\\
103.7	24.0\\
104.1	23.9\\
105.1	23.4\\
106.2	23.2\\
108.2	23.2\\
110.1	23.3\\
111.1	23.5\\
112.1	23.9\\
113.1	24.6\\
114.5	25.8\\
115.2	26.4\\
116.4	27.4\\
117.2	28.0\\
118.5	29.1\\
119.1	29.6\\
121.0	31.0\\
121.2	31.1\\
122.0	31.4\\
122.1	31.5\\
123.3	31.6\\
124.1	31.6\\
125.0	31.4\\
125.1	31.4\\
126.1	30.4\\
127.2	29.6\\
128.1	29.0\\
130.2	27.8\\
131.2	27.2\\
132.4	26.2\\
132.7	26.0\\
133.5	25.4\\
134.1	25.0\\
135.2	24.2\\
136.2	23.5\\
137.6	22.5\\
138.2	22.0\\
139.3	21.3\\
140.1	20.6\\
141.1	19.6\\
142.1	18.3\\
143.0	17.4\\
143.1	17.3\\
145.3	16.2\\
146.0	15.8\\
146.1	15.6\\
147.5	12.2\\
148.1	10.8\\
149.0	8.9\\
150.5	6.0\\
150.6	5.7\\
150.8	5.3\\
151.3	4.3\\
152.0	2.9\\
152.2	2.6\\
153.0	1.5\\
153.1	1.4\\
154.1	0.9\\
154.6	0.7\\
155.1	0.5\\
156.0	0.0\\
156.1	0.0\\
163.0	0.0\\
163.1	0.1\\
164.0	1.2\\
164.2	1.7\\
165.5	5.6\\
166.5	8.6\\
166.7	9.2\\
168.0	13.0\\
168.4	13.9\\
168.6	14.3\\
169.2	15.7\\
170.1	17.5\\
171.0	19.2\\
171.2	19.4\\
172.0	20.2\\
172.4	20.5\\
173.1	21.0\\
174.1	21.5\\
175.0	22.1\\
175.4	22.3\\
176.1	22.5\\
177.0	23.0\\
177.5	23.2\\
178.1	23.4\\
180.1	23.7\\
181.0	24.0\\
181.5	24.1\\
183.0	24.1\\
183.6	24.3\\
184.0	24.4\\
184.4	24.7\\
185.0	25.2\\
185.5	25.3\\
186.0	25.4\\
186.7	25.8\\
187.0	26.0\\
187.9	26.1\\
188.0	26.1\\
188.6	26.4\\
189.0	26.6\\
189.4	26.7\\
192.0	27.0\\
192.5	26.9\\
193.0	26.7\\
194.0	26.1\\
195.3	25.9\\
196.1	25.8\\
197.0	25.6\\
198.0	25.7\\
199.0	25.3\\
200.0	25.5\\
201.1	25.3\\
202.1	25.2\\
203.0	25.1\\
205.0	25.3\\
206.3	25.2\\
207.1	25.1\\
209.0	25.0\\
210.0	24.5\\
211.0	24.2\\
212.0	24.4\\
212.9	23.4\\
213.0	23.3\\
214.0	24.5\\
215.1	25.0\\
216.0	25.5\\
216.4	25.7\\
217.0	25.9\\
218.0	26.0\\
219.0	26.6\\
220.0	26.8\\
221.0	27.3\\
222.3	27.5\\
223.3	27.8\\
224.0	27.9\\
225.1	28.0\\
226.0	28.0\\
227.0	27.8\\
228.1	28.0\\
229.0	28.2\\
230.0	28.0\\
231.3	28.1\\
232.0	28.2\\
232.9	28.6\\
233.0	28.7\\
234.0	28.6\\
237.1	28.9\\
239.1	29.2\\
240.0	29.5\\
241.0	29.5\\
241.9	29.7\\
242.0	29.7\\
243.0	28.6\\
243.9	28.4\\
244.3	28.5\\
245.3	28.6\\
246.1	28.6\\
247.1	28.5\\
248.1	28.7\\
249.2	28.7\\
250.1	28.6\\
251.0	28.7\\
251.1	28.7\\
251.6	28.5\\
252.0	28.4\\
252.1	28.4\\
253.1	28.6\\
254.2	28.6\\
255.1	28.6\\
256.1	28.5\\
257.0	28.6\\
257.1	28.6\\
258.0	28.3\\
258.1	28.3\\
259.1	28.4\\
260.1	28.6\\
261.2	28.6\\
262.2	28.8\\
263.2	29.0\\
264.0	29.1\\
264.1	29.1\\
265.0	28.6\\
265.1	28.6\\
266.0	28.7\\
266.1	28.7\\
267.0	28.2\\
267.1	28.2\\
268.0	28.5\\
268.1	28.5\\
269.0	28.2\\
269.1	28.2\\
270.0	28.5\\
270.1	28.5\\
271.1	28.4\\
272.1	28.2\\
273.2	28.2\\
274.1	28.2\\
275.2	28.3\\
276.2	28.3\\
277.1	28.3\\
278.0	28.0\\
278.1	27.9\\
279.1	27.9\\
280.1	28.1\\
281.1	28.1\\
282.0	27.8\\
282.2	27.8\\
283.1	27.9\\
284.3	27.9\\
285.3	27.9\\
286.1	27.9\\
287.1	27.8\\
288.4	28.0\\
289.0	28.1\\
289.2	28.0\\
290.0	27.9\\
290.2	27.9\\
291.3	27.9\\
292.0	27.9\\
292.2	27.9\\
293.0	27.7\\
293.1	27.7\\
294.0	28.1\\
294.2	28.1\\
295.1	28.1\\
296.0	27.9\\
296.1	27.9\\
297.0	28.1\\
297.2	28.0\\
298.3	27.9\\
299.3	27.7\\
300.3	27.7\\
301.1	27.6\\
302.1	27.8\\
304.3	27.7\\
305.0	27.7\\
305.3	27.8\\
306.0	27.9\\
306.2	27.9\\
307.1	27.8\\
308.1	27.8\\
309.3	27.9\\
310.1	28.0\\
311.0	28.0\\
311.1	28.0\\
312.0	28.5\\
312.3	28.6\\
313.3	28.8\\
314.1	29.0\\
315.1	29.1\\
316.0	29.5\\
316.3	29.5\\
317.0	29.5\\
317.3	29.7\\
318.0	29.9\\
318.3	30.0\\
319.3	30.1\\
320.0	30.2\\
320.4	30.3\\
321.0	30.5\\
321.2	30.5\\
322.0	30.5\\
322.4	30.6\\
323.1	30.8\\
325.0	31.0\\
325.2	31.2\\
326.0	32.0\\
326.2	32.1\\
327.0	32.2\\
327.2	32.5\\
328.0	33.5\\
328.1	33.4\\
329.0	32.5\\
329.2	32.4\\
330.0	32.3\\
330.6	32.3\\
331.1	32.3\\
332.0	32.2\\
332.4	32.3\\
333.1	32.4\\
334.0	32.5\\
334.8	32.5\\
335.0	32.5\\
335.4	32.3\\
336.0	32.1\\
336.8	32.0\\
337.3	31.9\\
338.1	31.8\\
339.1	31.8\\
340.1	31.7\\
341.3	31.7\\
342.1	31.7\\
343.0	31.8\\
343.9	32.2\\
344.0	32.2\\
345.0	32.5\\
346.0	32.9\\
346.4	33.1\\
347.1	33.5\\
348.0	33.8\\
349.0	34.6\\
350.1	35.1\\
351.0	35.4\\
352.0	35.0\\
353.0	34.0\\
354.0	33.8\\
355.0	34.2\\
356.0	34.7\\
357.0	34.9\\
358.0	35.4\\
360.0	35.7\\
361.0	35.7\\
362.0	35.9\\
363.0	35.9\\
364.0	35.5\\
365.0	35.5\\
367.0	35.2\\
369.0	34.2\\
370.0	33.2\\
371.4	32.1\\
372.0	31.6\\
372.2	31.4\\
373.0	30.2\\
374.0	29.7\\
375.0	29.8\\
376.0	29.5\\
377.0	29.5\\
378.0	29.6\\
379.0	29.6\\
380.0	30.0\\
381.0	30.1\\
382.0	30.5\\
383.0	30.5\\
384.0	30.7\\
385.0	30.5\\
387.0	30.4\\
388.1	30.0\\
389.0	29.7\\
390.3	29.5\\
391.0	29.4\\
392.3	29.6\\
393.0	29.7\\
394.0	29.5\\
395.0	29.6\\
395.5	29.8\\
396.1	30.0\\
397.3	30.2\\
397.5	30.3\\
397.6	30.3\\
398.2	30.4\\
399.2	30.5\\
401.1	30.8\\
402.0	30.7\\
402.1	30.7\\
403.1	31.0\\
405.1	31.0\\
406.2	31.3\\
407.2	31.5\\
408.3	31.6\\
409.0	31.7\\
409.1	31.7\\
410.0	31.4\\
410.1	31.4\\
411.1	31.6\\
412.1	31.4\\
413.0	31.2\\
413.1	31.2\\
415.1	31.3\\
416.2	31.1\\
417.0	31.0\\
417.1	31.0\\
418.0	31.2\\
418.1	31.2\\
420.1	31.1\\
421.3	31.3\\
422.1	31.4\\
424.0	31.4\\
424.1	31.4\\
425.0	31.7\\
425.1	31.7\\
426.1	31.7\\
427.1	31.7\\
428.1	31.6\\
429.4	31.7\\
430.2	31.8\\
431.1	31.9\\
432.0	31.7\\
432.1	31.7\\
433.3	31.9\\
434.1	32.1\\
435.1	32.2\\
436.0	32.4\\
436.1	32.4\\
437.1	32.3\\
438.2	32.3\\
439.4	32.2\\
441.0	32.1\\
441.2	32.2\\
442.1	32.5\\
443.2	32.6\\
444.0	32.7\\
444.1	32.7\\
445.1	32.2\\
447.4	31.6\\
448.0	31.5\\
448.2	31.5\\
449.1	31.5\\
450.0	31.7\\
450.1	31.7\\
451.0	31.4\\
451.1	31.4\\
452.0	31.7\\
452.1	31.7\\
453.0	31.4\\
453.2	31.4\\
454.0	31.4\\
454.1	31.4\\
455.0	30.9\\
455.2	30.9\\
456.1	30.7\\
457.0	30.5\\
457.2	30.5\\
458.1	30.5\\
459.1	30.5\\
460.3	30.2\\
461.0	30.1\\
461.3	30.1\\
462.6	30.2\\
463.3	30.2\\
464.1	30.2\\
465.1	30.0\\
466.0	30.0\\
466.5	29.8\\
467.0	29.6\\
467.2	29.7\\
468.0	29.8\\
468.2	29.7\\
469.0	29.6\\
469.4	29.6\\
470.0	29.7\\
470.3	29.6\\
471.0	29.5\\
471.3	29.5\\
472.0	29.5\\
472.2	29.5\\
473.0	29.3\\
473.1	29.2\\
474.0	27.8\\
474.1	27.8\\
475.0	27.8\\
475.2	27.7\\
476.0	27.5\\
476.5	27.4\\
477.0	27.3\\
477.3	27.4\\
478.0	27.4\\
478.3	27.4\\
479.0	27.3\\
479.4	27.4\\
480.1	27.4\\
481.1	27.5\\
482.1	27.6\\
485.0	27.6\\
485.9	27.8\\
486.0	27.8\\
486.4	27.7\\
487.0	27.6\\
487.3	27.7\\
489.0	28.0\\
489.7	27.9\\
490.0	27.8\\
490.4	27.9\\
491.0	28.0\\
492.0	27.8\\
492.9	28.0\\
493.0	28.0\\
494.1	28.0\\
495.0	28.2\\
495.9	28.0\\
497.0	27.8\\
498.0	27.3\\
499.0	26.6\\
500.0	26.3\\
500.8	25.6\\
501.0	25.4\\
501.4	25.2\\
502.0	24.9\\
503.0	24.2\\
504.1	22.9\\
505.0	22.0\\
506.0	21.8\\
507.0	21.3\\
508.1	19.9\\
509.1	18.4\\
510.0	16.6\\
510.2	16.4\\
511.0	15.5\\
512.0	14.7\\
513.0	13.0\\
514.0	10.0\\
515.0	7.9\\
515.9	7.8\\
516.0	7.7\\
517.0	6.2\\
519.4	1.9\\
519.6	1.6\\
520.0	0.9\\
521.0	0.0\\
528.1	0.0\\
528.9	0.1\\
529.0	0.1\\
530.0	2.0\\
531.1	4.8\\
531.3	5.3\\
532.2	7.5\\
533.0	9.3\\
534.0	11.2\\
535.0	12.4\\
535.9	12.6\\
536.0	12.6\\
537.0	12.0\\
538.1	11.0\\
538.9	10.1\\
539.0	10.0\\
540.3	6.8\\
541.0	5.0\\
541.9	3.2\\
542.0	3.1\\
542.9	3.3\\
543.0	3.4\\
544.1	5.2\\
545.6	8.1\\
546.3	9.5\\
547.0	10.9\\
547.9	12.0\\
548.0	12.1\\
548.9	12.7\\
549.0	12.7\\
549.9	12.6\\
550.0	12.6\\
550.6	11.9\\
550.7	11.9\\
551.1	11.4\\
552.3	9.2\\
553.2	7.4\\
554.1	5.3\\
555.0	3.5\\
555.1	3.3\\
556.0	2.6\\
556.1	2.5\\
557.0	1.2\\
557.1	1.1\\
558.0	0.6\\
558.1	0.5\\
559.0	0.8\\
559.1	0.9\\
560.2	3.3\\
561.2	5.6\\
562.4	8.3\\
562.6	8.7\\
563.1	9.8\\
564.1	11.6\\
565.0	12.3\\
565.1	12.4\\
566.0	12.5\\
566.1	12.5\\
567.0	12.1\\
567.1	12.1\\
568.0	11.0\\
568.1	10.8\\
569.1	8.9\\
569.2	8.7\\
569.4	8.3\\
570.1	6.7\\
571.0	4.3\\
571.2	3.9\\
572.0	2.3\\
572.1	2.2\\
573.0	1.4\\
573.1	1.3\\
574.0	1.1\\
574.1	1.2\\
575.0	3.6\\
575.3	4.3\\
576.1	6.1\\
577.4	8.6\\
578.1	10.0\\
579.4	12.3\\
580.0	13.3\\
580.1	13.4\\
581.0	13.3\\
581.1	13.2\\
582.1	11.5\\
583.4	8.7\\
584.2	6.9\\
586.1	2.9\\
587.0	1.2\\
587.1	1.1\\
588.0	0.0\\
588.1	0.0\\
595.1	0.0\\
596.0	0.1\\
596.1	0.3\\
598.0	5.6\\
599.0	8.5\\
599.2	8.9\\
600.1	10.8\\
};
\addlegendentry{\scriptsize EPA US06}

\addplot [color=mycolor1, line width=1.3pt]
  table[row sep=crcr]{%
28.0	0.0\\
33.1	0.0\\
34.0	0.1\\
35.4	0.2\\
36.0	0.3\\
36.9	0.5\\
37.1	0.8\\
37.9	1.8\\
38.9	3.9\\
39.9	5.6\\
40.9	6.9\\
41.9	6.8\\
43.3	8.1\\
43.9	8.7\\
45.0	9.4\\
46.5	10.2\\
47.0	10.4\\
48.0	10.9\\
48.9	11.4\\
49.9	11.8\\
50.9	11.8\\
51.6	11.7\\
52.1	11.5\\
52.9	11.3\\
53.9	10.8\\
54.9	10.5\\
55.9	10.3\\
56.9	10.3\\
57.9	10.5\\
59.3	10.4\\
59.8	10.3\\
59.9	10.3\\
60.9	9.5\\
62.0	8.2\\
63.1	6.8\\
64.0	5.5\\
64.3	5.2\\
65.1	3.9\\
66.1	2.4\\
66.9	1.3\\
67.0	1.2\\
67.9	0.3\\
68.0	0.2\\
69.0	0.0\\
75.1	0.0\\
75.9	0.1\\
76.0	0.1\\
76.9	0.9\\
77.0	1.0\\
77.9	2.9\\
78.1	3.2\\
78.9	4.2\\
79.4	4.7\\
80.0	5.3\\
81.1	6.6\\
81.9	7.6\\
82.1	7.8\\
82.9	8.6\\
83.2	8.8\\
84.0	9.5\\
85.6	11.0\\
86.0	11.3\\
87.0	12.1\\
87.9	12.5\\
88.5	12.8\\
89.2	13.1\\
90.4	13.7\\
91.5	14.2\\
92.6	14.6\\
93.1	14.8\\
94.0	15.0\\
95.0	15.1\\
97.0	15.0\\
98.2	14.8\\
99.1	14.7\\
100.1	14.6\\
102.0	14.6\\
103.0	14.4\\
104.2	14.1\\
104.9	14.0\\
105.8	14.0\\
106.1	14.0\\
107.7	13.9\\
108.4	13.9\\
109.4	13.9\\
110.1	14.0\\
111.0	14.0\\
112.0	14.2\\
113.0	14.6\\
114.1	15.1\\
115.3	15.7\\
116.2	16.2\\
117.6	16.9\\
118.3	17.2\\
119.1	17.6\\
120.1	18.1\\
121.0	18.5\\
121.3	18.6\\
121.9	18.8\\
123.0	18.9\\
124.0	18.9\\
124.9	18.9\\
126.0	18.4\\
127.2	17.9\\
127.6	17.7\\
128.0	17.5\\
129.3	17.1\\
129.9	16.9\\
131.1	16.4\\
138.8	13.0\\
139.7	12.6\\
139.9	12.5\\
140.9	11.9\\
142.1	11.1\\
142.9	10.6\\
144.0	10.1\\
145.3	9.7\\
145.8	9.6\\
145.9	9.5\\
146.9	8.3\\
148.0	6.8\\
149.0	5.5\\
150.4	3.8\\
151.6	2.3\\
151.8	2.1\\
152.0	1.9\\
152.9	1.1\\
153.0	1.0\\
154.0	0.6\\
154.6	0.4\\
155.1	0.3\\
155.8	0.1\\
156.0	0.1\\
157.2	0.0\\
162.9	0.0\\
163.0	0.0\\
163.9	0.6\\
164.0	0.7\\
165.1	2.6\\
166.4	5.0\\
166.5	5.1\\
166.6	5.3\\
166.8	5.7\\
167.9	7.6\\
168.1	7.9\\
169.0	9.2\\
170.1	10.5\\
170.9	11.4\\
171.0	11.5\\
171.9	11.5\\
172.0	11.4\\
173.9	7.6\\
174.0	7.5\\
174.9	7.0\\
175.1	7.0\\
187.9	7.0\\
188.1	7.2\\
189.1	8.6\\
190.3	10.5\\
190.5	10.7\\
191.9	12.8\\
192.1	13.2\\
192.9	14.4\\
193.6	15.2\\
193.9	15.6\\
194.2	15.6\\
195.4	15.5\\
196.3	15.4\\
197.0	15.4\\
197.9	15.4\\
198.9	15.2\\
200.0	15.3\\
201.1	15.2\\
202.3	15.1\\
203.0	15.1\\
204.9	15.2\\
207.8	15.0\\
208.3	15.0\\
208.9	14.9\\
209.9	14.7\\
211.0	14.6\\
211.9	14.5\\
212.9	14.1\\
213.9	14.7\\
215.1	15.1\\
215.9	15.4\\
217.0	15.6\\
217.9	15.7\\
218.9	16.0\\
220.0	16.2\\
220.9	16.4\\
222.4	16.6\\
223.4	16.7\\
224.0	16.8\\
225.7	16.8\\
226.0	16.8\\
226.9	16.7\\
228.4	16.9\\
228.9	16.9\\
230.0	16.8\\
231.0	16.9\\
232.2	17.1\\
232.9	17.2\\
234.0	17.2\\
236.6	17.3\\
237.3	17.4\\
238.1	17.5\\
239.6	17.7\\
240.1	17.7\\
241.0	17.7\\
242.0	17.6\\
242.9	17.2\\
243.0	17.1\\
243.9	17.1\\
245.1	17.2\\
246.9	17.2\\
247.3	17.2\\
248.0	17.2\\
249.3	17.2\\
250.3	17.2\\
252.0	17.1\\
253.1	17.2\\
256.0	17.1\\
257.7	17.0\\
258.0	17.0\\
259.1	17.1\\
260.1	17.1\\
261.4	17.2\\
262.1	17.2\\
263.9	17.4\\
264.2	17.4\\
267.3	17.0\\
268.3	17.0\\
269.2	17.0\\
270.7	17.0\\
271.0	17.0\\
272.1	16.9\\
273.3	16.9\\
274.2	16.9\\
276.4	17.0\\
277.0	17.0\\
277.9	16.8\\
278.0	16.6\\
280.6	11.4\\
280.9	10.8\\
281.9	8.8\\
282.4	7.9\\
282.9	7.0\\
283.1	7.0\\
297.9	7.0\\
298.9	7.7\\
299.1	8.0\\
304.0	15.4\\
304.8	16.5\\
304.9	16.6\\
306.0	16.7\\
307.8	16.7\\
308.2	16.7\\
309.3	16.8\\
310.0	16.8\\
310.9	16.8\\
312.1	17.1\\
313.1	17.3\\
314.1	17.4\\
314.9	17.5\\
315.9	17.7\\
316.9	17.7\\
317.9	17.9\\
320.0	18.1\\
321.0	18.3\\
322.0	18.3\\
323.3	18.5\\
324.9	18.6\\
325.0	18.6\\
325.9	19.2\\
326.0	19.2\\
326.9	19.3\\
326.9	19.4\\
327.9	20.0\\
328.0	20.0\\
328.4	19.8\\
328.8	19.5\\
329.0	19.5\\
330.0	19.4\\
331.2	19.4\\
332.0	19.3\\
333.2	19.5\\
334.0	19.5\\
335.0	19.5\\
335.6	19.3\\
336.0	19.3\\
337.4	19.1\\
338.1	19.1\\
339.2	19.1\\
340.1	19.0\\
341.5	19.0\\
342.1	19.1\\
343.0	19.1\\
344.0	19.4\\
345.0	19.5\\
346.2	19.9\\
347.0	20.1\\
348.0	20.4\\
349.0	20.8\\
350.0	21.1\\
350.9	21.2\\
351.1	21.2\\
351.9	20.9\\
352.5	20.6\\
352.9	20.4\\
353.1	20.4\\
353.9	20.3\\
354.2	20.4\\
355.2	20.6\\
356.0	20.8\\
357.0	21.0\\
358.0	21.2\\
359.7	21.4\\
360.1	21.4\\
361.1	21.5\\
362.0	21.5\\
363.0	21.5\\
364.1	21.3\\
365.0	21.3\\
366.3	21.1\\
367.0	21.0\\
368.2	20.6\\
368.9	20.4\\
370.1	19.7\\
371.0	19.3\\
372.0	18.7\\
372.9	18.1\\
373.6	17.9\\
373.9	17.8\\
374.2	17.6\\
374.9	17.0\\
375.5	15.8\\
379.5	7.8\\
379.7	7.4\\
379.9	7.0\\
394.9	7.0\\
395.8	7.2\\
395.9	7.2\\
397.8	10.1\\
398.0	10.4\\
399.5	12.7\\
399.7	13.0\\
402.8	17.7\\
402.9	17.8\\
403.8	18.6\\
403.9	18.6\\
405.0	18.7\\
406.1	18.8\\
407.8	19.0\\
408.0	19.0\\
409.5	18.9\\
410.1	18.9\\
411.0	18.9\\
412.4	18.8\\
413.0	18.7\\
414.1	18.8\\
415.1	18.7\\
416.0	18.6\\
417.3	18.6\\
419.2	18.7\\
420.1	18.7\\
421.1	18.7\\
422.1	18.8\\
423.3	18.8\\
424.0	18.8\\
425.2	19.0\\
426.1	19.0\\
429.0	19.0\\
430.6	19.1\\
431.0	19.1\\
432.1	19.0\\
433.0	19.1\\
434.3	19.2\\
435.2	19.3\\
436.1	19.4\\
437.2	19.4\\
441.0	19.3\\
443.0	19.5\\
443.9	19.6\\
444.7	19.5\\
445.2	19.4\\
446.2	19.2\\
447.6	19.0\\
448.0	18.9\\
449.0	18.9\\
449.9	19.0\\
450.9	18.9\\
451.9	19.0\\
453.5	18.9\\
453.9	18.8\\
455.1	18.6\\
456.4	18.4\\
457.0	18.3\\
459.1	18.3\\
460.4	18.1\\
461.0	18.1\\
462.6	18.1\\
463.4	18.1\\
464.0	18.1\\
465.6	18.0\\
466.0	18.0\\
466.9	17.8\\
468.0	17.8\\
469.0	17.8\\
469.9	17.8\\
470.9	17.7\\
471.9	17.7\\
472.9	17.6\\
473.9	16.8\\
475.2	16.6\\
475.8	16.5\\
475.9	16.5\\
476.7	15.6\\
476.8	15.5\\
476.9	15.3\\
480.4	8.2\\
480.6	7.9\\
480.8	7.5\\
480.9	7.3\\
481.9	7.0\\
495.9	7.0\\
496.0	7.0\\
496.9	7.5\\
497.0	7.6\\
497.9	8.9\\
500.0	12.1\\
500.3	12.5\\
501.2	13.9\\
501.4	14.2\\
501.9	14.9\\
502.0	14.9\\
503.0	14.4\\
504.1	13.7\\
504.9	13.2\\
505.0	13.2\\
506.0	13.0\\
506.9	12.7\\
507.1	12.6\\
508.0	11.9\\
509.0	11.0\\
509.9	10.0\\
510.1	9.8\\
510.9	9.3\\
511.2	9.1\\
511.9	8.8\\
512.1	8.6\\
512.3	8.3\\
512.9	7.7\\
513.2	7.1\\
513.9	5.9\\
514.1	5.6\\
514.9	4.8\\
515.0	4.7\\
515.9	4.5\\
516.0	4.5\\
517.0	3.5\\
518.7	1.7\\
519.1	1.3\\
519.4	1.0\\
519.9	0.5\\
520.1	0.4\\
520.9	0.0\\
521.1	0.0\\
527.9	0.0\\
528.3	0.1\\
528.9	0.3\\
529.3	0.8\\
529.5	1.0\\
530.0	1.6\\
530.5	2.4\\
531.2	3.5\\
531.4	3.8\\
532.0	4.6\\
532.9	5.8\\
533.5	6.5\\
533.9	6.9\\
534.9	7.5\\
535.9	7.5\\
536.9	7.1\\
537.9	6.5\\
538.9	5.6\\
539.1	5.3\\
540.4	3.4\\
540.9	2.7\\
541.9	1.9\\
542.9	2.3\\
544.0	3.5\\
546.2	6.0\\
546.8	6.6\\
546.9	6.7\\
547.9	7.4\\
548.9	7.6\\
549.9	7.3\\
550.9	6.5\\
552.1	5.1\\
553.3	3.6\\
553.9	2.8\\
554.9	1.8\\
556.1	1.1\\
556.9	0.5\\
557.9	0.4\\
558.9	1.1\\
560.4	3.1\\
561.5	4.6\\
562.3	5.6\\
562.9	6.3\\
563.9	7.2\\
564.9	7.5\\
565.9	7.4\\
566.4	7.2\\
566.9	6.9\\
567.9	6.0\\
569.0	4.6\\
570.3	2.7\\
570.9	1.9\\
571.8	1.1\\
571.9	1.1\\
572.8	0.7\\
572.9	0.7\\
573.8	1.4\\
573.9	1.5\\
574.8	1.6\\
574.9	1.7\\
575.1	2.0\\
576.1	3.2\\
576.9	4.1\\
577.3	4.6\\
578.1	5.7\\
579.1	6.8\\
579.5	7.1\\
579.9	7.5\\
580.0	7.6\\
580.8	8.0\\
580.9	8.0\\
581.8	7.5\\
582.0	7.4\\
583.0	6.2\\
584.0	4.9\\
584.2	4.6\\
585.0	3.6\\
586.0	2.3\\
587.0	1.1\\
587.9	0.3\\
588.0	0.3\\
588.9	0.0\\
589.1	0.0\\
595.2	0.0\\
595.9	0.0\\
596.0	0.1\\
596.3	0.4\\
596.9	1.1\\
597.1	1.4\\
598.4	3.6\\
599.0	4.6\\
599.2	4.9\\
600.0	6.0\\
600.1	6.2\\
};
\addlegendentry{\scriptsize Modified EPA US06}

\addplot[area legend, draw=none, fill=white!10!black, fill opacity=0.2]
table[row sep=crcr] {%
x	y\\
177	0\\
177	40\\
192	40\\
192	0\\
}--cycle;
\addlegendentry{\scriptsize U-Turn}
\addplot[area legend, draw=none, fill=white!10!black, fill opacity=0.2]
table[row sep=crcr] {%
x	y\\
286	0\\
286	40\\
301	40\\
301	0\\
}--cycle;
\addplot[area legend, draw=none, fill=white!10!black, fill opacity=0.2]
table[row sep=crcr] {%
x	y\\
385	0\\
385	40\\
400	40\\
400	0\\
}--cycle;
\addplot[area legend, draw=none, fill=white!10!black, fill opacity=0.2]
table[row sep=crcr] {%
x	y\\
484	0\\
484	40\\
499	40\\
499	0\\
}--cycle;

\end{axis}
\end{tikzpicture}%

%% file: Images/EPA_UDDS_modified.tex
%
%
\definecolor{mycolor1}{rgb}{0.00000,0.44700,0.74100}
\begin{tikzpicture}

\begin{axis}[%
width=6.5cm,
height=3.5cm,
at={(0.0in,0.0in)},
scale only axis,
xmin=0.0,
xmax=1200.0,
xlabel style={font=\color{white!10!black} \scriptsize},
xlabel={Time [s]},
ticklabel style = {font = \scriptsize},
ymin=0.0,
ymax=36.0,
ylabel style={font=\color{white!10!black} \scriptsize},
ylabel={Drive Cycle [m/s]},
axis background/.style={fill=white},
xmajorgrids,
ymajorgrids,
legend style={anchor=south east, legend cell align=left, align=left, draw=white!12!black}
]

\addplot [color=mycolor1, dashed, line width=1.3pt]
  table[row sep=crcr]{%
0.0	0.0\\
48.0	0.0\\
49.2	1.6\\
49.4	1.9\\
50.1	2.8\\
51.4	4.4\\
52.6	5.9\\
52.8	6.1\\
53.1	6.5\\
54.0	7.6\\
55.0	7.7\\
56.0	8.1\\
57.0	9.3\\
58.0	9.7\\
59.0	10.0\\
60.0	10.1\\
61.1	9.9\\
63.1	9.3\\
64.3	9.0\\
64.9	8.9\\
65.0	8.8\\
66.0	7.6\\
66.9	6.7\\
67.0	6.7\\
68.0	6.7\\
70.1	7.0\\
71.0	7.2\\
72.0	7.7\\
74.0	9.4\\
75.0	10.1\\
76.0	10.2\\
77.2	10.1\\
77.9	10.1\\
78.0	10.1\\
79.0	9.5\\
80.0	8.5\\
81.0	7.6\\
81.9	7.1\\
82.0	7.1\\
82.9	7.1\\
83.0	7.1\\
84.1	8.0\\
85.0	8.9\\
86.0	9.7\\
86.2	9.8\\
87.0	10.4\\
87.9	10.8\\
88.0	10.8\\
89.1	11.0\\
90.0	11.1\\
91.0	11.2\\
92.0	11.0\\
93.0	11.0\\
94.2	11.1\\
95.1	11.1\\
97.3	11.0\\
98.1	11.0\\
100.0	11.4\\
100.2	11.5\\
101.1	11.5\\
102.1	11.3\\
103.0	11.1\\
103.1	11.1\\
104.1	11.2\\
105.1	11.4\\
106.1	11.6\\
107.1	11.6\\
108.0	11.5\\
108.1	11.5\\
109.2	11.7\\
110.2	12.0\\
111.1	12.3\\
112.0	12.8\\
112.3	12.9\\
113.1	13.1\\
114.1	13.3\\
117.1	13.7\\
118.1	13.7\\
119.2	13.6\\
121.1	13.5\\
122.1	13.6\\
123.0	13.8\\
123.1	13.8\\
124.2	13.6\\
125.2	13.3\\
126.0	13.2\\
126.1	13.2\\
127.1	13.3\\
128.3	13.6\\
129.1	13.7\\
130.1	13.8\\
131.1	13.9\\
132.1	13.8\\
133.3	13.5\\
134.0	13.3\\
134.1	13.3\\
135.1	13.4\\
136.1	13.5\\
138.5	14.1\\
139.2	14.2\\
140.1	14.4\\
141.0	14.5\\
141.2	14.5\\
142.1	14.4\\
143.0	14.2\\
143.1	14.1\\
144.1	12.7\\
145.0	11.4\\
152.0	1.0\\
152.2	0.8\\
153.0	0.0\\
153.1	0.0\\
191.0	0.0\\
191.2	0.3\\
197.0	8.8\\
198.0	9.9\\
199.0	10.9\\
200.0	11.5\\
201.0	11.8\\
202.1	11.5\\
203.1	11.2\\
204.0	11.0\\
205.2	11.2\\
207.1	11.4\\
208.0	11.5\\
209.0	12.2\\
209.9	11.9\\
210.0	11.8\\
211.0	10.7\\
211.9	10.2\\
212.0	10.1\\
213.0	8.7\\
214.0	7.9\\
215.0	7.7\\
216.0	8.1\\
216.5	8.2\\
217.0	8.3\\
218.0	9.0\\
219.6	10.6\\
220.0	11.0\\
220.2	11.2\\
221.0	12.2\\
222.1	13.8\\
223.0	15.0\\
224.0	16.2\\
224.3	16.3\\
225.0	16.7\\
225.9	17.5\\
226.0	17.6\\
227.0	18.1\\
228.0	18.8\\
229.1	19.5\\
230.0	20.2\\
231.1	20.6\\
231.5	20.7\\
232.2	21.0\\
233.0	21.2\\
234.1	21.2\\
235.2	21.1\\
236.2	21.1\\
237.1	21.0\\
241.0	21.0\\
243.0	21.2\\
244.3	21.5\\
246.0	22.0\\
246.7	22.1\\
247.2	22.2\\
248.2	22.4\\
248.3	22.4\\
249.1	22.6\\
250.3	22.9\\
251.1	23.0\\
252.1	23.4\\
253.3	23.9\\
254.1	24.2\\
255.2	24.4\\
256.1	24.5\\
257.1	24.6\\
258.1	24.5\\
259.1	24.4\\
260.1	24.4\\
261.3	24.5\\
262.3	24.7\\
263.1	24.8\\
264.1	24.9\\
265.1	25.1\\
266.3	25.2\\
267.1	25.3\\
268.3	25.3\\
269.1	25.3\\
270.1	25.3\\
275.3	25.2\\
276.1	25.2\\
278.1	24.9\\
279.1	24.6\\
280.2	24.4\\
281.1	24.2\\
282.3	24.1\\
283.1	24.0\\
284.1	24.0\\
285.1	24.1\\
287.2	24.2\\
288.1	24.2\\
289.2	24.0\\
292.4	23.4\\
293.0	23.3\\
293.1	23.3\\
294.0	23.4\\
294.1	23.4\\
295.1	23.2\\
296.3	23.2\\
298.1	23.0\\
299.2	23.1\\
300.3	23.2\\
301.3	23.3\\
302.3	23.5\\
305.0	24.1\\
305.3	24.2\\
306.1	24.6\\
307.0	24.8\\
307.3	24.8\\
308.1	24.9\\
309.0	25.0\\
309.3	25.0\\
310.1	25.0\\
311.0	24.9\\
311.4	24.8\\
312.3	24.6\\
313.2	24.3\\
314.2	23.9\\
315.2	23.4\\
316.0	23.0\\
316.1	23.0\\
318.0	23.0\\
318.3	23.0\\
319.0	22.8\\
319.3	22.7\\
320.0	22.4\\
320.2	22.4\\
321.1	22.4\\
322.1	22.4\\
323.1	22.3\\
324.0	22.2\\
324.5	22.2\\
325.2	22.1\\
327.0	22.1\\
327.4	22.1\\
328.2	21.9\\
330.0	21.5\\
330.8	21.2\\
331.1	21.1\\
333.5	19.9\\
335.0	19.0\\
336.5	18.3\\
337.0	18.0\\
337.2	17.9\\
338.1	17.2\\
339.1	16.5\\
340.0	15.7\\
340.6	15.4\\
341.3	14.9\\
341.6	14.8\\
342.1	14.5\\
343.1	14.0\\
344.0	13.7\\
345.1	13.6\\
346.0	13.4\\
346.3	13.3\\
347.0	13.0\\
348.0	12.3\\
348.3	11.9\\
349.0	11.1\\
349.7	10.0\\
349.9	9.8\\
350.0	9.6\\
350.3	9.4\\
351.0	9.0\\
352.0	8.5\\
353.0	8.3\\
354.8	7.1\\
355.0	6.9\\
356.0	5.6\\
357.0	4.8\\
357.9	3.7\\
358.0	3.6\\
360.0	0.6\\
361.0	0.0\\
374.0	0.0\\
374.9	0.4\\
375.0	0.5\\
376.0	1.9\\
379.2	6.6\\
380.1	7.9\\
381.1	9.1\\
382.0	10.1\\
383.1	10.7\\
384.2	11.4\\
385.4	12.2\\
386.0	12.6\\
386.9	13.3\\
387.0	13.4\\
389.0	14.1\\
390.1	14.4\\
391.2	14.7\\
392.2	15.1\\
392.9	15.4\\
393.0	15.4\\
394.0	15.5\\
395.0	15.6\\
396.1	15.5\\
397.0	15.4\\
398.0	15.5\\
399.0	15.9\\
400.0	16.1\\
405.3	16.1\\
406.1	16.1\\
407.1	16.3\\
408.1	16.3\\
409.1	16.3\\
410.1	16.1\\
411.5	15.5\\
412.1	15.2\\
413.0	15.0\\
413.1	14.9\\
414.1	14.0\\
415.1	12.9\\
416.1	11.4\\
417.5	9.7\\
418.5	8.5\\
419.1	7.8\\
420.0	6.6\\
420.2	6.3\\
421.1	5.3\\
421.8	4.3\\
424.1	0.9\\
425.0	0.1\\
425.1	0.0\\
430.0	0.0\\
430.1	0.1\\
431.1	1.2\\
432.4	3.2\\
432.5	3.3\\
432.7	3.6\\
432.9	3.9\\
433.7	5.1\\
434.9	6.9\\
435.1	7.1\\
436.6	9.4\\
436.8	9.6\\
437.1	10.1\\
438.0	11.1\\
438.1	11.2\\
439.0	11.4\\
439.2	11.6\\
440.1	12.3\\
441.0	12.9\\
441.2	13.0\\
442.0	13.4\\
442.1	13.4\\
443.1	13.5\\
444.1	13.4\\
445.2	13.2\\
446.3	13.0\\
447.1	12.9\\
448.0	12.5\\
448.1	12.4\\
449.3	10.8\\
455.5	1.6\\
456.0	0.9\\
456.1	0.8\\
457.0	0.0\\
457.1	0.0\\
475.0	0.0\\
475.1	0.1\\
483.0	11.8\\
483.2	11.9\\
484.1	12.5\\
485.0	13.0\\
485.2	13.2\\
486.0	14.1\\
486.4	14.3\\
487.0	14.7\\
487.3	14.8\\
488.0	15.0\\
488.8	15.4\\
489.0	15.6\\
489.3	15.6\\
490.1	15.7\\
492.0	16.1\\
492.4	16.1\\
493.1	16.1\\
495.0	16.2\\
495.9	16.1\\
496.2	16.1\\
497.0	16.0\\
497.5	16.0\\
498.0	16.1\\
499.0	16.1\\
500.1	15.9\\
502.2	15.8\\
503.2	15.7\\
508.1	15.7\\
509.0	15.6\\
511.0	15.7\\
512.0	15.9\\
513.2	15.7\\
514.1	15.6\\
516.1	15.6\\
518.2	15.5\\
519.0	15.4\\
520.0	15.0\\
521.0	14.3\\
522.1	13.4\\
523.0	12.5\\
524.1	11.3\\
525.0	10.0\\
526.0	8.8\\
528.1	5.8\\
529.3	4.2\\
530.5	2.5\\
531.0	1.8\\
532.0	0.4\\
533.0	0.0\\
538.0	0.0\\
539.0	0.6\\
539.2	0.8\\
540.1	1.7\\
541.0	2.5\\
542.0	2.9\\
542.2	3.1\\
543.0	3.8\\
544.0	4.3\\
545.0	4.7\\
545.3	4.9\\
546.0	5.3\\
547.0	6.3\\
547.2	6.5\\
548.1	7.2\\
549.0	7.9\\
550.0	8.5\\
551.0	9.0\\
552.3	9.5\\
554.1	10.3\\
555.1	10.7\\
556.0	11.0\\
556.2	11.0\\
556.3	11.0\\
557.1	11.1\\
558.3	11.2\\
563.1	11.2\\
564.1	11.4\\
566.1	11.6\\
568.1	11.3\\
569.1	11.2\\
571.0	11.2\\
571.1	11.2\\
572.1	10.9\\
573.0	10.4\\
573.1	10.3\\
574.4	8.3\\
580.0	0.1\\
580.1	0.0\\
596.0	0.0\\
596.1	0.1\\
596.3	0.4\\
599.4	4.9\\
600.0	5.8\\
600.1	5.9\\
601.2	6.6\\
602.0	7.1\\
602.2	7.2\\
603.0	7.6\\
603.1	7.6\\
605.0	7.6\\
605.2	7.6\\
606.1	7.8\\
607.1	7.9\\
608.1	7.9\\
609.1	7.8\\
610.0	7.6\\
610.3	7.6\\
611.1	7.5\\
612.0	7.4\\
612.2	7.5\\
613.1	7.6\\
614.1	7.6\\
615.1	7.6\\
616.1	7.4\\
617.2	7.4\\
618.2	7.4\\
619.1	7.4\\
620.1	7.6\\
621.1	7.9\\
622.1	8.3\\
623.1	8.6\\
624.1	9.1\\
625.0	9.4\\
625.2	9.4\\
627.0	9.5\\
627.4	9.5\\
630.0	10.0\\
630.6	10.0\\
631.2	10.1\\
633.1	10.1\\
634.0	10.1\\
634.2	10.2\\
635.0	10.6\\
635.5	10.9\\
636.0	11.2\\
636.4	11.4\\
637.0	11.6\\
637.5	11.7\\
639.0	12.1\\
639.2	12.0\\
640.0	11.7\\
640.2	11.4\\
647.2	1.1\\
648.0	0.0\\
648.1	0.0\\
673.0	0.0\\
674.0	0.9\\
675.0	2.0\\
675.9	3.4\\
676.0	3.5\\
677.0	4.6\\
677.2	4.8\\
678.0	5.6\\
679.1	6.3\\
680.0	6.9\\
681.2	8.0\\
682.0	8.8\\
683.2	9.5\\
684.2	10.0\\
685.3	10.6\\
686.0	11.0\\
687.0	11.3\\
688.2	11.5\\
689.0	11.6\\
691.2	11.7\\
692.1	11.7\\
693.1	11.8\\
694.2	11.8\\
695.0	11.8\\
697.0	11.4\\
698.1	10.4\\
699.0	9.5\\
699.6	8.8\\
700.0	8.2\\
701.1	7.2\\
702.0	6.5\\
702.2	6.2\\
703.0	5.1\\
703.9	4.0\\
704.1	3.7\\
705.0	2.6\\
706.0	1.5\\
706.2	1.4\\
707.0	0.9\\
707.9	0.1\\
708.0	0.0\\
721.0	0.0\\
721.1	0.0\\
722.1	0.7\\
723.0	1.4\\
723.1	1.5\\
724.0	1.9\\
724.1	2.0\\
725.0	2.9\\
725.2	3.1\\
726.0	4.0\\
726.2	4.2\\
727.7	5.7\\
728.0	6.0\\
728.1	6.1\\
729.0	6.5\\
729.2	6.6\\
730.0	7.3\\
730.1	7.3\\
731.1	7.5\\
732.1	7.4\\
733.0	7.4\\
733.1	7.4\\
734.1	8.2\\
735.2	8.6\\
736.1	9.0\\
736.4	9.2\\
737.1	9.6\\
738.0	10.0\\
738.1	10.1\\
739.1	10.0\\
740.0	9.9\\
740.1	9.9\\
742.0	10.4\\
742.2	10.4\\
743.0	10.5\\
743.1	10.5\\
743.4	10.3\\
744.2	10.0\\
745.1	9.6\\
746.0	9.2\\
746.1	9.1\\
747.1	8.0\\
750.2	3.8\\
751.0	2.8\\
751.1	2.7\\
752.1	2.0\\
753.1	1.3\\
754.0	1.0\\
754.2	0.8\\
755.0	0.3\\
755.1	0.2\\
756.0	0.2\\
756.1	0.3\\
756.3	0.5\\
757.1	1.5\\
757.8	2.6\\
758.0	2.9\\
758.2	3.1\\
759.2	4.5\\
759.9	5.4\\
760.0	5.5\\
760.1	5.6\\
761.0	6.2\\
761.4	6.6\\
763.1	8.1\\
764.2	8.9\\
765.0	9.6\\
765.4	9.9\\
766.1	10.4\\
767.1	11.0\\
769.0	11.8\\
769.3	11.9\\
770.2	12.2\\
771.1	12.3\\
772.2	12.5\\
773.2	12.7\\
774.1	12.8\\
775.1	12.8\\
776.3	12.6\\
777.2	12.6\\
778.0	12.5\\
778.6	12.4\\
779.1	12.3\\
780.0	12.0\\
780.3	11.8\\
781.0	11.4\\
781.2	11.2\\
783.1	9.5\\
784.2	8.3\\
784.4	8.1\\
785.0	7.4\\
785.3	7.2\\
786.0	6.7\\
786.3	6.4\\
787.0	5.6\\
787.9	4.4\\
788.8	3.1\\
790.0	1.4\\
790.2	1.2\\
791.0	0.7\\
791.2	0.7\\
792.0	0.7\\
792.2	0.6\\
793.0	0.2\\
793.5	0.1\\
794.0	0.0\\
794.2	0.3\\
795.2	1.6\\
796.4	3.4\\
798.1	5.9\\
799.0	7.1\\
799.3	7.3\\
800.0	7.8\\
800.5	8.0\\
801.1	8.3\\
802.4	8.9\\
803.4	9.5\\
804.2	9.9\\
805.0	10.4\\
806.1	11.2\\
807.0	11.8\\
808.0	12.3\\
809.1	12.5\\
810.0	12.7\\
811.0	12.9\\
812.5	12.9\\
813.2	12.9\\
814.1	12.9\\
815.2	12.7\\
816.1	12.7\\
818.2	12.6\\
819.0	12.6\\
820.0	12.3\\
821.2	12.3\\
826.2	12.3\\
827.0	12.3\\
828.2	12.6\\
829.0	12.7\\
830.0	13.4\\
832.4	14.5\\
832.7	14.6\\
833.0	14.8\\
834.0	14.8\\
835.1	15.0\\
836.2	15.2\\
837.0	15.3\\
838.2	15.3\\
839.1	15.2\\
840.2	15.2\\
841.1	15.1\\
842.1	15.0\\
843.0	14.8\\
844.0	14.7\\
846.0	14.3\\
847.0	14.3\\
848.0	14.1\\
849.0	14.1\\
850.0	13.7\\
851.0	13.4\\
852.2	13.4\\
855.0	13.4\\
856.0	13.2\\
857.2	13.2\\
858.0	13.2\\
859.0	13.1\\
860.0	12.9\\
861.0	12.6\\
862.1	12.3\\
862.9	12.1\\
863.0	12.0\\
863.2	11.9\\
863.3	11.9\\
864.2	11.3\\
865.2	10.5\\
865.4	10.3\\
866.2	9.7\\
867.1	9.1\\
868.0	8.6\\
868.1	8.6\\
869.0	8.6\\
869.1	8.6\\
870.4	9.1\\
871.1	9.4\\
872.3	9.6\\
875.2	10.4\\
876.2	10.8\\
878.1	11.6\\
878.6	11.8\\
879.0	11.9\\
879.1	11.9\\
880.1	11.9\\
883.1	12.2\\
884.1	12.4\\
885.1	12.6\\
886.1	12.9\\
889.1	13.0\\
890.0	13.0\\
890.1	12.9\\
891.1	12.5\\
892.2	12.3\\
893.0	12.1\\
893.1	12.0\\
894.0	11.6\\
894.2	11.5\\
895.1	11.2\\
896.0	11.0\\
896.1	11.0\\
898.4	11.3\\
899.1	11.4\\
900.1	11.5\\
901.1	11.7\\
902.2	12.1\\
903.1	12.3\\
904.1	12.4\\
906.1	13.0\\
907.1	13.1\\
910.0	12.9\\
910.3	12.9\\
912.1	12.6\\
913.2	12.5\\
914.0	12.5\\
914.2	12.5\\
916.2	12.1\\
917.0	11.9\\
917.1	11.9\\
918.3	12.1\\
919.1	12.3\\
920.2	12.4\\
921.0	12.5\\
921.3	12.5\\
922.0	12.4\\
922.3	12.5\\
923.1	12.5\\
925.1	12.5\\
927.1	12.2\\
928.1	12.0\\
929.1	11.9\\
930.3	11.8\\
932.1	11.8\\
933.2	11.7\\
934.2	11.7\\
935.1	11.7\\
937.0	11.4\\
937.5	11.4\\
938.1	11.5\\
939.0	11.6\\
939.5	11.6\\
940.1	11.5\\
941.0	11.4\\
941.3	11.3\\
942.1	11.0\\
943.1	10.5\\
944.0	9.9\\
944.4	9.8\\
945.0	9.7\\
945.3	9.7\\
946.2	9.7\\
947.0	9.7\\
947.9	10.1\\
948.2	10.2\\
949.1	10.5\\
950.0	10.7\\
950.6	10.8\\
952.0	10.9\\
952.7	11.1\\
953.0	11.1\\
954.0	11.2\\
956.5	11.4\\
957.0	11.4\\
958.4	11.2\\
959.1	11.2\\
961.0	11.2\\
962.5	11.0\\
964.1	10.9\\
965.1	10.9\\
966.0	11.0\\
967.0	11.2\\
968.0	11.2\\
969.0	11.0\\
970.0	11.0\\
971.0	10.8\\
972.1	11.0\\
973.2	11.3\\
974.0	11.4\\
975.0	11.2\\
976.0	10.7\\
977.2	9.7\\
978.0	9.0\\
979.9	6.2\\
981.3	4.1\\
982.2	2.8\\
983.5	0.9\\
984.0	0.2\\
985.0	0.0\\
986.9	0.0\\
987.0	0.0\\
987.9	0.8\\
988.0	0.9\\
988.2	1.2\\
990.4	4.5\\
992.0	6.8\\
992.9	7.7\\
993.0	7.8\\
994.0	8.3\\
995.0	8.9\\
996.0	9.4\\
997.4	10.0\\
998.0	10.3\\
999.1	11.1\\
1000.0	11.8\\
1001.0	12.3\\
1002.0	12.6\\
1003.0	12.7\\
1004.2	12.7\\
1006.0	12.7\\
1007.0	12.4\\
1008.2	12.3\\
1009.2	12.1\\
1010.1	12.0\\
1011.0	11.8\\
1012.0	11.6\\
1013.1	11.5\\
1013.9	11.3\\
1014.0	11.2\\
1014.3	11.1\\
1014.4	11.1\\
1015.1	10.7\\
1016.0	9.9\\
1016.1	9.8\\
1017.1	9.6\\
1018.1	9.6\\
1019.1	9.8\\
1020.1	10.1\\
1020.7	10.2\\
1021.1	10.3\\
1022.1	10.2\\
1023.1	10.2\\
1024.1	10.3\\
1025.1	10.1\\
1027.0	10.1\\
1027.1	10.2\\
1028.1	10.5\\
1029.3	10.8\\
1030.1	11.0\\
1031.3	11.1\\
1032.3	11.3\\
1033.1	11.4\\
1034.1	11.4\\
1035.1	11.4\\
1036.1	11.2\\
1037.1	10.8\\
1038.3	10.5\\
1039.1	10.4\\
1040.3	10.2\\
1041.3	10.0\\
1042.2	9.8\\
1043.0	9.7\\
1043.1	9.6\\
1044.0	9.2\\
1044.1	9.1\\
1045.2	7.6\\
1046.0	6.4\\
1050.0	0.5\\
1050.1	0.4\\
1051.0	0.0\\
1051.1	0.0\\
1080.0	0.0\\
1080.2	0.1\\
1081.0	0.5\\
1081.1	0.6\\
1082.0	1.8\\
1082.5	2.5\\
1082.8	2.9\\
1085.3	6.6\\
1086.0	7.6\\
1086.2	7.7\\
1087.9	8.9\\
1088.0	8.9\\
1088.7	9.5\\
1089.0	9.7\\
1089.3	9.9\\
1090.1	10.3\\
1091.1	10.8\\
1093.3	11.6\\
1094.0	11.8\\
1094.4	11.9\\
1095.0	12.0\\
1095.7	12.2\\
1096.2	12.3\\
1096.8	12.4\\
1097.2	12.5\\
1098.0	12.6\\
1098.3	12.6\\
1099.1	12.5\\
1101.0	12.1\\
1102.0	12.1\\
1102.5	11.9\\
1103.0	11.8\\
1103.3	11.5\\
1104.1	10.9\\
1105.0	10.1\\
1105.4	9.9\\
1106.2	9.5\\
1107.0	9.2\\
1107.4	8.7\\
1108.0	8.0\\
1109.0	6.7\\
1109.8	5.7\\
1110.0	5.5\\
1110.3	5.3\\
1111.0	5.0\\
1112.5	4.6\\
1113.2	4.4\\
1113.8	4.3\\
1114.2	4.2\\
1116.1	3.9\\
1117.0	3.8\\
1119.0	4.0\\
1120.1	3.9\\
1121.0	3.8\\
1122.0	3.6\\
1123.0	3.1\\
1124.0	2.2\\
1125.0	1.9\\
1127.0	0.4\\
1128.0	0.0\\
1129.0	0.0\\
1130.0	0.3\\
1131.0	0.7\\
1132.0	1.6\\
1133.0	3.1\\
1134.0	4.5\\
1134.8	5.5\\
1135.0	5.7\\
1136.0	6.3\\
1137.0	6.5\\
1138.0	7.2\\
1139.1	8.2\\
1139.9	8.9\\
1140.0	8.9\\
1141.0	9.4\\
1142.5	9.5\\
1144.1	9.6\\
1145.0	9.7\\
1146.0	10.1\\
1147.0	10.3\\
1148.2	10.7\\
1149.1	11.0\\
1150.0	11.2\\
1152.0	11.1\\
1153.0	11.2\\
1155.0	11.5\\
1156.0	11.6\\
1157.0	11.8\\
1158.0	11.9\\
1159.0	12.0\\
1160.2	12.1\\
1162.2	12.1\\
1164.2	12.0\\
1165.1	12.0\\
1166.2	11.8\\
1167.1	11.8\\
1168.3	11.6\\
1169.1	11.4\\
1170.2	10.9\\
1171.0	10.5\\
1171.1	10.5\\
1172.0	9.7\\
1172.2	9.5\\
1173.0	9.0\\
1173.1	8.9\\
1174.1	7.8\\
1175.1	7.1\\
1176.0	6.3\\
1176.1	6.2\\
1180.0	0.4\\
1180.1	0.3\\
1181.0	0.0\\
1181.1	0.0\\
1196.0	0.0\\
1196.1	0.1\\
1197.0	0.9\\
1197.2	1.2\\
1200.1	5.4\\
};
\addlegendentry{\scriptsize EPA UDDS}

\addplot [color=mycolor1, line width=1.3pt]
  table[row sep=crcr]{%
28.0	0.0\\
46.9	0.0\\
47.8	0.1\\
47.9	0.1\\
48.9	1.3\\
49.9	2.4\\
50.9	3.4\\
51.4	4.0\\
51.6	4.2\\
52.1	4.7\\
52.8	5.5\\
53.0	5.7\\
53.8	6.4\\
53.9	6.4\\
54.6	6.6\\
54.9	6.6\\
55.8	7.0\\
55.9	7.0\\
56.8	7.9\\
56.9	7.9\\
57.9	8.3\\
58.9	8.5\\
59.9	8.5\\
61.0	8.3\\
62.8	7.9\\
63.7	7.8\\
64.0	7.7\\
64.8	7.3\\
65.0	7.3\\
65.2	7.1\\
66.0	6.3\\
66.8	5.7\\
67.0	5.7\\
68.0	5.7\\
69.9	5.9\\
70.2	6.0\\
70.9	6.2\\
71.1	6.2\\
71.9	6.6\\
72.0	6.7\\
73.1	7.5\\
73.9	8.1\\
74.1	8.2\\
74.9	8.6\\
75.0	8.6\\
76.0	8.7\\
77.0	8.6\\
77.9	8.5\\
78.0	8.4\\
79.0	7.8\\
80.0	7.0\\
80.2	6.9\\
80.9	6.4\\
81.1	6.3\\
81.9	6.0\\
82.0	6.0\\
82.9	6.2\\
83.1	6.3\\
85.1	7.8\\
86.0	8.4\\
86.9	8.9\\
87.1	9.0\\
87.5	9.1\\
87.9	9.2\\
88.2	9.3\\
89.4	9.4\\
90.0	9.5\\
91.0	9.4\\
92.0	9.3\\
93.0	9.3\\
94.1	9.4\\
95.1	9.4\\
96.9	9.3\\
97.7	9.4\\
98.0	9.4\\
99.0	9.6\\
99.9	9.7\\
100.4	9.7\\
101.0	9.7\\
102.1	9.6\\
102.9	9.5\\
103.9	9.6\\
105.3	9.8\\
105.9	9.9\\
107.4	9.8\\
107.9	9.8\\
109.0	10.0\\
110.0	10.3\\
110.4	10.4\\
112.0	11.0\\
113.0	11.2\\
114.0	11.4\\
116.1	11.6\\
117.0	11.7\\
118.4	11.6\\
119.8	11.5\\
120.2	11.5\\
121.0	11.5\\
122.0	11.6\\
123.9	11.6\\
125.9	11.3\\
126.9	11.3\\
128.7	11.6\\
129.0	11.6\\
130.0	11.7\\
131.0	11.8\\
131.9	11.8\\
133.0	11.6\\
133.9	11.4\\
134.9	11.3\\
136.0	11.4\\
137.2	11.6\\
137.8	11.8\\
138.5	11.9\\
139.2	12.0\\
139.9	12.2\\
140.9	12.3\\
141.9	12.3\\
142.8	12.1\\
142.9	12.1\\
143.9	11.3\\
145.4	9.4\\
147.4	6.9\\
147.6	6.6\\
149.2	4.7\\
151.3	2.0\\
151.5	1.7\\
152.0	1.2\\
152.4	0.8\\
152.6	0.6\\
152.9	0.3\\
153.0	0.3\\
153.9	0.0\\
154.0	0.0\\
190.9	0.0\\
192.1	1.3\\
196.9	7.4\\
197.9	8.4\\
198.9	9.2\\
199.9	9.8\\
200.9	10.0\\
201.6	9.9\\
203.0	9.5\\
203.9	9.4\\
205.3	9.5\\
207.0	9.7\\
207.9	9.8\\
208.8	10.3\\
208.9	10.3\\
209.9	10.1\\
210.8	9.2\\
210.9	9.1\\
211.8	8.7\\
211.9	8.6\\
212.8	7.5\\
212.9	7.3\\
213.8	6.8\\
213.9	6.7\\
214.8	6.6\\
214.9	6.6\\
215.9	6.9\\
216.8	7.1\\
216.9	7.1\\
217.9	7.7\\
219.2	8.8\\
219.9	9.4\\
221.0	10.6\\
221.9	11.7\\
222.2	11.9\\
222.8	12.7\\
224.0	13.8\\
225.0	14.3\\
226.0	15.0\\
227.0	15.5\\
227.8	16.0\\
228.0	16.0\\
229.2	13.6\\
231.8	8.3\\
231.9	8.1\\
232.0	8.0\\
232.9	7.0\\
233.0	7.0\\
246.9	7.0\\
247.0	7.0\\
247.9	7.5\\
248.0	7.6\\
249.3	9.6\\
256.0	19.7\\
256.9	20.9\\
257.0	20.9\\
258.5	20.8\\
259.0	20.7\\
260.1	20.8\\
261.2	20.9\\
263.3	21.2\\
264.4	21.3\\
265.0	21.4\\
267.0	21.5\\
268.1	21.5\\
269.4	21.5\\
270.1	21.5\\
274.6	21.5\\
275.1	21.4\\
276.1	21.4\\
277.0	21.2\\
278.1	21.0\\
278.7	20.9\\
279.0	20.8\\
280.0	20.7\\
281.2	20.5\\
282.7	20.4\\
282.9	20.4\\
284.1	20.4\\
285.1	20.5\\
286.5	20.5\\
287.0	20.6\\
288.0	20.5\\
289.6	20.3\\
289.9	20.2\\
290.9	20.2\\
292.5	20.0\\
293.0	19.9\\
294.7	19.8\\
295.0	19.8\\
296.1	19.7\\
297.6	19.6\\
298.1	19.6\\
299.0	19.6\\
300.1	19.7\\
301.1	19.8\\
302.2	19.9\\
303.8	20.2\\
305.0	20.5\\
306.2	20.8\\
307.0	21.0\\
309.0	21.2\\
310.0	21.3\\
311.0	21.2\\
311.7	21.1\\
312.0	21.0\\
313.0	20.8\\
314.1	20.4\\
315.9	19.7\\
316.1	19.7\\
317.0	19.6\\
318.0	19.6\\
318.9	19.5\\
319.1	19.4\\
319.4	19.3\\
319.9	19.1\\
320.0	18.9\\
320.7	17.5\\
322.5	13.9\\
322.7	13.5\\
323.9	11.1\\
324.1	10.7\\
325.9	7.1\\
326.0	7.0\\
337.2	7.0\\
341.9	7.0\\
342.1	7.1\\
342.9	7.5\\
343.1	7.8\\
344.2	9.4\\
344.9	10.5\\
345.1	10.7\\
345.3	10.9\\
345.9	11.4\\
346.0	11.4\\
346.9	11.1\\
347.9	10.6\\
348.9	9.6\\
349.4	9.0\\
349.9	8.4\\
350.3	8.1\\
350.9	7.7\\
351.9	7.3\\
352.9	7.1\\
353.9	6.6\\
354.9	6.0\\
355.9	4.9\\
356.9	4.2\\
357.5	3.6\\
357.7	3.4\\
357.9	3.2\\
359.7	0.9\\
359.9	0.7\\
360.9	0.1\\
362.0	0.0\\
373.9	0.0\\
374.8	0.3\\
374.9	0.4\\
377.8	4.0\\
379.0	5.5\\
379.9	6.6\\
381.0	7.7\\
381.9	8.6\\
383.0	9.1\\
384.1	9.7\\
385.3	10.4\\
386.0	10.8\\
386.8	11.4\\
386.9	11.4\\
388.9	12.0\\
390.0	12.2\\
391.2	12.6\\
392.7	13.1\\
392.9	13.1\\
394.0	13.2\\
395.0	13.3\\
396.0	13.2\\
397.0	13.1\\
398.0	13.2\\
399.0	13.5\\
400.0	13.7\\
405.0	13.7\\
406.1	13.7\\
407.0	13.8\\
408.0	13.9\\
409.0	13.8\\
410.0	13.6\\
412.8	12.6\\
412.9	12.6\\
413.0	12.5\\
414.0	11.7\\
414.9	10.8\\
415.4	10.2\\
416.0	9.5\\
418.4	7.0\\
420.5	4.7\\
420.8	4.4\\
421.0	4.2\\
421.6	3.4\\
421.8	3.2\\
423.0	1.7\\
423.9	0.6\\
424.1	0.5\\
424.9	0.0\\
425.1	0.0\\
428.9	0.0\\
429.2	0.1\\
429.9	0.3\\
430.1	0.5\\
430.9	1.3\\
433.1	4.1\\
434.4	5.8\\
436.0	7.8\\
436.9	8.8\\
437.5	9.3\\
437.9	9.6\\
438.5	9.8\\
438.9	10.0\\
439.4	10.3\\
439.9	10.6\\
440.1	10.7\\
440.9	11.2\\
441.9	11.4\\
442.9	11.4\\
444.0	11.3\\
445.2	11.2\\
446.0	11.0\\
446.9	10.8\\
447.8	10.3\\
447.9	10.2\\
449.4	8.4\\
452.1	5.0\\
452.3	4.7\\
453.8	2.8\\
455.0	1.3\\
455.8	0.5\\
455.9	0.4\\
456.8	0.0\\
456.9	0.0\\
474.8	0.0\\
474.9	0.0\\
475.2	0.2\\
475.8	0.6\\
475.9	0.7\\
476.2	1.1\\
476.4	1.3\\
479.0	4.6\\
479.2	4.9\\
481.3	7.5\\
482.8	9.4\\
482.9	9.5\\
483.8	10.3\\
483.9	10.4\\
484.9	10.9\\
485.4	11.2\\
485.7	11.4\\
486.8	12.2\\
486.9	12.3\\
487.0	12.3\\
488.8	13.0\\
489.0	13.1\\
490.0	13.3\\
491.2	13.5\\
492.0	13.6\\
493.2	13.7\\
494.3	13.7\\
495.1	13.7\\
496.1	13.7\\
497.0	13.6\\
499.1	13.7\\
500.4	13.5\\
501.2	13.5\\
502.4	13.4\\
503.3	13.4\\
504.3	13.4\\
508.1	13.4\\
509.1	13.3\\
510.2	13.3\\
511.1	13.4\\
511.9	13.5\\
512.3	13.4\\
513.2	13.4\\
514.0	13.3\\
515.4	13.3\\
516.1	13.3\\
517.5	13.2\\
518.3	13.1\\
518.9	13.1\\
519.3	13.0\\
519.9	12.8\\
520.1	12.7\\
520.9	12.3\\
522.0	11.5\\
523.0	10.7\\
524.1	9.7\\
525.0	8.7\\
526.0	7.6\\
526.5	7.0\\
526.7	6.8\\
527.3	6.0\\
528.0	5.2\\
528.2	4.9\\
529.9	3.0\\
531.0	1.6\\
531.9	0.6\\
532.9	0.1\\
534.0	0.0\\
537.9	0.0\\
538.9	0.4\\
540.0	1.3\\
540.9	2.0\\
541.9	2.4\\
542.9	3.1\\
543.2	3.3\\
544.0	3.6\\
544.9	4.0\\
545.9	4.5\\
546.9	5.2\\
547.1	5.4\\
548.0	6.1\\
548.9	6.7\\
549.9	7.2\\
550.9	7.6\\
552.5	8.2\\
552.8	8.3\\
552.9	8.3\\
553.8	8.1\\
553.9	8.1\\
554.8	7.1\\
554.9	7.0\\
566.8	7.0\\
566.9	7.0\\
567.8	7.6\\
567.9	7.7\\
568.9	9.1\\
569.0	9.2\\
569.9	9.5\\
570.0	9.5\\
571.0	9.5\\
571.9	9.3\\
572.0	9.2\\
572.9	8.7\\
573.0	8.6\\
574.2	7.1\\
574.4	6.9\\
575.8	5.1\\
577.7	2.7\\
577.9	2.5\\
578.9	1.2\\
579.2	0.9\\
579.9	0.0\\
580.0	0.0\\
594.9	0.0\\
595.1	0.0\\
595.9	0.2\\
596.0	0.3\\
597.6	2.4\\
598.9	4.0\\
599.2	4.3\\
599.4	4.5\\
599.9	5.1\\
600.1	5.2\\
600.9	5.6\\
601.6	6.0\\
601.9	6.2\\
602.3	6.3\\
602.9	6.5\\
603.2	6.5\\
604.1	6.5\\
605.0	6.5\\
606.0	6.7\\
607.0	6.7\\
608.0	6.7\\
609.1	6.6\\
610.0	6.4\\
611.7	6.4\\
611.9	6.3\\
612.6	6.4\\
613.0	6.5\\
614.0	6.5\\
615.1	6.4\\
616.0	6.3\\
617.1	6.3\\
618.0	6.3\\
619.0	6.4\\
620.0	6.6\\
621.1	6.9\\
622.3	7.2\\
623.1	7.5\\
623.9	7.8\\
624.9	8.0\\
626.1	8.0\\
627.0	8.1\\
629.0	8.4\\
630.0	8.5\\
631.3	8.5\\
632.2	8.6\\
632.9	8.6\\
633.9	8.8\\
634.2	8.9\\
636.0	9.7\\
637.0	10.0\\
637.6	10.1\\
637.9	10.1\\
638.8	10.1\\
638.9	10.1\\
639.2	9.9\\
639.9	9.3\\
641.2	7.7\\
641.4	7.5\\
644.6	3.4\\
644.8	3.2\\
646.1	1.5\\
646.8	0.7\\
646.9	0.6\\
647.8	0.0\\
647.9	0.0\\
672.9	0.0\\
673.0	0.0\\
673.9	0.4\\
674.0	0.5\\
674.5	0.9\\
674.7	1.1\\
674.9	1.3\\
675.1	1.5\\
675.9	2.4\\
676.9	3.5\\
677.3	3.8\\
677.5	4.0\\
678.0	4.5\\
678.9	5.1\\
679.2	5.2\\
680.0	5.7\\
681.0	6.4\\
681.9	7.1\\
682.3	7.4\\
682.5	7.5\\
683.0	7.8\\
684.0	8.3\\
685.6	9.0\\
686.0	9.2\\
686.9	9.5\\
687.3	9.6\\
688.1	9.7\\
689.0	9.8\\
690.4	9.9\\
691.3	9.9\\
692.2	10.0\\
694.1	10.1\\
694.9	10.1\\
695.5	10.0\\
696.1	9.9\\
696.9	9.7\\
697.9	9.2\\
698.9	8.4\\
699.1	8.2\\
699.9	7.3\\
700.9	6.4\\
701.9	5.7\\
703.0	4.6\\
704.8	2.6\\
705.0	2.4\\
705.5	1.9\\
705.9	1.5\\
707.1	0.8\\
707.9	0.2\\
708.9	0.0\\
720.9	0.0\\
721.9	0.5\\
722.9	1.1\\
723.9	1.6\\
724.9	2.4\\
725.6	3.1\\
725.8	3.3\\
726.0	3.5\\
727.9	5.0\\
728.9	5.5\\
729.8	6.1\\
729.9	6.2\\
730.9	6.3\\
732.1	6.3\\
732.8	6.3\\
732.9	6.3\\
733.2	6.5\\
733.9	6.9\\
734.9	7.3\\
735.9	7.6\\
736.9	8.1\\
737.8	8.5\\
737.9	8.5\\
738.9	8.5\\
739.8	8.4\\
739.9	8.4\\
741.2	8.7\\
742.0	8.8\\
742.9	8.9\\
743.0	8.9\\
743.4	8.8\\
745.0	8.2\\
745.9	7.8\\
746.0	7.8\\
747.0	6.8\\
747.2	6.6\\
750.2	3.2\\
750.9	2.4\\
751.0	2.4\\
751.8	1.8\\
752.1	1.6\\
752.9	1.2\\
753.1	1.1\\
754.0	0.8\\
754.9	0.2\\
755.0	0.2\\
755.9	0.2\\
756.0	0.3\\
757.0	1.3\\
758.1	2.7\\
758.3	2.9\\
759.0	3.7\\
759.9	4.7\\
760.0	4.8\\
761.0	5.4\\
762.1	6.2\\
763.0	6.9\\
763.3	7.1\\
764.1	7.6\\
765.0	8.2\\
765.3	8.4\\
766.0	8.8\\
766.9	9.3\\
767.3	9.5\\
768.7	10.0\\
769.0	10.1\\
770.1	10.3\\
771.0	10.5\\
772.3	10.7\\
773.2	10.8\\
773.9	10.9\\
774.9	10.9\\
775.8	10.8\\
776.0	10.7\\
777.1	10.7\\
778.0	10.6\\
778.7	10.5\\
779.0	10.4\\
779.9	10.1\\
780.3	9.9\\
780.9	9.6\\
781.2	9.4\\
781.4	9.2\\
782.2	8.6\\
783.0	8.0\\
784.2	6.8\\
784.9	6.2\\
785.9	5.5\\
787.0	4.5\\
788.1	3.1\\
788.9	2.2\\
789.6	1.4\\
789.8	1.2\\
789.9	1.0\\
790.9	0.6\\
791.9	0.5\\
792.9	0.2\\
793.8	0.2\\
793.9	0.2\\
795.2	1.8\\
797.2	4.3\\
797.9	5.2\\
798.8	6.1\\
798.9	6.2\\
799.9	6.7\\
801.1	7.2\\
802.2	7.7\\
803.7	8.4\\
804.2	8.6\\
804.9	9.0\\
805.9	9.7\\
806.9	10.2\\
807.9	10.5\\
809.4	10.8\\
811.0	11.0\\
812.8	11.0\\
813.2	10.9\\
814.1	10.9\\
815.0	10.8\\
816.0	10.8\\
817.7	10.7\\
817.9	10.7\\
819.1	10.6\\
819.9	10.5\\
821.3	10.4\\
825.5	10.5\\
826.0	10.5\\
827.0	10.6\\
828.0	10.7\\
829.0	11.1\\
830.0	11.6\\
831.9	12.3\\
832.2	12.4\\
833.0	12.5\\
833.9	12.6\\
834.0	12.6\\
834.9	12.1\\
835.0	11.9\\
836.5	8.9\\
836.7	8.5\\
836.9	8.1\\
837.0	8.0\\
837.9	7.0\\
838.0	7.0\\
851.9	7.0\\
852.2	7.1\\
852.9	7.4\\
853.0	7.5\\
854.9	10.4\\
855.1	10.6\\
855.9	11.3\\
856.0	11.3\\
857.1	11.2\\
858.2	11.2\\
859.0	11.2\\
860.0	11.0\\
862.8	10.4\\
862.9	10.4\\
863.9	9.9\\
865.7	8.7\\
866.0	8.5\\
866.9	8.0\\
867.1	7.9\\
867.9	7.5\\
868.9	7.3\\
870.0	7.6\\
870.9	7.8\\
872.5	8.2\\
873.9	8.5\\
875.1	8.8\\
876.0	9.1\\
877.7	9.7\\
877.9	9.8\\
878.9	10.0\\
882.3	10.3\\
882.9	10.3\\
884.1	10.5\\
885.0	10.7\\
885.9	10.9\\
887.0	11.0\\
889.0	11.0\\
889.9	11.0\\
891.2	10.7\\
892.1	10.5\\
892.9	10.3\\
893.9	9.9\\
894.9	9.6\\
895.8	9.4\\
895.9	9.3\\
897.0	9.4\\
898.1	9.6\\
899.0	9.7\\
900.0	9.8\\
901.0	10.0\\
902.5	10.3\\
903.0	10.4\\
904.0	10.6\\
905.8	11.0\\
906.0	11.0\\
907.0	11.1\\
908.7	11.0\\
910.0	11.0\\
912.1	10.7\\
913.1	10.6\\
914.0	10.6\\
916.0	10.3\\
916.9	10.1\\
917.0	10.1\\
918.1	10.3\\
918.8	10.4\\
919.0	10.4\\
920.2	10.6\\
921.0	10.6\\
922.0	10.6\\
923.0	10.6\\
925.0	10.6\\
927.1	10.4\\
928.0	10.2\\
929.2	10.1\\
930.2	10.1\\
932.0	10.1\\
933.2	10.0\\
934.2	10.0\\
935.0	9.9\\
936.9	9.7\\
937.6	9.7\\
938.0	9.7\\
938.9	9.8\\
939.7	9.8\\
940.0	9.8\\
940.9	9.7\\
941.3	9.5\\
942.0	9.3\\
943.1	8.8\\
943.9	8.4\\
944.3	8.3\\
944.9	8.2\\
945.9	8.2\\
946.9	8.3\\
947.5	8.5\\
948.1	8.7\\
949.0	8.9\\
949.9	9.1\\
951.7	9.3\\
952.0	9.3\\
952.9	9.5\\
954.2	9.6\\
956.6	9.7\\
956.9	9.7\\
958.1	9.5\\
959.1	9.5\\
960.4	9.5\\
961.0	9.5\\
962.2	9.3\\
963.5	9.3\\
964.0	9.2\\
965.0	9.3\\
966.1	9.4\\
966.9	9.5\\
968.0	9.5\\
969.1	9.3\\
970.0	9.3\\
970.9	9.2\\
972.1	9.4\\
972.9	9.6\\
973.8	9.7\\
973.9	9.7\\
974.9	9.4\\
975.2	9.3\\
975.9	8.9\\
977.0	8.1\\
977.9	7.3\\
979.3	5.5\\
980.0	4.8\\
980.2	4.5\\
983.0	1.0\\
983.9	0.2\\
984.0	0.1\\
985.0	0.0\\
985.9	0.0\\
986.0	0.0\\
986.9	0.2\\
987.0	0.3\\
987.9	1.1\\
988.0	1.4\\
988.5	1.9\\
988.7	2.1\\
990.0	3.8\\
990.2	4.0\\
991.1	5.1\\
991.6	5.7\\
991.8	5.9\\
991.9	6.0\\
992.0	6.1\\
992.9	6.8\\
993.2	6.9\\
994.5	7.5\\
995.0	7.8\\
996.3	8.3\\
997.1	8.6\\
998.0	9.0\\
999.3	9.8\\
999.6	10.0\\
999.9	10.2\\
1000.1	10.2\\
1000.9	10.5\\
1001.1	10.6\\
1002.0	10.7\\
1003.1	10.8\\
1004.4	10.8\\
1005.0	10.8\\
1005.9	10.7\\
1006.8	10.5\\
1007.0	10.5\\
1008.2	10.4\\
1010.6	10.0\\
1012.9	9.7\\
1013.3	9.6\\
1013.9	9.4\\
1014.2	9.2\\
1014.9	8.8\\
1015.7	8.4\\
1015.9	8.3\\
1016.1	8.3\\
1017.0	8.3\\
1017.9	8.2\\
1018.1	8.0\\
1018.9	7.4\\
1019.2	7.3\\
1019.9	7.0\\
1020.2	7.0\\
1031.9	7.0\\
1032.3	7.5\\
1033.1	8.6\\
1033.9	9.7\\
1034.9	9.7\\
1035.9	9.6\\
1037.0	9.3\\
1038.0	9.0\\
1039.2	8.8\\
1041.4	8.5\\
1043.0	8.2\\
1043.4	8.1\\
1043.9	7.9\\
1044.9	7.0\\
1046.1	5.5\\
1046.5	5.0\\
1046.7	4.8\\
1049.9	0.8\\
1050.8	0.2\\
1050.9	0.1\\
1051.9	0.0\\
1079.9	0.0\\
1080.0	0.0\\
1080.9	0.4\\
1081.0	0.4\\
1082.0	1.4\\
1084.0	3.9\\
1085.0	5.2\\
1085.9	6.3\\
1086.0	6.3\\
1086.3	6.5\\
1087.1	7.1\\
1088.0	7.6\\
1088.4	7.8\\
1088.9	8.2\\
1089.3	8.4\\
1090.0	8.7\\
1091.1	9.1\\
1092.6	9.6\\
1092.9	9.7\\
1094.0	10.0\\
1094.9	10.2\\
1097.1	10.6\\
1097.9	10.7\\
1098.2	10.7\\
1099.0	10.6\\
1101.0	10.3\\
1101.9	10.3\\
1102.6	10.1\\
1102.9	10.0\\
1104.1	9.2\\
1104.9	8.6\\
1105.6	8.3\\
1106.0	8.1\\
1106.4	8.0\\
1106.9	7.8\\
1108.0	6.7\\
1109.0	5.6\\
1109.9	4.7\\
1110.9	4.2\\
1112.1	4.0\\
1113.1	3.8\\
1114.1	3.6\\
1115.9	3.3\\
1116.9	3.3\\
1118.9	3.4\\
1120.0	3.3\\
1120.9	3.3\\
1121.9	3.0\\
1122.9	2.6\\
1123.1	2.5\\
1123.9	1.9\\
1124.9	1.5\\
1126.3	0.7\\
1126.8	0.4\\
1126.9	0.3\\
1127.3	0.2\\
1127.9	0.0\\
1128.9	0.1\\
1129.9	0.3\\
1130.8	0.7\\
1130.9	0.7\\
1131.1	0.9\\
1131.8	1.4\\
1131.9	1.5\\
1132.1	1.8\\
1133.2	3.1\\
1133.9	3.9\\
1134.8	4.8\\
1134.9	4.9\\
1135.9	5.4\\
1136.9	5.6\\
1137.9	6.2\\
1138.9	7.0\\
1139.8	7.6\\
1139.9	7.7\\
1140.8	8.0\\
1140.9	8.0\\
1142.2	8.1\\
1143.5	8.1\\
1144.0	8.2\\
1145.0	8.3\\
1146.0	8.6\\
1147.0	8.8\\
1148.2	9.2\\
1149.0	9.3\\
1150.0	9.5\\
1151.3	9.4\\
1152.0	9.4\\
1153.0	9.5\\
1154.7	9.8\\
1155.1	9.8\\
1156.1	9.9\\
1157.1	10.1\\
1158.7	10.2\\
1159.1	10.2\\
1160.3	10.3\\
1162.0	10.2\\
1164.0	10.2\\
1165.0	10.1\\
1166.8	10.0\\
1167.0	10.0\\
1168.0	9.8\\
1169.0	9.6\\
1170.0	9.2\\
1170.9	8.7\\
1171.3	8.5\\
1172.2	7.8\\
1173.0	7.3\\
1173.9	6.5\\
1174.4	6.2\\
1174.9	5.9\\
1175.3	5.5\\
1175.9	5.0\\
1176.9	3.7\\
1177.4	3.1\\
1178.9	1.2\\
1179.4	0.7\\
1179.9	0.2\\
1180.2	0.1\\
1180.8	0.0\\
1180.9	0.0\\
1181.6	0.0\\
1194.9	0.0\\
1195.9	0.3\\
1196.9	1.3\\
1200.0	5.2\\
};
\addlegendentry{\scriptsize Modified EPA UDDS}

\addplot[area legend, draw=none, fill=white!10!black, fill opacity=0.2]
table[row sep=crcr] {%
x	y\\
237	0\\
237	36\\
252	36\\
252	0\\
}--cycle;
\addlegendentry{\scriptsize U-Turn}
\addplot[area legend, draw=none, fill=white!10!black, fill opacity=0.2]
table[row sep=crcr] {%
x	y\\
557	0\\
557	36\\
572	36\\
572	0\\
}--cycle;
\addplot[area legend, draw=none, fill=white!10!black, fill opacity=0.2]
table[row sep=crcr] {%
x	y\\
331	0\\
331	36\\
346	36\\
346	0\\
}--cycle;
\addplot[area legend, draw=none, fill=white!10!black, fill opacity=0.2]
table[row sep=crcr] {%
x	y\\
842	0\\
842	36\\
857	36\\
857	0\\
}--cycle;
\addplot[area legend, draw=none, fill=white!10!black, fill opacity=0.2]
table[row sep=crcr] {%
x	y\\
1022	0\\
1022	36\\
1037	36\\
1037	0\\
}--cycle;

\end{axis}
\end{tikzpicture}%

%% file: Images/VIL_mf_Mazda.tex
%
%
\definecolor{mycolor1}{rgb}{0.00000,0.44700,0.74100}%
\definecolor{mycolor2}{rgb}{0.85000,0.32500,0.09800}%
\definecolor{mycolor3}{rgb}{0.92900,0.69400,0.12500}%
\definecolor{mycolor4}{rgb}{0.49400,0.18400,0.55600}%
\begin{tikzpicture}

\begin{axis}[%
width=6.5cm,
height=4.0cm,
at={(0cm,4.4cm)},
scale only axis,
xmin=505,
xmax=2800,
xlabel style={font=\color{white!15!black}},
xlabel={},
xticklabels={,,},
ymin=0,
ymax=11,
ylabel style={font=\color{white!15!black} \scriptsize},
ylabel={$\dot{m}_f$ [g/s]},
ticklabel style = {font = \scriptsize},
axis background/.style={fill=white},
xmajorgrids,
ymajorgrids,
legend style={legend pos=north west, legend cell align=left, align=left, draw=white!15!black}
]
\addplot [color=mycolor1, line width=1.3pt]
  table[row sep=crcr]{%
494.233894499133	1.32142426993414\\
514.134329212761	1.2068266305605\\
538.459178415803	1.2247949027892\\
558.343714468361	1.21170689288829\\
582.630888294954	1.25794422415447\\
604.715084769789	1.28849243899322\\
624.612025238661	1.40430823436373\\
646.729199302196	1.36201583325692\\
668.85534449756	0.713561471945923\\
691.006583351501	0.300322598429375\\
713.181334428823	0.269199354794637\\
735.401243638431	0.272139400921787\\
757.666063877218	0.261560948979877\\
779.947160689435	0.2623584341477\\
802.218460741121	0.243854681516041\\
824.441355513287	0.242351460486134\\
846.631072094701	0.24552378610224\\
890.879490336088	0.235405152968269\\
957.213817275288	0.223773217678627\\
1000.62126069603	0.227611698033343\\
1021.65524740837	0.219473936003396\\
1042.07657905245	0.227302453171433\\
1061.74248455328	0.221970049705305\\
1080.50429161652	0.226338916234909\\
1098.29410885091	0.223686434498632\\
1115.13277993203	0.227113942049073\\
1131.04281405984	0.218230525573745\\
1145.99773677834	0.223140132058688\\
1185.07726815229	0.219033007812868\\
1196.24715307353	0.229741109004863\\
1206.59668289203	0.296038423548453\\
1216.13975038371	0.232540722169233\\
1225.80383549008	0.225506939520074\\
1233.20747376319	0.253334421367526\\
1241.88981888446	0.239928164603498\\
1248.73493135687	0.234221454968065\\
1256.77948123288	0.263817326983826\\
1263.06875983526	0.258732569941458\\
1276.31470319317	0.272719060824329\\
1282.59202510648	0.263327821256553\\
1288.64833270297	0.26151576088796\\
1295.07962656365	0.274007451414946\\
1300.72999946124	0.273226069688462\\
1306.31428639481	0.265836291303003\\
1311.27707791748	0.238290045881968\\
1317.15082567352	0.239327284869887\\
1326.97240023499	0.250234580146298\\
1331.26384041515	0.24905703978493\\
1339.0945650725	0.237139318287973\\
1345.65544650096	0.246597413874497\\
1349.02904946211	0.252871321678413\\
1351.81900972577	0.24132283335075\\
1356.60602932508	0.24213215786267\\
1360.35871456353	0.262512769108071\\
1365.15023137244	0.3187783714543\\
1366.7699439341	0.348249744832628\\
1368.38711864842	0.384159041789189\\
1370.00547362415	0.386433616149588\\
1371.62696821884	0.352527180832112\\
1373.26217826905	0.357350370797121\\
1374.90700775369	0.379623985620128\\
1378.25237744391	0.392464049806676\\
1379.95781602797	0.346083269444534\\
1381.68324929733	0.41924896076398\\
1385.18020712558	0.387207091072014\\
1387.1876649943	0.330209745072352\\
1389.32337611313	0.307904891035832\\
1391.61069863265	0.308618948972253\\
1396.701342226	0.296406311639657\\
1399.40308514612	0.299654768729852\\
1408.05666502183	0.288746451296447\\
1414.21226004424	0.292752593252771\\
1417.41018414905	0.285067375516974\\
1420.73696326121	0.299882621789038\\
1424.22532786746	0.279306220377293\\
1427.80077607542	0.313209058011125\\
1431.43387108856	0.297148430022844\\
1435.11521733533	0.301384586240601\\
1442.60360195287	0.289948083359377\\
1446.40611985741	0.29260959893054\\
1450.23157147377	0.313357267282754\\
1454.07134031885	0.298916341447239\\
1465.46043424616	0.289094224831842\\
1473.05429245317	0.297501648516572\\
1477.08704294332	0.335313927479547\\
1480.42083384801	0.336602058232984\\
1484.16470200714	0.346306467425165\\
1487.91994836162	0.416676965364786\\
1492.06919598729	0.514890639655732\\
1495.46613383626	0.649065129761766\\
1499.1864665212	0.777606941207978\\
1502.91445927788	0.921417485291386\\
1507.0401714248	1.0577799564694\\
1510.43261207613	0.983530325992433\\
1514.58718251251	0.866693735444187\\
1518.35679262641	0.586604012179578\\
1521.97287270206	0.501328930235559\\
1524.68627096507	0.443337489961777\\
1528.0380721503	0.666465905827408\\
1532.16916555265	1.25342999753639\\
1535.94000841412	1.72200679996786\\
1538.96351360965	2.35876445791064\\
1543.22991825528	2.93520937495532\\
1547.90859031743	3.42629014083923\\
1553.18095646497	3.61439382011167\\
1557.30282708839	3.7077367033371\\
1563.20256301045	3.56956930807428\\
1569.61916366717	3.63902054507844\\
1575.478280501	3.79966090435892\\
1581.08191590623	3.81164530977367\\
1587.31923848113	3.75462113226786\\
1593.22369512032	3.81696570149143\\
1600.46155695501	3.73243989691355\\
1607.11614496026	3.67797051308025\\
1615.89463948876	3.5770140444065\\
1625.19958768063	3.36125833213555\\
1635.74026742871	3.15512379682741\\
1646.31879993707	2.92084219835988\\
1660.52922311507	2.79284104204362\\
1672.9959777294	2.28824525407708\\
1689.22151900984	2.19997208519499\\
1704.88597361351	1.72900124711668\\
1721.37706236492	1.68786559020373\\
1736.87759541466	1.77076186749628\\
1756.60330551526	1.77529516422874\\
1775.21106600858	1.79101055196816\\
1794.39463184781	1.757141695678\\
1814.05159815798	1.56283464546232\\
1834.18201321549	1.48167101402032\\
1854.73584287605	1.38277368198123\\
1875.66390436	1.32045103135488\\
1896.90574074926	1.38004351788004\\
1918.39978704331	1.51734308435289\\
1940.07053851826	1.55695878141159\\
1961.86391411375	1.69767287862987\\
1983.72842617346	1.81359599994084\\
2005.65997943733	1.80961505703272\\
2027.67175918055	1.7262828818034\\
2049.7488456951	1.52520156467517\\
2071.89611324774	1.54766423156116\\
2091.89965648629	1.5793848797307\\
2114.17210968823	1.6189007806297\\
2136.47216942574	1.69397968376506\\
2158.72068556822	1.75432714638509\\
2180.92034799855	1.80008973551958\\
2203.07843980497	1.75952406002716\\
2225.22103400021	1.73754796580124\\
2247.37558465671	1.77080203072455\\
2269.55623235227	1.84453989024269\\
2291.77440373885	1.83675044310576\\
2314.02432513572	1.73865738650602\\
2336.30263202727	1.63444732797416\\
2380.79434047147	1.58320840157876\\
2403.00400005741	1.63999412496469\\
2425.20657725931	1.49156032366454\\
2447.40819896794	0.541761119995499\\
2469.63827852704	0.354964488641144\\
2491.90048550672	0.276171032764523\\
2514.16914847948	0.287081273358581\\
2558.69238205698	0.277677143953042\\
2580.98050950646	0.260430316778638\\
2603.29943145405	0.262664245371525\\
2625.61323033047	0.259721146143875\\
2647.91389235382	0.252386674493209\\
2670.18162393969	0.250515157949849\\
2716.84963380094	0.239203883706978\\
2800.44259606472	0.240192485440275\\
};
\addlegendentry{\scriptsize MPC-C}

\addplot [color=mycolor2, dashed, line width=1.3pt]
  table[row sep=crcr]{%
490.307287242085	1.28665038170993\\
512.656228776507	1.22539079099624\\
534.98951879672	0.604943913312127\\
557.270060369384	1.20281733044885\\
579.511345121682	1.29835753127509\\
601.753208371534	1.31858335296965\\
624.006071910749	1.27416521139821\\
646.279331323674	1.286600142319\\
668.561808986349	1.23039054205765\\
690.859159106618	1.20270713231275\\
713.163366957292	1.02454774148646\\
735.489552118605	1.10256795046053\\
757.820682179167	1.16919123619209\\
780.088086801442	1.18735007264468\\
802.321734735516	1.28432575500483\\
824.553245995397	0.275864661031392\\
849.035962575656	0.294689287926758\\
869.099741544153	2.11958682799195\\
893.653195350053	1.51191685404001\\
915.999398475548	0.33729210512638\\
938.359340306492	0.299730401016404\\
960.702403806112	0.28515114734364\\
1003.04795115139	0.249725212252088\\
1025.30531188276	0.251826258264828\\
1047.54953566828	1.07999139404819\\
1071.09637969161	0.298850296554519\\
1091.53022326166	0.249755280349291\\
1107.92149632973	0.239596549445196\\
1130.6046324156	0.254541142954622\\
1149.86907717269	0.708794625781138\\
1166.49022817787	0.324620027296078\\
1181.20084819788	0.379339672311744\\
1194.33350380499	0.690806556483039\\
1205.56929738432	0.369511133485958\\
1214.88743186458	0.242634130787792\\
1223.87174972284	0.238605205647218\\
1231.17336212516	0.362717383826748\\
1237.36086201158	0.580622956906609\\
1242.85038199299	0.358606881253309\\
1247.54210267276	0.836776811831896\\
1252.27069638414	0.30572351910223\\
1257.28425370926	0.253696774712807\\
1262.24214685826	0.234164254814004\\
1267.33947085908	0.240731492605391\\
1272.58622306754	0.273491008159453\\
1277.55017158394	0.240088606731206\\
1281.85987095223	0.285434003949831\\
1286.04855272649	0.247937705344611\\
1305.17919756686	0.21505357816568\\
1315.53691943392	0.213387213360875\\
1319.05173528075	0.220850872010487\\
1319.69460622867	0.277038126096613\\
1320.099985694	0.292066948288721\\
1320.46535251373	0.315549850036405\\
1320.72816150657	0.268040165533876\\
1320.74566152057	0.685935648234135\\
1321.06139961882	1.17997654695773\\
1321.07195518283	0.985594499561103\\
1321.08028852282	0.352549214163446\\
1321.08751075083	0.740493122732914\\
1321.32759534527	0.545009087963081\\
1321.35537314527	0.607291128450925\\
1321.96893138295	0.995452772585395\\
1323.00288783474	0.64189118210561\\
1324.42102553587	1.2115560538432\\
1326.09177429451	0.946872616108976\\
1328.1953334376	1.10959722433927\\
1331.40576947072	0.290256718172259\\
1335.62386477683	0.253998454161319\\
1340.06985539488	0.251794767138108\\
1344.73628182619	0.294794472408739\\
1349.76905476472	0.264396660288639\\
1355.05599694279	0.272246565930345\\
1360.72300105348	0.242540724780611\\
1366.77066890117	0.267034613205851\\
1373.32823338743	0.25721447462638\\
1387.87307885393	0.22597341474966\\
1400.69407247849	0.224127300871714\\
1405.59954857059	0.305861734466816\\
1409.58289083657	0.320953903134978\\
1412.65859379434	0.298245563082219\\
1414.87706863479	0.300698691579328\\
1416.14791481473	0.332337631106839\\
1416.73999668306	0.362647688616107\\
1417.23401792719	0.459552235473893\\
1417.63331874958	0.519665656706366\\
1417.89004910018	0.628766132142118\\
1417.90949356018	0.687799071378777\\
1418.27445419955	0.557203220440442\\
1418.70492731975	0.578879782250169\\
1419.5852897491	0.563651763013695\\
1420.81592269045	0.3919950880545\\
1422.34347594622	0.308645593194342\\
1423.94751370128	0.240134649161519\\
1426.19560266365	0.247582115666319\\
1428.60979758047	0.231719080814401\\
1431.09871822975	0.234367221587945\\
1437.86044196344	0.229614559485981\\
1441.86887787668	0.240115453442741\\
1446.255616981	0.241199781984051\\
1450.9588931944	0.259578964141383\\
1455.80672248377	0.250303095585878\\
1460.69730286252	0.279783803179726\\
1465.43001586017	0.297926361713508\\
1469.84367717743	0.284718663993772\\
1473.91337048105	0.336072936492201\\
1477.66024746199	0.343494427682344\\
1481.11668395972	0.271848177961147\\
1484.31011085391	0.248601796740786\\
1487.28036350008	0.248878508332382\\
1490.07587043458	0.311952598556672\\
1492.89586925388	0.312400209293628\\
1495.71564422412	0.297974708580568\\
1498.59188574084	0.336229597932743\\
1501.56540878698	0.29988350853273\\
1507.87603968043	0.331509005035969\\
1511.23530214949	0.305128505863195\\
1514.66419139048	0.418426854218524\\
1517.99974105528	0.540906054203788\\
1521.29606597005	0.620364304872965\\
1524.03472451903	0.357879358850369\\
1526.65642997041	0.31830152826069\\
1530.08603088581	0.274419977904927\\
1533.26270094537	0.33028111361773\\
1535.9948423756	0.507155861265346\\
1538.2756223564	0.390750026184833\\
1541.91900194038	0.293592567118139\\
1545.17501248125	0.728324787627571\\
1550.85749737338	2.15989362140681\\
1554.1393700291	3.26759962498136\\
1557.54289897974	5.08810173747952\\
1560.91620700227	6.09181315837895\\
1564.67218703588	6.44272365204688\\
1567.13765310279	6.05244865937493\\
1570.10788413885	6.15238207838865\\
1573.19216744606	4.87897799373559\\
1576.49784289217	3.92555701992205\\
1579.79353836734	3.65741401557352\\
1583.81048581972	3.21154698210421\\
1588.93077516874	2.76477091193874\\
1596.50638075899	3.18973148700843\\
1605.74137287828	4.07738575120311\\
1617.46802322232	4.1545959644468\\
1630.61906671518	3.94399313437543\\
1645.32742968781	3.58421944389966\\
1661.50765835593	2.45819664088958\\
1678.81576156319	1.8911032684432\\
1696.83656319705	1.44516104932609\\
1715.52005332565	1.27350077101619\\
1734.66893017253	1.82418782545574\\
1754.20316810515	1.90243310306414\\
1774.05728370381	1.86110906940576\\
1796.47208218557	1.89528034996829\\
1817.40147652274	1.82450253080151\\
1838.82195321404	1.73269995663441\\
1860.64896573525	1.66883115866813\\
1882.70741501413	1.38016016095344\\
1904.84980843035	1.26883437512424\\
1924.78545913674	1.45302957671174\\
1946.83428855095	1.58881951095464\\
1971.08413541323	1.77260828598537\\
1993.19415244857	1.81904973940527\\
2015.36947624988	1.85174777785051\\
2037.61520599569	1.75035925736574\\
2059.93209624467	1.64393554076332\\
2082.29271717081	1.51262623406774\\
2104.67237694362	1.41569045624419\\
2127.0196843924	1.49038107837578\\
2149.27679241013	1.17183206804293\\
2167.01732774343	0.456481069878919\\
2191.39545162054	1.68126031791962\\
2213.58293272762	2.03632094237582\\
2235.81748485803	2.16145660554275\\
2258.09438645579	2.05290005301504\\
2280.4218761756	1.81639276540454\\
2302.79078399245	1.74662656415194\\
2325.1569295323	0.880029062765061\\
2347.50149878136	0.364579190102177\\
2369.80999794865	0.309236881785182\\
2392.00750032656	0.296967762478744\\
2413.97416826017	1.49060259784801\\
2435.81220584437	2.03156382028965\\
2457.73943907982	2.46001318255685\\
2479.80334054893	2.4590684483187\\
2502.00709815679	2.43142019291236\\
2524.33300598376	0.771179897797538\\
2546.74147682599	0.463015542432004\\
2569.11018390288	0.356013798696495\\
2591.21017960957	0.317130468627056\\
2612.98992030023	1.27096401579547\\
2634.42569387182	1.29962945180159\\
2655.69910365402	1.07831209216465\\
2677.07288665349	0.309227514962004\\
2698.61689471621	0.254938354635669\\
2720.36194645102	0.276233779902668\\
2764.29781969505	0.273901080232918\\
2786.0029068882	0.25399391813653\\
2807.37106106198	0.249689841959935\\
};
\addlegendentry{\scriptsize MPC-U}

\addplot [color=mycolor3, dashdotted, line width=1.3pt]
  table[row sep=crcr]{%
486.485510338253	1.28106311578813\\
506.524974725858	1.29198994814988\\
528.803309837383	1.22097780984632\\
551.0694876583	1.16312509848012\\
575.575539124666	1.18424980254622\\
595.629544610726	1.19312645521268\\
617.913403121842	1.2095459668476\\
640.206279309231	0.885665651186173\\
662.503274786523	0.822049503245125\\
684.794168352287	1.55394808664369\\
709.333929970231	1.59443581796677\\
729.421645596276	0.778664924522218\\
753.978486142169	0.487329858418889\\
776.310348766963	0.469324001493987\\
798.627161608593	0.438240831617804\\
818.700536741053	1.61194802542514\\
843.217187316065	1.90136123748971\\
863.280765712897	1.9595023504462\\
885.518532940517	1.44333507310921\\
909.883371532103	0.502555731999109\\
929.871468334519	1.42096434153973\\
952.176100642043	0.257058688906454\\
976.677076403129	0.300355367284283\\
998.748313240179	0.294242924974697\\
1020.60639642322	0.264572103388673\\
1042.33170588054	0.258989098575057\\
1064.17173047093	0.312666667093254\\
1086.20531521272	0.272910874416084\\
1108.43775961458	0.289658963382863\\
1130.77474445929	0.242949168741688\\
1152.9732781303	0.235630282209968\\
1174.8984036784	0.39120350686153\\
1194.63210744447	1.00205128342168\\
1208.04179620409	0.30827035849461\\
1223.97943859488	0.270283502581151\\
1233.37993561808	0.411817190816237\\
1239.49471564163	0.359279038234945\\
1244.80855675755	2.3898406131957\\
1249.0293904948	3.81731001823027\\
1250.96880304178	0.492883226720096\\
1252.54538660831	1.33260450186572\\
1254.17218075641	2.44261676596807\\
1256.32318877133	1.21854766390607\\
1259.60950753713	0.304645604689085\\
1262.22648084858	0.251774938942617\\
1264.74009398189	0.26257415679629\\
1267.30509795007	0.282485122816524\\
1270.19231648107	0.264678031266158\\
1274.88911286671	0.256709592443713\\
1280.96679600689	0.282574293918969\\
1288.8677957207	0.280904599519999\\
1295.90197115681	0.254514840111369\\
1315.40001195593	0.237521616877075\\
1336.86097754027	0.220052807428601\\
1340.05789869054	0.230961908328936\\
1342.3242468504	0.234410078781366\\
1343.90025192558	0.218907087476509\\
1344.94458363377	0.218568579308794\\
1345.31485836193	0.229576409285073\\
1345.31485836193	0.605356215717165\\
1345.31485836193	0.313682795821023\\
1345.31485836193	0.771758264779237\\
1345.31485836193	0.362484496831257\\
1345.31485836193	1.00957190135659\\
1345.31485836193	0.390466483384898\\
1345.31485836193	1.17170112921849\\
1345.31930280993	0.789737666931615\\
1345.55470964129	0.699529969061587\\
1346.21857821967	0.710098705881137\\
1347.81676118525	0.6592660803949\\
1349.89678484581	0.746381928125629\\
1352.71284169814	0.756811381584612\\
1356.03703643644	0.322833949523556\\
1359.89858199729	0.295669425942833\\
1364.67756361918	0.314573723168451\\
1369.54480909593	0.280159290506163\\
1375.08335937516	0.246381231513624\\
1381.21791024547	0.328891332768762\\
1387.53916350378	0.281707646494851\\
1393.46797368033	0.317321381571219\\
1400.32755745544	0.277284138609957\\
1414.9376865423	0.242704176156622\\
1422.28636991596	0.230559938194801\\
1427.75175264314	0.231455946887309\\
1431.21673581504	0.377901758120515\\
1433.4689440757	0.307294714408272\\
1435.10613674313	0.299596855314121\\
1436.09299896963	0.668561711875554\\
1436.70881690885	0.315572171614804\\
1436.99498405903	0.713223055789058\\
1437.01887296703	0.393392961816971\\
1437.28313382613	0.386636793813523\\
1437.58379552263	0.590873220309277\\
1438.00153285948	0.638030123295721\\
1438.58891694001	0.750210223724935\\
1439.5669620871	0.392274886194173\\
1440.80191490404	0.316476719102866\\
1442.31437184981	0.253809955630913\\
1444.2610432925	0.24071417481764\\
1447.02533139618	0.235040077703616\\
1456.8233937526	0.233157301132451\\
1460.74888862032	0.242731928652574\\
1465.60143896365	0.389309143054106\\
1470.6846230685	0.250790781088654\\
1475.76132734807	0.264014684686117\\
1480.82665342144	0.341090617271675\\
1485.72095191673	0.449483336020421\\
1494.58447601947	0.614114507477098\\
1498.15636018032	0.601748368175777\\
1501.49643031772	0.561325716314968\\
1504.84180000193	0.345321598899318\\
1507.57567163568	0.316803277288273\\
1509.25639447377	0.308263172274565\\
1511.30562374138	0.351170928110605\\
1513.16619125893	0.322050172383115\\
1515.18253600082	0.432006966057997\\
1517.50322283147	0.622607353268904\\
1520.16325846803	0.416262872751304\\
1523.1426482464	0.438997531023688\\
1525.66526848057	0.694291340862037\\
1529.60122598433	0.707943128085844\\
1532.72777614831	0.529844938795122\\
1537.45574254975	0.546633230603675\\
1540.64322283395	0.5882615278706\\
1542.76297709475	0.640531136831669\\
1545.928728653	1.35601969589879\\
1548.17136694985	2.37849546141251\\
1550.92170531873	3.17295976158039\\
1553.61231164591	5.65375658624089\\
1555.65855133525	7.73609833036926\\
1559.06000077043	9.42340607374899\\
1562.39920491425	11.4374730005052\\
1566.46246858572	11.7256102557244\\
1570.26810246036	9.35304698495975\\
1573.38024951476	10.3275070321074\\
1578.67985191989	8.40145969983496\\
1583.44365773564	2.69743361296514\\
1589.67453261702	0.614235506188379\\
1596.96943083263	1.77117552567688\\
1607.42653605283	2.20180032233839\\
1619.39775993867	1.58108054282184\\
1633.67431713871	1.67814118660226\\
1650.17130790951	1.64364057306921\\
1668.39408460587	1.58948240910559\\
1688.19934648841	1.61534840015111\\
1709.94223817802	1.63426718742858\\
1734.87826077374	1.62415130765157\\
1754.79235938254	1.5906038995322\\
1777.29288469579	1.3238262267987\\
1799.67435121921	0.892054693053524\\
1822.12446503649	1.17091267550722\\
1844.53785823199	0.726047230895347\\
1864.68080791634	0.695471302401529\\
1887.0275854695	0.674711494158601\\
1909.34236400995	0.670600590183312\\
1931.65965577798	0.624374808121502\\
1953.94833721776	0.615949530661965\\
1976.22958302205	2.91383456077347\\
1998.49276968742	5.3771795646021\\
2020.69044866883	5.64832432930234\\
2042.71478708318	3.52655386263677\\
2064.58145917108	1.59136480543202\\
2086.21928019625	2.12376474804933\\
2107.62370524472	1.92801044664884\\
2128.78845492246	0.923913765841462\\
2151.7868717817	0.486618219414595\\
2168.31890473341	0.374420527542497\\
2190.82900158727	0.287304833696908\\
2209.32536374583	1.67796007665493\\
2232.66229557739	2.59053548948941\\
2254.97061807439	4.62185987189196\\
2279.22440760581	4.28033736667294\\
2301.11044761125	4.54427322533456\\
2323.43404291084	4.51977741986366\\
2345.80902934464	4.27956536072588\\
2365.74294566752	1.60931421869964\\
2393.83612550125	0.815250238131739\\
2412.60353568989	0.680399388464593\\
2450.46963613075	0.645886573395273\\
2471.5728208577	0.624563708441656\\
2491.40865237311	0.625066443845299\\
2509.89989144166	0.585203026088038\\
2554.96873812465	0.53642319877099\\
2575.01759391917	0.53320067484492\\
2599.48850290198	2.03243015911539\\
2619.29881547777	4.99795664748444\\
2643.2238864406	7.98710538480736\\
2664.7014555181	3.91876995599068\\
2685.92583494254	1.79675718534418\\
2706.8856223447	0.318052929748319\\
2727.60811930466	0.23994549639292\\
2748.0776274254	0.234770628389015\\
2807.99923817975	0.237666156289833\\
};
\addlegendentry{\scriptsize WIE}

\addplot [color=mycolor4, dotted, line width=1.3pt]
  table[row sep=crcr]{%
481.172480360261	1.10369171339516\\
502.912284376304	1.14086158899636\\
546.500883536622	1.06104828861498\\
568.351059513288	1.02513876378544\\
590.241117350087	1.01712761190356\\
612.13783232353	1.03348961247502\\
634.008072796331	1.0115743148026\\
655.845358941192	1.04237289862203\\
677.656571754413	1.09401544565162\\
699.435007558284	1.09638739529646\\
721.213443365717	1.11081924914379\\
742.997780493888	1.14733658152454\\
764.788975353404	1.15257093926675\\
786.581173151497	1.13816079723756\\
808.362146809367	0.908901914949183\\
830.126242126343	0.362173747988436\\
851.882322720679	0.334108742032186\\
873.641967361353	0.278441530360851\\
917.126497045493	0.263643521313952\\
938.852007094875	0.263104096440657\\
960.575333950035	0.238609449517753\\
982.298259622277	0.22857150699474\\
1004.02276675476	0.234973881334554\\
1025.74769829604	0.227303803538689\\
1047.46888846606	0.224778720772974\\
1069.12417002562	0.234328143425046\\
1090.57823130996	0.334595707268818\\
1111.68282446664	0.292395931416195\\
1132.31811213421	0.297031035414875\\
1152.15597836695	0.345908272918223\\
1170.65121177223	0.373289486866724\\
1186.84030824261	0.537698310605265\\
1201.54400073148	0.277912675923289\\
1211.53090215918	0.244300317372108\\
1221.09122591403	0.327777303830771\\
1229.71371775064	0.243532471768503\\
1237.95594704055	0.283679040518564\\
1245.9891419749	0.275296946970229\\
1261.62164873506	0.246173758744135\\
1269.34574279932	0.237315825861515\\
1277.03198416407	0.221605129825093\\
1284.71608885356	0.216283537808977\\
1306.36900984413	0.233885366099003\\
1309.96148475773	0.225789926196285\\
1309.97037365373	0.244876418301374\\
1309.97481810173	0.256609301921344\\
1309.98342921973	0.240325706620752\\
1309.98676255573	0.423919886058684\\
1309.98676255573	1.16141395868681\\
1309.98676255573	1.12060762809733\\
1309.98676255573	1.56584763318051\\
1309.98676255573	0.965520271058722\\
1309.99231811573	0.552511628161483\\
1310.2842514498	0.338621570310806\\
1311.31896372433	0.300828228233513\\
1313.3140637189	0.242204309314275\\
1315.86301496516	0.26727249357964\\
1320.04757744957	0.319905770721562\\
1324.93032041567	0.258956555240275\\
1330.78344844562	0.27965965409885\\
1337.51807263583	0.274048998276612\\
1345.0885494214	0.261468129750483\\
1353.32839492901	0.241572071448445\\
1379.81262152589	0.23145428098087\\
1388.61613829494	0.21807113631894\\
1405.22186160899	0.22011829206167\\
1405.22686161299	0.320643687148277\\
1405.23047272699	0.621163149246968\\
1405.23047272699	1.02653807717161\\
1405.23047272699	0.820741797465416\\
1405.23047272699	0.843481105070168\\
1405.23047272699	0.298797480615121\\
1405.27658387499	0.240760765738742\\
1406.05766766245	0.227547884528121\\
1407.54442918901	0.245256594321745\\
1409.87392429091	0.263004190471293\\
1416.99789849323	0.274914043275203\\
1421.68060150614	0.269719011092093\\
1426.92655136348	0.281271762411961\\
1432.65730419588	0.243706269808627\\
1438.79265740894	0.28383432468263\\
1445.10979138822	0.268069166582336\\
1451.44300135229	0.279202903187979\\
1457.50890511565	0.272736730275483\\
1463.10198586864	0.258958060947407\\
1468.11298208034	0.320167644977573\\
1472.50464283724	0.48020336410309\\
1476.27068889156	0.54644456262622\\
1479.46858972642	0.662318708527437\\
1482.14933501333	0.720786337305981\\
1484.1997445494	0.587111595793431\\
1485.71300384874	0.331083080226563\\
1487.10504494995	0.29769213641066\\
1488.50454554919	0.298043783372577\\
1490.00702841677	0.239425746662619\\
1491.64346526449	0.450279446014974\\
1493.32029265951	0.302741733815765\\
1495.13185313489	0.320761174309155\\
1497.30270177835	0.315607639927748\\
1499.9658129069	0.259366393890559\\
1503.10882864412	0.304553829219003\\
1506.71251011929	0.398543159186374\\
1510.72646945183	0.387657931542435\\
1514.90223810429	0.481501851736084\\
1519.0322499992	0.524446656509099\\
1522.97584746796	0.576045080040331\\
1526.10807101565	0.450952812957439\\
1529.76969754799	0.348341378906298\\
1536.0381750883	0.344712430678555\\
1538.59454219878	0.701139502280967\\
1541.5347403078	1.35927723744089\\
1544.2205819987	1.354435991424\\
1547.23731769904	1.04390649417655\\
1549.7140958079	2.39673524957743\\
1552.92417744615	2.9329150725639\\
1555.9753478776	3.60569147650904\\
1559.28905516754	4.24616045431094\\
1562.8737414385	4.88490394444079\\
1565.97282308456	5.2876469942812\\
1569.40575512532	5.40289863797852\\
1573.22492742685	5.563976086205\\
1577.05531258741	4.9990023605742\\
1581.11489921004	4.65289572549727\\
1587.06862402465	4.26975337781869\\
1593.16967494964	3.91086424156583\\
1601.3464523738	3.53802429365442\\
1609.89627248455	3.22990621684949\\
1620.34896111692	2.92045490478631\\
1632.03366349853	2.61453666472516\\
1645.18350245751	2.25944436175996\\
1659.73791690052	2.16116498407973\\
1692.4674829843	1.86488005860383\\
1710.3319512086	1.75019065675042\\
1748.47590889772	1.6175925288785\\
1768.46217181532	1.42015096139266\\
1788.88187075541	1.33951024677117\\
1809.66573185634	1.34096798637938\\
1830.72769930451	1.31052101190016\\
1851.98932921247	1.39660626429259\\
1873.39254974005	1.44068074006964\\
1894.92150992336	1.45626281850355\\
1916.55460937824	1.52479120881071\\
1938.26811184649	1.52930139641967\\
1960.02574644466	1.56099138088985\\
1981.82907094286	1.54955581102422\\
2003.65512790023	1.4818143284574\\
2025.50138003941	1.51404985268618\\
2047.32213731357	1.46600094683845\\
2069.09778286032	1.47287646348559\\
2090.81887340233	1.37958252188992\\
2112.48000509357	1.32082983214104\\
2134.11772964079	1.19771819616471\\
2155.75685944862	1.20730368946715\\
2177.43072645513	1.28448554090573\\
2199.14206887706	1.42196003779327\\
2220.89166557847	1.45878789460994\\
2242.68732958253	1.59309244062888\\
2264.50376677781	1.70480329879001\\
2286.31354724817	1.71507518301496\\
2308.11571549874	1.69153064565171\\
2329.95159306459	1.65024003801318\\
2351.80392377504	1.58218651737661\\
2395.3640315781	1.459500467412\\
2416.98944640572	1.40703768366166\\
2438.54621390606	1.44581918562608\\
2460.07258840159	1.41287236360085\\
2481.57288994732	1.40222567026467\\
2503.10963898436	1.42855331417968\\
2524.69820479887	1.41395930161661\\
2546.34562150309	1.41881485389104\\
2589.80985279186	1.49839943126835\\
2611.5866787083	1.52289339342724\\
2633.35958606263	1.34776666849666\\
2655.10621969158	0.940393783100717\\
2676.81010237869	0.486457771562527\\
2698.46553283719	0.334715703520942\\
2720.0921039194	0.381834131468622\\
2748.16504041581	0.288251970364854\\
2769.74848312733	0.280142680941935\\
2791.31685384545	0.279139070861675\\
2812.87032917451	0.265599802185079\\
};
\addlegendentry{\scriptsize IDM}

\addplot[area legend, draw=none, fill=white!10!black, fill opacity=0.2]
table[row sep=crcr] {%
x	y\\
1454.4	0\\
1454.4	11\\
1532	11\\
1532	0\\
}--cycle;
\addlegendentry{\scriptsize U-Turn}

\end{axis}

\input{Images/VIL_Vel_Mazda}

\end{tikzpicture}%

%% file: Images/VIL_Vel_Mazda.tex
%

\begin{axis}[%
width=6.5cm,
height=4.0cm,
at={(0cm,0cm)},
scale only axis,
xmin=505,
xmax=2800,
xlabel style={font=\color{white!15!black} \scriptsize},
xlabel={Position [km]},
xtick={1000, 1500, 2000, 2500},
xticklabels={1.0, 1.5, 2.0, 2.5},
ymin=0,
ymax=24,
ylabel style={font=\color{white!15!black} \scriptsize},
ylabel={Velocity [m/s]},
ticklabel style = {font = \scriptsize},
axis background/.style={fill=white},
xmajorgrids,
ymajorgrids,
legend style={legend pos=south east, legend cell align=left, align=left, draw=white!15!black}
]
\addplot [color=mycolor1, line width=1.3pt]
  table[row sep=crcr]{%
196.06920504614	21.8969113393432\\
213.525606852582	22.00684623797\\
235.419583427471	22.0368270967165\\
253.736978012829	22.0768041166566\\
276.555826818642	22.1467619953432\\
297.682574598626	22.1067868816122\\
319.080146093077	22.0967926266271\\
340.699935442263	22.0668098616716\\
362.453077021353	22.2367083839999\\
382.112525989068	22.1767447602988\\
406.195008597293	22.1367677403582\\
425.939735874274	22.2666911489555\\
450.11787206711	22.1267734853732\\
472.15205854986	22.2067256190448\\
494.233894499133	22.2367083839999\\
514.134329212761	22.2267141290149\\
538.459178415803	22.1067868816122\\
558.343714468361	22.3966126513433\\
582.630888294954	22.2267141290149\\
604.715084769789	22.0368270967165\\
624.612025238661	22.0568156066865\\
646.729199302196	21.9868577279999\\
668.85534449756	21.9768634730149\\
691.006583351501	22.0168404929555\\
713.181334428823	22.0668098616716\\
735.401243638431	21.6570549383282\\
757.666063877218	21.3072636386864\\
779.947160689435	20.8175565816714\\
802.218460741121	19.9680639700296\\
824.441355513287	19.2085177470449\\
846.631072094701	18.2790749079404\\
868.777837023662	17.2996588877013\\
890.879490336088	16.3002562637012\\
912.983504168829	15.530716738836\\
935.114948919619	14.5413083698209\\
957.213817275288	13.5119239339701\\
979.103651233995	12.4825385450149\\
1000.62126069603	11.5730823096715\\
1021.65524740837	10.8035427848058\\
1042.07657905245	10.0140147499701\\
1061.74248455328	9.1745163933133\\
1098.29410885091	8.08516738065691\\
1115.13277993203	7.81532869123885\\
1131.04281405984	7.54549000182078\\
1145.99773677834	7.1057527900748\\
1159.98639371355	6.96583607959701\\
1173.02290153094	6.53609312283561\\
1185.07726815229	6.33621278865667\\
1196.24715307353	6.22627836658194\\
1206.59668289203	5.99641574399993\\
1216.13975038371	5.82651722167157\\
1225.80383549008	5.62663688749262\\
1233.20747376319	5.49671443199986\\
1241.88981888446	5.36679197650756\\
1248.73493135687	5.24686377599983\\
1256.77948123288	4.85976960752396\\
1263.06875983526	4.36738935250742\\
1269.81420700478	4.09755066308935\\
1276.31470319317	3.85769402379856\\
1282.59202510648	3.64781967291037\\
1288.64833270297	3.64781967291037\\
1295.07962656365	3.43111155786528\\
1300.72999946124	3.11813607250724\\
1306.31428639481	2.99820787199997\\
1311.27707791748	2.35859008779835\\
1317.15082567352	2.35859008779835\\
1322.36834165758	1.86888291164541\\
1326.97240023499	1.63687871405455\\
1331.26384041515	1.56827125726431\\
1335.29402977834	1.51909196941779\\
1339.0945650725	1.60903823893659\\
1345.65544650096	1.58905020551856\\
1351.81900972577	1.63277402862013\\
1354.35382107432	1.65900832248144\\
1356.60602932508	1.65900832248144\\
1358.58151254228	1.66900233919023\\
1360.35871456353	1.68899049174615\\
1362.00766334214	1.71897254187297\\
1363.55855148902	1.76894262541782\\
1365.15023137244	1.84971202531597\\
1366.7699439341	1.95882918116422\\
1368.38711864842	2.07875738167149\\
1371.62696821884	2.41855442632823\\
1373.26217826905	2.56846467696278\\
1374.90700775369	2.65841106561948\\
1376.56929407772	2.76859509983569\\
1378.25237744391	2.83830336638039\\
1379.95781602797	2.96822582187315\\
1381.68324929733	3.03818393883603\\
1383.2914063414	3.0581719722536\\
1385.18020712558	3.18809442774636\\
1387.1876649943	3.26804656141803\\
1389.32337611313	3.39696961523941\\
1391.61069863265	3.60712955371355\\
1394.08590382216	3.64781967291037\\
1396.701342226	3.6678077063284\\
1399.40308514612	3.72777180658204\\
1405.0752580326	3.78173949681059\\
1408.05666502183	3.80772394025371\\
1411.10847786361	3.79772992354492\\
1414.21226004424	3.80772394025371\\
1417.41018414905	3.77774189012689\\
1420.73696326121	3.7793409328001\\
1427.80077607542	3.80772394025371\\
1431.43387108856	3.78773590683568\\
1435.11521733533	3.67780172303719\\
1438.84009374302	3.69129364559421\\
1442.60360195287	3.73776582329083\\
1446.40611985741	3.75775385670886\\
1450.23157147377	3.7677478734181\\
1454.07134031885	3.73776582329083\\
1457.89993156146	3.72352434948061\\
1461.67557377841	3.77174548010134\\
1465.46043424616	3.74775984000007\\
1469.2539344908	3.78773590683568\\
1473.05429245317	3.74775984000007\\
1477.08704294332	3.78288168036033\\
1480.42083384801	3.90509445560838\\
1484.16470200714	3.93764639574647\\
1487.91994836162	3.87768229549238\\
1492.06919598729	3.92765237903723\\
1495.46613383626	4.07756262967177\\
1499.1864665212	4.01759852941768\\
1502.91445927788	4.29943005931\\
1507.0401714248	4.65635977211832\\
1510.43261207613	5.04698344182088\\
1514.58718251251	5.59665459908956\\
1518.35679262641	6.266254433418\\
1521.97287270206	6.22627836658194\\
1528.0380721503	6.496117056\\
1532.16916555265	6.79593803382068\\
1535.94000841412	6.36619460050724\\
1538.96351360965	6.46613476759694\\
1543.22991825528	7.69540001417909\\
1547.90859031743	8.70479689316426\\
1553.18095646497	9.98403198501501\\
1557.30282708839	11.2732613318508\\
1563.20256301045	12.4325682231938\\
1569.61916366717	13.5618933026867\\
1575.478280501	14.4113859143285\\
1581.08191590623	15.3108478946865\\
1587.31923848113	16.1103711376122\\
1593.22369512032	16.9298809842985\\
1600.46155695501	17.6694386973136\\
1607.11614496026	18.3690193903881\\
1625.19958768063	19.5083396779701\\
1635.74026742871	19.8981060913434\\
1646.31879993707	20.3178552696718\\
1660.52922311507	20.7276101930147\\
1672.9959777294	21.097390002627\\
1689.22151900984	21.2872751287164\\
1704.88597361351	21.4871564159998\\
1721.37706236492	21.6470606833432\\
1736.87759541466	21.8069649506865\\
1756.60330551526	21.8069649506865\\
1775.21106600858	21.9368883592838\\
1794.39463184781	21.9568749630448\\
1814.05159815798	21.9868577279999\\
1834.18201321549	22.0368270967165\\
1854.73584287605	22.1567562503283\\
1875.66390436	22.1667505053133\\
1918.39978704331	22.3266547726566\\
1940.07053851826	22.2167198740299\\
1961.86391411375	22.0967926266271\\
1983.72842617346	22.1267734853732\\
2005.65997943733	22.0568156066865\\
2027.67175918055	22.1467619953432\\
2049.7488456951	22.1067868816122\\
2071.89611324774	22.1367677403582\\
2091.89965648629	22.2167198740299\\
2114.17210968823	22.1567562503283\\
2136.47216942574	22.1367677403582\\
2158.72068556822	22.1567562503283\\
2180.92034799855	22.1167792303881\\
2203.07843980497	22.1567562503283\\
2225.22103400021	22.2067256190448\\
2247.37558465671	22.2067256190448\\
2269.55623235227	22.1867390152838\\
2291.77440373885	22.2067256190448\\
2314.02432513572	22.1867390152838\\
2336.30263202727	22.25669689397\\
2358.55660039648	22.2766854039405\\
2380.79434047147	22.25669689397\\
2403.00400005741	22.2666911489555\\
2425.20657725931	22.1967313640594\\
2447.40819896794	22.2167198740299\\
2469.63827852704	22.1667505053133\\
2491.90048550672	21.8469419706271\\
2514.16914847948	21.5071449259704\\
2536.43213434758	20.9774608490147\\
2558.69238205698	20.3678246383879\\
2580.98050950646	19.8181539576717\\
2603.29943145405	18.8587283536122\\
2647.91389235382	17.139754620358\\
2670.18162393969	16.3102505186866\\
2694.6395727061	15.5606985506865\\
2716.84963380094	14.7311944490148\\
2736.72100556663	13.9616549241491\\
2758.47097962771	13.2920550898207\\
2779.77302391359	12.5724849336716\\
2800.44259606472	11.9428611661792\\
2820.48595141333	11.3532134655225\\
2839.82369059538	10.8035427848058\\
2860.00532018655	10.3238290296717\\
2877.48934962684	9.7941459058211\\
2892.50583450708	9.03459968283596\\
2908.3832650945	8.28504771483586\\
2924.92552015043	7.84531050308942\\
2937.76645832073	7.42556132476102\\
2951.32732987767	7.04578868982071\\
2965.43635190523	6.66601557832837\\
2977.62386813582	6.42615870076133\\
2988.07829117224	5.82651722167157\\
2999.16714555158	5.54668475382095\\
3010.73160873578	5.43675033174623\\
};
\addlegendentry{\scriptsize MPC-C}

\addplot [color=mycolor2, dashed, line width=1.3pt]
  table[row sep=crcr]{%
179.509123373835	22.3466432826272\\
201.553153155956	22.3066662626866\\
245.622953516447	22.3266547726566\\
267.654625400475	22.3566375376122\\
289.712771794682	22.3466432826272\\
311.82032644433	22.3566375376122\\
334.001225952778	22.3266547726566\\
356.23469655544	22.246702638985\\
378.520915607532	22.2167198740299\\
423.213321828197	22.2766854039405\\
445.588938467483	22.2866777527161\\
490.307287242085	22.2666911489555\\
512.656228776507	22.3066662626866\\
534.98951879672	22.2866777527161\\
557.270060369384	22.2367083839999\\
579.511345121682	22.1467619953432\\
601.753208371534	22.1867390152838\\
624.006071910749	22.2866777527161\\
646.279331323674	22.3266547726566\\
713.163366957292	22.3566375376122\\
735.489552118605	22.3166605176716\\
757.820682179167	22.2666911489555\\
780.088086801442	22.1967313640594\\
802.321734735516	22.1967313640594\\
824.553245995397	22.2267141290149\\
849.035962575656	21.1973287400597\\
869.099741544153	20.4377844232836\\
893.653195350053	20.5377231607163\\
915.999398475548	20.6676465693135\\
938.359340306492	18.8987034673432\\
960.702403806112	16.5501069197012\\
980.779767440607	14.7466611867458\\
1003.04795115139	12.6624313223283\\
1025.30531188276	10.643637564358\\
1047.54953566828	9.3244264056716\\
1071.09637969161	9.05458819280602\\
1091.53022326166	7.26565705741768\\
1107.92149632973	6.22414937337999\\
1130.6046324156	5.53669049883592\\
1149.86907717269	4.67720410876109\\
1166.49022817787	4.82711459767143\\
1181.20084819788	4.88707869792552\\
1194.33350380499	4.91706098632812\\
1205.56929738432	5.13692935392555\\
1214.88743186458	5.22687574258225\\
1223.87174972284	4.81712081923888\\
1231.17336212516	4.14752098491044\\
1237.36086201158	4.23746689701511\\
1242.85038199299	4.60724623007445\\
1247.54210267276	4.65721607534306\\
1252.27069638414	5.03698918683585\\
1257.28425370926	4.88707869792552\\
1262.24214685826	4.06756885123878\\
1267.33947085908	3.30802262825364\\
1272.58622306754	2.66840508232826\\
1277.55017158394	2.00879926470907\\
1281.85987095223	1.35918758293656\\
1286.04855272649	0.659605758050475\\
1290.61515977185	0.409755072265853\\
1295.26568754637	0.32980287902501\\
1300.15624465363	0.219868576088629\\
1305.17919756686	0.159904416265817\\
1309.39865046339	0\\
1319.69460622867	0\\
1320.099985694	0.629623648354482\\
1320.46535251373	0\\
1320.72816150657	1.48910991929097\\
1320.74566152057	1.66900233919023\\
1321.06139961882	2.29862598754471\\
1321.08751075083	4.43734770774608\\
1321.32759534527	4.74716246400021\\
1321.35537314527	5.06697147523892\\
1321.96893138295	5.3268159096715\\
1323.00288783474	5.78654115483596\\
1324.42102553587	6.0763678776716\\
1326.09177429451	6.63603328992531\\
1328.1953334376	7.09575853508932\\
1331.40576947072	7.59545984708939\\
1335.62386477683	6.78594377883564\\
1340.06985539488	5.76655312141793\\
1344.73628182619	4.80712656425385\\
1349.76905476472	3.86768804050735\\
1355.05599694279	2.98821385529118\\
1360.72300105348	2.09874541508952\\
1366.77066890117	1.18928917974608\\
1373.32823338743	0.599641598227663\\
1380.33746922671	0.41974908897464\\
1387.87307885393	0.379772992354447\\
1394.79799277161	0.269838719202653\\
1405.59954857059	0.159904416265817\\
1409.58289083657	0.479713248797452\\
1412.65859379434	0.899462337772547\\
1414.87706863479	1.23925926329093\\
1416.14791481473	1.52908598612703\\
1417.23401792719	2.05876934825392\\
1417.63331874958	2.41855442632823\\
1417.89004910018	2.78833328283599\\
1417.90949356018	3.20808246116439\\
1418.27445419955	3.64781967291037\\
1418.70492731975	4.06756885123878\\
1419.5852897491	4.43734770774608\\
1420.81592269045	4.71718017559715\\
1422.34347594622	4.83710885265691\\
1423.94751370128	4.88707869792552\\
1426.19560266365	4.71718017559715\\
1428.60979758047	4.37738360749245\\
1431.09871822975	4.03758656283571\\
1437.86044196344	3.42795106703716\\
1441.86887787668	3.17810041103712\\
1446.255616981	2.94823778845512\\
1450.9588931944	2.78833328283599\\
1455.80672248377	2.8283093496716\\
1460.69730286252	2.81831533296281\\
1465.43001586017	2.85829139979842\\
1469.84367717743	2.96822582187315\\
1473.91337048105	3.11813607250724\\
1477.66024746199	3.23806451129121\\
1481.11668395972	3.37798098349276\\
1484.31011085391	3.37798098349276\\
1487.28036350008	3.30802262825364\\
1490.07587043458	3.29802861154485\\
1492.89586925388	3.3479986950897\\
1495.71564422412	3.32801066167167\\
1498.59188574084	3.32801066167167\\
1501.56540878698	3.3479986950897\\
1504.66130703499	3.31801664496288\\
1507.87603968043	3.1481183609103\\
1511.23530214949	3.06816598896285\\
1514.66419139048	3.08815402238042\\
1517.99974105528	2.96822582187315\\
1521.29606597005	3.18809442774636\\
1524.03472451903	3.54787926754489\\
1526.65642997041	3.5778613176717\\
1530.08603088581	3.5778613176717\\
1533.26270094537	3.36798672850728\\
1535.9948423756	3.35799271179849\\
1538.2756223564	3.52789123412686\\
1541.91900194038	3.58785533438049\\
1545.17501248125	3.58785533438049\\
1547.53942464668	3.92765237903723\\
1550.85749737338	4.72717443058218\\
1554.1393700291	5.9864214890149\\
1557.54289897974	7.58546606865684\\
1560.91620700227	9.63424163847776\\
1564.67218703588	11.7329865770148\\
1567.13765310279	13.2720665798506\\
1570.10788413885	14.8111465826864\\
1573.19216744606	16.2802696599406\\
1576.49784289217	17.2496895189852\\
1579.79353836734	18.0292242519404\\
1583.81048581972	18.6788355762988\\
1588.93077516874	19.2085177470449\\
1596.50638075899	19.6082803216118\\
1605.74137287828	19.9280888562985\\
1617.46802322232	20.4078016583285\\
1630.61906671518	20.8975087153431\\
1645.32742968781	21.367227262388\\
1661.50765835593	21.7869764407164\\
1678.81576156319	22.0268347479405\\
1696.83656319705	22.1067868816122\\
1715.52005332565	22.1667505053133\\
1734.66893017253	22.0368270967165\\
1754.20316810515	22.0468213517015\\
1774.05728370381	22.1067868816122\\
1796.47208218557	22.1567562503283\\
1817.40147652274	22.2167198740299\\
1838.82195321404	22.2966720077015\\
1860.64896573525	22.3666298863877\\
1882.70741501413	22.3766241413732\\
1904.84980843035	22.3266547726566\\
1924.78545913674	22.2067256190448\\
1946.83428855095	22.1867390152838\\
1971.08413541323	22.25669689397\\
1993.19415244857	22.2655360340577\\
2015.36947624988	22.3366490276417\\
2037.61520599569	22.3966126513433\\
2059.93209624467	22.4166011613133\\
2082.29271717081	22.3266547726566\\
2127.0196843924	22.3866183963582\\
2149.27679241013	22.3566375376122\\
2191.39545162054	21.8769228293731\\
2213.58293272762	21.8169592056715\\
2258.09438645579	22.0368270967165\\
2280.4218761756	22.1867390152838\\
2302.79078399245	22.3066662626866\\
2325.1569295323	22.3966126513433\\
2347.50149878136	22.2966720077015\\
2369.80999794865	21.9868577279999\\
2392.00750032656	21.6670491933132\\
2413.97416826017	21.3872157723581\\
2435.81220584437	21.367227262388\\
2457.73943907982	21.4571736510447\\
2479.80334054893	21.5571142946865\\
2502.00709815679	21.7569955819704\\
2524.33300598376	22.0168404929555\\
2546.74147682599	21.9268941042988\\
2569.11018390288	21.6470606833432\\
2591.21017960957	21.347240658627\\
2612.98992030023	20.9674665940297\\
2634.42569387182	20.7775814679403\\
2655.69910365402	20.6976274280596\\
2677.07288665349	20.5577116706863\\
2698.61689471621	20.0580103586867\\
2720.36194645102	19.3484354106267\\
2742.32663094652	18.3690193903881\\
2764.29781969505	17.4095937863285\\
2786.0029068882	16.3802083973728\\
2807.37106106198	15.290860337821\\
2828.39012498162	14.1115649365074\\
2849.19520910449	12.7923537778211\\
2869.9406634784	11.5830765646565\\
2890.61922905358	10.3638050965073\\
2910.97967553778	9.03459968283596\\
2930.62335252823	7.92526263676109\\
2949.3057105159	6.93585426774644\\
2966.98528318198	5.89647557691023\\
2983.62343885334	5.28683984283589\\
2999.18606669837	4.80712656425385\\
3013.58221253518	4.34740131908939\\
};
\addlegendentry{\scriptsize MPC-U}

\addplot [color=mycolor3, dashdotted, line width=1.3pt]
  table[row sep=crcr]{%
196.324919218047	22.25669689397\\
218.754487806495	22.2766854039405\\
241.148340362229	22.2866777527161\\
263.513959682334	22.246702638985\\
285.846424061699	22.2367083839999\\
308.154172708307	22.246702638985\\
330.437228852822	22.246702638985\\
352.711668533617	22.2367083839999\\
374.987866956727	22.2666911489555\\
397.277001759985	22.25669689397\\
419.574976938005	22.3466432826272\\
441.888025120895	22.3466432826272\\
486.485510338253	22.3866183963582\\
506.524974725858	22.3366490276417\\
528.803309837383	22.3566375376122\\
551.0694876583	22.3466432826272\\
575.575539124666	22.3466432826272\\
595.629544610726	22.3066662626866\\
617.913403121842	22.246702638985\\
662.503274786523	22.246702638985\\
684.794168352287	22.1167792303881\\
709.333929970231	22.0668098616716\\
729.421645596276	22.1767447602988\\
753.978486142169	22.2866777527161\\
776.310348766963	22.1967313640594\\
798.627161608593	22.00684623797\\
818.700536741053	21.8069649506865\\
843.217187316065	21.7170185620298\\
863.280765712897	21.8969113393432\\
885.518532940517	22.1067868816122\\
909.883371532103	22.2666911489555\\
929.871468334519	22.3566375376122\\
952.176100642043	22.1767447602988\\
976.677076403129	21.7370070719999\\
998.748313240179	19.2684832769555\\
1020.60639642322	16.2702754049551\\
1042.33170588054	12.76237196597\\
1064.17173047093	9.25446852698497\\
1086.20531521272	5.78468164773312\\
1108.43775961458	5.36679197650756\\
1130.77474445929	3.80772394025371\\
1152.9732781303	1.76894262541782\\
1174.8984036784	1.6016140552224\\
1194.63210744447	1.65900832248144\\
1208.04179620409	2.46852450987308\\
1223.97943859488	3.1481183609103\\
1233.37993561808	2.592878632066\\
1239.49471564163	2.55926853496203\\
1244.80855675755	2.88356300506985\\
1249.0293904948	3.10814205579845\\
1250.96880304178	5.16691164232816\\
1252.54538660831	6.98582411301504\\
1254.17218075641	7.27565131240317\\
1256.32318877133	8.09516163564194\\
1259.60950753713	9.3244264056716\\
1262.22648084858	9.77415739585058\\
1264.74009398189	8.79474328182096\\
1267.30509795007	6.58642017469356\\
1270.19231648107	4.20377286693565\\
1274.88911286671	2.66590647948988\\
1280.96679600689	2.15870975361941\\
1288.8677957207	1.48910991929097\\
1295.90197115681	0.869480287645729\\
1305.72178696488	0.249850656000035\\
1315.40001195593	0\\
1345.31485836193	0\\
1345.31930280993	4.90706673134309\\
1345.55470964129	5.81652344323902\\
1346.21857821967	6.24626639999997\\
1347.81676118525	6.55608115625364\\
1349.89678484581	6.73079951171712\\
1352.71284169814	6.95584230116401\\
1356.03703643644	7.29563934582075\\
1359.89858199729	7.59545984708939\\
1364.67756361918	7.03579443483568\\
1369.54480909593	4.85709688607449\\
1375.08335937516	2.93824377174633\\
1381.21791024547	2.05762715789569\\
1387.53916350378	1.27635940423625\\
1393.46797368033	0.786463224490035\\
1400.32755745544	0.338369180763493\\
1407.48895991034	0.146483876138518\\
1414.9376865423	0.161903231521592\\
1422.28636991596	0.321807650340361\\
1427.75175264314	0.339796895734253\\
1431.21673581504	0.483710869380502\\
1433.4689440757	0.723995807220945\\
1435.10613674313	1.16501789589393\\
1436.09299896963	1.49182258777728\\
1436.70881690885	2.15585428453278\\
1436.99498405903	2.54062420184482\\
1437.01887296703	3.32415579602002\\
1437.28313382613	3.47792115058201\\
1437.58379552263	3.53374493035199\\
1438.00153285948	3.81457709601682\\
1438.58891694001	4.2978596424532\\
1439.5669620871	4.78556818205743\\
1440.80191490404	5.06068952187888\\
1442.31437184981	5.04626938945967\\
1444.2610432925	4.80576071391488\\
1447.02533139618	4.76515188469693\\
1453.55096777047	4.00018029410421\\
1456.8233937526	3.52882591433763\\
1460.74888862032	3.16339476932308\\
1465.60143896365	2.25305037364342\\
1470.6846230685	1.87607285600961\\
1480.82665342144	1.88750485644869\\
1485.72095191673	1.98262780139839\\
1489.42051080974	2.02051881218176\\
1494.58447601947	2.46852441456303\\
1498.15636018032	3.02319285420344\\
1501.49643031772	3.41597236881989\\
1504.84180000193	3.74775984000007\\
1507.57567163568	3.74421586244353\\
1509.25639447377	3.60570187788835\\
1511.30562374138	2.86614393510672\\
1513.16619125893	2.15014336848435\\
1515.18253600082	2.27928500587859\\
1517.50322283147	2.56989238512551\\
1520.16325846803	2.96822582187315\\
1523.1426482464	2.98321684693656\\
1525.66526848057	3.08315710763463\\
1529.60122598433	3.16810639432833\\
1532.72777614831	3.83147458960957\\
1537.45574254975	3.93764645531519\\
1540.64322283395	4.10611720140196\\
1542.76297709475	4.28636655897935\\
1545.928728653	4.63580064862754\\
1548.17136694985	4.77714475240282\\
1550.92170531873	5.82651722167157\\
1553.61231164591	7.05256999511175\\
1555.65855133525	8.7972413690145\\
1559.06000077043	10.3638050965073\\
1562.39920491425	12.6224552554927\\
1566.46246858572	14.9710508500298\\
1570.26810246036	17.0298216279402\\
1573.38024951476	18.6788355762988\\
1578.67985191989	20.4081585279273\\
1583.44365773564	21.987571671435\\
1589.67453261702	22.7746010063665\\
1596.96943083263	22.9962546069851\\
1607.42653605283	22.4665705300299\\
1619.39775993867	22.4465820200594\\
1633.67431713871	22.4565762750449\\
1650.17130790951	22.3766241413732\\
1668.39408460587	22.3134481466482\\
1688.19934648841	22.2337101075045\\
1709.94223817802	22.25669689397\\
1754.79235938254	22.2067256190448\\
1777.29288469579	22.2267141290149\\
1799.67435121921	22.2367083839999\\
1822.12446503649	22.113924382521\\
1844.53785823199	21.9334953068569\\
1864.68080791634	21.8269534606566\\
1887.0275854695	21.5471200397014\\
1909.34236400995	21.3172578936715\\
1931.65965577798	20.9913540048715\\
1976.22958302205	20.5876944356419\\
1998.49276968742	20.4277901682985\\
2020.69044866883	20.6976274280596\\
2042.71478708318	21.2273115050148\\
2064.58145917108	21.9368883592838\\
2086.21928019625	22.2067256190448\\
2107.62370524472	22.236414267129\\
2128.78845492246	22.3366490276417\\
2151.7868717817	22.3366490276417\\
2168.31890473341	22.0568156066865\\
2190.82900158727	21.5371257847164\\
2209.32536374583	21.097390002627\\
2232.66229557739	19.0286268759405\\
2254.97061807439	19.0386192247161\\
2279.22440760581	19.248494766985\\
2301.11044761125	19.8081597026867\\
2323.43404291084	20.6112517505294\\
2345.80902934464	21.1373651163581\\
2365.74294566752	21.7470013269849\\
2393.83612550125	22.2367083839999\\
2412.60353568989	22.1467619953432\\
2431.71749869276	21.9468807080593\\
2491.40865237311	21.2173172500297\\
2509.89989144166	20.9274914802986\\
2533.1691555704	20.6676465693135\\
2554.96873812465	20.387813148358\\
2599.48850290198	19.9580697150445\\
2619.29881547777	19.728207569015\\
2643.2238864406	19.6882324552835\\
2664.7014555181	20.4677652820296\\
2685.92583494254	21.3972100273431\\
2706.8856223447	22.0168404929555\\
2727.60811930466	21.8360898670262\\
2748.0776274254	20.9774608490147\\
2768.2941467113	19.5483147917012\\
2788.26084096971	18.2690806529554\\
2807.99923817975	16.7899633207162\\
2827.78920314994	15.2708718278504\\
2848.28498495885	13.9416664141791\\
2869.84860998692	12.4325682231938\\
2892.03531217434	10.5037218069851\\
2914.05357199653	9.02460542785093\\
2935.14359577117	7.55548378025378\\
2954.77041804587	6.73597345701501\\
2973.00869242439	5.90646935534323\\
2989.74885142919	5.27684606440289\\
3005.03872102302	3.41103249681828\\
};
\addlegendentry{\scriptsize WIE}

\addplot [color=mycolor4, dotted, line width=1.3pt]
  table[row sep=crcr]{%
199.041650893305	21.6870377032837\\
220.073678265228	21.7170185620298\\
241.244156810219	21.7769840919404\\
262.521890950029	21.8569362256121\\
283.89476997527	21.7969706957015\\
303.206239169029	21.7969706957015\\
326.926046868143	21.8469419706271\\
348.593840310409	21.7170185620298\\
370.318617830417	21.8569362256121\\
392.09115232277	21.8969113393432\\
413.845096224118	22.0668098616716\\
437.737890889835	21.8569362256121\\
481.172480360261	21.7170185620298\\
502.912284376304	21.7569955819704\\
524.680522257661	21.7070243070448\\
546.500883536622	21.7669898369554\\
568.351059513288	21.7370070719999\\
590.241117350087	21.7470013269849\\
612.13783232353	21.7270128170148\\
634.008072796331	21.7470013269849\\
655.845358941192	21.7470013269849\\
677.656571754413	21.8369477156421\\
699.435007558284	21.8469419706271\\
721.213443365717	21.8169592056715\\
742.997780493888	21.7969706957015\\
764.788975353404	21.7869764407164\\
786.581173151497	21.6970300520597\\
808.362146809367	21.6770434482987\\
830.126242126343	21.6170779183881\\
851.882322720679	21.367227262388\\
873.641967361353	20.9874551040002\\
895.388629111146	20.5477174157013\\
917.126497045493	19.7082190590445\\
938.852007094875	18.339038531642\\
960.575333950035	16.0803883726567\\
982.298259622277	13.681821503194\\
1004.02276675476	11.1633273863285\\
1025.74769829604	9.504319182985\\
1047.46888846606	8.58486869265653\\
1069.12417002562	8.35500559352249\\
1090.57823130996	7.99522099200021\\
1111.68282446664	7.75536459098521\\
1132.31811213421	7.77535262440279\\
1152.15597836695	7.65542394734348\\
1170.65121177223	7.60545410207442\\
1186.84030824261	7.81532869123885\\
1201.54400073148	7.92526263676109\\
1211.53090215918	7.63543591392545\\
1221.09122591403	5.80652918825353\\
1229.71371775064	3.32801066167167\\
1237.95594704055	1.03937869083575\\
1245.9891419749	0\\
1284.71608885356	0\\
1292.49978907371	0.179892479468435\\
1300.80147046186	0\\
1309.98676255573	0\\
1309.98676255573	0.199880527778532\\
1309.98676255573	0\\
1309.99231811573	8.7247844500298\\
1310.2842514498	9.02460542785093\\
1311.31896372433	8.94465329417926\\
1313.3140637189	8.92466573731326\\
1315.86301496516	7.97523295858218\\
1320.04757744957	5.71658279959684\\
1324.93032041567	2.05876934825392\\
1330.78344844562	0.799522111113674\\
1337.51807263583	0.239856624398726\\
1345.0885494214	0.249850656000035\\
1353.32839492901	0.279832735911441\\
1362.04598507956	0.20987454448732\\
1370.04010053505	0\\
1405.23047272699	0\\
1405.23047272699	6.256260654985\\
1405.27658387499	6.38618263392527\\
1406.05766766245	6.11634394450721\\
1407.54442918901	5.7465650879999\\
1409.87392429091	5.03698918683585\\
1413.04278951331	4.49731180800018\\
1416.99789849323	3.83770599038053\\
1421.68060150614	3.25805254470879\\
1426.92655136348	3.25805254470879\\
1432.65730419588	2.09874541508952\\
1438.79265740894	1.48910991929097\\
1445.10979138822	1.37917561635459\\
1451.44300135229	1.33919954951853\\
1457.50890511565	1.33919954951853\\
1463.10198586864	1.57905618880977\\
1468.11298208034	1.8389008615186\\
1472.50464283724	1.91885299519026\\
1476.27068889156	2.2586499207091\\
1479.46858972642	2.64841704891069\\
1482.14933501333	3.07816000567163\\
1484.1997445494	3.44793910045519\\
1485.71300384874	3.98761647929086\\
1487.10504494995	4.18749705174605\\
1488.50454554919	4.20748508516408\\
1490.00702841677	4.16750901832847\\
1491.64346526449	3.95763442916405\\
1493.32029265951	3.6678077063284\\
1497.30270177835	3.04817795554482\\
1499.9658129069	2.87827967149269\\
1503.10882864412	2.80832131625357\\
1506.71251011929	2.77833926612675\\
1510.72646945183	2.83830336638039\\
1514.90223810429	2.9182557383283\\
1519.0322499992	3.10814205579845\\
1522.97584746796	3.26804656141803\\
1526.10807101565	3.48791516729125\\
1529.76969754799	3.70778377316401\\
1533.35832480938	3.77774189012689\\
1536.0381750883	3.87768229549238\\
1538.59454219878	3.95763442916405\\
1541.5347403078	4.07756262967177\\
1544.2205819987	4.95703705316419\\
1547.23731769904	5.92645738876126\\
1549.7140958079	6.46613476759694\\
1552.92417744615	7.28564509083571\\
1555.9753478776	8.46494049214925\\
1559.28905516754	9.84411622764173\\
1562.8737414385	11.2532737749852\\
1565.97282308456	12.8823001664778\\
1569.40575512532	14.3614155925075\\
1573.22492742685	15.6306573824777\\
1577.05531258741	16.6800303282985\\
1581.11489921004	17.7393965760002\\
1587.06862402465	18.5788949326566\\
1593.16967494964	19.3284469006567\\
1601.3464523738	19.8381424676418\\
1609.89627248455	20.2179146260296\\
1620.34896111692	20.627669549373\\
1632.03366349853	20.9474780840596\\
1645.18350245751	21.107384257612\\
1659.73791690052	21.367227262388\\
1692.4674829843	21.5771028046565\\
1710.3319512086	21.5970913146271\\
1729.04817107007	21.6570549383282\\
1748.47590889772	21.6870377032837\\
1768.46217181532	21.7070243070448\\
1788.88187075541	21.6970300520597\\
1809.66573185634	21.7370070719999\\
1830.72769930451	21.6770434482987\\
1851.98932921247	21.7070243070448\\
1873.39254974005	21.6370664283581\\
1894.92150992336	21.6470606833432\\
1916.55460937824	21.6970300520597\\
1938.26811184649	21.7869764407164\\
1960.02574644466	21.6970300520597\\
1981.82907094286	21.6870377032837\\
2003.65512790023	21.7270128170148\\
2025.50138003941	21.6570549383282\\
2047.32213731357	21.6070855696121\\
2069.09778286032	21.5970913146271\\
2090.81887340233	21.6370664283581\\
2112.48000509357	21.8269534606566\\
2134.11772964079	21.8669285743881\\
2155.75685944862	21.6870377032837\\
2177.43072645513	21.6170779183881\\
2199.14206887706	21.4571736510447\\
2220.89166557847	21.4371870472837\\
2242.68732958253	21.5071449259704\\
2264.50376677781	21.5571142946865\\
2286.31354724817	21.6770434482987\\
2308.11571549874	21.7370070719999\\
2329.95159306459	21.7270128170148\\
2351.80392377504	21.7569955819704\\
2373.63407597548	21.7669898369554\\
2395.3640315781	21.8469419706271\\
2416.98944640572	21.8469419706271\\
2438.54621390606	21.8069649506865\\
2460.07258840159	21.7370070719999\\
2481.57288994732	21.6070855696121\\
2503.10963898436	21.5371257847164\\
2524.69820479887	21.5970913146271\\
2546.34562150309	21.5970913146271\\
2568.05342274896	21.6570549383282\\
2589.80985279186	21.5671085496715\\
2611.5866787083	21.5571142946865\\
2633.35958606263	21.8169592056715\\
2655.10621969158	21.7969706957015\\
2676.81010237869	21.1973287400597\\
2698.46553283719	21.2572942699703\\
2720.0921039194	20.8975087153431\\
2748.16504041581	20.4877537920001\\
2769.74848312733	19.9880524800001\\
2791.31685384545	19.3084583906866\\
2812.87032917451	17.879312333373\\
2834.41222546661	16.8699154543879\\
2855.97235752977	15.1409493723581\\
2877.54680685192	13.0621919906866\\
2899.07476327056	10.893488220358\\
2922.57566483748	7.62544213549245\\
2943.63647596568	5.66661295432823\\
2964.32486215968	4.24746115200014\\
2982.52882488928	2.96822582187315\\
3004.05266197719	2.19868582045501\\
};
\addlegendentry{\scriptsize IDM}

\addplot[area legend, draw=none, fill=white!10!black, fill opacity=0.2]
table[row sep=crcr] {%
x	y\\
1454.4	0\\
1454.4	24\\
1532	24\\
1532	0\\
}--cycle;
\addlegendentry{\scriptsize U-Turn}

\end{axis}
%

%% file: Images/US06_pout_Nissan.tex
%
%
\definecolor{mycolor1}{rgb}{0.00000,0.44700,0.74100}%
\definecolor{mycolor2}{rgb}{0.85000,0.32500,0.09800}%
\definecolor{mycolor3}{rgb}{0.92900,0.69400,0.12500}%
\definecolor{mycolor4}{rgb}{0.49400,0.18400,0.55600}%
\begin{tikzpicture}

\begin{axis}[%
width=6.5cm,
height=4.0cm,
at={(0cm,8.8cm)},
scale only axis,
xmin=505,
xmax=2800,
xlabel style={font=\color{white!15!black} \scriptsize},
xlabel={},
xticklabels={,,},
ymin=0,
ymax=65,
ylabel style={font=\color{white!15!black} \scriptsize},
ylabel={Power Out [KW]},
ytick={0,20,40,60},
yticklabels={0,+20,+40,+60},
ticklabel style = {font = \scriptsize},
axis background/.style={fill=white},
xmajorgrids,
ymajorgrids,
legend style={legend cell align=left, align=left, draw=white!15!black}
]

\addplot [color=mycolor1, line width=1.3pt]
  table[row sep=crcr]{%
503.6	13.6\\
506.5	12.4\\
510.6	13.6\\
516.2	12.5\\
518.6	13.4\\
523.9	12.7\\
526.8	13.2\\
531.9	12.8\\
536.5	12.8\\
538.7	11.9\\
542.7	12.1\\
547.2	11.6\\
552.4	12.4\\
555.5	11.7\\
559.3	11.4\\
565.5	11.1\\
568.5	11.1\\
574.2	10.8\\
576.3	11.0\\
581.1	10.6\\
584.9	10.9\\
589.4	11.0\\
593.8	11.0\\
599.0	10.8\\
603.3	10.6\\
607.2	10.8\\
610.7	10.5\\
614.4	10.5\\
618.6	10.3\\
623.7	10.7\\
628.6	10.1\\
632.4	10.8\\
637.8	10.3\\
641.7	10.6\\
645.1	10.2\\
650.1	10.7\\
654.7	10.6\\
659.2	10.9\\
661.2	10.8\\
668.2	10.9\\
677.6	10.5\\
682.4	11.3\\
685.5	10.9\\
690.8	11.3\\
695.2	10.5\\
698.1	11.3\\
703.9	11.0\\
708.1	11.3\\
712.6	10.7\\
716.9	11.1\\
719.6	10.7\\
725.6	11.0\\
730.0	10.7\\
734.4	11.3\\
738.9	11.8\\
742.2	12.1\\
747.8	12.0\\
751.8	12.3\\
757.0	12.1\\
761.4	12.3\\
765.9	12.4\\
770.0	12.2\\
775.3	12.6\\
779.0	12.0\\
783.8	12.1\\
788.7	12.3\\
792.9	12.1\\
797.6	11.5\\
801.9	11.9\\
806.4	11.6\\
811.0	10.5\\
815.5	10.1\\
820.1	10.2\\
824.6	10.1\\
829.3	8.0\\
833.7	6.6\\
838.5	9.1\\
842.4	9.9\\
847.1	6.1\\
851.6	6.3\\
856.3	6.3\\
861.0	6.6\\
865.4	6.4\\
870.3	6.3\\
874.6	6.5\\
878.6	6.1\\
884.0	6.0\\
887.9	4.9\\
892.9	4.2\\
896.9	4.8\\
901.3	3.7\\
906.0	0.5\\
910.2	0.2\\
914.9	1.9\\
919.1	3.8\\
923.5	1.0\\
928.1	0.3\\
932.3	0.8\\
936.6	1.0\\
941.5	0.4\\
945.0	1.1\\
949.8	0.0\\
954.0	0.8\\
958.3	0.0\\
1232.8	0.0\\
1233.5	0.3\\
1234.0	0.3\\
1235.0	0.5\\
1236.5	0.5\\
1237.0	0.7\\
1240.1	0.7\\
1240.5	0.8\\
1240.9	0.8\\
1241.3	0.9\\
1241.7	0.8\\
1242.1	1.0\\
1242.5	0.9\\
1242.8	1.4\\
1243.2	0.9\\
1243.7	0.8\\
1244.0	1.1\\
1244.4	1.4\\
1244.8	2.8\\
1245.2	3.8\\
1245.7	4.2\\
1246.3	4.3\\
1246.9	4.5\\
1247.5	4.6\\
1248.5	4.9\\
1249.0	4.8\\
1249.7	4.8\\
1250.6	4.7\\
1251.4	4.9\\
1253.6	5.1\\
1254.2	5.0\\
1255.3	5.1\\
1256.7	5.2\\
1257.2	5.5\\
1259.6	5.5\\
1261.0	5.8\\
1262.4	5.7\\
1263.1	5.7\\
1264.9	5.9\\
1270.0	5.9\\
1271.7	6.0\\
1273.2	6.0\\
1274.5	5.8\\
1276.1	5.8\\
1277.4	5.9\\
1279.2	5.5\\
1280.5	5.4\\
1283.2	5.3\\
1283.4	5.6\\
1285.8	5.4\\
1288.3	4.8\\
1289.3	4.6\\
1291.7	4.4\\
1293.7	4.4\\
1294.8	4.7\\
1296.9	4.5\\
1298.5	4.5\\
1300.0	4.0\\
1301.6	3.5\\
1305.0	3.5\\
1307.5	3.6\\
1309.3	3.8\\
1311.1	3.4\\
1312.4	3.4\\
1314.7	3.5\\
1316.2	2.8\\
1318.4	2.9\\
1319.6	2.7\\
1321.5	2.2\\
1324.0	2.1\\
1325.8	1.8\\
1327.5	1.7\\
1329.5	0.7\\
1331.2	0.0\\
1333.1	0.9\\
1335.0	1.6\\
1336.6	0.7\\
1338.5	0.8\\
1340.4	0.6\\
1342.1	0.7\\
1344.0	1.6\\
1345.0	1.5\\
1348.7	2.7\\
1350.8	4.8\\
1352.6	5.2\\
1357.5	5.2\\
1359.0	5.3\\
1361.9	5.1\\
1364.1	5.5\\
1366.3	5.3\\
1367.9	5.4\\
1370.6	5.7\\
1372.1	5.6\\
1375.0	5.9\\
1376.4	6.2\\
1378.3	5.9\\
1380.5	5.9\\
1383.7	5.3\\
1385.7	3.3\\
1388.0	3.4\\
1389.6	3.0\\
1391.7	2.8\\
1393.5	2.0\\
1396.3	1.9\\
1398.1	1.0\\
1399.8	0.0\\
1402.3	0.3\\
1405.0	1.5\\
1406.9	1.4\\
1409.3	1.5\\
1410.7	1.6\\
1413.5	1.5\\
1415.1	1.7\\
1417.4	1.2\\
1419.3	1.0\\
1421.3	1.3\\
1422.9	1.7\\
1424.9	2.0\\
1426.9	20.3\\
1429.1	20.5\\
1431.5	20.6\\
1435.9	21.1\\
1438.3	21.6\\
1440.7	21.7\\
1443.2	22.3\\
1445.6	22.7\\
1448.0	23.2\\
1450.5	23.6\\
1453.7	23.9\\
1456.2	25.6\\
1459.1	25.9\\
1462.0	25.7\\
1464.3	26.0\\
1467.8	25.9\\
1470.7	26.3\\
1473.6	26.4\\
1476.9	27.1\\
1480.0	27.8\\
1483.3	27.8\\
1486.6	28.2\\
1489.8	28.5\\
1493.3	28.4\\
1496.3	28.8\\
1500.2	29.3\\
1503.4	29.4\\
1507.2	29.7\\
1511.0	29.9\\
1515.0	30.1\\
1518.7	30.5\\
1522.3	30.4\\
1526.5	30.9\\
1530.8	31.3\\
1534.8	31.2\\
1538.3	31.4\\
1542.8	31.4\\
1546.3	31.3\\
1551.0	32.0\\
1555.2	31.1\\
1559.1	30.3\\
1563.5	30.2\\
1567.4	30.7\\
1572.2	30.7\\
1581.9	28.7\\
1585.9	29.0\\
1595.6	29.1\\
1600.2	28.7\\
1604.8	28.0\\
1609.3	27.7\\
1614.0	26.9\\
1618.7	26.3\\
1623.3	23.9\\
1628.2	23.9\\
1632.7	25.0\\
1637.7	25.3\\
1642.6	25.1\\
1647.2	24.0\\
1652.2	23.7\\
1656.7	23.6\\
1662.0	23.7\\
1666.9	22.5\\
1671.6	22.0\\
1676.8	21.8\\
1681.5	18.1\\
1686.5	17.9\\
1691.4	18.1\\
1696.0	17.9\\
1701.0	18.5\\
1706.4	17.7\\
1711.9	18.4\\
1716.4	16.2\\
1721.0	16.3\\
1726.4	15.7\\
1730.9	12.6\\
1736.5	11.5\\
1741.7	12.5\\
1746.6	13.7\\
1751.6	12.5\\
1756.5	11.7\\
1761.5	11.3\\
1766.6	12.0\\
1771.4	11.5\\
1776.6	11.2\\
1781.0	10.6\\
1786.8	10.4\\
1791.6	9.8\\
1796.4	8.6\\
1801.6	7.6\\
1807.0	8.0\\
1811.6	7.7\\
1816.9	7.7\\
1821.5	7.4\\
1826.9	8.0\\
1832.2	7.6\\
1836.2	6.3\\
1841.9	5.4\\
1846.1	6.1\\
1851.4	6.2\\
1856.5	6.6\\
1861.3	5.9\\
1866.4	6.5\\
1871.2	6.3\\
1875.9	6.4\\
1880.8	6.2\\
1885.6	6.3\\
1895.3	6.3\\
1900.2	6.1\\
1905.0	6.5\\
1909.8	6.0\\
1919.6	5.9\\
1924.6	6.1\\
1929.1	6.2\\
1934.7	5.9\\
1938.6	5.9\\
1943.2	5.6\\
1948.1	6.4\\
1952.9	5.9\\
1957.7	6.3\\
1962.0	5.9\\
1966.5	5.1\\
1972.4	5.7\\
1976.5	5.5\\
1981.5	5.6\\
1985.4	5.5\\
1990.7	5.4\\
1995.4	5.6\\
2000.0	5.3\\
2004.6	5.4\\
2009.2	6.2\\
2018.5	6.2\\
2023.1	6.1\\
2027.6	6.1\\
2032.3	6.0\\
2036.6	6.2\\
2042.8	6.4\\
2046.3	9.3\\
2050.8	8.9\\
2055.4	6.3\\
2060.0	6.2\\
2064.4	6.4\\
2069.2	7.0\\
2073.7	8.1\\
2079.1	8.2\\
2083.7	8.5\\
2088.2	7.1\\
2092.7	6.2\\
2096.8	6.0\\
2101.1	8.1\\
2106.0	7.6\\
2110.4	8.6\\
2115.5	7.4\\
2121.1	8.4\\
2124.7	8.1\\
2128.4	7.6\\
2134.5	8.4\\
2137.6	7.8\\
2142.0	8.4\\
2149.3	8.0\\
2154.2	8.2\\
2157.1	7.9\\
2161.5	7.7\\
2166.5	8.2\\
2171.2	7.8\\
2175.1	8.2\\
2179.4	7.6\\
2183.9	8.4\\
2188.4	7.5\\
2193.5	8.8\\
2197.5	7.4\\
2202.0	8.6\\
2208.3	6.6\\
2211.2	6.8\\
2215.6	6.5\\
2220.1	8.0\\
2225.8	7.7\\
2230.8	7.7\\
2233.2	8.0\\
2238.9	7.9\\
2242.5	7.9\\
2248.4	7.9\\
2253.0	7.6\\
2257.9	7.8\\
2262.8	7.6\\
2267.3	8.0\\
2271.8	8.1\\
2276.5	7.6\\
2280.5	7.9\\
2283.3	7.8\\
2287.0	8.0\\
2298.3	8.1\\
2301.4	7.7\\
2306.1	8.1\\
2309.4	7.6\\
2315.3	8.1\\
2319.5	7.7\\
2324.1	8.2\\
2328.5	7.8\\
2333.1	8.9\\
2336.2	9.5\\
2340.1	9.1\\
2346.4	8.5\\
2349.2	8.6\\
2355.1	8.7\\
2359.8	8.6\\
2363.9	8.8\\
2368.5	8.7\\
2372.9	8.8\\
2375.9	8.2\\
2381.7	8.7\\
2384.5	8.3\\
2391.2	9.1\\
2394.4	8.2\\
2399.9	9.2\\
2402.6	8.5\\
2408.5	8.8\\
2413.2	8.5\\
2417.4	8.4\\
2421.8	8.6\\
2427.2	8.2\\
2436.4	8.4\\
2439.9	9.2\\
2448.0	9.0\\
2451.7	9.9\\
2458.7	9.6\\
2463.5	9.1\\
2474.2	9.4\\
2476.5	8.5\\
2483.6	8.5\\
2490.0	8.6\\
2494.9	9.0\\
2499.2	9.4\\
2502.9	9.4\\
2507.9	9.0\\
2510.6	8.9\\
2516.8	9.0\\
2521.0	9.1\\
2525.2	8.8\\
2529.8	8.8\\
2538.0	9.3\\
2540.2	9.0\\
2545.3	8.9\\
2548.2	9.0\\
2554.1	8.8\\
2557.6	9.2\\
2563.4	8.5\\
2567.8	9.4\\
2572.2	8.6\\
2576.4	9.2\\
2580.8	9.1\\
2585.2	9.2\\
2589.6	9.1\\
2594.0	8.6\\
2598.4	8.7\\
2602.8	8.5\\
2607.3	9.3\\
2611.7	8.9\\
2616.1	9.1\\
2620.6	8.7\\
2625.0	8.6\\
2629.4	8.3\\
2633.9	8.2\\
2638.3	6.2\\
2642.7	6.2\\
2647.1	6.3\\
2651.6	5.9\\
2656.0	5.3\\
2660.4	4.8\\
2665.7	3.8\\
2669.5	1.0\\
2674.1	0.7\\
2682.6	0.9\\
2686.9	1.3\\
2691.0	1.2\\
2695.3	0.9\\
2699.6	0.3\\
2703.8	0.3\\
2708.0	0.1\\
2712.1	0.0\\
2812.1	0.0\\
};
\addlegendentry{\scriptsize MPC-C }

\addplot [color=mycolor2, dashed, line width=1.3pt]
  table[row sep=crcr]{%
500.9	5.1\\
505.3	9.3\\
509.6	9.2\\
513.4	7.1\\
517.3	5.9\\
521.1	7.7\\
525.1	8.9\\
528.8	7.9\\
533.5	5.5\\
536.5	4.8\\
541.6	4.8\\
544.4	4.9\\
548.0	0.9\\
551.7	0.0\\
559.2	0.0\\
562.8	0.9\\
566.7	1.4\\
570.2	2.2\\
575.6	2.4\\
578.0	4.3\\
582.8	4.8\\
585.3	5.0\\
589.0	5.7\\
592.9	5.3\\
596.7	5.3\\
600.1	5.1\\
604.7	4.9\\
608.9	4.6\\
611.6	4.6\\
615.6	4.3\\
618.9	5.0\\
623.6	4.4\\
626.3	5.0\\
630.3	4.6\\
634.3	5.5\\
637.4	5.4\\
641.0	5.6\\
644.6	5.2\\
648.8	6.6\\
652.3	8.8\\
657.6	7.0\\
661.4	7.1\\
665.2	10.6\\
669.1	9.7\\
672.5	10.2\\
677.8	10.8\\
681.0	14.2\\
684.1	15.0\\
689.1	14.3\\
692.8	14.7\\
695.5	15.2\\
700.3	16.1\\
703.7	16.6\\
707.2	19.8\\
711.9	20.4\\
715.4	20.2\\
720.2	19.9\\
728.2	19.8\\
731.5	18.2\\
737.2	19.8\\
741.1	20.3\\
745.3	20.1\\
748.9	20.3\\
760.6	20.3\\
765.5	20.5\\
771.7	20.8\\
775.7	20.8\\
778.6	21.7\\
784.3	22.1\\
787.5	22.3\\
791.7	22.5\\
796.0	22.2\\
801.0	22.3\\
812.0	22.3\\
817.0	22.2\\
821.0	22.4\\
826.3	22.0\\
831.3	22.5\\
835.7	21.8\\
840.6	21.3\\
844.7	15.7\\
849.8	15.8\\
853.3	16.6\\
859.3	16.5\\
864.4	16.3\\
869.3	15.7\\
872.5	15.1\\
878.2	12.7\\
882.0	10.1\\
887.4	11.1\\
893.5	10.6\\
898.2	4.8\\
901.3	1.1\\
907.6	1.7\\
912.6	1.2\\
917.2	0.0\\
1171.6	0.0\\
1176.2	2.7\\
1180.3	0.9\\
1184.6	0.7\\
1186.4	0.5\\
1189.8	0.7\\
1192.5	0.0\\
1239.5	0.0\\
1240.2	0.1\\
1240.7	0.4\\
1241.1	0.5\\
1241.5	0.7\\
1241.9	0.7\\
1242.3	0.8\\
1242.6	1.0\\
1242.9	1.0\\
1243.1	0.9\\
1243.3	0.9\\
1243.4	1.1\\
1243.5	0.9\\
1243.5	1.1\\
1243.5	0.5\\
1243.5	0.6\\
1243.5	0.5\\
1243.5	0.7\\
1243.5	0.4\\
1243.5	0.7\\
1243.5	0.4\\
1243.5	0.6\\
1243.5	0.5\\
1243.5	0.6\\
1243.5	0.5\\
1243.6	0.8\\
1243.7	1.8\\
1243.9	2.4\\
1244.1	2.3\\
1244.5	3.8\\
1244.9	5.6\\
1245.5	6.7\\
1246.0	5.9\\
1246.7	5.7\\
1247.4	6.4\\
1248.2	6.9\\
1249.1	7.1\\
1249.9	7.4\\
1250.9	7.8\\
1253.3	8.1\\
1254.5	8.2\\
1255.7	8.5\\
1256.9	8.6\\
1258.2	8.6\\
1260.9	8.4\\
1262.3	7.2\\
1263.7	6.3\\
1265.2	7.6\\
1266.7	8.2\\
1268.3	8.4\\
1269.9	7.4\\
1271.5	6.7\\
1273.1	7.9\\
1274.8	8.5\\
1276.4	6.7\\
1278.2	5.8\\
1279.9	5.9\\
1281.6	5.1\\
1284.0	0.0\\
1294.0	0.0\\
1295.5	4.4\\
1296.9	4.6\\
1298.4	3.4\\
1300.0	5.7\\
1301.6	8.3\\
1303.2	8.3\\
1304.8	7.0\\
1306.5	6.6\\
1308.1	6.6\\
1309.9	6.4\\
1311.6	5.8\\
1313.3	5.8\\
1314.6	5.2\\
1316.7	5.1\\
1318.8	3.4\\
1319.8	3.3\\
1321.7	4.3\\
1324.0	3.6\\
1325.4	2.6\\
1327.0	3.1\\
1329.5	2.7\\
1331.2	2.6\\
1333.3	2.4\\
1334.9	2.2\\
1336.8	2.1\\
1338.6	1.4\\
1340.4	1.5\\
1343.2	1.5\\
1345.6	1.2\\
1347.4	1.4\\
1349.1	1.3\\
1350.8	1.4\\
1352.6	1.9\\
1355.7	2.6\\
1357.4	4.2\\
1359.6	4.6\\
1361.8	3.3\\
1364.0	2.7\\
1366.5	2.8\\
1368.2	3.7\\
1370.9	4.7\\
1372.6	4.6\\
1374.4	3.9\\
1375.7	3.9\\
1378.0	4.1\\
1380.6	3.8\\
1381.3	3.9\\
1383.4	4.0\\
1385.3	3.7\\
1387.3	4.0\\
1389.7	3.8\\
1390.9	3.0\\
1392.7	2.4\\
1394.3	2.6\\
1397.1	2.2\\
1399.5	1.7\\
1401.2	1.5\\
1403.8	1.5\\
1404.8	1.3\\
1406.6	1.3\\
1409.4	1.4\\
1412.1	1.5\\
1413.9	1.4\\
1415.0	1.3\\
1416.9	1.3\\
1418.8	0.8\\
1420.5	1.9\\
1422.6	4.1\\
1425.3	4.5\\
1428.1	25.1\\
1431.0	26.3\\
1432.8	27.7\\
1435.6	28.9\\
1437.5	32.0\\
1440.1	30.0\\
1443.6	29.8\\
1445.2	30.0\\
1448.4	31.5\\
1450.6	33.0\\
1454.7	33.8\\
1456.9	34.6\\
1459.6	35.8\\
1463.3	35.0\\
1465.8	35.5\\
1469.3	36.4\\
1472.9	38.6\\
1477.3	40.9\\
1481.0	40.9\\
1482.9	39.9\\
1487.0	40.4\\
1490.9	39.1\\
1495.7	35.7\\
1497.8	35.5\\
1501.9	39.6\\
1506.1	41.9\\
1511.1	41.9\\
1516.0	38.9\\
1520.4	36.9\\
1523.0	36.7\\
1526.9	32.2\\
1531.6	31.2\\
1537.1	31.1\\
1542.1	32.0\\
1546.1	31.9\\
1550.3	32.2\\
1555.5	31.2\\
1557.8	29.9\\
1564.2	24.2\\
1568.8	23.7\\
1573.7	29.0\\
1576.1	28.7\\
1581.2	25.7\\
1587.7	25.2\\
1590.7	24.4\\
1601.9	22.6\\
1604.9	15.3\\
1611.6	14.5\\
1616.4	14.6\\
1620.7	15.0\\
1625.9	14.4\\
1628.5	13.5\\
1634.3	11.9\\
1638.5	8.0\\
1643.7	7.4\\
1649.4	8.2\\
1654.5	6.2\\
1659.2	5.0\\
1663.9	5.2\\
1667.8	3.8\\
1671.8	3.1\\
1675.9	0.9\\
1682.4	0.0\\
1695.7	0.0\\
1700.8	2.1\\
1705.4	2.7\\
1709.5	2.7\\
1714.0	3.0\\
1718.4	3.2\\
1722.9	4.2\\
1726.8	4.1\\
1731.6	0.0\\
1745.0	0.0\\
1749.3	17.2\\
1751.8	16.7\\
1757.8	18.2\\
1762.3	19.2\\
1766.7	14.0\\
1770.3	12.4\\
1773.8	11.8\\
1779.7	12.5\\
1783.9	14.3\\
1788.5	14.7\\
1793.0	14.7\\
1795.8	14.5\\
1801.8	10.6\\
1806.3	10.4\\
1810.8	10.6\\
1815.3	12.8\\
1819.2	10.9\\
1824.2	10.2\\
1828.8	10.1\\
1833.1	10.8\\
1837.6	14.5\\
1842.2	14.4\\
1846.6	14.4\\
1851.1	11.5\\
1855.7	10.7\\
1860.0	11.0\\
1864.6	11.0\\
1868.6	13.0\\
1872.1	14.8\\
1878.2	14.6\\
1882.7	12.2\\
1887.5	11.1\\
1892.0	10.5\\
1896.2	10.3\\
1901.0	10.0\\
1905.6	9.9\\
1910.5	10.4\\
1915.6	10.1\\
1917.9	10.2\\
1925.5	10.1\\
1934.0	10.2\\
1938.7	9.7\\
1943.1	6.4\\
1947.7	5.0\\
1952.2	7.9\\
1956.8	8.3\\
1961.5	6.4\\
1965.8	5.3\\
1970.3	6.0\\
1975.0	5.8\\
1979.5	5.9\\
1984.0	5.8\\
1988.4	5.2\\
1993.0	9.2\\
1997.0	9.5\\
2001.9	9.5\\
2006.3	8.8\\
2010.7	9.4\\
2015.3	9.2\\
2019.3	9.6\\
2024.3	5.4\\
2028.8	4.7\\
2033.1	4.7\\
2037.7	5.1\\
2042.2	7.8\\
2046.6	9.4\\
2051.1	9.2\\
2055.4	9.2\\
2060.0	9.4\\
2064.5	9.4\\
2068.9	10.3\\
2073.4	10.3\\
2077.7	13.5\\
2082.3	14.6\\
2090.9	11.9\\
2095.9	9.1\\
2099.9	6.8\\
2104.8	6.8\\
2109.2	8.8\\
2113.7	9.5\\
2118.2	9.1\\
2122.5	9.1\\
2127.2	9.8\\
2131.6	9.5\\
2136.1	10.1\\
2140.6	10.0\\
2144.9	10.3\\
2149.6	10.4\\
2153.6	10.8\\
2158.4	10.5\\
2162.9	12.9\\
2167.2	14.0\\
2171.9	13.2\\
2176.4	11.3\\
2180.9	11.8\\
2185.5	14.1\\
2189.7	14.4\\
2194.6	14.1\\
2199.1	11.6\\
2203.7	11.7\\
2208.2	12.7\\
2212.6	13.3\\
2217.4	11.3\\
2225.5	11.3\\
2230.2	11.0\\
2235.0	4.5\\
2239.0	4.0\\
2244.1	4.1\\
2248.6	4.1\\
2252.9	0.0\\
2262.0	0.0\\
2266.5	1.1\\
2270.9	9.2\\
2275.2	9.3\\
2279.8	8.2\\
2283.7	8.9\\
2288.7	9.3\\
2293.3	10.2\\
2297.4	10.1\\
2302.0	9.5\\
2306.0	9.9\\
2310.9	9.9\\
2315.2	9.7\\
2319.3	9.0\\
2324.1	9.3\\
2328.1	9.1\\
2333.0	9.2\\
2337.5	9.4\\
2341.8	10.1\\
2346.4	10.6\\
2350.4	10.8\\
2355.3	9.8\\
2359.8	7.7\\
2364.1	5.8\\
2368.7	8.3\\
2372.6	9.4\\
2377.6	9.0\\
2382.0	9.0\\
2386.3	8.7\\
2390.9	7.4\\
2395.4	6.6\\
2399.8	7.8\\
2404.2	7.3\\
2408.5	7.8\\
2413.1	7.7\\
2417.5	7.2\\
2421.9	9.6\\
2426.4	10.7\\
2430.6	10.9\\
2435.2	11.2\\
2439.7	10.9\\
2444.1	10.8\\
2449.6	10.5\\
2454.4	10.1\\
2458.9	10.0\\
2463.0	10.1\\
2467.6	9.3\\
2471.6	9.1\\
2476.5	9.6\\
2481.0	9.3\\
2485.3	9.4\\
2489.9	7.4\\
2494.3	5.7\\
2498.8	7.8\\
2503.2	8.6\\
2507.5	7.1\\
2512.1	6.8\\
2516.0	6.0\\
2520.9	7.1\\
2525.4	10.8\\
2529.4	10.7\\
2534.2	10.7\\
2538.7	10.7\\
2543.1	10.8\\
2547.5	10.1\\
2551.8	10.1\\
2556.4	9.8\\
2560.9	10.2\\
2565.3	10.2\\
2569.8	10.5\\
2574.1	10.7\\
2582.9	10.5\\
2587.7	11.0\\
2592.1	11.1\\
2596.4	14.6\\
2601.1	12.9\\
2605.6	11.1\\
2610.1	12.9\\
2614.5	14.0\\
2618.8	14.0\\
2623.5	11.7\\
2627.6	7.6\\
2632.5	5.7\\
2636.8	5.7\\
2641.4	8.2\\
2645.9	6.5\\
2650.1	5.5\\
2654.9	6.0\\
2660.3	5.5\\
2663.9	6.3\\
2668.0	5.6\\
2672.7	6.1\\
2677.2	5.4\\
2681.6	6.0\\
2686.0	5.8\\
2690.4	6.1\\
2694.5	5.4\\
2699.1	5.9\\
2703.7	6.1\\
2708.1	5.9\\
2712.5	5.8\\
2716.5	6.0\\
2721.0	6.0\\
2725.6	5.5\\
2729.9	5.9\\
2737.0	5.1\\
2743.7	4.0\\
2750.6	1.5\\
2755.1	0.0\\
2769.5	0.0\\
2773.3	0.3\\
2777.7	0.0\\
2780.7	0.0\\
2784.5	0.2\\
2788.2	0.0\\
2800.5	0.0\\
};
\addlegendentry{\scriptsize MPC-U }

\addplot [color=mycolor3, dashdotted, line width=1.3pt]
  table[row sep=crcr]{%
503.2	0.6\\
507.3	3.6\\
511.3	3.1\\
515.9	0.5\\
519.4	0.0\\
553.0	0.0\\
556.9	13.7\\
560.2	13.4\\
563.7	12.1\\
567.4	11.3\\
572.0	13.8\\
574.7	13.6\\
578.3	13.7\\
582.0	13.4\\
585.7	13.6\\
589.4	13.6\\
593.2	13.8\\
596.9	13.7\\
600.7	13.8\\
604.2	13.7\\
608.3	13.7\\
612.2	13.6\\
616.0	13.9\\
619.8	13.2\\
624.3	13.5\\
628.8	6.9\\
632.6	1.6\\
636.4	1.3\\
640.1	2.4\\
647.7	0.0\\
694.9	0.0\\
701.1	1.4\\
704.0	7.1\\
708.0	7.1\\
712.7	9.4\\
715.2	9.5\\
719.0	13.6\\
722.7	13.9\\
727.9	14.5\\
731.7	15.7\\
734.2	17.9\\
739.7	19.2\\
741.9	19.7\\
745.9	19.7\\
749.7	20.8\\
753.8	20.9\\
758.2	21.8\\
762.0	24.6\\
766.4	22.7\\
772.5	25.7\\
774.9	29.4\\
779.3	25.7\\
783.6	25.9\\
789.6	26.9\\
794.5	30.9\\
797.1	29.0\\
801.5	25.9\\
805.8	30.6\\
810.7	29.4\\
815.1	26.8\\
821.5	20.0\\
826.1	19.8\\
829.0	19.5\\
833.8	20.1\\
839.4	19.7\\
843.2	20.2\\
847.9	19.6\\
852.7	19.7\\
862.1	20.1\\
867.1	42.3\\
872.8	38.4\\
877.6	16.2\\
884.1	19.7\\
886.8	20.5\\
892.0	20.5\\
896.9	20.4\\
902.2	20.9\\
906.8	20.8\\
913.1	21.3\\
919.2	21.1\\
923.1	44.2\\
927.4	45.6\\
933.0	4.6\\
938.8	3.1\\
944.6	2.3\\
947.7	1.4\\
952.4	0.0\\
979.0	0.0\\
983.3	3.1\\
987.8	0.7\\
993.0	0.0\\
997.8	0.0\\
1000.3	0.7\\
1006.5	0.0\\
1013.3	0.0\\
1018.9	5.2\\
1021.8	0.0\\
1027.7	0.0\\
1031.1	3.1\\
1034.7	0.7\\
1040.7	0.0\\
1095.3	0.0\\
1099.3	13.1\\
1102.4	10.9\\
1105.2	13.9\\
1108.8	13.9\\
1113.8	14.0\\
1115.9	12.9\\
1119.9	12.6\\
1123.1	12.7\\
1127.6	12.9\\
1130.2	4.6\\
1134.2	0.4\\
1138.4	0.0\\
1183.2	0.0\\
1184.7	7.2\\
1188.2	7.0\\
1190.9	9.3\\
1193.6	10.0\\
1196.4	9.6\\
1198.8	9.7\\
1201.3	10.0\\
1203.8	8.7\\
1206.1	4.8\\
1208.5	0.2\\
1210.6	0.0\\
1241.7	0.0\\
1242.1	0.7\\
1242.6	0.8\\
1242.7	1.1\\
1242.7	0.9\\
1242.7	1.3\\
1242.7	0.6\\
1242.7	0.7\\
1242.7	0.5\\
1242.7	0.6\\
1242.7	0.5\\
1242.7	0.7\\
1242.7	0.5\\
1242.7	0.7\\
1242.7	0.6\\
1242.7	0.8\\
1242.7	0.6\\
1242.8	1.0\\
1242.8	0.9\\
1242.9	1.2\\
1243.1	0.9\\
1243.4	0.9\\
1243.8	1.1\\
1244.3	2.2\\
1244.9	3.8\\
1245.7	5.7\\
1246.5	7.7\\
1247.7	9.1\\
1248.7	10.5\\
1249.9	12.2\\
1251.2	14.0\\
1252.6	14.9\\
1254.1	16.2\\
1255.8	18.5\\
1257.4	20.1\\
1259.3	21.6\\
1261.2	22.8\\
1263.3	23.5\\
1265.5	24.3\\
1267.7	26.3\\
1270.0	26.6\\
1272.5	27.0\\
1275.0	27.5\\
1277.7	27.6\\
1280.2	27.4\\
1283.0	28.5\\
1285.8	29.1\\
1288.6	28.2\\
1291.4	22.8\\
1294.2	18.9\\
1297.0	14.7\\
1300.0	9.3\\
1302.7	9.9\\
1305.3	7.8\\
1310.4	0.0\\
1342.6	0.0\\
1344.5	3.9\\
1346.4	3.5\\
1348.3	2.7\\
1350.0	2.0\\
1351.9	2.0\\
1355.1	1.7\\
1357.1	1.3\\
1359.5	0.9\\
1361.7	0.0\\
1364.2	0.0\\
1366.4	0.6\\
1368.7	0.0\\
1385.7	0.0\\
1387.6	4.4\\
1389.4	4.2\\
1391.7	3.1\\
1393.0	2.7\\
1394.5	2.8\\
1396.1	2.8\\
1397.9	2.7\\
1400.6	3.0\\
1402.5	2.6\\
1404.2	3.0\\
1406.2	3.1\\
1408.1	2.6\\
1410.1	2.3\\
1411.7	2.5\\
1413.9	2.4\\
1415.9	2.6\\
1417.8	2.1\\
1419.7	2.2\\
1423.3	2.2\\
1425.2	2.2\\
1426.7	2.1\\
1428.6	2.1\\
1430.1	2.2\\
1432.5	2.1\\
1434.0	2.1\\
1435.7	2.2\\
1437.9	2.1\\
1440.0	2.4\\
1442.2	2.2\\
1444.5	20.3\\
1446.9	38.9\\
1449.3	43.7\\
1451.9	44.6\\
1454.5	35.3\\
1457.5	39.4\\
1460.6	41.6\\
1463.7	52.0\\
1467.0	57.8\\
1470.2	61.0\\
1473.8	63.3\\
1477.4	52.3\\
1481.0	47.4\\
1484.7	53.7\\
1488.3	57.2\\
1492.3	53.6\\
1496.2	45.1\\
1500.1	32.9\\
1504.1	30.9\\
1509.4	29.6\\
1513.1	29.5\\
1515.8	29.4\\
1521.0	29.6\\
1524.6	30.2\\
1528.4	29.6\\
1532.6	29.1\\
1537.2	28.2\\
1542.8	28.6\\
1545.8	29.1\\
1549.6	23.8\\
1553.9	22.2\\
1558.3	3.3\\
1562.4	0.3\\
1566.5	0.0\\
1640.0	0.0\\
1643.4	15.5\\
1646.4	15.1\\
1651.6	14.8\\
1655.9	14.4\\
1660.4	14.7\\
1663.9	14.7\\
1668.0	14.8\\
1672.8	14.7\\
1675.8	15.2\\
1681.0	15.1\\
1683.5	15.0\\
1687.5	15.5\\
1691.4	16.2\\
1696.4	18.7\\
1699.3	17.0\\
1704.3	16.1\\
1707.2	16.5\\
1712.8	18.8\\
1716.9	18.9\\
1719.1	17.0\\
1724.1	16.5\\
1727.8	17.9\\
1731.3	19.0\\
1736.1	19.2\\
1739.4	18.8\\
1744.2	5.9\\
1749.1	7.8\\
1751.2	9.1\\
1756.1	8.1\\
1759.3	7.0\\
1764.7	7.0\\
1766.8	6.3\\
1772.8	5.3\\
1776.8	5.8\\
1780.4	5.2\\
1782.9	5.9\\
1788.0	4.3\\
1790.8	4.6\\
1794.2	3.5\\
1798.4	3.9\\
1803.3	3.5\\
1805.9	1.7\\
1810.2	0.0\\
1815.6	4.1\\
1819.0	2.8\\
1822.4	3.0\\
1825.1	2.1\\
1829.1	11.8\\
1832.7	12.3\\
1838.3	10.6\\
1841.3	12.6\\
1846.2	13.6\\
1850.0	11.5\\
1854.1	13.8\\
1857.9	13.5\\
1862.3	12.5\\
1868.2	14.0\\
1870.2	15.0\\
1874.7	14.7\\
1879.2	14.6\\
1883.2	14.9\\
1887.0	14.9\\
1891.6	15.9\\
1893.9	16.7\\
1899.7	20.1\\
1903.2	18.4\\
1906.2	15.7\\
1912.2	16.6\\
1914.4	19.0\\
1920.4	19.2\\
1924.6	18.8\\
1928.9	16.8\\
1933.3	18.3\\
1936.5	19.9\\
1941.6	19.4\\
1944.8	19.8\\
1950.3	19.8\\
1954.6	19.6\\
1957.0	19.6\\
1963.3	19.5\\
1967.9	19.5\\
1970.9	19.8\\
1976.4	19.7\\
1981.1	19.8\\
1985.4	20.0\\
1989.5	20.1\\
1994.7	19.7\\
1998.8	19.8\\
2007.9	19.8\\
2012.8	19.7\\
2017.2	20.4\\
2021.8	19.9\\
2026.4	20.3\\
2030.8	20.3\\
2035.4	21.1\\
2038.4	21.2\\
2044.7	21.4\\
2048.1	21.6\\
2054.3	21.6\\
2056.6	21.5\\
2063.3	9.9\\
2067.9	8.6\\
2072.5	8.4\\
2077.1	6.3\\
2081.7	3.6\\
2085.9	4.2\\
2090.9	3.6\\
2095.4	4.0\\
2099.7	4.0\\
2104.3	3.9\\
2108.8	4.3\\
2113.2	3.7\\
2117.7	4.1\\
2122.1	4.4\\
2126.8	4.4\\
2131.1	4.0\\
2135.3	4.1\\
2139.7	3.6\\
2142.4	4.8\\
2148.4	3.3\\
2155.1	4.0\\
2157.8	3.1\\
2162.1	4.2\\
2166.3	2.8\\
2170.6	4.0\\
2174.8	3.0\\
2179.0	3.7\\
2183.4	2.8\\
2187.7	11.1\\
2192.0	16.8\\
2195.4	16.9\\
2200.7	16.3\\
2205.1	19.6\\
2209.5	20.1\\
2213.9	17.0\\
2218.4	16.2\\
2222.8	19.7\\
2227.3	19.7\\
2231.9	19.8\\
2236.2	16.7\\
2240.7	18.0\\
2245.2	19.6\\
2249.8	19.7\\
2254.3	19.5\\
2258.9	19.7\\
2263.5	19.1\\
2268.0	20.0\\
2272.6	19.2\\
2277.4	19.6\\
2281.9	19.3\\
2286.5	19.5\\
2291.1	19.9\\
2295.8	19.7\\
2300.6	19.7\\
2305.2	20.1\\
2309.9	20.1\\
2314.8	20.5\\
2319.4	20.7\\
2324.1	20.6\\
2328.7	21.7\\
2333.4	21.6\\
2338.1	17.4\\
2341.9	10.6\\
2347.5	9.6\\
2352.2	9.0\\
2356.8	5.4\\
2361.5	4.2\\
2366.1	5.1\\
2370.4	4.6\\
2375.4	4.5\\
2379.9	4.6\\
2384.5	4.5\\
2387.8	4.7\\
2393.5	4.5\\
2398.1	4.5\\
2402.6	4.7\\
2407.1	4.8\\
2411.6	3.9\\
2416.1	3.2\\
2420.5	1.9\\
2424.9	1.6\\
2429.3	0.0\\
2433.7	3.0\\
2438.1	2.2\\
2442.0	0.0\\
2480.3	0.0\\
2484.6	16.8\\
2488.8	15.3\\
2493.1	15.4\\
2497.4	15.4\\
2501.7	15.7\\
2506.1	16.5\\
2510.4	19.3\\
2514.7	17.8\\
2519.1	16.7\\
2523.5	16.8\\
2527.9	19.1\\
2532.3	19.7\\
2536.7	17.1\\
2541.1	16.3\\
2545.6	19.4\\
2550.1	20.0\\
2554.6	20.1\\
2559.1	19.7\\
2563.6	19.7\\
2568.1	17.1\\
2572.7	18.3\\
2577.3	19.7\\
2581.9	19.6\\
2586.4	20.0\\
2591.1	19.5\\
2595.7	20.0\\
2600.5	19.4\\
2604.7	19.7\\
2609.7	19.5\\
2614.4	19.4\\
2619.1	19.9\\
2624.0	19.5\\
2628.4	19.7\\
2633.2	19.6\\
2637.9	20.1\\
2642.6	19.8\\
2647.4	20.1\\
2652.0	20.0\\
2656.7	3.7\\
2661.3	0.0\\
2720.3	0.0\\
2724.6	1.8\\
2728.9	2.1\\
2733.3	0.1\\
2737.4	0.2\\
2741.6	0.7\\
2745.8	1.0\\
2750.0	1.4\\
2754.1	1.3\\
2758.3	0.9\\
2762.4	0.7\\
2766.7	0.8\\
2770.4	0.5\\
2774.9	18.2\\
2779.2	15.9\\
2783.4	15.7\\
2787.8	15.7\\
2796.2	15.5\\
2800.7	15.9\\
};
\addlegendentry{\scriptsize WIE }

\addplot [color=mycolor4, dotted, line width=1.3pt]
  table[row sep=crcr]{%
502.0	0.0\\
605.9	0.0\\
609.4	1.7\\
613.4	2.0\\
616.6	0.0\\
620.5	1.1\\
624.2	3.8\\
627.5	3.7\\
631.4	1.1\\
635.2	5.1\\
638.7	3.9\\
645.8	4.7\\
649.5	5.6\\
653.2	6.1\\
656.8	9.1\\
660.2	8.1\\
664.1	7.8\\
667.7	7.6\\
671.4	8.2\\
677.2	8.9\\
679.5	9.7\\
683.5	10.5\\
687.2	14.1\\
690.7	14.1\\
694.7	14.1\\
698.4	14.3\\
701.9	13.8\\
706.0	14.8\\
709.5	15.8\\
713.7	19.3\\
717.6	20.4\\
721.4	17.8\\
725.4	16.3\\
729.0	19.5\\
733.1	19.7\\
737.3	19.7\\
741.2	19.6\\
745.3	19.8\\
749.0	19.7\\
753.6	20.1\\
757.9	19.6\\
761.8	20.2\\
766.0	20.2\\
770.2	20.7\\
774.4	21.6\\
778.7	21.9\\
783.0	22.4\\
787.2	22.2\\
791.7	22.6\\
796.3	22.6\\
800.3	22.4\\
805.0	22.5\\
809.3	23.2\\
814.0	23.4\\
818.3	23.8\\
827.6	23.2\\
832.1	23.1\\
836.9	22.6\\
841.4	22.5\\
846.1	22.0\\
850.8	20.7\\
855.5	20.6\\
860.4	16.8\\
865.2	17.6\\
869.8	17.3\\
874.5	18.0\\
879.3	17.1\\
884.1	17.0\\
888.8	15.3\\
893.6	11.9\\
898.4	10.5\\
903.2	10.8\\
908.2	10.3\\
912.9	9.2\\
917.7	4.5\\
922.5	4.6\\
927.4	4.5\\
932.1	4.8\\
937.3	4.4\\
941.6	3.5\\
946.3	2.1\\
951.0	0.0\\
955.9	0.0\\
960.1	1.3\\
965.3	0.1\\
969.8	0.0\\
1072.6	0.0\\
1076.1	1.0\\
1081.5	0.0\\
1192.8	0.0\\
1195.5	1.2\\
1197.2	0.0\\
1235.1	0.0\\
1235.7	0.5\\
1235.9	0.7\\
1236.2	1.1\\
1236.3	1.2\\
1236.4	1.1\\
1236.4	0.9\\
1236.4	0.5\\
1236.4	0.6\\
1236.4	0.5\\
1236.4	0.8\\
1236.4	0.4\\
1236.5	0.7\\
1236.5	0.4\\
1236.5	0.5\\
1236.5	0.4\\
1236.5	0.3\\
1236.5	0.5\\
1236.5	0.4\\
1236.5	0.5\\
1236.5	0.4\\
1236.7	1.6\\
1236.8	2.3\\
1237.1	2.9\\
1237.4	4.2\\
1237.8	5.2\\
1238.3	6.2\\
1238.9	7.0\\
1240.3	8.3\\
1241.1	9.8\\
1242.1	11.3\\
1243.4	12.2\\
1244.6	13.1\\
1245.6	13.5\\
1247.0	14.1\\
1248.4	14.7\\
1249.8	15.5\\
1251.4	16.4\\
1253.0	18.2\\
1254.7	19.1\\
1256.5	19.4\\
1258.4	19.7\\
1260.3	19.9\\
1262.4	20.3\\
1264.5	20.5\\
1266.7	21.5\\
1269.0	22.0\\
1271.2	22.2\\
1273.6	22.8\\
1276.1	23.1\\
1278.6	23.2\\
1281.1	23.7\\
1283.8	22.9\\
1286.5	22.6\\
1289.2	19.9\\
1292.1	19.6\\
1295.0	18.7\\
1297.8	15.8\\
1300.7	14.4\\
1303.8	12.3\\
1306.8	9.9\\
1309.6	8.0\\
1312.6	5.4\\
1315.5	4.0\\
1318.5	0.5\\
1321.1	0.0\\
1342.8	0.0\\
1344.4	2.8\\
1346.0	2.1\\
1347.6	2.2\\
1350.6	1.7\\
1352.1	1.9\\
1355.4	1.7\\
1357.0	2.0\\
1359.0	2.1\\
1360.7	2.1\\
1362.7	2.3\\
1364.8	3.6\\
1366.5	4.6\\
1370.6	2.9\\
1372.4	2.9\\
1374.4	3.9\\
1375.8	4.6\\
1377.8	4.4\\
1379.1	4.2\\
1380.8	4.5\\
1383.2	4.6\\
1384.3	4.1\\
1388.1	4.1\\
1389.6	3.9\\
1391.3	2.9\\
1392.7	2.5\\
1395.7	2.2\\
1397.5	2.1\\
1399.3	1.4\\
1402.1	1.9\\
1405.3	1.9\\
1406.8	2.0\\
1409.1	2.5\\
1411.1	4.5\\
1412.7	4.8\\
1414.8	4.7\\
1421.5	4.7\\
1422.9	4.7\\
1424.7	4.6\\
1426.5	17.7\\
1428.1	21.0\\
1430.3	21.1\\
1432.3	22.2\\
1434.9	23.8\\
1441.0	24.3\\
1444.1	24.8\\
1445.7	25.6\\
1449.1	26.2\\
1452.6	29.2\\
1455.6	29.6\\
1458.1	29.6\\
1461.5	29.5\\
1464.2	29.5\\
1467.6	30.7\\
1470.7	32.0\\
1473.7	34.0\\
1477.2	35.4\\
1480.4	35.2\\
1483.8	34.5\\
1487.3	35.2\\
1491.0	35.7\\
1494.4	35.9\\
1497.7	35.9\\
1502.0	36.3\\
1505.6	37.1\\
1509.5	37.2\\
1513.1	37.8\\
1517.4	38.5\\
1521.4	38.5\\
1525.5	39.1\\
1529.5	39.1\\
1533.7	39.2\\
1537.5	39.1\\
1542.5	37.8\\
1546.7	36.9\\
1551.0	37.9\\
1555.6	37.5\\
1560.3	37.2\\
1564.9	36.1\\
1569.4	33.4\\
1573.8	33.1\\
1578.9	32.5\\
1583.4	31.4\\
1588.2	27.7\\
1593.0	27.7\\
1597.4	26.3\\
1602.6	22.9\\
1606.8	22.4\\
1612.4	20.7\\
1617.1	16.7\\
1621.7	15.6\\
1626.8	15.7\\
1631.1	12.3\\
1636.5	10.1\\
1641.3	10.3\\
1646.3	9.3\\
1651.1	4.2\\
1655.7	4.7\\
1660.2	4.3\\
1665.3	5.1\\
1669.9	3.6\\
1674.8	3.1\\
1678.9	0.0\\
1684.3	0.0\\
1688.9	1.5\\
1693.1	1.0\\
1698.1	0.0\\
1751.2	0.0\\
1755.5	0.8\\
1759.4	0.0\\
1763.8	0.0\\
1769.0	1.1\\
1772.2	3.2\\
1776.4	3.1\\
1781.1	3.1\\
1784.6	3.5\\
1789.8	4.4\\
1792.9	5.5\\
1797.5	8.0\\
1800.8	7.3\\
1805.1	6.7\\
1808.9	8.7\\
1813.4	8.6\\
1818.0	8.8\\
1822.3	9.3\\
1825.8	9.1\\
1830.8	9.3\\
1833.9	9.3\\
1837.7	9.4\\
1842.0	9.6\\
1846.2	10.7\\
1851.4	14.1\\
1856.2	14.6\\
1858.5	9.9\\
1862.6	10.8\\
1867.1	13.6\\
1871.0	11.8\\
1875.4	12.5\\
1879.3	14.5\\
1883.6	13.9\\
1887.8	11.0\\
1892.0	11.2\\
1896.6	14.4\\
1901.4	11.7\\
1904.6	13.6\\
1911.1	12.4\\
1913.0	11.5\\
1917.9	14.4\\
1923.1	11.7\\
1926.3	10.7\\
1930.9	10.9\\
1936.2	12.7\\
1940.0	12.2\\
1946.0	11.8\\
1954.9	11.8\\
1957.5	11.7\\
1963.2	10.6\\
1966.0	9.9\\
1972.5	9.5\\
1975.1	9.6\\
1981.0	9.6\\
1983.5	9.9\\
1989.2	9.9\\
1994.2	9.8\\
1996.7	10.9\\
2002.7	10.8\\
2007.4	10.6\\
2019.3	10.9\\
2024.4	10.5\\
2029.0	11.2\\
2033.4	10.7\\
2041.7	10.9\\
2044.4	10.8\\
2049.2	10.8\\
2053.2	11.0\\
2059.3	10.7\\
2062.9	10.9\\
2068.1	11.9\\
2071.8	13.7\\
2077.0	11.9\\
2081.0	11.5\\
2085.1	14.5\\
2090.0	14.3\\
2094.4	11.0\\
2097.2	11.0\\
2104.0	13.6\\
2108.2	11.5\\
2112.6	11.2\\
2117.2	10.6\\
2126.3	11.2\\
2132.4	10.1\\
2134.6	11.4\\
2139.0	10.5\\
2143.8	11.5\\
2147.9	10.2\\
2152.5	14.5\\
2157.0	13.7\\
2161.6	11.7\\
2165.9	11.1\\
2170.5	11.6\\
2175.0	11.7\\
2177.7	13.0\\
2184.2	13.2\\
2188.4	13.9\\
2192.7	14.5\\
2197.0	12.8\\
2200.0	13.5\\
2206.4	12.6\\
2210.8	11.3\\
2215.4	12.3\\
2219.9	14.6\\
2224.5	10.8\\
2229.2	11.2\\
2233.7	10.5\\
2237.9	10.2\\
2241.4	9.0\\
2247.2	4.9\\
2252.1	4.3\\
2256.6	8.5\\
2260.5	9.9\\
2265.1	6.2\\
2269.8	4.4\\
2273.0	9.0\\
2278.8	9.5\\
2283.3	9.0\\
2287.8	6.4\\
2292.3	6.8\\
2297.0	7.2\\
2301.3	7.4\\
2305.8	7.1\\
2310.1	7.5\\
2314.7	7.3\\
2319.4	8.2\\
2323.7	8.6\\
2328.2	8.6\\
2332.6	8.5\\
2339.4	8.7\\
2341.5	8.9\\
2346.2	9.9\\
2350.7	9.7\\
2355.3	9.7\\
2359.6	10.0\\
2364.1	10.8\\
2368.6	10.8\\
2372.9	11.2\\
2377.9	10.8\\
2382.1	11.4\\
2386.5	11.1\\
2391.0	11.1\\
2395.5	10.6\\
2400.1	10.7\\
2404.3	10.5\\
2409.0	10.1\\
2412.8	10.3\\
2418.0	9.9\\
2422.6	10.0\\
2426.4	9.9\\
2431.4	10.0\\
2435.9	9.9\\
2440.4	9.4\\
2445.0	9.3\\
2447.7	8.9\\
2453.9	6.3\\
2458.4	7.6\\
2462.9	10.2\\
2470.8	10.0\\
2473.2	10.4\\
2477.9	8.8\\
2482.2	7.5\\
2486.7	6.5\\
2490.9	7.5\\
2495.4	7.0\\
2500.1	7.3\\
2504.4	7.1\\
2508.8	7.1\\
2517.7	7.0\\
2526.6	8.5\\
2535.3	8.7\\
2539.9	8.4\\
2543.8	8.9\\
2548.8	8.4\\
2553.2	8.3\\
2557.6	8.6\\
2566.5	8.6\\
2570.9	9.0\\
2575.3	9.6\\
2579.8	9.9\\
2584.2	10.8\\
2588.6	10.8\\
2593.1	10.9\\
2597.5	11.3\\
2602.1	13.9\\
2606.4	13.8\\
2610.9	11.4\\
2615.4	11.5\\
2618.5	12.0\\
2624.1	11.3\\
2628.6	11.1\\
2633.3	10.7\\
2637.2	10.6\\
2642.2	10.9\\
2648.6	9.8\\
2650.6	10.0\\
2655.5	9.4\\
2659.9	9.3\\
2664.4	7.1\\
2668.9	7.3\\
2673.4	8.0\\
2677.3	8.4\\
2682.1	8.3\\
2686.8	8.5\\
2691.2	8.3\\
2695.8	7.5\\
2700.3	7.6\\
2704.6	7.8\\
2709.0	7.4\\
2713.3	7.4\\
2717.9	8.0\\
2722.3	7.8\\
2726.8	7.9\\
2731.4	7.8\\
2735.6	7.9\\
2740.1	7.6\\
2744.0	7.7\\
2748.9	7.5\\
2757.7	7.5\\
2762.1	7.7\\
2766.2	8.0\\
2771.0	7.8\\
2775.4	7.9\\
2779.7	8.6\\
2782.7	8.5\\
2788.2	9.8\\
2792.8	9.8\\
2797.2	11.2\\
2801.3	13.6\\
};
\addlegendentry{\scriptsize IDM }

\addplot[area legend, draw=none, fill=white!10!black, fill opacity=0.2]
table[row sep=crcr] {%
x	y\\
1354.4	0\\
1354.4	65\\
1432	65\\
1432	0\\
}--cycle;
\addlegendentry{\scriptsize U-Turn}

\end{axis}
     
\input{Images/US06_pin_Nissan}
\input{Images/US06_vel_Nissan}
     
\end{tikzpicture}%

%% file: Images/US06_pin_Nissan.tex
%
%

\begin{axis}[%
width=6.5cm,
height=4.0cm,
at={(0cm,4.4cm)},
scale only axis,
xmin=505,
xmax=2800,
xlabel style={font=\color{white!15!black} \scriptsize},
xlabel={},
xticklabels={,,},
ymin=-25,
ymax=0,
ylabel style={font=\color{white!15!black} \scriptsize},
ylabel={Power In [KW]},
ticklabel style = {font = \scriptsize},
axis background/.style={fill=white},
xmajorgrids,
ymajorgrids,
legend style={at={(0.97,0.03)}, anchor=south east, legend cell align=left, align=left, draw=white!15!black}
]

\addplot [color=mycolor1, line width=1.3pt]
  table[row sep=crcr]{%
503.6	0.0\\
945.0	0.0\\
949.8	-0.2\\
954.0	0.0\\
958.3	-0.9\\
962.7	-0.7\\
966.8	-1.3\\
971.2	-1.5\\
975.3	-3.4\\
979.4	-3.9\\
982.1	-4.0\\
987.6	-2.9\\
991.9	-2.7\\
996.0	-3.2\\
1000.0	-4.0\\
1003.7	-3.7\\
1008.1	-3.9\\
1011.8	-4.2\\
1015.9	-4.7\\
1020.1	-4.7\\
1024.5	-4.6\\
1028.4	-5.3\\
1032.2	-4.8\\
1036.3	-4.9\\
1040.2	-4.9\\
1044.3	-5.0\\
1048.3	-4.5\\
1051.9	-5.1\\
1055.7	-5.0\\
1059.8	-5.4\\
1063.1	-4.5\\
1066.9	-5.2\\
1070.4	-4.4\\
1074.0	-5.1\\
1077.6	-4.6\\
1081.4	-4.7\\
1085.3	-4.3\\
1088.0	-5.0\\
1092.0	-4.8\\
1098.6	-4.9\\
1102.0	-5.0\\
1105.3	-5.0\\
1108.6	-4.8\\
1111.9	-5.0\\
1115.1	-11.5\\
1120.2	-11.5\\
1122.2	-11.5\\
1125.3	-11.6\\
1128.6	-11.9\\
1131.5	-11.7\\
1134.4	-11.7\\
1137.3	-11.6\\
1140.3	-11.4\\
1142.9	-10.9\\
1145.7	-11.0\\
1148.4	-11.0\\
1151.5	-10.6\\
1154.2	-10.3\\
1156.9	-10.2\\
1159.7	-10.3\\
1161.8	-9.8\\
1164.5	-9.5\\
1166.6	-9.5\\
1169.1	-9.3\\
1171.3	-8.8\\
1173.3	-9.0\\
1175.5	-8.4\\
1178.4	-7.7\\
1179.7	-7.8\\
1181.6	-7.4\\
1183.5	-7.4\\
1185.4	-6.9\\
1187.4	-6.7\\
1189.2	-6.5\\
1191.0	-6.1\\
1192.6	-5.9\\
1194.5	-5.7\\
1196.1	-5.6\\
1197.7	-5.3\\
1199.4	-5.2\\
1200.9	-5.1\\
1202.4	-5.0\\
1203.9	-4.5\\
1205.3	-4.6\\
1206.7	-4.1\\
1208.0	-4.3\\
1209.4	-4.1\\
1210.5	-3.8\\
1211.9	-3.8\\
1213.0	-3.6\\
1214.3	-3.1\\
1215.6	-3.0\\
1216.6	-3.0\\
1217.7	-2.7\\
1218.7	-2.6\\
1219.6	-2.6\\
1220.6	-2.5\\
1221.5	-2.3\\
1222.4	-1.7\\
1223.3	-1.8\\
1224.1	-1.9\\
1224.9	-1.5\\
1225.7	-1.5\\
1226.5	-1.5\\
1227.2	-1.4\\
1227.9	-1.2\\
1228.6	-0.9\\
1229.3	-1.2\\
1229.9	-0.7\\
1230.5	-0.9\\
1231.1	-0.6\\
1231.7	-0.1\\
1232.3	-0.5\\
1232.8	-0.2\\
1233.5	0.0\\
1329.5	0.0\\
1331.2	-0.4\\
1333.1	0.0\\
2708.0	0.0\\
2712.1	-1.1\\
2716.2	-0.9\\
2720.4	-0.9\\
2724.6	-0.8\\
2728.7	-0.9\\
2732.8	-1.2\\
2737.2	-2.2\\
2741.0	-3.9\\
2745.3	-1.9\\
2749.8	-1.8\\
2753.7	-2.1\\
2757.7	-2.8\\
2761.6	-2.5\\
2765.5	-2.6\\
2769.3	-2.6\\
2773.2	-3.0\\
2777.3	-3.9\\
2780.8	-3.9\\
2784.5	-3.5\\
2812.1	-3.7\\
};
\addlegendentry{\scriptsize MPC-C }

\addplot [color=mycolor2, dashed, line width=1.3pt]
  table[row sep=crcr]{%
500.9	0.0\\
548.0	0.0\\
551.7	-0.5\\
555.6	-0.5\\
559.2	-0.1\\
562.8	0.0\\
912.6	0.0\\
917.2	-2.3\\
921.9	-4.9\\
926.6	-6.6\\
931.7	-5.5\\
938.4	-2.9\\
942.3	-2.7\\
947.7	-3.5\\
952.7	-4.0\\
957.0	-2.6\\
961.5	-1.9\\
965.9	-2.3\\
970.0	-1.7\\
974.9	-1.9\\
980.5	-1.9\\
984.1	-1.6\\
988.1	-1.7\\
992.6	-3.1\\
996.7	-4.6\\
1002.5	-5.0\\
1006.7	-5.2\\
1010.7	-5.5\\
1015.3	-5.8\\
1019.5	-5.9\\
1023.8	-5.4\\
1027.9	-5.9\\
1032.3	-6.1\\
1036.4	-5.3\\
1040.4	-5.3\\
1044.6	-6.2\\
1048.6	-5.1\\
1052.6	-5.7\\
1056.7	-5.0\\
1060.6	-5.5\\
1064.6	-5.3\\
1068.3	-5.5\\
1071.9	-5.1\\
1076.0	-5.3\\
1079.7	-5.0\\
1083.7	-5.4\\
1087.5	-5.3\\
1092.4	-5.1\\
1096.1	-4.9\\
1099.2	-4.9\\
1102.9	-5.0\\
1106.7	-4.5\\
1110.4	-5.3\\
1114.0	-4.6\\
1116.5	-5.3\\
1121.2	-5.0\\
1124.7	-5.6\\
1128.2	-4.5\\
1131.7	-5.4\\
1135.1	-4.1\\
1138.5	-9.1\\
1141.8	-14.6\\
1145.0	-14.8\\
1148.1	-14.9\\
1151.2	-15.0\\
1154.2	-15.3\\
1157.1	-15.6\\
1160.0	-15.7\\
1162.7	-16.3\\
1165.5	-15.3\\
1168.7	-13.3\\
1171.6	-7.3\\
1176.2	0.0\\
1189.8	0.0\\
1192.5	-1.5\\
1195.0	-6.2\\
1197.6	-6.8\\
1200.0	-7.3\\
1202.4	-4.3\\
1204.7	-9.8\\
1206.8	-14.8\\
1208.7	-17.5\\
1210.8	-17.5\\
1212.6	-17.9\\
1214.3	-16.2\\
1215.9	-14.8\\
1217.5	-12.9\\
1219.0	-11.5\\
1221.2	-10.1\\
1222.2	-7.2\\
1223.5	-6.5\\
1224.7	-6.3\\
1225.8	-5.2\\
1226.9	-5.0\\
1228.0	-4.9\\
1229.0	-4.5\\
1229.9	-4.0\\
1230.8	-3.6\\
1231.7	-3.6\\
1232.5	-3.1\\
1233.3	-2.8\\
1234.0	-1.9\\
1234.7	-1.8\\
1235.4	-1.8\\
1236.1	-1.6\\
1236.7	-1.4\\
1237.3	-1.2\\
1237.9	-0.7\\
1238.4	-0.6\\
1239.0	-0.4\\
1239.5	-0.1\\
1240.2	0.0\\
1243.5	0.0\\
1281.6	0.0\\
1284.0	-1.8\\
1285.0	-4.5\\
1286.6	-5.7\\
1289.6	-5.7\\
1291.1	-5.1\\
1292.5	-4.8\\
1294.0	-1.1\\
1295.5	0.0\\
1675.9	0.0\\
1682.4	-5.8\\
1685.1	-6.0\\
1691.4	-2.9\\
1695.7	-2.7\\
1700.8	0.0\\
1726.8	0.0\\
1731.6	-0.2\\
1735.2	-5.9\\
1740.4	-5.7\\
1745.0	-5.7\\
1749.3	0.0\\
2248.6	0.0\\
2252.9	-2.0\\
2257.5	-2.2\\
2262.0	-2.4\\
2266.5	0.0\\
2750.6	0.0\\
2755.1	-1.6\\
2760.6	-0.5\\
2763.7	-0.5\\
2769.5	-0.1\\
2773.3	0.0\\
2784.5	0.0\\
2792.8	-0.4\\
2796.6	-1.7\\
2800.5	-2.1\\
};
\addlegendentry{\scriptsize MPC-U }

\addplot [color=mycolor3, dashdotted, line width=1.3pt]
  table[row sep=crcr]{%
503.2	0.0\\
515.9	0.0\\
519.4	-0.1\\
523.3	-1.8\\
527.2	-2.9\\
530.9	-5.0\\
534.5	-1.8\\
538.1	-1.8\\
541.5	-9.3\\
544.9	-17.7\\
548.4	-18.9\\
553.0	-17.8\\
556.9	0.0\\
643.9	0.0\\
647.7	-0.6\\
651.4	-0.5\\
655.1	-0.6\\
658.8	-0.5\\
662.2	-0.5\\
665.9	-0.3\\
669.6	-0.6\\
673.2	-1.1\\
676.9	-0.7\\
680.5	-1.0\\
684.1	-0.8\\
687.7	-1.1\\
691.3	-0.9\\
694.9	-1.4\\
701.1	0.0\\
947.7	0.0\\
952.4	-3.2\\
958.1	-2.2\\
963.9	-3.9\\
969.3	-3.8\\
972.3	-9.0\\
979.0	-4.6\\
983.3	0.0\\
987.8	0.0\\
993.0	-3.5\\
997.8	-5.2\\
1000.3	0.0\\
1006.5	-6.1\\
1009.6	-15.1\\
1013.3	-15.3\\
1018.9	0.0\\
1021.8	-3.9\\
1027.7	-12.1\\
1031.1	0.0\\
1034.7	0.0\\
1040.7	-1.5\\
1043.3	-3.4\\
1049.0	-4.2\\
1053.0	-4.4\\
1056.7	-6.1\\
1060.4	-11.2\\
1062.7	-8.9\\
1067.2	-13.9\\
1071.4	-17.9\\
1073.7	-10.7\\
1077.7	-14.2\\
1081.6	-19.4\\
1085.4	-19.6\\
1088.6	-19.7\\
1091.5	-20.1\\
1095.3	-20.4\\
1099.3	0.0\\
1134.2	0.0\\
1138.4	-0.6\\
1142.2	-1.2\\
1144.6	-4.1\\
1149.1	-1.6\\
1151.9	-3.2\\
1155.2	-4.2\\
1158.8	-3.7\\
1161.9	-10.4\\
1165.3	-9.7\\
1167.6	-11.8\\
1169.3	-11.5\\
1172.2	-5.5\\
1175.7	-6.7\\
1178.1	-12.6\\
1180.6	-18.1\\
1183.2	-4.8\\
1184.7	0.0\\
1208.5	0.0\\
1210.6	-4.7\\
1212.7	-6.0\\
1214.8	-11.6\\
1216.7	-5.5\\
1218.1	-7.1\\
1220.7	-11.5\\
1222.1	-15.3\\
1223.8	-16.1\\
1225.3	-10.2\\
1226.7	-8.2\\
1228.1	-8.8\\
1229.4	-10.6\\
1230.7	-12.0\\
1231.8	-12.3\\
1232.8	-8.3\\
1233.8	-7.1\\
1234.6	-7.2\\
1235.4	-8.5\\
1236.2	-7.9\\
1237.0	-7.1\\
1237.7	-6.0\\
1238.4	-3.9\\
1239.0	-0.5\\
1239.6	-0.7\\
1240.1	-1.8\\
1240.6	-1.7\\
1241.0	-1.4\\
1241.7	-0.1\\
1242.1	0.0\\
1242.7	0.0\\
1242.7	0.0\\
1307.8	0.0\\
1310.4	-0.4\\
1312.8	-2.5\\
1315.8	-4.7\\
1318.8	-8.4\\
1321.7	-6.9\\
1324.5	-7.8\\
1327.5	-7.9\\
1330.7	-7.9\\
1331.6	-5.9\\
1334.9	-6.1\\
1336.7	-5.9\\
1338.7	-5.8\\
1340.5	-1.4\\
1342.6	-1.1\\
1344.5	0.0\\
1359.5	0.0\\
1361.7	-1.1\\
1364.2	-0.7\\
1366.4	0.0\\
1368.7	0.0\\
1370.8	-0.5\\
1372.7	-0.6\\
1378.3	-0.4\\
1380.1	-0.6\\
1381.7	-2.8\\
1383.8	-4.0\\
1385.7	-2.1\\
1387.6	0.0\\
1562.4	0.0\\
1566.5	-1.4\\
1571.8	-0.6\\
1574.5	-0.8\\
1578.5	-1.1\\
1582.6	-2.5\\
1587.0	-2.4\\
1593.7	-2.6\\
1595.2	-2.2\\
1602.4	-2.1\\
1608.6	-1.5\\
1612.1	-1.5\\
1615.9	-1.1\\
1619.5	-1.2\\
1623.4	-0.2\\
1627.1	-0.9\\
1630.8	-0.2\\
1634.4	-0.7\\
1640.0	-0.4\\
1643.4	0.0\\
1805.9	0.0\\
1810.2	-0.1\\
1815.6	0.0\\
2424.9	0.0\\
2429.3	-1.5\\
2433.7	0.0\\
2438.1	0.0\\
2442.0	-2.2\\
2446.7	-0.9\\
2450.9	-0.9\\
2455.2	-1.2\\
2459.4	-0.9\\
2463.6	-1.0\\
2467.8	-1.0\\
2472.0	-0.8\\
2476.3	-0.7\\
2480.3	-0.8\\
2484.6	0.0\\
2656.7	0.0\\
2661.3	-1.4\\
2665.9	-2.8\\
2670.4	-0.4\\
2675.0	-2.3\\
2679.5	-5.1\\
2684.0	-5.5\\
2689.0	-3.3\\
2692.8	-2.0\\
2700.8	-1.8\\
2702.6	-1.5\\
2707.3	-1.2\\
2711.7	-3.0\\
2716.0	-1.7\\
2720.3	-0.7\\
2724.6	0.0\\
2800.7	0.0\\
};
\addlegendentry{\scriptsize WIE }

\addplot [color=mycolor4, dotted, line width=1.3pt]
  table[row sep=crcr]{%
502.0	-0.3\\
507.4	-0.7\\
512.9	-0.9\\
518.3	-1.5\\
523.0	-1.3\\
527.1	-1.1\\
530.8	-1.5\\
538.9	-1.4\\
543.3	-0.7\\
547.0	-0.5\\
551.2	-1.3\\
555.2	-1.7\\
559.1	-4.5\\
562.8	-3.1\\
567.0	-2.0\\
570.9	-2.8\\
574.4	-4.0\\
578.5	-2.3\\
582.0	-1.4\\
586.1	-1.9\\
590.5	-1.3\\
594.8	-1.8\\
597.3	-1.1\\
602.2	-1.1\\
605.9	-1.0\\
609.4	0.0\\
613.4	0.0\\
616.6	-0.3\\
620.5	0.0\\
946.3	0.0\\
951.0	-1.8\\
955.9	-0.5\\
960.1	0.0\\
965.3	0.0\\
969.8	-0.9\\
974.4	-1.1\\
979.2	-1.1\\
983.7	-2.0\\
988.2	-2.5\\
993.3	-3.1\\
996.8	-5.7\\
1004.4	-6.0\\
1010.6	-5.9\\
1012.9	-4.8\\
1017.3	-4.0\\
1022.1	-5.6\\
1027.1	-5.4\\
1030.5	-6.5\\
1035.5	-5.8\\
1039.0	-6.0\\
1045.2	-5.8\\
1047.7	-6.1\\
1052.0	-5.8\\
1056.1	-8.0\\
1061.5	-11.5\\
1068.0	-11.6\\
1072.6	-11.7\\
1076.1	0.0\\
1081.5	-1.0\\
1084.2	-2.8\\
1088.8	-7.1\\
1091.7	-7.4\\
1096.5	-6.6\\
1099.2	-5.7\\
1103.3	-5.2\\
1106.9	-5.2\\
1111.8	-4.9\\
1116.1	-4.9\\
1120.1	-4.7\\
1122.9	-4.7\\
1125.8	-4.8\\
1129.7	-4.6\\
1132.4	-4.5\\
1136.1	-4.2\\
1138.9	-4.3\\
1142.7	-4.1\\
1145.5	-3.9\\
1148.8	-4.1\\
1152.3	-4.1\\
1156.4	-3.8\\
1158.5	-3.5\\
1162.0	-3.2\\
1165.4	-3.3\\
1167.1	-3.5\\
1171.1	-3.2\\
1174.0	-3.3\\
1176.6	-3.3\\
1178.4	-3.0\\
1183.6	-3.0\\
1185.1	-2.8\\
1186.7	-2.7\\
1190.3	-2.7\\
1192.8	-2.3\\
1195.5	0.0\\
1197.2	-0.1\\
1199.8	-7.7\\
1202.7	-12.9\\
1204.3	-12.1\\
1207.0	-11.8\\
1209.5	-13.6\\
1211.5	-14.3\\
1213.4	-13.7\\
1215.4	-13.1\\
1216.7	-15.6\\
1218.9	-14.6\\
1220.2	-13.2\\
1221.4	-11.7\\
1223.3	-10.8\\
1224.9	-9.4\\
1226.2	-8.9\\
1227.4	-8.4\\
1229.2	-8.8\\
1229.7	-8.6\\
1230.7	-7.5\\
1231.6	-6.8\\
1232.0	-5.0\\
1233.1	-3.7\\
1233.5	-2.6\\
1234.1	-1.5\\
1234.7	-0.7\\
1235.1	0.0\\
1236.5	0.0\\
1236.5	0.0\\
1318.5	0.0\\
1321.1	-5.0\\
1323.7	-10.4\\
1326.1	-9.9\\
1328.4	-8.7\\
1330.6	-7.7\\
1332.5	-6.9\\
1334.4	-7.0\\
1336.1	-10.5\\
1337.8	-12.4\\
1339.5	-2.6\\
1341.1	-1.1\\
1342.8	-0.9\\
1344.4	0.0\\
1674.8	0.0\\
1678.9	-1.7\\
1684.3	-1.8\\
1688.9	0.0\\
1693.1	0.0\\
1698.1	-1.7\\
1702.5	-1.1\\
1707.0	-2.5\\
1711.7	-2.1\\
1716.2	-2.6\\
1720.7	-2.3\\
1725.0	-2.5\\
1729.4	-4.5\\
1734.0	-1.1\\
1738.4	-2.3\\
1742.6	-5.1\\
1747.9	-6.3\\
1751.2	-10.8\\
1755.5	0.0\\
1759.4	-0.3\\
1763.8	-0.3\\
1769.0	0.0\\
2801.3	0.0\\
};
\addlegendentry{\scriptsize IDM }

\addplot[area legend, draw=none, fill=white!10!black, fill opacity=0.2]
table[row sep=crcr] {%
x	y\\
1354.4	-25\\
1354.4	0\\
1432	0\\
1432	-25\\
}--cycle;
\addlegendentry{\scriptsize U-Turn}

\end{axis}

%% file: Images/UDDS_pout_Nissan.tex
%
%
\definecolor{mycolor1}{rgb}{0.00000,0.44700,0.74100}%
\definecolor{mycolor2}{rgb}{0.85000,0.32500,0.09800}%
\definecolor{mycolor3}{rgb}{0.92900,0.69400,0.12500}%
\definecolor{mycolor4}{rgb}{0.49400,0.18400,0.55600}%
\begin{tikzpicture}

\begin{axis}[%
width=6.5cm,
height=4.0cm,
at={(0cm,8.8cm)},
scale only axis,
xmin=505,
xmax=2800,
xlabel style={font=\color{white!15!black} \scriptsize},
xlabel={},
xticklabels={,,},
ymin=0,
ymax=65,
ylabel style={font=\color{white!15!black} \scriptsize},
ylabel={Power Out [KW]},
ytick={0,20,40,60},
yticklabels={0,+20,+40,+60},
ticklabel style = {font = \scriptsize},
axis background/.style={fill=white},
xmajorgrids,
ymajorgrids,
legend style={legend cell align=left, align=left, draw=white!15!black}
]

\addplot [color=mycolor1, line width=1.3pt]
  table[row sep=crcr]{%
504.8	6.3\\
507.5	6.1\\
510.5	6.8\\
513.3	7.5\\
515.9	6.8\\
518.6	6.1\\
521.4	6.6\\
527.4	6.2\\
532.7	6.4\\
535.3	6.4\\
537.9	6.2\\
540.8	6.4\\
543.6	6.3\\
546.6	6.3\\
549.6	6.4\\
552.4	6.2\\
555.3	6.3\\
558.2	6.2\\
561.0	6.2\\
563.6	6.4\\
566.4	6.2\\
569.6	6.5\\
572.6	6.3\\
578.2	6.3\\
581.0	6.4\\
584.2	6.2\\
587.1	6.2\\
589.9	6.2\\
592.7	6.4\\
598.9	6.2\\
601.8	6.1\\
604.7	6.3\\
607.7	5.5\\
613.8	5.5\\
619.4	5.3\\
622.4	5.1\\
625.4	5.5\\
631.4	5.3\\
634.4	5.4\\
637.5	5.3\\
640.4	5.1\\
643.5	5.4\\
646.4	5.4\\
649.5	5.6\\
652.5	5.3\\
655.5	5.7\\
661.3	5.7\\
664.2	5.5\\
667.4	5.9\\
670.3	5.6\\
676.1	5.6\\
678.9	5.2\\
682.1	5.5\\
685.3	5.1\\
688.6	5.5\\
691.6	5.3\\
694.7	5.2\\
697.8	5.3\\
703.8	5.3\\
706.5	5.4\\
709.6	5.2\\
712.7	5.1\\
717.2	5.2\\
719.5	5.0\\
722.0	5.0\\
725.0	4.2\\
727.9	3.0\\
730.9	3.2\\
734.0	3.1\\
736.9	2.8\\
740.1	3.1\\
744.3	3.0\\
746.3	2.2\\
752.6	1.6\\
758.5	1.6\\
761.6	1.8\\
765.0	1.2\\
767.9	1.1\\
773.8	0.5\\
776.7	0.3\\
779.7	0.3\\
782.4	0.0\\
930.2	0.0\\
930.7	0.5\\
931.0	0.5\\
931.7	0.6\\
932.2	0.6\\
932.7	0.7\\
932.9	0.8\\
933.5	0.7\\
934.0	0.7\\
934.7	0.6\\
935.3	0.6\\
935.7	0.8\\
936.1	0.8\\
936.3	0.9\\
937.1	0.9\\
937.6	1.1\\
938.2	0.9\\
938.6	0.9\\
939.0	1.0\\
939.1	1.1\\
939.5	1.1\\
939.9	1.2\\
940.2	1.2\\
940.5	1.0\\
940.7	1.0\\
940.7	1.1\\
940.8	1.0\\
940.8	1.0\\
940.8	0.9\\
940.9	0.9\\
940.9	1.0\\
941.1	1.1\\
941.1	0.9\\
941.2	1.0\\
941.2	0.9\\
941.4	0.9\\
941.4	1.0\\
941.5	0.9\\
941.6	1.0\\
941.8	1.0\\
941.9	1.1\\
941.9	0.9\\
942.3	0.9\\
942.5	1.0\\
943.1	1.0\\
943.2	0.9\\
943.3	1.1\\
943.5	0.9\\
943.5	1.0\\
943.7	0.9\\
943.8	0.9\\
944.0	1.1\\
944.0	1.0\\
944.3	0.9\\
944.5	1.0\\
944.6	0.9\\
944.7	0.9\\
944.9	1.0\\
945.0	1.1\\
945.0	0.9\\
945.2	1.1\\
945.3	0.9\\
945.5	0.9\\
945.6	1.1\\
945.7	1.1\\
945.7	0.9\\
945.9	1.0\\
945.9	1.0\\
946.1	0.9\\
946.4	0.9\\
946.5	1.0\\
946.5	1.0\\
946.6	0.9\\
946.7	0.9\\
946.8	0.8\\
946.9	1.0\\
947.0	1.0\\
947.0	1.1\\
947.1	1.0\\
947.1	1.1\\
947.2	0.8\\
947.4	0.9\\
947.7	0.9\\
947.9	1.0\\
948.0	1.1\\
948.1	0.9\\
948.2	0.9\\
948.2	1.1\\
948.3	0.9\\
948.5	1.0\\
948.7	0.9\\
948.8	0.9\\
948.9	1.1\\
949.1	0.9\\
949.2	0.9\\
949.2	1.1\\
949.3	0.9\\
949.5	0.9\\
949.7	1.0\\
949.9	1.0\\
949.9	1.2\\
950.0	1.0\\
950.1	1.0\\
950.1	0.9\\
950.2	1.0\\
950.2	0.8\\
950.4	1.2\\
950.5	2.5\\
950.5	3.7\\
950.6	4.4\\
951.1	4.3\\
951.7	4.1\\
952.1	3.8\\
953.2	4.1\\
954.2	4.3\\
954.9	4.4\\
955.8	4.9\\
956.6	5.1\\
957.5	6.1\\
958.0	6.3\\
958.9	5.8\\
959.8	6.0\\
960.9	6.5\\
962.1	7.0\\
962.9	6.7\\
964.1	7.0\\
967.4	7.0\\
968.6	7.3\\
969.9	7.3\\
971.0	7.5\\
972.5	7.5\\
974.0	7.7\\
975.4	8.2\\
979.6	8.2\\
982.3	8.3\\
983.9	8.2\\
985.2	8.4\\
986.8	8.3\\
988.9	8.4\\
990.0	8.4\\
991.5	8.5\\
993.5	8.3\\
995.4	8.2\\
998.4	8.1\\
999.4	8.2\\
1001.9	8.2\\
1003.5	8.3\\
1004.9	8.2\\
1014.4	8.2\\
1016.0	8.1\\
1017.5	8.1\\
1019.8	8.0\\
1022.8	7.8\\
1025.1	7.9\\
1031.0	7.9\\
1033.3	8.0\\
1036.5	7.9\\
1041.1	6.7\\
1043.3	6.8\\
1045.6	6.8\\
1047.9	8.4\\
1050.4	7.9\\
1052.0	6.7\\
1055.5	6.9\\
1057.4	7.1\\
1059.7	8.0\\
1062.7	8.1\\
1065.4	8.1\\
1067.2	7.3\\
1070.6	6.7\\
1073.1	6.7\\
1076.3	8.1\\
1078.6	8.2\\
1081.2	7.9\\
1083.1	7.8\\
1089.0	7.8\\
1091.6	7.9\\
1094.5	7.8\\
1096.1	7.7\\
1099.8	7.8\\
1101.7	7.7\\
1105.6	7.6\\
1107.7	6.7\\
1111.0	6.6\\
1113.8	6.6\\
1115.5	8.1\\
1118.6	8.1\\
1122.7	7.9\\
1124.6	7.4\\
1128.2	6.7\\
1129.8	6.6\\
1133.8	6.4\\
1135.3	8.0\\
1139.9	7.8\\
1141.7	6.7\\
1145.9	6.2\\
1148.6	6.6\\
1151.6	6.4\\
1154.5	6.5\\
1157.4	5.9\\
1160.4	5.3\\
1163.6	3.9\\
1166.6	2.0\\
1168.3	2.0\\
1171.6	2.4\\
1178.9	2.8\\
1180.4	2.5\\
1184.8	2.1\\
1188.0	1.9\\
1190.9	1.9\\
1193.7	2.0\\
1196.6	1.9\\
1201.6	1.9\\
1203.2	2.0\\
1206.1	1.8\\
1209.1	1.9\\
1212.1	1.6\\
1215.1	1.2\\
1218.2	1.2\\
1221.2	0.6\\
1224.2	0.7\\
1232.9	0.4\\
1235.9	0.5\\
1238.8	0.2\\
1241.8	0.0\\
1354.3	0.0\\
1356.2	5.9\\
1358.1	6.3\\
1360.0	4.6\\
1362.0	2.5\\
1364.0	1.5\\
1366.0	0.0\\
1367.9	1.4\\
1369.5	2.7\\
1371.4	2.3\\
1374.0	1.5\\
1376.2	0.2\\
1378.5	0.0\\
1380.7	0.1\\
1382.9	0.7\\
1390.9	0.7\\
1393.0	1.3\\
1395.1	2.0\\
1397.1	3.3\\
1398.5	5.1\\
1401.3	5.0\\
1402.9	4.3\\
1404.6	4.0\\
1406.4	4.0\\
1408.2	4.1\\
1410.2	3.7\\
1412.6	2.9\\
1414.6	2.9\\
1416.4	2.1\\
1418.6	2.0\\
1420.6	1.7\\
1422.7	0.6\\
1424.7	0.0\\
1426.7	0.0\\
1428.6	0.7\\
1430.5	1.0\\
1432.4	0.0\\
1434.7	0.0\\
1436.8	0.3\\
1438.6	1.3\\
1440.5	1.2\\
1442.3	1.2\\
1444.4	15.3\\
1445.8	19.4\\
1447.9	19.4\\
1449.9	19.7\\
1451.8	20.3\\
1453.9	21.1\\
1455.9	21.8\\
1457.9	23.1\\
1460.1	25.0\\
1461.6	25.4\\
1463.5	25.3\\
1468.1	25.9\\
1470.2	25.9\\
1473.5	27.2\\
1476.3	29.0\\
1479.2	29.9\\
1481.5	29.5\\
1484.5	29.9\\
1487.2	30.5\\
1490.0	31.7\\
1493.0	32.6\\
1496.2	35.4\\
1499.5	35.5\\
1504.0	36.0\\
1506.2	36.1\\
1509.7	36.0\\
1513.1	36.4\\
1516.9	36.7\\
1520.6	37.8\\
1524.3	38.2\\
1529.6	39.0\\
1531.9	41.0\\
1536.1	42.0\\
1541.4	41.0\\
1544.3	41.6\\
1548.3	41.1\\
1552.5	41.8\\
1557.0	42.1\\
1561.3	42.3\\
1565.5	42.7\\
1572.4	42.7\\
1577.1	43.0\\
1583.8	43.2\\
1587.9	42.8\\
1593.2	42.7\\
1596.3	42.8\\
1601.0	42.7\\
1606.0	42.7\\
1610.4	42.6\\
1617.3	42.9\\
1621.1	42.6\\
1625.7	42.4\\
1631.0	41.9\\
1636.8	41.9\\
1646.7	42.1\\
1652.2	42.0\\
1657.4	41.5\\
1665.6	41.0\\
1670.6	40.8\\
1674.0	39.1\\
1681.1	38.6\\
1685.2	38.9\\
1691.9	38.4\\
1698.0	37.9\\
1701.9	35.1\\
1707.4	33.2\\
1716.0	32.9\\
1720.8	27.1\\
1727.3	26.8\\
1733.8	22.6\\
1739.1	22.6\\
1745.2	21.4\\
1754.5	15.9\\
1761.1	17.2\\
1767.7	16.0\\
1772.4	16.9\\
1777.9	16.2\\
1786.0	17.0\\
1789.3	15.8\\
1797.4	16.1\\
1801.3	14.0\\
1806.8	14.4\\
1813.0	13.6\\
1819.4	13.9\\
1824.9	13.9\\
1833.2	13.6\\
1836.5	10.6\\
1842.3	11.5\\
1848.5	13.3\\
1854.1	13.8\\
1861.9	14.2\\
1867.7	14.4\\
1873.5	14.5\\
1879.0	14.2\\
1885.2	14.7\\
1891.2	14.6\\
1897.1	14.9\\
1903.0	14.3\\
1908.5	15.1\\
1914.6	15.4\\
1920.2	15.8\\
1924.9	15.2\\
1937.1	15.4\\
1943.4	15.6\\
1949.1	15.0\\
1954.6	15.3\\
1960.4	15.3\\
1965.9	15.0\\
1972.2	15.3\\
1978.1	15.1\\
1984.0	15.3\\
1989.9	15.0\\
1995.1	15.2\\
2001.2	14.7\\
2007.1	14.4\\
2012.9	14.3\\
2018.9	14.1\\
2024.3	14.4\\
2030.5	14.1\\
2036.3	14.3\\
2041.8	14.1\\
2048.1	14.4\\
2053.7	14.3\\
2060.5	14.5\\
2065.9	14.2\\
2071.9	14.7\\
2081.3	14.9\\
2088.7	14.7\\
2096.3	14.9\\
2101.0	14.7\\
2105.0	14.4\\
2112.6	14.8\\
2118.0	14.6\\
2129.9	14.4\\
2135.9	14.5\\
2141.1	14.2\\
2146.8	14.4\\
2152.9	14.9\\
2158.7	14.3\\
2164.7	14.6\\
2170.6	14.5\\
2175.9	15.2\\
2182.2	15.3\\
2199.0	15.3\\
2205.3	15.5\\
2211.1	14.9\\
2216.8	15.4\\
2222.7	14.6\\
2228.6	14.7\\
2233.9	14.0\\
2240.1	13.5\\
2245.5	12.9\\
2251.7	13.4\\
2257.0	13.2\\
2263.3	13.6\\
2269.2	12.9\\
2274.5	13.9\\
2280.8	12.5\\
2286.1	13.9\\
2292.5	13.2\\
2298.4	14.4\\
2304.1	13.7\\
2309.6	15.5\\
2313.7	14.1\\
2321.3	15.2\\
2324.8	14.2\\
2332.8	14.7\\
2338.3	14.0\\
2344.4	14.7\\
2350.3	13.5\\
2355.9	14.4\\
2360.2	13.6\\
2367.5	14.2\\
2371.6	13.4\\
2378.5	13.8\\
2384.9	14.6\\
2389.6	14.0\\
2396.5	13.0\\
2401.4	12.1\\
2408.4	12.0\\
2414.0	12.9\\
2420.2	12.2\\
2425.3	11.9\\
2431.3	12.1\\
2437.2	12.9\\
2442.7	11.8\\
2448.6	10.1\\
2454.1	10.4\\
2466.1	10.6\\
2471.8	10.8\\
2477.7	10.7\\
2483.3	10.4\\
2489.1	11.2\\
2495.0	10.5\\
2500.4	10.8\\
2506.1	10.8\\
2511.8	11.5\\
2517.5	10.2\\
2523.4	10.0\\
2528.9	10.0\\
2534.6	10.3\\
2540.0	8.9\\
2545.6	9.4\\
2551.7	9.0\\
2557.3	8.1\\
2563.0	3.8\\
2568.6	3.5\\
2574.0	4.0\\
2580.0	4.3\\
2585.3	3.7\\
2590.8	4.1\\
2596.3	3.8\\
2602.0	3.3\\
2608.0	1.6\\
2613.2	0.0\\
2803.5	0.0\\
};
\addlegendentry{ \scriptsize MPC-C}

\addplot [color=mycolor2, dashed, line width=1.3pt]
  table[row sep=crcr]{%
503.1	10.2\\
505.2	9.6\\
508.0	9.1\\
510.5	9.2\\
513.7	6.5\\
517.5	6.4\\
519.5	6.4\\
522.4	7.9\\
526.1	7.4\\
529.4	6.2\\
531.4	6.3\\
534.1	6.5\\
538.1	6.4\\
540.1	6.6\\
544.1	6.2\\
549.0	6.6\\
552.1	6.4\\
555.0	6.5\\
559.0	5.9\\
561.0	5.0\\
563.8	5.2\\
567.3	4.6\\
570.5	3.6\\
573.9	3.1\\
577.0	2.9\\
579.0	2.9\\
581.9	3.1\\
584.7	3.4\\
588.0	3.5\\
591.0	3.3\\
593.9	3.6\\
597.3	4.0\\
599.9	4.0\\
602.9	4.3\\
605.7	4.9\\
608.8	5.2\\
613.2	5.7\\
616.3	5.6\\
617.8	5.4\\
621.5	5.3\\
624.0	5.1\\
627.8	4.9\\
630.1	4.9\\
633.0	4.7\\
636.9	4.6\\
641.5	1.8\\
642.4	0.9\\
648.2	1.0\\
651.6	1.1\\
655.2	1.0\\
658.3	1.1\\
660.0	1.0\\
663.2	1.1\\
667.3	1.6\\
668.9	4.0\\
673.0	4.0\\
676.0	4.5\\
677.9	5.0\\
682.3	5.3\\
683.8	5.6\\
687.6	7.0\\
690.9	7.6\\
692.8	7.9\\
696.3	6.4\\
698.7	6.3\\
702.1	6.2\\
708.9	6.2\\
711.4	5.9\\
714.8	5.9\\
718.1	5.6\\
719.9	5.1\\
724.1	4.3\\
727.5	3.4\\
730.3	3.5\\
733.1	2.2\\
736.3	1.4\\
739.4	1.6\\
741.9	1.5\\
745.5	1.4\\
748.6	1.0\\
750.2	0.7\\
754.6	0.8\\
757.6	0.7\\
760.4	0.9\\
762.6	1.0\\
766.2	2.2\\
770.3	4.9\\
775.8	4.7\\
778.8	4.8\\
784.5	5.8\\
787.8	6.2\\
790.1	8.5\\
793.8	8.6\\
796.8	8.2\\
799.6	7.9\\
803.0	7.8\\
806.0	7.8\\
809.1	8.0\\
812.3	8.1\\
815.4	8.6\\
818.1	8.1\\
821.5	8.1\\
824.6	7.8\\
827.9	6.7\\
830.9	6.5\\
833.9	6.6\\
837.1	6.4\\
840.2	6.3\\
843.4	5.3\\
846.3	5.3\\
849.7	4.2\\
852.9	1.4\\
856.0	1.0\\
859.1	1.3\\
862.0	0.2\\
865.2	0.0\\
926.5	0.0\\
927.2	0.2\\
927.8	0.7\\
928.2	0.7\\
928.4	1.0\\
929.0	1.0\\
929.2	1.1\\
929.3	0.9\\
929.3	1.0\\
929.3	0.9\\
929.3	1.1\\
929.4	0.5\\
929.4	0.6\\
929.4	0.5\\
929.4	0.6\\
929.4	0.4\\
929.4	0.6\\
929.4	0.4\\
929.4	0.5\\
929.4	0.4\\
929.4	0.5\\
929.4	0.4\\
929.4	0.5\\
929.4	0.4\\
929.4	0.7\\
929.4	0.4\\
929.4	0.5\\
929.4	0.4\\
929.4	0.5\\
929.4	0.4\\
929.4	0.5\\
929.4	0.3\\
929.5	0.6\\
929.5	0.4\\
929.5	0.5\\
929.5	0.4\\
929.5	0.6\\
929.5	0.4\\
929.5	0.6\\
929.5	0.4\\
929.5	0.5\\
929.5	0.3\\
929.5	0.6\\
929.5	0.4\\
929.5	0.5\\
929.5	0.4\\
929.5	0.7\\
929.5	0.5\\
929.5	0.4\\
929.5	0.5\\
929.5	0.4\\
929.5	0.5\\
929.5	0.4\\
929.5	0.5\\
929.5	0.4\\
929.5	0.6\\
929.5	0.4\\
929.5	0.6\\
929.5	0.4\\
929.5	0.7\\
929.6	0.4\\
929.6	0.5\\
929.6	0.3\\
929.6	0.4\\
929.6	0.3\\
929.6	0.5\\
929.6	0.4\\
929.6	0.6\\
929.6	0.4\\
929.6	0.5\\
929.6	0.4\\
929.6	0.5\\
929.7	1.7\\
929.9	3.2\\
930.2	4.5\\
930.6	5.8\\
931.3	7.1\\
931.8	8.5\\
932.4	9.5\\
933.1	10.7\\
934.0	10.3\\
935.4	10.8\\
936.1	11.6\\
937.3	12.5\\
938.4	13.4\\
940.0	14.3\\
941.0	14.9\\
944.5	15.4\\
945.6	16.2\\
947.3	16.5\\
949.5	17.2\\
951.3	17.9\\
952.6	18.6\\
954.6	18.6\\
957.8	9.4\\
959.6	3.4\\
962.6	4.9\\
965.6	5.6\\
967.6	4.8\\
969.3	5.1\\
971.4	5.6\\
976.0	5.4\\
977.8	6.4\\
979.7	6.2\\
982.2	3.6\\
984.1	2.7\\
986.6	4.5\\
988.3	1.0\\
990.5	0.0\\
997.7	0.0\\
999.6	0.6\\
1001.9	0.6\\
1003.9	0.7\\
1004.9	0.6\\
1007.8	1.0\\
1009.9	1.1\\
1011.4	3.1\\
1013.9	5.6\\
1015.2	5.3\\
1017.9	5.8\\
1020.3	5.9\\
1022.0	6.0\\
1023.9	6.2\\
1026.1	5.9\\
1029.1	6.1\\
1031.1	6.6\\
1033.6	6.5\\
1035.7	6.6\\
1038.1	8.7\\
1040.4	9.3\\
1042.5	9.0\\
1044.6	8.2\\
1047.5	13.3\\
1049.8	14.6\\
1051.6	14.9\\
1054.8	13.5\\
1056.9	3.4\\
1060.1	1.9\\
1061.5	2.2\\
1064.9	0.6\\
1067.2	0.0\\
1094.4	0.0\\
1096.1	4.1\\
1097.7	3.7\\
1099.7	2.6\\
1101.4	2.6\\
1105.0	8.8\\
1107.6	8.0\\
1109.5	6.4\\
1110.7	5.9\\
1115.3	5.4\\
1117.3	8.4\\
1119.3	10.2\\
1121.4	10.2\\
1123.5	11.2\\
1125.6	14.4\\
1129.1	16.3\\
1130.1	16.5\\
1133.9	14.9\\
1136.3	15.2\\
1138.8	15.3\\
1141.3	16.2\\
1143.8	17.4\\
1146.4	17.5\\
1149.1	17.5\\
1151.8	16.9\\
1157.3	16.9\\
1160.2	17.1\\
1162.8	15.9\\
1165.2	15.0\\
1168.8	15.4\\
1171.7	15.3\\
1174.8	15.5\\
1176.5	15.1\\
1180.1	14.3\\
1182.9	14.0\\
1185.6	11.1\\
1188.6	9.2\\
1192.1	9.7\\
1195.7	9.3\\
1199.5	9.5\\
1203.0	8.9\\
1205.8	9.5\\
1209.3	6.8\\
1212.5	5.6\\
1215.8	5.6\\
1222.0	5.2\\
1225.5	4.8\\
1228.7	4.0\\
1231.9	2.1\\
1235.8	3.1\\
1238.5	0.0\\
1283.9	0.0\\
1286.2	0.9\\
1288.2	5.8\\
1290.6	5.5\\
1292.7	5.1\\
1294.9	3.1\\
1297.1	2.6\\
1299.2	2.3\\
1301.5	2.2\\
1303.7	1.5\\
1305.9	1.2\\
1308.1	0.8\\
1310.1	0.6\\
1312.4	0.0\\
1325.0	0.0\\
1327.0	2.6\\
1329.0	1.0\\
1331.0	0.0\\
1334.7	0.0\\
1336.5	0.9\\
1338.5	0.0\\
1340.2	0.5\\
1342.2	2.5\\
1344.1	2.6\\
1345.9	2.2\\
1347.8	1.5\\
1351.2	0.0\\
1353.2	0.6\\
1356.0	1.2\\
1358.2	1.0\\
1361.1	1.2\\
1362.8	1.4\\
1364.9	2.0\\
1367.0	2.1\\
1369.1	2.3\\
1371.0	2.6\\
1373.1	3.1\\
1375.1	4.1\\
1376.9	4.0\\
1381.9	4.0\\
1383.8	3.9\\
1385.6	3.1\\
1387.7	2.9\\
1389.3	2.8\\
1391.1	3.2\\
1392.8	3.2\\
1395.4	2.8\\
1397.5	2.3\\
1399.3	1.4\\
1401.4	1.4\\
1403.6	0.7\\
1405.5	0.5\\
1407.6	0.8\\
1409.5	1.0\\
1411.4	1.3\\
1413.4	1.0\\
1415.0	1.1\\
1418.9	1.1\\
1420.8	1.2\\
1422.4	2.1\\
1424.0	3.5\\
1425.8	14.0\\
1428.0	23.6\\
1429.9	24.0\\
1432.3	25.0\\
1434.4	26.6\\
1437.1	28.5\\
1439.4	29.8\\
1441.8	32.4\\
1444.5	35.4\\
1447.1	37.0\\
1450.0	38.3\\
1452.7	40.1\\
1455.7	42.3\\
1458.8	46.1\\
1461.8	49.5\\
1465.3	51.7\\
1468.8	52.9\\
1472.2	55.2\\
1476.0	56.6\\
1479.7	58.4\\
1484.4	59.9\\
1487.7	61.8\\
1491.7	63.0\\
1496.0	61.9\\
1500.4	63.2\\
1504.6	64.7\\
1509.4	64.6\\
1513.8	64.6\\
1518.5	62.2\\
1523.1	61.3\\
1530.7	63.0\\
1535.4	64.5\\
1539.9	64.9\\
1545.2	63.6\\
1550.5	61.0\\
1555.8	59.4\\
1561.0	60.1\\
1566.3	59.2\\
1572.2	56.8\\
1577.8	51.9\\
1583.4	51.8\\
1589.1	50.3\\
1598.4	41.2\\
1604.1	40.4\\
1609.9	36.1\\
1615.7	30.6\\
1621.6	27.9\\
1627.4	20.2\\
1633.0	18.7\\
1638.8	20.0\\
1644.6	18.4\\
1650.5	18.7\\
1656.3	17.0\\
1664.1	16.3\\
1668.6	16.3\\
1673.8	15.9\\
1679.6	14.5\\
1685.4	15.4\\
1691.2	14.7\\
1696.8	15.5\\
1702.9	15.8\\
1708.1	16.1\\
1714.1	16.4\\
1721.0	16.5\\
1727.6	17.2\\
1731.9	17.4\\
1738.5	16.9\\
1743.2	17.0\\
1748.7	17.4\\
1757.1	17.1\\
1760.5	17.6\\
1767.9	16.7\\
1772.2	17.7\\
1778.0	16.5\\
1783.6	17.6\\
1790.0	16.8\\
1797.4	17.2\\
1803.5	15.7\\
1806.9	16.0\\
1812.7	12.3\\
1819.0	9.5\\
1824.4	9.5\\
1830.2	9.3\\
1838.4	9.8\\
1843.4	13.9\\
1847.7	13.2\\
1859.0	10.6\\
1864.8	8.7\\
1870.5	8.8\\
1878.3	8.9\\
1882.1	14.0\\
1887.7	13.2\\
1893.2	13.5\\
1898.5	12.9\\
1910.2	14.1\\
1916.0	15.1\\
1921.5	15.8\\
1927.8	15.8\\
1932.8	16.2\\
1938.7	18.6\\
1944.7	19.6\\
1953.1	16.5\\
1956.5	16.5\\
1964.1	16.9\\
1968.3	17.9\\
1975.5	18.3\\
1980.0	17.8\\
1985.0	17.9\\
1991.0	17.7\\
1998.0	17.9\\
2004.4	17.8\\
2009.2	17.2\\
2016.0	17.0\\
2019.9	18.0\\
2027.3	17.9\\
2033.5	17.9\\
2038.4	17.1\\
2044.8	17.3\\
2049.0	17.0\\
2056.1	17.3\\
2062.2	16.8\\
2068.2	16.9\\
2074.0	16.7\\
2079.2	16.9\\
2084.9	16.6\\
2088.4	16.4\\
2096.7	15.7\\
2102.6	16.2\\
2107.8	15.9\\
2113.7	15.1\\
2119.3	15.4\\
2123.3	15.5\\
2129.5	15.3\\
2136.6	15.3\\
2143.1	15.4\\
2149.4	15.2\\
2153.7	15.4\\
2158.5	15.3\\
2166.6	15.3\\
2172.4	15.4\\
2178.0	15.4\\
2183.2	14.6\\
2189.6	14.2\\
2195.6	11.5\\
2201.4	9.6\\
2207.2	13.0\\
2214.1	12.9\\
2221.3	14.2\\
2227.9	14.6\\
2236.3	13.2\\
2243.7	9.8\\
2250.6	9.9\\
2256.9	9.5\\
2262.6	8.7\\
2268.4	8.1\\
2274.1	8.0\\
2280.1	6.7\\
2285.7	8.2\\
2291.1	9.1\\
2296.8	9.6\\
2302.4	9.6\\
2307.8	9.9\\
2313.4	9.6\\
2319.1	9.9\\
2324.6	9.2\\
2328.4	8.7\\
2336.0	7.7\\
2339.7	4.6\\
2347.0	7.5\\
2352.6	8.8\\
2358.4	8.4\\
2363.7	8.1\\
2368.7	4.0\\
2375.0	4.1\\
2380.4	6.5\\
2383.9	7.1\\
2391.1	7.7\\
2396.8	7.8\\
2401.7	8.1\\
2407.6	8.2\\
2413.2	8.6\\
2418.5	9.6\\
2423.9	9.8\\
2429.3	9.8\\
2434.8	11.1\\
2440.4	12.7\\
2445.6	10.7\\
2450.8	10.1\\
2456.3	13.6\\
2461.7	13.8\\
2467.3	13.6\\
2472.5	13.1\\
2478.0	13.6\\
2483.4	13.7\\
2488.8	13.3\\
2494.4	13.5\\
2499.7	13.5\\
2505.1	14.0\\
2510.5	13.2\\
2515.9	14.3\\
2521.6	14.4\\
2526.8	15.2\\
2532.2	14.6\\
2537.7	15.4\\
2543.1	15.8\\
2548.8	16.1\\
2553.6	16.4\\
2559.5	16.3\\
2570.4	16.5\\
2575.9	16.3\\
2580.7	15.9\\
2586.6	15.6\\
2592.1	16.0\\
2597.6	14.6\\
2603.3	15.1\\
2608.6	14.3\\
2614.1	14.2\\
2619.6	10.2\\
2625.0	9.5\\
2630.7	10.5\\
2636.0	10.7\\
2641.4	9.3\\
2646.9	9.0\\
2652.3	7.8\\
2658.0	5.6\\
2663.2	2.4\\
2668.6	2.9\\
2673.9	2.7\\
2679.3	2.6\\
2684.8	2.4\\
2690.0	2.0\\
2695.0	0.0\\
2803.9	0.0\\
};
\addlegendentry{\scriptsize MPC-U }

\addplot [color=mycolor3, dashdotted, line width=1.3pt]
  table[row sep=crcr]{%
503.1	15.4\\
506.6	15.6\\
509.3	16.4\\
512.1	17.8\\
515.1	19.1\\
517.7	18.5\\
522.3	18.1\\
523.8	18.6\\
528.5	19.5\\
531.3	19.7\\
534.7	20.6\\
537.9	20.1\\
540.1	20.5\\
544.1	3.7\\
547.3	1.6\\
550.4	2.2\\
553.5	2.5\\
556.7	2.8\\
560.0	4.5\\
563.3	4.4\\
566.5	4.1\\
569.5	3.4\\
572.7	3.2\\
575.8	3.0\\
578.8	3.0\\
582.1	3.3\\
585.3	3.5\\
588.6	2.9\\
591.5	3.5\\
594.6	2.8\\
597.7	3.2\\
600.9	1.2\\
604.0	0.0\\
679.1	0.0\\
681.8	5.4\\
684.3	10.2\\
687.4	15.7\\
690.0	13.4\\
692.8	11.9\\
695.6	12.4\\
698.4	12.3\\
701.6	12.9\\
704.6	12.2\\
707.2	12.6\\
710.4	12.4\\
713.1	14.1\\
716.2	13.9\\
719.3	13.9\\
722.5	14.0\\
725.7	13.9\\
728.6	14.2\\
731.7	14.3\\
734.8	15.0\\
737.8	15.1\\
741.4	4.2\\
744.4	1.4\\
747.8	1.8\\
751.0	2.4\\
754.2	2.7\\
757.4	2.5\\
760.3	2.7\\
763.6	2.7\\
766.7	2.8\\
770.0	2.8\\
773.2	2.6\\
779.4	3.0\\
782.3	2.6\\
785.4	2.3\\
788.6	1.5\\
791.4	0.0\\
824.9	0.0\\
827.9	0.2\\
830.6	8.7\\
833.4	8.5\\
836.1	8.8\\
839.3	10.4\\
842.1	9.3\\
845.0	9.5\\
848.2	9.4\\
850.9	9.6\\
854.1	9.6\\
857.2	9.5\\
860.1	9.5\\
863.5	9.9\\
866.4	10.1\\
869.8	10.3\\
872.9	4.3\\
876.0	1.4\\
879.1	1.1\\
882.9	0.0\\
934.7	0.0\\
935.5	0.1\\
936.0	0.0\\
939.8	0.0\\
939.9	0.8\\
940.2	0.7\\
940.5	0.8\\
940.8	1.0\\
941.0	1.3\\
941.1	1.3\\
941.3	0.9\\
941.4	0.9\\
941.5	1.0\\
941.5	0.9\\
941.7	1.1\\
941.8	0.6\\
942.0	0.7\\
942.1	0.8\\
942.3	0.6\\
942.3	0.7\\
942.5	0.6\\
942.5	0.7\\
942.6	0.7\\
942.7	0.8\\
942.8	0.7\\
942.8	0.8\\
942.8	0.7\\
942.9	0.8\\
942.9	0.8\\
943.1	0.7\\
943.3	0.7\\
943.3	0.5\\
943.5	0.7\\
943.6	0.6\\
943.7	0.8\\
943.8	0.7\\
944.0	0.7\\
944.1	0.8\\
944.6	0.6\\
944.7	0.7\\
944.7	0.6\\
944.9	0.7\\
944.9	0.6\\
945.1	0.6\\
945.2	0.9\\
945.3	0.7\\
945.5	0.7\\
945.8	0.8\\
945.8	0.6\\
946.1	0.6\\
946.2	0.8\\
946.5	0.8\\
946.7	0.9\\
946.8	0.9\\
946.9	0.8\\
947.0	0.5\\
947.2	0.6\\
947.3	0.6\\
947.4	0.7\\
947.5	0.7\\
947.6	0.8\\
947.7	0.8\\
947.7	0.6\\
948.0	0.9\\
948.1	0.6\\
948.3	0.8\\
948.3	0.7\\
948.4	0.8\\
948.4	0.7\\
948.6	0.8\\
948.7	0.7\\
949.2	0.7\\
949.4	0.6\\
949.6	0.7\\
949.8	0.6\\
949.9	0.7\\
950.2	0.7\\
950.6	0.6\\
950.7	0.7\\
950.9	0.6\\
951.0	0.8\\
951.1	0.8\\
951.2	0.7\\
951.5	0.7\\
951.5	0.6\\
951.7	0.8\\
951.9	0.7\\
951.9	0.7\\
952.0	0.8\\
952.0	0.7\\
952.1	0.6\\
952.2	0.8\\
952.3	0.8\\
952.3	0.7\\
952.7	0.7\\
952.8	0.8\\
953.0	0.7\\
953.0	0.8\\
953.1	0.8\\
953.2	0.7\\
953.2	0.7\\
953.3	0.8\\
953.4	0.7\\
953.5	0.7\\
953.7	0.6\\
953.7	0.8\\
953.9	0.6\\
953.9	0.7\\
954.2	0.7\\
954.3	0.8\\
954.3	0.8\\
954.6	0.7\\
954.6	0.7\\
954.6	0.6\\
954.7	0.6\\
954.9	0.7\\
955.1	0.8\\
955.2	0.6\\
955.4	3.6\\
955.6	4.6\\
955.9	4.7\\
956.2	4.4\\
956.6	4.4\\
957.1	5.4\\
957.7	7.4\\
958.2	9.0\\
959.0	9.9\\
959.9	10.8\\
960.8	11.7\\
961.8	12.9\\
962.9	13.7\\
964.0	14.1\\
965.2	15.9\\
966.5	16.1\\
967.9	16.9\\
969.5	17.5\\
971.0	17.7\\
972.2	17.8\\
974.8	16.9\\
976.7	17.5\\
978.4	15.4\\
980.4	15.9\\
982.5	18.7\\
984.7	16.2\\
986.1	15.1\\
988.6	15.5\\
990.9	16.4\\
993.3	14.4\\
995.7	10.3\\
998.0	10.2\\
1000.5	7.7\\
1003.0	6.5\\
1005.5	6.5\\
1008.0	6.4\\
1010.2	6.4\\
1012.6	6.5\\
1017.6	6.5\\
1020.2	6.4\\
1022.8	7.0\\
1025.4	6.6\\
1027.8	7.1\\
1030.3	7.4\\
1032.9	7.3\\
1035.2	7.3\\
1037.9	7.4\\
1043.4	7.4\\
1046.2	4.9\\
1048.7	3.1\\
1051.2	2.5\\
1055.5	1.0\\
1058.8	0.0\\
1136.0	0.0\\
1138.0	7.0\\
1139.9	6.6\\
1141.9	4.1\\
1144.0	0.0\\
1145.9	0.0\\
1148.0	5.5\\
1149.9	4.7\\
1151.9	3.2\\
1154.1	2.5\\
1156.6	1.0\\
1159.9	0.4\\
1163.2	1.0\\
1164.1	3.2\\
1167.2	4.4\\
1169.5	6.3\\
1171.8	7.2\\
1173.9	10.5\\
1175.7	13.5\\
1177.9	15.2\\
1180.3	18.8\\
1182.7	20.3\\
1185.0	22.5\\
1187.6	23.2\\
1189.9	23.3\\
1192.6	25.1\\
1195.5	27.1\\
1198.2	27.4\\
1200.9	27.9\\
1203.7	29.0\\
1206.7	31.7\\
1210.0	32.4\\
1213.1	32.5\\
1219.5	30.3\\
1222.6	28.2\\
1226.2	29.2\\
1229.5	29.7\\
1233.0	30.4\\
1236.6	30.3\\
1244.5	24.8\\
1248.2	22.7\\
1251.9	18.4\\
1255.7	18.7\\
1259.7	25.7\\
1263.7	27.1\\
1267.4	27.7\\
1271.4	27.3\\
1275.4	11.3\\
1279.6	1.4\\
1283.8	0.0\\
1334.8	0.0\\
1337.2	3.4\\
1339.6	6.5\\
1341.6	1.5\\
1344.0	0.0\\
1354.5	0.0\\
1356.2	7.3\\
1358.0	7.3\\
1360.4	1.5\\
1361.9	4.9\\
1363.7	3.9\\
1366.1	2.7\\
1368.1	1.4\\
1372.1	1.2\\
1374.1	1.2\\
1377.6	0.5\\
1379.9	0.5\\
1382.5	0.6\\
1384.5	4.8\\
1386.8	4.2\\
1391.2	3.0\\
1395.7	1.9\\
1399.0	3.2\\
1400.4	3.6\\
1402.4	3.5\\
1404.5	3.3\\
1408.9	3.3\\
1410.5	2.6\\
1412.1	2.7\\
1413.5	1.3\\
1415.5	0.6\\
1417.8	0.7\\
1421.9	3.5\\
1422.8	3.5\\
1426.0	2.7\\
1428.6	1.8\\
1429.8	1.3\\
1433.0	1.3\\
1434.5	1.0\\
1437.4	0.7\\
1438.9	0.8\\
1441.6	0.6\\
1442.9	1.0\\
1444.5	2.6\\
1446.7	2.5\\
1449.1	2.5\\
1450.8	2.7\\
1452.5	2.6\\
1453.6	2.7\\
1456.4	2.5\\
1460.0	2.5\\
1461.0	2.4\\
1461.9	2.4\\
1465.2	2.3\\
1467.3	2.4\\
1468.2	6.9\\
1470.2	44.6\\
1472.6	47.7\\
1475.1	51.9\\
1477.3	52.4\\
1480.1	54.8\\
1482.0	57.6\\
1486.0	60.6\\
1488.3	64.3\\
1492.5	66.1\\
1495.9	66.9\\
1498.6	67.9\\
1503.3	69.9\\
1505.7	72.8\\
1510.9	72.5\\
1513.5	61.6\\
1518.5	71.5\\
1523.5	75.0\\
1527.1	73.2\\
1532.6	69.4\\
1537.2	69.1\\
1540.7	70.0\\
1545.0	65.0\\
1551.9	63.6\\
1557.1	56.2\\
1562.1	49.6\\
1567.7	51.1\\
1572.3	52.3\\
1577.9	53.1\\
1583.2	53.2\\
1588.6	52.8\\
1594.5	51.5\\
1599.7	52.2\\
1605.4	47.0\\
1611.0	41.1\\
1616.9	2.5\\
1622.5	0.2\\
1627.7	1.8\\
1633.5	1.0\\
1639.0	0.0\\
1725.5	0.0\\
1730.6	25.1\\
1735.5	27.2\\
1740.8	25.0\\
1746.0	23.0\\
1756.2	23.2\\
1761.2	23.7\\
1766.8	24.7\\
1771.9	22.8\\
1778.1	22.7\\
1783.1	40.0\\
1787.8	41.6\\
1793.7	40.2\\
1799.1	39.3\\
1804.6	37.8\\
1810.6	36.5\\
1816.2	36.5\\
1821.6	34.5\\
1825.9	31.9\\
1832.6	29.5\\
1838.7	29.1\\
1844.5	26.1\\
1850.2	25.0\\
1855.8	21.6\\
1861.2	12.5\\
1866.8	10.0\\
1872.6	22.0\\
1878.6	22.0\\
1884.7	23.4\\
1890.2	11.2\\
1896.0	9.2\\
1901.7	8.4\\
1907.5	7.8\\
1913.2	8.9\\
1918.5	8.3\\
1924.2	7.3\\
1929.9	8.1\\
1935.6	8.2\\
1941.5	6.2\\
1947.1	5.9\\
1953.0	6.1\\
1958.6	6.0\\
1964.1	6.6\\
1969.9	6.0\\
1975.3	6.1\\
1981.2	6.0\\
1986.6	6.2\\
1992.2	5.5\\
1997.7	5.3\\
2003.1	4.8\\
2008.6	5.1\\
2014.3	4.2\\
2019.9	5.0\\
2025.4	4.3\\
2036.4	4.5\\
2041.7	4.0\\
2047.1	3.3\\
2052.5	3.7\\
2057.6	3.6\\
2063.2	3.7\\
2068.5	3.4\\
2073.8	3.9\\
2079.2	3.6\\
2084.2	3.4\\
2089.9	3.8\\
2094.9	2.5\\
2100.0	1.9\\
2105.3	1.9\\
2110.5	1.8\\
2115.7	1.9\\
2125.8	1.9\\
2130.9	2.1\\
2136.1	2.2\\
2141.1	2.0\\
2146.1	2.3\\
2151.4	2.1\\
2156.5	2.2\\
2161.7	2.1\\
2166.7	16.4\\
2171.8	18.4\\
2176.5	18.7\\
2182.0	18.2\\
2186.8	18.9\\
2192.1	18.5\\
2197.0	18.7\\
2202.5	25.2\\
2207.6	44.4\\
2218.0	45.3\\
2223.3	35.7\\
2228.6	42.1\\
2234.6	41.4\\
2240.1	41.6\\
2245.6	38.4\\
2251.1	38.3\\
2256.3	12.9\\
2262.1	10.6\\
2267.6	11.3\\
2273.1	12.4\\
2278.7	11.3\\
2283.8	7.2\\
2289.5	6.1\\
2295.5	7.4\\
2300.9	6.2\\
2306.6	4.5\\
2311.5	0.0\\
2317.0	0.2\\
2322.5	0.2\\
2328.1	0.0\\
2333.6	4.4\\
2339.1	3.4\\
2344.4	3.3\\
2349.6	0.6\\
2355.2	0.0\\
2370.9	0.0\\
2376.4	0.2\\
2381.7	0.0\\
2386.8	0.2\\
2392.2	0.6\\
2399.9	0.4\\
2403.7	0.0\\
2474.0	0.0\\
2478.8	7.5\\
2483.3	17.7\\
2488.3	17.1\\
2493.2	17.6\\
2498.1	21.2\\
2502.9	22.1\\
2507.8	19.7\\
2512.8	16.0\\
2517.7	15.7\\
2522.4	16.3\\
2527.1	16.3\\
2532.1	17.4\\
2537.3	17.0\\
2542.3	17.6\\
2547.2	17.7\\
2552.2	18.2\\
2557.0	18.1\\
2561.9	20.5\\
2567.1	43.9\\
2572.3	44.5\\
2577.6	43.2\\
2583.1	44.0\\
2587.8	44.3\\
2593.5	44.3\\
2598.8	45.1\\
2604.2	41.8\\
2609.5	42.4\\
2614.9	41.4\\
2620.4	39.3\\
2626.1	37.3\\
2631.9	30.6\\
2637.8	29.7\\
2643.5	31.5\\
2649.2	26.3\\
2654.8	24.7\\
2660.6	10.7\\
2666.3	11.0\\
2672.0	10.6\\
2677.7	11.1\\
2683.4	10.6\\
2689.0	11.7\\
2694.7	10.5\\
2700.5	9.6\\
2706.3	8.8\\
2712.3	11.2\\
2718.2	11.7\\
2723.9	6.8\\
2729.4	5.0\\
2735.1	4.7\\
2740.8	5.0\\
2746.3	5.3\\
2751.7	5.3\\
2757.1	1.6\\
2762.7	0.0\\
2784.9	0.0\\
2790.2	1.3\\
2795.6	0.0\\
2801.0	0.0\\
};
\addlegendentry{\scriptsize WIE }

\addplot [color=mycolor4, dotted, line width=1.3pt]
  table[row sep=crcr]{%
504.0	10.5\\
506.8	10.4\\
509.7	10.4\\
512.4	10.5\\
515.4	10.8\\
521.9	10.8\\
525.2	10.9\\
528.7	9.7\\
532.1	9.7\\
534.6	9.5\\
536.7	9.7\\
540.3	6.9\\
543.7	9.2\\
545.9	9.9\\
549.7	8.6\\
551.6	8.0\\
558.1	7.7\\
564.8	5.0\\
567.0	4.9\\
570.2	5.1\\
573.2	5.1\\
577.6	5.2\\
579.7	5.1\\
583.9	2.3\\
585.5	4.2\\
588.5	5.1\\
592.1	4.3\\
595.6	3.8\\
599.2	1.6\\
601.8	1.5\\
603.9	4.5\\
608.3	5.1\\
610.9	4.5\\
614.3	3.3\\
617.6	2.9\\
620.3	3.7\\
622.4	4.1\\
627.9	3.4\\
632.8	4.4\\
635.1	4.3\\
637.8	3.7\\
641.7	2.7\\
644.7	2.2\\
646.8	4.4\\
650.4	3.8\\
653.7	2.2\\
656.0	1.2\\
659.3	1.8\\
662.9	1.7\\
665.3	3.8\\
668.6	4.4\\
671.6	2.3\\
675.0	2.3\\
677.7	3.3\\
680.1	2.5\\
682.5	3.9\\
686.7	4.5\\
689.3	4.8\\
692.5	4.8\\
694.8	4.9\\
697.5	5.1\\
700.4	6.2\\
703.9	5.9\\
706.2	6.0\\
710.3	5.8\\
713.1	6.0\\
716.2	8.3\\
718.2	6.5\\
722.2	5.6\\
725.2	5.8\\
731.4	6.0\\
734.0	5.2\\
737.4	5.2\\
740.4	3.6\\
743.3	4.2\\
746.4	5.0\\
749.6	3.3\\
752.2	2.0\\
755.5	1.6\\
758.4	1.7\\
761.3	3.7\\
763.5	5.1\\
767.3	2.2\\
770.4	3.8\\
773.3	5.1\\
776.4	4.7\\
779.5	4.9\\
782.2	5.0\\
785.3	4.9\\
788.3	5.4\\
791.2	5.4\\
794.3	5.5\\
797.2	5.5\\
800.3	5.6\\
803.5	6.9\\
806.2	9.2\\
808.3	9.2\\
812.4	9.0\\
815.5	9.0\\
818.3	8.8\\
821.5	9.3\\
824.7	8.8\\
826.7	9.4\\
831.6	7.5\\
835.4	7.0\\
837.6	10.3\\
841.6	9.3\\
844.8	6.3\\
848.0	6.3\\
851.1	7.1\\
854.3	6.2\\
857.5	5.1\\
860.6	1.6\\
863.8	1.1\\
866.9	0.1\\
870.0	0.0\\
923.4	0.0\\
923.8	0.7\\
924.0	0.8\\
924.2	1.0\\
924.2	1.3\\
924.2	0.7\\
924.2	0.8\\
924.2	0.7\\
924.2	0.8\\
924.2	0.7\\
924.2	0.8\\
924.2	0.7\\
924.2	0.8\\
924.2	0.7\\
924.2	0.8\\
924.2	0.7\\
924.2	0.8\\
924.2	0.7\\
924.2	0.8\\
924.2	0.7\\
924.2	0.8\\
924.2	0.7\\
924.2	0.8\\
924.2	0.7\\
924.2	0.8\\
924.2	0.7\\
924.2	0.8\\
924.3	0.7\\
924.3	0.8\\
924.3	0.7\\
924.3	0.8\\
924.3	0.7\\
924.3	0.8\\
924.3	0.7\\
924.3	0.8\\
924.3	0.7\\
924.3	0.8\\
924.3	0.7\\
924.3	0.8\\
924.3	0.6\\
924.3	0.8\\
924.3	0.7\\
924.3	0.8\\
924.3	0.7\\
924.3	0.8\\
924.3	0.7\\
924.3	0.8\\
924.3	0.7\\
924.3	0.8\\
924.3	0.7\\
924.3	0.9\\
924.3	0.8\\
924.3	0.8\\
924.3	0.8\\
924.3	0.8\\
924.4	0.7\\
924.4	0.8\\
924.4	0.7\\
924.4	0.8\\
924.4	0.6\\
924.4	0.8\\
924.4	0.6\\
924.4	0.8\\
924.4	0.8\\
924.4	0.7\\
924.4	0.8\\
924.4	0.7\\
924.4	0.8\\
924.4	0.7\\
924.4	0.8\\
924.4	0.6\\
924.4	0.8\\
924.4	0.7\\
924.4	0.8\\
924.4	0.7\\
924.4	0.8\\
924.4	0.7\\
924.4	0.8\\
924.4	0.7\\
924.4	0.8\\
924.4	0.7\\
924.4	0.8\\
924.4	0.7\\
924.5	0.8\\
924.6	2.0\\
925.1	3.7\\
925.5	4.5\\
926.0	5.5\\
926.5	6.6\\
927.5	7.8\\
928.6	8.8\\
929.6	9.6\\
930.4	11.4\\
931.2	12.5\\
932.5	13.4\\
933.9	13.8\\
935.4	14.4\\
936.4	14.4\\
937.6	15.1\\
939.7	15.6\\
941.3	16.7\\
942.4	17.4\\
944.9	19.1\\
946.7	19.7\\
948.7	19.7\\
950.5	19.6\\
952.8	20.1\\
954.8	20.6\\
957.0	21.0\\
958.9	21.2\\
960.7	21.9\\
964.2	22.0\\
965.4	22.6\\
968.2	22.5\\
971.1	23.0\\
974.4	22.8\\
977.9	23.0\\
981.2	23.1\\
985.3	22.9\\
989.4	20.7\\
991.1	20.1\\
995.2	20.5\\
997.9	19.8\\
1001.1	17.6\\
1004.1	16.1\\
1007.1	14.5\\
1010.3	13.6\\
1013.6	10.3\\
1016.7	9.4\\
1018.7	5.6\\
1023.2	5.4\\
1025.3	5.0\\
1029.5	3.3\\
1031.5	1.4\\
1036.0	1.5\\
1038.6	0.6\\
1042.3	0.3\\
1045.6	0.0\\
1051.7	0.0\\
1054.1	0.1\\
1056.6	0.0\\
1115.1	0.0\\
1116.8	0.7\\
1118.5	1.5\\
1120.2	2.1\\
1121.9	1.5\\
1123.6	0.8\\
1125.3	0.9\\
1127.0	2.3\\
1128.7	3.2\\
1130.2	5.6\\
1132.3	6.2\\
1134.1	9.1\\
1135.9	9.8\\
1137.8	10.1\\
1139.8	12.7\\
1141.9	13.3\\
1143.8	13.9\\
1146.0	15.2\\
1148.2	17.6\\
1150.5	17.7\\
1152.7	17.8\\
1155.9	18.4\\
1159.9	20.9\\
1163.8	22.2\\
1168.2	22.2\\
1172.5	22.3\\
1177.1	22.9\\
1181.6	24.1\\
1184.9	26.4\\
1187.9	27.1\\
1191.2	26.5\\
1194.4	26.5\\
1197.7	26.8\\
1201.1	27.2\\
1204.4	27.9\\
1208.0	28.3\\
1211.5	28.3\\
1215.1	28.8\\
1218.7	29.0\\
1225.9	29.0\\
1229.5	29.7\\
1233.5	29.6\\
1237.5	30.3\\
1241.4	29.6\\
1245.4	26.8\\
1249.3	21.4\\
1253.5	16.5\\
1257.0	9.9\\
1261.5	5.2\\
1265.4	0.9\\
1269.1	0.0\\
1343.4	0.0\\
1344.8	2.8\\
1346.2	3.3\\
1347.6	4.8\\
1349.0	4.1\\
1350.1	3.5\\
1351.7	3.4\\
1353.9	3.3\\
1355.9	3.0\\
1357.7	3.4\\
1359.7	3.7\\
1361.6	4.4\\
1364.0	2.8\\
1368.4	2.4\\
1373.6	2.2\\
1375.6	2.0\\
1378.8	1.6\\
1383.2	1.9\\
1384.8	1.9\\
1386.4	2.2\\
1387.6	2.4\\
1389.4	2.4\\
1390.7	2.3\\
1392.3	2.6\\
1394.0	2.7\\
1395.9	3.8\\
1397.2	5.2\\
1399.1	5.1\\
1401.9	5.0\\
1403.1	3.6\\
1404.8	3.3\\
1407.4	3.2\\
1409.0	2.6\\
1410.8	2.4\\
1412.3	2.5\\
1416.9	2.2\\
1418.6	2.2\\
1421.4	2.4\\
1422.8	3.9\\
1424.9	5.3\\
1426.4	5.2\\
1428.1	5.3\\
1430.0	5.4\\
1435.5	5.4\\
1438.0	7.5\\
1439.8	14.9\\
1441.3	21.6\\
1443.2	21.6\\
1445.5	22.2\\
1447.7	23.5\\
1450.3	24.4\\
1454.9	24.6\\
1457.3	25.3\\
1459.8	25.9\\
1462.9	26.8\\
1465.0	29.2\\
1468.7	29.9\\
1473.2	29.9\\
1476.7	30.0\\
1479.5	30.9\\
1482.6	32.4\\
1486.9	34.8\\
1489.0	35.4\\
1492.0	35.3\\
1495.7	36.2\\
1498.7	35.9\\
1502.8	37.1\\
1506.2	37.4\\
1509.8	39.6\\
1513.6	40.5\\
1517.5	41.0\\
1521.1	40.8\\
1525.3	41.8\\
1529.2	41.6\\
1533.1	41.9\\
1537.0	42.1\\
1550.2	42.9\\
1554.8	43.1\\
1560.4	43.8\\
1563.8	43.6\\
1568.4	43.6\\
1572.9	43.7\\
1577.8	44.0\\
1582.6	44.2\\
1589.1	43.8\\
1592.1	43.9\\
1603.5	43.5\\
1606.5	42.9\\
1612.6	42.5\\
1617.4	41.9\\
1622.2	41.9\\
1629.1	41.7\\
1634.7	41.9\\
1640.9	41.0\\
1645.8	39.6\\
1653.2	38.3\\
1657.7	38.9\\
1663.9	38.5\\
1668.1	38.5\\
1674.7	37.5\\
1680.8	34.5\\
1685.3	34.1\\
1689.8	33.4\\
1695.4	32.5\\
1701.5	29.1\\
1706.7	28.0\\
1711.5	28.9\\
1720.0	28.9\\
1726.2	27.1\\
1729.4	25.4\\
1735.1	22.1\\
1741.0	22.6\\
1746.2	23.2\\
1751.7	22.6\\
1757.7	21.5\\
1763.5	20.9\\
1768.6	16.7\\
1775.0	16.7\\
1783.3	16.4\\
1786.4	17.1\\
1794.9	16.8\\
1797.7	17.3\\
1803.9	16.4\\
1809.8	16.9\\
1818.1	15.7\\
1820.8	15.7\\
1829.4	14.6\\
1832.8	14.5\\
1839.4	14.1\\
1846.6	14.4\\
1851.5	14.0\\
1856.4	14.4\\
1861.3	11.5\\
1867.7	10.6\\
1874.9	13.9\\
1878.8	13.6\\
1885.0	13.6\\
1892.6	13.4\\
1897.9	13.4\\
1901.6	12.9\\
1909.6	13.4\\
1914.8	13.1\\
1918.6	12.1\\
1926.7	12.9\\
1930.0	12.7\\
1936.1	13.0\\
1943.7	12.9\\
1946.7	12.9\\
1955.0	12.6\\
1960.4	11.9\\
1964.5	12.0\\
1969.9	12.0\\
1977.4	12.5\\
1982.1	11.9\\
1989.0	12.4\\
1994.4	11.9\\
1998.7	12.0\\
2005.8	12.2\\
2011.3	11.5\\
2014.9	12.5\\
2022.6	12.3\\
2027.6	12.4\\
2034.0	12.4\\
2039.5	12.6\\
2043.4	12.3\\
2051.1	12.2\\
2056.7	12.2\\
2062.2	12.0\\
2068.1	12.1\\
2071.2	11.8\\
2078.7	12.1\\
2082.6	12.6\\
2087.9	13.5\\
2094.5	12.3\\
2101.2	13.2\\
2105.8	13.4\\
2112.2	12.9\\
2117.7	12.7\\
2120.8	13.4\\
2128.9	13.4\\
2134.5	13.3\\
2140.0	13.7\\
2143.6	13.4\\
2149.5	12.9\\
2156.7	14.0\\
2162.3	13.2\\
2167.8	14.0\\
2171.5	13.3\\
2179.1	13.3\\
2184.9	13.4\\
2195.9	13.4\\
2199.3	12.9\\
2206.7	13.9\\
2212.3	13.8\\
2217.1	14.0\\
2223.4	12.9\\
2229.2	13.7\\
2234.5	13.5\\
2240.1	12.8\\
2245.6	11.8\\
2256.7	12.0\\
2262.2	12.0\\
2267.7	12.4\\
2273.3	12.0\\
2278.8	12.3\\
2284.3	12.1\\
2289.8	12.6\\
2295.1	12.2\\
2300.6	12.5\\
2307.1	12.0\\
2312.1	12.3\\
2318.7	11.6\\
2323.5	11.9\\
2329.2	12.0\\
2335.0	11.9\\
2346.0	12.2\\
2351.3	11.8\\
2356.8	12.2\\
2362.2	12.1\\
2367.7	11.9\\
2373.2	12.2\\
2378.6	12.2\\
2384.1	11.9\\
2389.5	12.2\\
2393.2	13.0\\
2400.5	14.0\\
2405.9	13.1\\
2411.4	12.8\\
2416.8	13.2\\
2427.7	12.7\\
2433.2	13.0\\
2438.6	12.9\\
2444.1	13.0\\
2449.5	12.9\\
2455.0	13.4\\
2460.4	13.2\\
2465.8	13.1\\
2471.3	13.8\\
2476.7	13.1\\
2482.0	13.7\\
2487.4	13.5\\
2492.8	13.8\\
2498.3	13.5\\
2503.7	13.6\\
2509.2	13.9\\
2514.6	14.6\\
2520.1	14.6\\
2525.5	14.8\\
2530.9	15.5\\
2536.4	15.8\\
2541.8	15.9\\
2547.3	16.0\\
2552.8	15.1\\
2558.2	16.4\\
2563.7	14.7\\
2569.2	16.4\\
2574.6	15.2\\
2580.1	16.0\\
2585.6	15.4\\
2591.1	16.1\\
2596.6	15.1\\
2602.1	15.7\\
2607.5	15.8\\
2613.0	15.5\\
2618.5	15.7\\
2624.0	15.6\\
2629.5	15.7\\
2635.0	15.0\\
2640.5	14.8\\
2651.5	14.6\\
2657.1	13.5\\
2662.3	11.9\\
2667.8	11.3\\
2673.3	12.9\\
2678.8	13.2\\
2684.3	13.5\\
2689.8	11.1\\
2695.3	10.8\\
2700.7	10.6\\
2706.2	10.8\\
2717.1	10.7\\
2722.6	11.1\\
2728.0	10.1\\
2733.5	10.8\\
2738.9	9.3\\
2744.3	10.0\\
2749.8	8.8\\
2755.2	10.6\\
2760.6	8.7\\
2766.0	9.8\\
2771.4	8.6\\
2776.8	9.1\\
2782.1	8.7\\
2787.5	9.5\\
2792.9	9.0\\
2798.3	9.3\\
2803.6	9.4\\
};
\addlegendentry{\scriptsize IDM }

\addplot[area legend, draw=none, fill=white!10!black, fill opacity=0.2]
table[row sep=crcr] {%
x	y\\
1354.4	0\\
1354.4	80\\
1432	80\\
1432	0\\
}--cycle;
\addlegendentry{\scriptsize U-Turn}

\end{axis}

\input{Images/UDDS_pin_Nissan}
\input{Images/UDDS_vel_Nissan}
     
\end{tikzpicture}%

%% file: Images/UDDS_pin_Nissan.tex
%
%

\begin{axis}[%
width=6.5cm,
height=4.0cm,
at={(0cm,4.4cm)},
scale only axis,
xmin=505,
xmax=2800,
xlabel style={font=\color{white!15!black} \scriptsize},
xlabel={},
xticklabels={,,},
ymin=-25,
ymax=0,
ylabel style={font=\color{white!15!black} \scriptsize},
ylabel={Power In [KW]},
ticklabel style = {font = \scriptsize},
axis background/.style={fill=white},
xmajorgrids,
ymajorgrids,
legend style={at={(0.97,0.03)}, anchor=south east, legend cell align=left, align=left, draw=white!15!black}
]

\addplot [color=mycolor1, line width=1.3pt]
  table[row sep=crcr]{%
504.8	0.0\\
779.7	0.0\\
782.4	-1.9\\
785.3	-2.6\\
788.2	-1.4\\
791.0	-0.8\\
793.8	-1.4\\
796.7	-2.1\\
799.5	-2.5\\
802.2	-2.4\\
805.0	-2.2\\
807.8	-2.0\\
810.5	-3.2\\
813.3	-2.5\\
816.0	-2.6\\
818.7	-2.5\\
821.5	-2.7\\
826.6	-2.5\\
829.3	-2.2\\
831.7	-2.5\\
834.3	-2.4\\
836.8	-2.3\\
839.4	-2.4\\
841.9	-2.0\\
844.5	-2.5\\
847.1	-2.3\\
849.4	-1.9\\
851.7	-2.4\\
853.8	-2.4\\
856.1	-2.4\\
858.5	-2.5\\
860.8	-2.0\\
863.0	-2.2\\
865.0	-2.3\\
867.3	-2.2\\
869.5	-6.0\\
872.1	-6.1\\
873.7	-5.8\\
875.9	-5.8\\
877.7	-5.8\\
879.7	-5.6\\
882.2	-5.3\\
883.8	-4.9\\
884.8	-4.9\\
886.5	-4.6\\
888.3	-4.7\\
890.2	-4.4\\
891.9	-4.0\\
893.4	-4.2\\
894.9	-4.0\\
897.3	-3.8\\
898.6	-3.8\\
900.0	-3.3\\
901.4	-3.5\\
902.3	-3.2\\
904.1	-2.9\\
905.5	-2.9\\
906.2	-2.8\\
907.4	-2.5\\
909.2	-2.5\\
909.9	-2.3\\
911.2	-2.1\\
912.7	-2.0\\
913.3	-1.8\\
914.3	-1.6\\
915.7	-1.6\\
916.7	-1.4\\
918.5	-1.3\\
919.3	-1.4\\
919.8	-1.3\\
920.9	-0.9\\
922.2	-1.2\\
922.9	-0.9\\
924.2	-0.8\\
925.4	-0.9\\
926.0	-0.5\\
926.6	-0.7\\
927.3	-0.6\\
927.6	-0.5\\
928.3	-0.2\\
929.2	-0.2\\
929.6	-0.1\\
930.2	0.0\\
1238.8	0.0\\
1241.8	-0.2\\
1244.6	-1.3\\
1247.3	-1.3\\
1250.1	-1.5\\
1253.0	-0.9\\
1255.6	-0.3\\
1258.3	-0.7\\
1261.2	-1.8\\
1264.1	-1.5\\
1266.7	-1.7\\
1269.5	-1.0\\
1272.2	-1.5\\
1275.1	-1.6\\
1278.0	-1.9\\
1280.6	-1.9\\
1283.3	-1.5\\
1285.9	-1.6\\
1288.4	-1.8\\
1291.1	-1.9\\
1294.0	-2.2\\
1296.6	-2.1\\
1299.2	-2.4\\
1301.5	-2.4\\
1303.9	-1.7\\
1306.5	-2.8\\
1308.9	-2.3\\
1311.4	-1.9\\
1313.8	-2.3\\
1316.2	-2.0\\
1318.6	-2.2\\
1321.0	-2.0\\
1323.2	-1.8\\
1325.7	-2.0\\
1328.3	-2.0\\
1330.9	-2.0\\
1333.4	-1.9\\
1335.9	-2.1\\
1338.1	-2.0\\
1340.4	-1.8\\
1342.3	-1.8\\
1344.4	-2.0\\
1346.5	-6.7\\
1348.5	-7.3\\
1350.6	-7.3\\
1352.5	-7.0\\
1354.3	-5.2\\
1356.2	0.0\\
1364.0	0.0\\
1366.0	-1.4\\
1367.9	0.0\\
1376.2	0.0\\
1378.5	-0.3\\
1380.7	0.0\\
1422.7	0.0\\
1424.7	-0.8\\
1426.7	-0.6\\
1428.6	0.0\\
1430.5	0.0\\
1432.4	-0.5\\
1434.7	-0.5\\
1436.8	0.0\\
2608.0	0.0\\
2613.2	-2.1\\
2618.4	-2.6\\
2623.8	-2.4\\
2629.4	-2.6\\
2635.1	-1.6\\
2640.2	-3.2\\
2645.3	-3.3\\
2650.6	-4.5\\
2656.1	-3.0\\
2662.0	-4.0\\
2667.2	-5.0\\
2672.6	-6.0\\
2677.7	-5.8\\
2685.1	-5.6\\
2688.0	-5.5\\
2692.9	-5.4\\
2698.2	-6.0\\
2703.4	-6.6\\
2708.5	-6.6\\
2713.4	-6.8\\
2718.0	-6.6\\
2722.9	-7.0\\
2727.9	-6.9\\
2733.0	-7.4\\
2738.6	-7.0\\
2743.5	-7.3\\
2748.5	-7.5\\
2753.8	-6.9\\
2758.6	-6.9\\
2763.1	-6.4\\
2767.7	-6.6\\
2772.4	-6.4\\
2776.9	-6.8\\
2781.5	-6.0\\
2785.9	-6.9\\
2790.4	-5.9\\
2795.1	-6.4\\
2799.4	-6.0\\
2803.5	-6.6\\
};
\addlegendentry{\scriptsize MPC-C }

\addplot [color=mycolor2, dashed, line width=1.3pt]
  table[row sep=crcr]{%
503.1	0.0\\
862.0	0.0\\
865.2	-8.5\\
868.1	-13.9\\
870.9	-14.7\\
873.7	-17.4\\
876.2	-17.5\\
879.0	-17.9\\
881.5	-18.0\\
883.9	-18.1\\
886.2	-18.5\\
888.3	-17.9\\
890.6	-15.7\\
892.3	-15.0\\
894.7	-14.3\\
896.3	-13.8\\
898.4	-13.0\\
900.1	-11.9\\
901.8	-11.2\\
903.4	-10.1\\
904.8	-9.4\\
906.3	-8.6\\
907.7	-7.9\\
909.0	-7.4\\
910.4	-6.3\\
911.6	-5.7\\
912.8	-5.7\\
913.9	-4.9\\
914.6	-4.9\\
916.2	-4.5\\
917.2	-3.9\\
917.7	-3.9\\
919.0	-2.4\\
919.5	-2.4\\
920.7	-2.3\\
921.5	-2.4\\
922.3	-2.3\\
923.0	-2.3\\
923.6	-2.0\\
924.3	-1.9\\
924.9	-1.3\\
925.5	-1.2\\
926.0	-0.7\\
926.5	-0.2\\
927.2	0.0\\
929.4	0.0\\
929.4	0.0\\
929.4	0.0\\
929.6	0.0\\
988.3	0.0\\
990.5	-1.6\\
993.4	-1.3\\
994.8	-1.4\\
997.7	-0.7\\
999.6	0.0\\
1064.9	0.0\\
1067.2	-4.3\\
1069.4	-10.1\\
1071.6	-14.6\\
1073.7	-3.6\\
1075.4	-1.9\\
1078.0	-2.1\\
1080.0	-2.1\\
1082.0	-3.1\\
1083.9	-7.6\\
1085.7	-11.5\\
1087.4	-13.1\\
1089.2	-1.2\\
1090.9	-1.4\\
1092.6	-1.4\\
1094.4	-1.3\\
1096.1	0.0\\
1235.8	0.0\\
1238.5	-5.4\\
1243.5	-12.3\\
1246.2	-12.6\\
1249.2	-13.0\\
1251.9	-13.2\\
1254.9	-13.2\\
1257.6	-14.5\\
1260.3	-14.2\\
1265.3	-14.2\\
1268.0	-14.2\\
1270.5	-13.6\\
1272.9	-13.2\\
1275.2	-13.1\\
1277.3	-12.3\\
1279.7	-9.7\\
1281.9	-9.3\\
1283.9	-7.7\\
1286.2	0.0\\
1310.1	0.0\\
1312.4	-0.7\\
1314.6	-1.1\\
1316.7	-1.2\\
1318.8	-1.2\\
1320.7	-2.0\\
1322.9	-5.5\\
1325.0	-5.8\\
1327.0	0.0\\
1329.0	0.0\\
1331.0	-5.4\\
1332.8	-6.2\\
1334.7	-0.8\\
1336.5	0.0\\
1338.5	-5.4\\
1340.2	0.0\\
1349.3	0.0\\
1351.2	-0.6\\
1353.2	0.0\\
2690.0	0.0\\
2695.0	-2.7\\
2700.3	-1.9\\
2705.5	-1.3\\
2711.0	-2.0\\
2716.0	-1.1\\
2721.1	-3.4\\
2726.3	-3.5\\
2731.4	-4.5\\
2736.6	-7.1\\
2741.5	-7.3\\
2746.5	-7.2\\
2751.5	-7.1\\
2756.4	-7.0\\
2761.5	-6.7\\
2766.2	-5.6\\
2771.0	-4.8\\
2775.8	-6.5\\
2780.6	-6.9\\
2785.5	-7.0\\
2790.1	-6.4\\
2794.7	-8.0\\
2799.4	-7.5\\
2803.9	-7.7\\
};
\addlegendentry{\scriptsize MPC-U }

\addplot [color=mycolor3, dashdotted, line width=1.3pt]
  table[row sep=crcr]{%
503.1	0.0\\
600.9	0.0\\
604.0	-1.0\\
607.0	-0.9\\
610.1	-0.2\\
613.3	-0.3\\
616.3	-0.1\\
618.7	-0.2\\
622.4	-0.1\\
625.3	-0.3\\
628.4	0.0\\
631.4	-0.1\\
634.2	-0.2\\
637.1	-0.5\\
639.8	-0.3\\
643.0	-0.4\\
645.6	-0.1\\
648.5	-0.2\\
651.2	-0.5\\
653.9	-0.0\\
656.8	-0.5\\
659.6	-0.1\\
662.6	-0.6\\
668.4	-0.2\\
671.3	-0.4\\
674.0	-0.6\\
676.4	-0.4\\
679.1	-0.6\\
681.8	0.0\\
788.6	0.0\\
791.4	-0.8\\
794.9	-0.9\\
798.0	-0.4\\
801.0	-0.1\\
803.8	-0.2\\
806.6	-0.0\\
810.0	-0.1\\
813.1	-0.2\\
815.8	-0.2\\
818.7	-0.2\\
821.5	-0.6\\
824.9	-0.6\\
827.9	0.0\\
879.1	0.0\\
882.9	-1.3\\
885.6	-3.5\\
888.7	-3.5\\
891.7	-5.4\\
894.4	-7.5\\
897.5	-7.7\\
900.1	-7.6\\
902.6	-8.2\\
905.2	-8.2\\
907.4	-8.6\\
912.1	-8.9\\
913.9	-9.2\\
916.0	-9.5\\
917.7	-9.9\\
919.7	-10.1\\
921.7	-10.1\\
923.5	-10.3\\
924.9	-10.7\\
926.0	-11.3\\
927.4	-10.2\\
928.7	-8.3\\
930.2	-6.3\\
931.4	-6.1\\
932.2	-6.4\\
933.1	-6.5\\
933.9	-5.0\\
934.7	-3.4\\
935.5	0.0\\
936.0	-0.6\\
936.6	-0.7\\
937.2	-0.7\\
938.3	-0.8\\
939.2	-0.4\\
939.9	0.0\\
1055.5	0.0\\
1058.8	-0.4\\
1062.3	-0.6\\
1065.6	-0.4\\
1068.8	-0.5\\
1070.5	-0.7\\
1072.9	-0.7\\
1075.6	-0.7\\
1078.2	-1.3\\
1080.8	-0.8\\
1083.2	-1.1\\
1085.6	-1.1\\
1088.1	-1.3\\
1091.6	-1.2\\
1095.3	-1.2\\
1096.6	-1.3\\
1099.9	-1.2\\
1102.6	-1.3\\
1105.0	-1.2\\
1107.4	-1.2\\
1109.7	-1.0\\
1112.1	-0.6\\
1114.4	-1.3\\
1116.8	-1.3\\
1118.7	-1.4\\
1121.0	-2.3\\
1123.2	-4.4\\
1125.5	-6.7\\
1127.8	-7.0\\
1129.7	-6.9\\
1132.0	-6.5\\
1134.1	-5.8\\
1136.0	-5.1\\
1138.0	0.0\\
1141.9	0.0\\
1144.0	-1.0\\
1145.9	-0.8\\
1148.0	0.0\\
1279.6	0.0\\
1283.8	-3.9\\
1287.5	-4.7\\
1291.4	-4.8\\
1295.4	-5.2\\
1299.1	-4.5\\
1302.4	-6.3\\
1306.6	-7.1\\
1309.5	-7.0\\
1312.5	-7.7\\
1315.5	-7.9\\
1318.2	-8.0\\
1320.8	-8.0\\
1323.3	-8.9\\
1325.7	-9.2\\
1327.9	-3.1\\
1330.2	-3.6\\
1332.4	-3.4\\
1334.8	-3.2\\
1337.2	0.0\\
1341.6	0.0\\
1344.0	-1.9\\
1346.2	-4.2\\
1348.3	-8.6\\
1350.6	-9.7\\
1352.5	-8.8\\
1354.5	-4.8\\
1356.2	0.0\\
1633.5	0.0\\
1639.0	-0.1\\
1642.6	-0.6\\
1648.5	-1.4\\
1655.9	-2.0\\
1661.6	-2.3\\
1665.5	-2.6\\
1670.4	-2.5\\
1677.8	-2.5\\
1683.3	-3.1\\
1693.9	-2.0\\
1699.2	-1.8\\
1704.3	-1.7\\
1709.2	-1.5\\
1714.6	-1.4\\
1719.5	-0.8\\
1725.5	-0.7\\
1730.6	0.0\\
2306.6	0.0\\
2311.5	-1.1\\
2317.0	0.0\\
2322.5	0.0\\
2328.1	-0.1\\
2333.6	0.0\\
2349.6	0.0\\
2355.2	-1.9\\
2360.4	-1.0\\
2365.6	-0.8\\
2370.9	-0.2\\
2376.4	0.0\\
2381.7	-0.2\\
2386.8	0.0\\
2399.9	0.0\\
2403.7	-0.4\\
2409.9	-0.1\\
2416.8	-0.7\\
2423.5	-0.7\\
2429.8	-1.5\\
2434.2	-4.3\\
2439.3	-4.6\\
2444.6	-3.7\\
2449.7	-4.6\\
2454.7	-4.7\\
2459.7	-4.5\\
2464.5	-4.7\\
2469.3	-4.5\\
2474.0	-4.7\\
2478.8	0.0\\
2757.1	0.0\\
2762.7	-2.8\\
2768.3	-2.5\\
2774.0	-2.9\\
2779.4	-3.4\\
2784.9	-3.8\\
2790.2	0.0\\
2795.6	-2.5\\
2801.0	-2.0\\
};
\addlegendentry{\scriptsize WIE }

\addplot [color=mycolor4, dotted, line width=1.3pt]
  table[row sep=crcr]{%
504.0	0.0\\
866.9	0.0\\
870.0	-2.5\\
873.0	-10.7\\
875.9	-13.5\\
878.8	-13.9\\
881.6	-17.4\\
884.4	-17.1\\
887.0	-17.2\\
889.6	-18.3\\
892.1	-18.9\\
894.5	-19.0\\
896.8	-18.1\\
899.0	-17.2\\
901.1	-12.7\\
903.2	-16.3\\
905.1	-15.7\\
906.9	-15.0\\
908.7	-14.1\\
910.3	-14.9\\
911.9	-14.0\\
913.4	-12.8\\
914.8	-12.3\\
916.1	-11.2\\
917.2	-10.7\\
918.3	-9.4\\
919.3	-8.6\\
920.2	-7.6\\
921.0	-6.6\\
921.8	-4.5\\
922.4	-3.2\\
923.0	-1.5\\
923.4	-0.6\\
923.8	0.0\\
924.3	0.0\\
924.4	0.0\\
924.4	0.0\\
1042.3	0.0\\
1045.6	-1.3\\
1048.6	-2.6\\
1051.7	-1.0\\
1054.1	0.0\\
1056.6	-5.5\\
1059.8	-10.9\\
1062.8	-11.3\\
1066.5	-11.6\\
1069.4	-11.8\\
1072.2	-11.8\\
1074.9	-12.4\\
1077.6	-13.2\\
1080.2	-12.3\\
1082.7	-11.6\\
1085.3	-12.7\\
1087.7	-3.8\\
1090.1	-3.1\\
1092.4	-1.9\\
1094.7	-2.0\\
1096.9	-1.7\\
1098.9	-1.8\\
1100.7	-1.6\\
1102.8	-1.4\\
1104.6	-1.4\\
1106.3	-1.3\\
1108.2	-1.4\\
1109.9	-1.3\\
1111.7	-1.2\\
1113.4	-1.1\\
1115.1	-0.9\\
1116.8	0.0\\
1265.4	0.0\\
1269.1	-4.5\\
1273.0	-15.2\\
1276.7	-18.7\\
1281.0	-19.1\\
1284.0	-19.4\\
1287.3	-19.9\\
1290.5	-20.0\\
1293.5	-20.3\\
1296.3	-20.5\\
1299.3	-20.7\\
1302.0	-20.7\\
1304.5	-20.3\\
1307.0	-21.9\\
1309.2	-22.4\\
1311.8	-20.4\\
1313.7	-13.0\\
1316.4	-1.4\\
1318.5	-1.6\\
1320.6	-2.7\\
1322.6	-7.5\\
1324.6	-11.8\\
1326.3	-10.3\\
1328.5	-1.0\\
1330.4	-1.3\\
1332.2	-1.3\\
1334.9	-2.1\\
1336.0	-2.7\\
1337.6	-0.5\\
1339.2	-0.6\\
1340.5	-0.9\\
1342.0	-0.8\\
1343.4	-0.2\\
1344.8	0.0\\
2803.6	0.0\\
};
\addlegendentry{\scriptsize IDM }

\addplot[area legend, draw=none, fill=white!10!black, fill opacity=0.2]
table[row sep=crcr] {%
x	y\\
1354.4	-25\\
1354.4	0\\
1432	0\\
1432	-25\\
}--cycle;
\addlegendentry{\scriptsize U-Turn}

\end{axis}

%% file: Images/Protobuf.tex
\newcommand{\plus}{0.10cm}
\newcommand{\lene}{0.50cm}
\newcommand{\lent}{0.75cm}
\newcommand{\side}{1.50cm}
\newcommand{\below}{4.50cm}
\newcommand{\xcen}{0.5*\x2-0.5*\x1}
\newcommand{\ycen}{0.5*\y2-0.5*\y1}

\tikzstyle{process} = 
    [
        rectangle
      , draw
      , top color=white, bottom color=white!10,
      , text centered
      , node distance=1.5cm
      , inner sep=4pt
      , minimum height = 1.2cm
      , font = \ttfamily
      ]
\tikzstyle{simulation} =
    [
        rectangle
      , draw=red!50
      , top color=white, bottom color=red!10,
      , text centered
      , node distance=1.5cm
      , minimum height=2em
      , inner sep=8pt
    ]
\tikzstyle{database} =
    [
        cylinder
      , draw
      , top color = white, bottom color=white
      , shape border rotate=90
      , minimum height=1.8cm
      , minimum width=0.8cm
      , inner sep=10pt
      , shape aspect=0.25
    ]
\tikzstyle{line} = 
    [
        draw
      , -latex'
    ]
\tikzstyle{dash} =
    [
        draw
      , -latex'
      , dashed
    ]
    
\begin{tikzpicture}[node distance = 1cm, auto, every node/.style={black,scale=0.80}]]
    \node [] (Subscription) {\large Subscription/Unsubscription Message:};
    \node [below right = \lene and \lene of Subscription.west, process, align = center] (1) {Preamble\\ \#1};
    \node [right = \plus of 1] (p) {+};
    \node [right = \plus of p, process, align = center] (Identification) {Identification};
    \node [right = 0 of Identification.east, process, align = center, anchor=west] (Flag) {(Un)Subscription\\ Flag};
    \node [right = 0 of Flag.east, process, align=center, anchor=west] (Probe1) {Probe Data};
    \node [right = 0 of Probe1.east, process, align=center, anchor=west] (error1) {Error\\ Flag};
    
    \node [below = \lene of Identification, process, align = left] (IDMessage) {Sim/Physical Flag\\ Physical Vehicle ID};
    \draw (IDMessage.north west) -- (Identification.south west);
    \draw (IDMessage.north east) -- (Identification.south east);
    
    \node [below = \lene of Probe1, process, align = left] (IDProbe) {Velocity ($v$)\\ Location ($x,\ y$)\\ Heading ($\theta$)\\ Brake On/Off\\ Timestamp (Unix)};
    \draw (IDProbe.north west) -- (Probe1.south west);
    \draw (IDProbe.north east) -- (Probe1.south east);
    
    \node [below = 0.02cm of 1, align=left] {\scriptsize Preamble byte\\ \scriptsize = 0x16};
    
    \node [below right = 0.02cm and -2.1cm of Flag, align=left] {\scriptsize Unsubscription = 0\\ \scriptsize Subscription = 1};
    
    \node [below = \below of Subscription.west, anchor = west] (V2Sim) {\large V2Sim Message:};
    \node [below right = \lene and \lene of V2Sim.west, process, align=center] (2) {Preamble\\ \#2};
    \node [right = \plus of 2] (p2) {+};
    \node [right = \plus of p2, process, align = center] (ID) {Physical Vehicle ID};
    \node [right = 0 of ID.east, process, align=center, anchor=west] (Probe2) {Probe Data};
    \node [right = 0 of Probe2.east, process, align=center, anchor=west] (error2) {Error\\ Flag};
    
    \node [below = \lene of Probe2, process, align = left] (VProbe) {Velocity ($v$)\\ Location ($x,\ y$)\\ Heading ($\theta$)\\ Brake On/Off\\ Timestamp (Unix)};
    \draw (VProbe.north west) -- (Probe2.south west);
    \draw (VProbe.north east) -- (Probe2.south east);
    
    \node [below = 0.02cm of 2, align=left] {\scriptsize Preamble byte\\ \scriptsize = 0x43};
    \node [below right = 0.02cm and -0.50cm of error2, align=left] {\scriptsize OK=1\\ \scriptsize Error = 0};
    
    \node [below = \below of V2Sim.west, anchor = west] (Sim2V) {\large Sim2V Message:};
    \node [below right = \lene and \lene of Sim2V.west, process, align = center] (3) {Preamble\\ \#3};
    \node [right = \plus of 3] (p3) {+};
    \node [right = \plus of p3, process, align = center] (Probe3) {Probe Data 1};
    \node [right = 0 of Probe3.east, process, align=center, anchor=west] (dot) {$\cdots$};
    \node [right = 0 of dot.east, process, align=center, anchor=west] (Probek) {Probe Data k};
    \node [right = 0 of Probek.east, process, align=center, anchor=west] (error3) {Error\\ Flag};
    
    \node [below right = \lene and -2*\side of Probe3, process, align = left] (SProbe) {Simulated Vehicle ID\\ Simulated Vehicle Type\\ Velocity ($v$)\\ Longitudinal Distance Away ($dx$)\\ Lateral Distance Away ($dy$)\\ Heading ($\theta$)\\ Brake On/Off\\ Timestamp (Unix)};
    \draw (SProbe.north west) -- (Probe3.south west);
    \draw (SProbe.north east) -- (Probe3.south east);
    
    \node [below left = 0.02cm and -1cm of 3, align=left] {\scriptsize Preamble byte\\ \scriptsize = 0xEC};
    
    \node [below = \below+1.0cm of Sim2V.west, anchor=west] (V2V) {\large V2V Message:};
    \node [below right = \lene and \lene of V2V.west, process, align = center] (4) {Preamble \\ \#4};
    \node [right = \plus of 4] (p4) {+};
    \node [right = \plus of p4, process, align=center] (ID2) {Physical Vehicle ID};
    \node [right = 0 of ID2.east, process, align=center, anchor=west] (Time) {Timestamp};
    \node [right = 0 of Time.east, process, align=center, anchor=west] (MPC) {MPC Message};
    \node [right = 0 of MPC.east, process, align=center, anchor=west] (error4) {Error\\ Flag};
    
    \node [below left = \lene and -2*\side of MPC, process, align = left] (MPCMessage) {Longitudinal Distance ($s_i \ | \ i=1,...,N$)\\ Lateral Distance ($l_i \ | \ i=1,...,N$)\\ Terminal Velocity ($v_N$)};
    \draw (MPCMessage.north west) -- (MPC.south west);
    \draw (MPCMessage.north east) -- (MPC.south east);
    
    \node [below = 0.02cm of 4, align=left] {\scriptsize Preamble byte\\ \scriptsize = 0x6B};
    
\end{tikzpicture}

%% file: main.bbl
\begin{thebibliography}{50}
\expandafter\ifx\csname natexlab\endcsname\relax\def\natexlab#1{#1}\fi
\providecommand{\url}[1]{\texttt{#1}}
\providecommand{\href}[2]{#2}
\providecommand{\path}[1]{#1}
\providecommand{\DOIprefix}{doi:}
\providecommand{\ArXivprefix}{arXiv:}
\providecommand{\URLprefix}{URL: }
\providecommand{\Pubmedprefix}{pmid:}
\providecommand{\doi}[1]{\href{http://dx.doi.org/#1}{\path{#1}}}
\providecommand{\Pubmed}[1]{\href{pmid:#1}{\path{#1}}}
\providecommand{\bibinfo}[2]{#2}
\ifx\xfnm\relax \def\xfnm[#1]{\unskip,\space#1}\fi
\bibitem[{Alexander and Lunenfeld(1986)}]{alexanderExpectancy}
\bibinfo{author}{Alexander, G.}, \bibinfo{author}{Lunenfeld, H.},
  \bibinfo{year}{1986}.
\newblock \bibinfo{title}{Driver expectancy in highway design and traffic
  operations}.
\newblock \bibinfo{journal}{{U.S. Department of Transportation Federal Highway
  Administration}} .
\bibitem[{Amidi and Thorpe(1991)}]{Amidi1991}
\bibinfo{author}{Amidi, O.}, \bibinfo{author}{Thorpe, C.E.},
  \bibinfo{year}{1991}.
\newblock \bibinfo{title}{{Integrated mobile robot control}}.
\newblock \bibinfo{journal}{Mobile Robots V} \bibinfo{volume}{1388},
  \bibinfo{pages}{504--523}.
\newblock \DOIprefix\doi{10.1117/12.25494}.
\bibitem[{Ard et~al.(2020)Ard, Ashtiani, Vahidi and Borhan}]{ard2019platooning}
\bibinfo{author}{Ard, T.}, \bibinfo{author}{Ashtiani, F.},
  \bibinfo{author}{Vahidi, A.}, \bibinfo{author}{Borhan, H.},
  \bibinfo{year}{2020}.
\newblock \bibinfo{title}{Optimizing gap tracking subject to dynamic losses via
  connected and anticipative {MPC} in truck platooning}.
\newblock \bibinfo{journal}{Proceedings of the American Control Conference} .
\bibitem[{Ard et~al.(2019)Ard, Dollar, Vahidi, Zhang and
  Karbowski}]{ard2019microsimulation}
\bibinfo{author}{Ard, T.}, \bibinfo{author}{Dollar, R.A.},
  \bibinfo{author}{Vahidi, A.}, \bibinfo{author}{Zhang, Y.},
  \bibinfo{author}{Karbowski, D.}, \bibinfo{year}{2019}.
\newblock \bibinfo{title}{Microsimulation of energy and flow effects from
  optimal automated driving in mixed traffic}.
\newblock \bibinfo{journal}{arXiv preprint arXiv:1911.06818} .
\bibitem[{Bhangu et~al.(2005)Bhangu, Bentley, Stone and
  Bingham}]{bhangu2005nonlinear}
\bibinfo{author}{Bhangu, B.S.}, \bibinfo{author}{Bentley, P.},
  \bibinfo{author}{Stone, D.A.}, \bibinfo{author}{Bingham, C.M.},
  \bibinfo{year}{2005}.
\newblock \bibinfo{title}{Nonlinear observers for predicting state-of-charge
  and state-of-health of lead-acid batteries for hybrid-electric vehicles}.
\newblock \bibinfo{journal}{IEEE transactions on vehicular technology}
  \bibinfo{volume}{54}, \bibinfo{pages}{783--794}.
\bibitem[{Chen et~al.(2010)Chen, Li and Zhang}]{Chen2010ATraffic}
\bibinfo{author}{Chen, X.}, \bibinfo{author}{Li, L.}, \bibinfo{author}{Zhang,
  Y.}, \bibinfo{year}{2010}.
\newblock \bibinfo{title}{{A markov model for headway/spacing distribution of
  road traffic}}.
\newblock \bibinfo{journal}{IEEE Transactions on Intelligent Transportation
  Systems} \bibinfo{volume}{11}, \bibinfo{pages}{773--785}.
\newblock \DOIprefix\doi{10.1109/TITS.2010.2050141}.
\bibitem[{Dollar and Vahidi(2018)}]{dollar2018efficient}
\bibinfo{author}{Dollar, R.A.}, \bibinfo{author}{Vahidi, A.},
  \bibinfo{year}{2018}.
\newblock \bibinfo{title}{Efficient and collision-free anticipative cruise
  control in randomly mixed strings}.
\newblock \bibinfo{journal}{IEEE Transactions on Intelligent Vehicles}
  \bibinfo{volume}{3}, \bibinfo{pages}{439--452}.
\bibitem[{Dollar and Vahidi(2019)}]{dollar2019automated}
\bibinfo{author}{Dollar, R.A.}, \bibinfo{author}{Vahidi, A.},
  \bibinfo{year}{2019}.
\newblock \bibinfo{title}{Automated vehicles in hazardous merging traffic: A
  chance-constrained approach}.
\newblock \bibinfo{journal}{IFAC-PapersOnLine} \bibinfo{volume}{52},
  \bibinfo{pages}{218--223}.
\bibitem[{{ELM Electronics}(2012)}]{elm327}
\bibinfo{author}{{ELM Electronics}}, \bibinfo{year}{2012}.
\newblock \bibinfo{title}{{ELM327, OBD to RS232} interpreter}.
\newblock \URLprefix
  \url{https://www.elmelectronics.com/wp-content/uploads/2017/01/ELM327DS.pdf}.
\bibitem[{Fayazi and Vahidi(2017)}]{fayazi2017vehicle}
\bibinfo{author}{Fayazi, S.A.}, \bibinfo{author}{Vahidi, A.},
  \bibinfo{year}{2017}.
\newblock \bibinfo{title}{Vehicle-in-the-loop {(VIL)} verification of a smart
  city intersection control scheme for autonomous vehicles}, in:
  \bibinfo{booktitle}{2017 IEEE Conference on Control Technology and
  Applications (CCTA)}, \bibinfo{organization}{IEEE}. pp.
  \bibinfo{pages}{1575--1580}.
\bibitem[{Fayazi et~al.(2019)Fayazi, Vahidi and Luckow}]{fayazi2019}
\bibinfo{author}{Fayazi, S.A.}, \bibinfo{author}{Vahidi, A.},
  \bibinfo{author}{Luckow, A.}, \bibinfo{year}{2019}.
\newblock \bibinfo{title}{A vehicle-in-the-loop {(VIL)} verification of an
  all-autonomous intersection control scheme}.
\newblock \bibinfo{journal}{Transportation Research Part C: Emerging
  Technologies} \bibinfo{volume}{107}, \bibinfo{pages}{193 -- 210}.
\newblock \URLprefix
  \url{http://www.sciencedirect.com/science/article/pii/S0968090X19310745},
  \DOIprefix\doi{https://doi.org/10.1016/j.trc.2019.07.027}.
\bibitem[{Fellendorf and Vortisch(2011)}]{FellendorfVISSIM}
\bibinfo{author}{Fellendorf, M.}, \bibinfo{author}{Vortisch, P.},
  \bibinfo{year}{2011}.
\newblock \bibinfo{title}{{Microscopic traffic flow simulator VISSIM}}.
\newblock \bibinfo{journal}{Fundamentals of Traffic Simulation. International
  Series in Operations Research and Management Science} ,
  \bibinfo{pages}{63--93}\DOIprefix\doi{10.1007/978-1-4419-6142-6_2}.
\bibitem[{{Gao} et~al.(2014){Gao}, {Gray}, {Carvalho}, {Tseng} and
  {Borrelli}}]{Gao2014}
\bibinfo{author}{{Gao}, Y.}, \bibinfo{author}{{Gray}, A.},
  \bibinfo{author}{{Carvalho}, A.}, \bibinfo{author}{{Tseng}, H.E.},
  \bibinfo{author}{{Borrelli}, F.}, \bibinfo{year}{2014}.
\newblock \bibinfo{title}{Robust nonlinear predictive control for
  semiautonomous ground vehicles}, in: \bibinfo{booktitle}{2014 American
  Control Conference}, pp. \bibinfo{pages}{4913--4918}.
\bibitem[{{Google Inc.}(2020)}]{protobuf}
\bibinfo{author}{{Google Inc.}}, \bibinfo{year}{2020}.
\newblock \bibinfo{title}{Protocol buffers}.
\newblock \URLprefix \url{https://developers.google.com/protocol-buffers}.
\bibitem[{Gunter et~al.(2019)Gunter, Gloudemans, Stern, McQuade, Bhadani,
  Bunting, Monache, Lysecky, Seibold, Sprinkle, Piccoli and Work}]{Gunter2019}
\bibinfo{author}{Gunter, G.}, \bibinfo{author}{Gloudemans, D.},
  \bibinfo{author}{Stern, R.E.}, \bibinfo{author}{McQuade, S.},
  \bibinfo{author}{Bhadani, R.}, \bibinfo{author}{Bunting, M.},
  \bibinfo{author}{Monache, M.L.D.}, \bibinfo{author}{Lysecky, R.},
  \bibinfo{author}{Seibold, B.}, \bibinfo{author}{Sprinkle, J.},
  \bibinfo{author}{Piccoli, B.}, \bibinfo{author}{Work, D.B.},
  \bibinfo{year}{2019}.
\newblock \bibinfo{title}{{Are commercially implemented adaptive cruise control
  systems string stable?}} \bibinfo{volume}{19122}, \bibinfo{pages}{1--22}.
\newblock \URLprefix \url{http://arxiv.org/abs/1905.02108},
  \href{http://arxiv.org/abs/1905.02108}{\tt arXiv:1905.02108}.
\bibitem[{{Gurobi Optimization}(2020)}]{gurobiref}
\bibinfo{author}{{Gurobi Optimization}}, \bibinfo{year}{2020}.
\newblock \bibinfo{title}{Gurobi optimizer reference manual}.
\newblock \URLprefix \url{https://www.gurobi.com/documentation/}.
\bibitem[{{Hajdu} et~al.(2020){Hajdu}, {Ge}, {Insperger} and
  {Orosz}}]{Hajdu2020}
\bibinfo{author}{{Hajdu}, D.}, \bibinfo{author}{{Ge}, J.I.},
  \bibinfo{author}{{Insperger}, T.}, \bibinfo{author}{{Orosz}, G.},
  \bibinfo{year}{2020}.
\newblock \bibinfo{title}{Robust design of connected cruise control among
  human-driven vehicles}.
\newblock \bibinfo{journal}{IEEE Transactions on Intelligent Transportation
  Systems} \bibinfo{volume}{21}, \bibinfo{pages}{749--761}.
\bibitem[{Han et~al.(2019)Han, Vahidi and Sciarretta}]{han2019fundamentals}
\bibinfo{author}{Han, J.}, \bibinfo{author}{Vahidi, A.},
  \bibinfo{author}{Sciarretta, A.}, \bibinfo{year}{2019}.
\newblock \bibinfo{title}{Fundamentals of energy efficient driving for
  combustion engine and electric vehicles: An optimal control perspective}.
\newblock \bibinfo{journal}{Automatica} \bibinfo{volume}{103},
  \bibinfo{pages}{558--572}.
\bibitem[{Kamal et~al.(2011)Kamal, Mukai, Murata and Kawabe}]{Kamal2011}
\bibinfo{author}{Kamal, M.A.}, \bibinfo{author}{Mukai, M.},
  \bibinfo{author}{Murata, J.}, \bibinfo{author}{Kawabe, T.},
  \bibinfo{year}{2011}.
\newblock \bibinfo{title}{Ecological driving based on preceding vehicle
  prediction using {MPC}}.
\newblock \bibinfo{journal}{IFAC Proceedings Volumes (IFAC-PapersOnline)}
  \bibinfo{volume}{44}, \bibinfo{pages}{3843--3848}.
\newblock \URLprefix \url{http://dx.doi.org/10.3182/20110828-6-IT-1002.02748},
  \DOIprefix\doi{10.3182/20110828-6-IT-1002.02748}.
\bibitem[{Li et~al.(2010)Li, Li, Rajamani and Wang}]{li2010model}
\bibinfo{author}{Li, S.}, \bibinfo{author}{Li, K.}, \bibinfo{author}{Rajamani,
  R.}, \bibinfo{author}{Wang, J.}, \bibinfo{year}{2010}.
\newblock \bibinfo{title}{Model predictive multi-objective vehicular adaptive
  cruise control}.
\newblock \bibinfo{journal}{IEEE Transactions on Control Systems Technology}
  \bibinfo{volume}{19}, \bibinfo{pages}{556--566}.
\bibitem[{Liu et~al.(2020)Liu, Shladover, Lu and Kan}]{Liu2020}
\bibinfo{author}{Liu, H.}, \bibinfo{author}{Shladover, S.E.},
  \bibinfo{author}{Lu, X.Y.}, \bibinfo{author}{Kan, X.}, \bibinfo{year}{2020}.
\newblock \bibinfo{title}{{Freeway vehicle fuel efficiency improvement via
  cooperative adaptive cruise control}}.
\newblock \bibinfo{journal}{Journal of Intelligent Transportation Systems:
  Technology, Planning, and Operations} \bibinfo{volume}{0},
  \bibinfo{pages}{1--13}.
\newblock \DOIprefix\doi{10.1080/15472450.2020.1720673}.
\bibitem[{McMahon et~al.(1990)McMahon, Hedrick and
  Shladover}]{mcmahon1990vehicle}
\bibinfo{author}{McMahon, D.H.}, \bibinfo{author}{Hedrick, J.K.},
  \bibinfo{author}{Shladover, S.E.}, \bibinfo{year}{1990}.
\newblock \bibinfo{title}{Vehicle modelling and control for automated highway
  systems}, in: \bibinfo{booktitle}{1990 American Control Conference},
  \bibinfo{organization}{IEEE}. pp. \bibinfo{pages}{297--303}.
\bibitem[{Mills(2010)}]{ntpmills}
\bibinfo{author}{Mills, D.L.}, \bibinfo{year}{2010}.
\newblock \bibinfo{title}{Computer Network Time Synchronization: The Network
  Time Protocol on Earth and in Space, Second Edition}.
\newblock \bibinfo{edition}{2nd} ed., \bibinfo{publisher}{CRC Press, Inc.},
  \bibinfo{address}{USA}.
\bibitem[{{NTSB Media Relations}(2020)}]{TeslaCrash}
\bibinfo{author}{{NTSB Media Relations}}, \bibinfo{year}{2020}.
\newblock \bibinfo{title}{Tesla crash investigation yields 9 {NTSB} safety
  recommendations}.
\newblock \URLprefix
  \url{https://www.ntsb.gov/news/press-releases/Pages/NR20200225.aspx}.
\bibitem[{{Online}(2020a)}]{avlkma}
\bibinfo{author}{{Online}}, \bibinfo{year}{2020}a.
\newblock \bibinfo{title}{{AVL KMA} mobile fuel consumption measurement
  system}.
\newblock \URLprefix \url{https://www.avl.com/web/guest/-/avl-kma-mobile}.
\bibitem[{{Online}(2020b)}]{leafspy}
\bibinfo{author}{{Online}}, \bibinfo{year}{2020}b.
\newblock \bibinfo{title}{Leafspy pro}.
\newblock \URLprefix
  \url{https://apps.apple.com/us/app/leafspy-pro/id967376861}.
\bibitem[{{Online}(2020c)}]{somateDaq}
\bibinfo{author}{{Online}}, \bibinfo{year}{2020}c.
\newblock \bibinfo{title}{{Somat eDAQ and Somat eDAQXR} data acquisition
  systems}.
\newblock \URLprefix
  \url{https://www.hbm.com/en/7775/edaqxr‐edaq‐mobile‐rugged‐data‐acquisition‐systems/}.
\bibitem[{Parvini et~al.(2017)Parvini, Vahidi and
  Fayazi}]{parvini2017heuristic}
\bibinfo{author}{Parvini, Y.}, \bibinfo{author}{Vahidi, A.},
  \bibinfo{author}{Fayazi, S.A.}, \bibinfo{year}{2017}.
\newblock \bibinfo{title}{Heuristic versus optimal charging of supercapacitors,
  lithium-ion, and lead-acid batteries: An efficiency point of view}.
\newblock \bibinfo{journal}{IEEE Transactions on Control Systems Technology}
  \bibinfo{volume}{26}, \bibinfo{pages}{167--180}.
\bibitem[{Ploeg et~al.(2011)Ploeg, Scheepers, {Van Nunen}, {Van De Wouw} and
  Nijmeijer}]{Ploeg2011}
\bibinfo{author}{Ploeg, J.}, \bibinfo{author}{Scheepers, B.T.},
  \bibinfo{author}{{Van Nunen}, E.}, \bibinfo{author}{{Van De Wouw}, N.},
  \bibinfo{author}{Nijmeijer, H.}, \bibinfo{year}{2011}.
\newblock \bibinfo{title}{{Design and experimental evaluation of cooperative
  adaptive cruise control}}.
\newblock \bibinfo{journal}{IEEE Conference on Intelligent Transportation
  Systems, Proceedings, ITSC} ,
  \bibinfo{pages}{260--265}\DOIprefix\doi{10.1109/ITSC.2011.6082981}.
\bibitem[{Pomerleau(1990)}]{Pomerleau1990}
\bibinfo{author}{Pomerleau, D.A.}, \bibinfo{year}{1990}.
\newblock \bibinfo{title}{Alvinn, an autonomous land vehicle in a neural
  network} \URLprefix
  \url{{https://kilthub.cmu.edu/articles/ALVINN_an_autonomous_land_vehicle_in_a_neural_network/6603146}},
  \DOIprefix\doi{10.1184/R1/6603146.v1}.
\bibitem[{{Pourabdollah} et~al.(2017){Pourabdollah}, {Bjärkvik}, {Fürer},
  {Lindenberg} and {Burgdorf}}]{pourabdollahcalibration}
\bibinfo{author}{{Pourabdollah}, M.}, \bibinfo{author}{{Bjärkvik}, E.},
  \bibinfo{author}{{Fürer}, F.}, \bibinfo{author}{{Lindenberg}, B.},
  \bibinfo{author}{{Burgdorf}, K.}, \bibinfo{year}{2017}.
\newblock \bibinfo{title}{Calibration and evaluation of car following models
  using real-world driving data}, in: \bibinfo{booktitle}{2017 IEEE 20th
  International Conference on Intelligent Transportation Systems (ITSC)}, pp.
  \bibinfo{pages}{1--6}.
\bibitem[{Rios-Torres and Malikopoulos(2017)}]{Rios-Torres2017}
\bibinfo{author}{Rios-Torres, J.}, \bibinfo{author}{Malikopoulos, A.A.},
  \bibinfo{year}{2017}.
\newblock \bibinfo{title}{Automated and cooperative vehicle merging at highway
  on-ramps}.
\newblock \bibinfo{journal}{IEEE Transactions on Intelligent Transportation
  Systems} \bibinfo{volume}{18}, \bibinfo{pages}{780--789}.
\newblock \DOIprefix\doi{10.1109/TITS.2016.2587582}.
\bibitem[{Rios-Torres and Malikopoulos(2018)}]{Rios-Torres2018}
\bibinfo{author}{Rios-Torres, J.}, \bibinfo{author}{Malikopoulos, A.A.},
  \bibinfo{year}{2018}.
\newblock \bibinfo{title}{Impact of partial penetrations of connected and
  automated vehicles on fuel consumption and traffic flow}.
\newblock \bibinfo{journal}{IEEE Transactions on Intelligent Vehicles}
  \bibinfo{volume}{3}, \bibinfo{pages}{453--462}.
\newblock \DOIprefix\doi{10.1109/TIV.2018.2873899}.
\bibitem[{Schmied et~al.(2015)Schmied, Waschl, Quirynen, Diehl and {Del
  Re}}]{Schmied2015}
\bibinfo{author}{Schmied, R.}, \bibinfo{author}{Waschl, H.},
  \bibinfo{author}{Quirynen, R.}, \bibinfo{author}{Diehl, M.},
  \bibinfo{author}{{Del Re}, L.}, \bibinfo{year}{2015}.
\newblock \bibinfo{title}{Nonlinear {MPC} for emission efficient cooperative
  adaptive cruise control}.
\newblock \bibinfo{journal}{IFAC-PapersOnline} \bibinfo{volume}{48},
  \bibinfo{pages}{160--165}.
\newblock \URLprefix \url{http://dx.doi.org/10.1016/j.ifacol.2015.11.277},
  \DOIprefix\doi{10.1016/j.ifacol.2015.11.277}.
\bibitem[{Stern et~al.(2018)Stern, Cui, {Delle Monache}, Bhadani, Bunting,
  Churchill, Hamilton, Haulcy, Pohlmann, Wu, Piccoli, Seibold, Sprinkle and
  Work}]{Stern2018}
\bibinfo{author}{Stern, R.E.}, \bibinfo{author}{Cui, S.},
  \bibinfo{author}{{Delle Monache}, M.L.}, \bibinfo{author}{Bhadani, R.},
  \bibinfo{author}{Bunting, M.}, \bibinfo{author}{Churchill, M.},
  \bibinfo{author}{Hamilton, N.}, \bibinfo{author}{Haulcy, R.},
  \bibinfo{author}{Pohlmann, H.}, \bibinfo{author}{Wu, F.},
  \bibinfo{author}{Piccoli, B.}, \bibinfo{author}{Seibold, B.},
  \bibinfo{author}{Sprinkle, J.}, \bibinfo{author}{Work, D.B.},
  \bibinfo{year}{2018}.
\newblock \bibinfo{title}{{Dissipation of stop-and-go waves via control of
  autonomous vehicles: Field experiments}}.
\newblock \bibinfo{journal}{Transportation Research Part C: Emerging
  Technologies} \bibinfo{volume}{89}, \bibinfo{pages}{205--221}.
\newblock \URLprefix \url{https://doi.org/10.1016/j.trc.2018.02.005},
  \DOIprefix\doi{10.1016/j.trc.2018.02.005},
  \href{http://arxiv.org/abs/1705.01693}{\tt arXiv:1705.01693}.
\bibitem[{Szumanowski and Chang(2008)}]{szumanowski2008battery}
\bibinfo{author}{Szumanowski, A.}, \bibinfo{author}{Chang, Y.},
  \bibinfo{year}{2008}.
\newblock \bibinfo{title}{Battery management system based on battery nonlinear
  dynamics modeling}.
\newblock \bibinfo{journal}{IEEE transactions on vehicular technology}
  \bibinfo{volume}{57}, \bibinfo{pages}{1425--1432}.
\bibitem[{Talebpour and Mahmassani(2016)}]{Talebpour2016}
\bibinfo{author}{Talebpour, A.}, \bibinfo{author}{Mahmassani, H.S.},
  \bibinfo{year}{2016}.
\newblock \bibinfo{title}{{Influence of connected and autonomous vehicles on
  traffic flow stability and throughput}}.
\newblock \bibinfo{journal}{Transportation Research Part C: Emerging
  Technologies} \bibinfo{volume}{71}, \bibinfo{pages}{143--163}.
\newblock \URLprefix \url{http://dx.doi.org/10.1016/j.trc.2016.07.007},
  \DOIprefix\doi{10.1016/j.trc.2016.07.007}.
\bibitem[{Thrun et~al.(2006)}]{stanleyrobot}
\bibinfo{author}{Thrun, S.}, et~al., \bibinfo{year}{2006}.
\newblock \bibinfo{title}{Stanley: The robot that won the {DARPA} grand
  challenge}.
\newblock \bibinfo{journal}{Journal of Field Robotics} \bibinfo{volume}{23},
  \bibinfo{pages}{661--692}.
\newblock \DOIprefix\doi{10.1002/rob.20147}.
\bibitem[{Treiber et~al.(2000)Treiber, Hennecke and
  Helbing}]{treiber2000congested}
\bibinfo{author}{Treiber, M.}, \bibinfo{author}{Hennecke, A.},
  \bibinfo{author}{Helbing, D.}, \bibinfo{year}{2000}.
\newblock \bibinfo{title}{Congested traffic states in empirical observations
  and microscopic simulations}.
\newblock \bibinfo{journal}{Physical review E} \bibinfo{volume}{62},
  \bibinfo{pages}{1805}.
\bibitem[{Tseng et~al.(2016)Tseng, Zhou, Al~Hashmi, Chau, Song and
  Wilhelm}]{tseng2016data}
\bibinfo{author}{Tseng, C.M.}, \bibinfo{author}{Zhou, W.},
  \bibinfo{author}{Al~Hashmi, M.}, \bibinfo{author}{Chau, C.K.},
  \bibinfo{author}{Song, S.G.}, \bibinfo{author}{Wilhelm, E.},
  \bibinfo{year}{2016}.
\newblock \bibinfo{title}{Data extraction from electric vehicles through {OBD}
  and application of carbon footprint evaluation}, in:
  \bibinfo{booktitle}{Proceedings of the Workshop on Electric Vehicle Systems,
  Data, and Applications}, pp. \bibinfo{pages}{1--6}.
\bibitem[{Turri et~al.(2017)Turri, Kim, Guanetti, Johansson and
  Borrelli}]{Turri2017}
\bibinfo{author}{Turri, V.}, \bibinfo{author}{Kim, Y.},
  \bibinfo{author}{Guanetti, J.}, \bibinfo{author}{Johansson, K.H.},
  \bibinfo{author}{Borrelli, F.}, \bibinfo{year}{2017}.
\newblock \bibinfo{title}{{A model predictive controller for non-cooperative
  eco-platooning}}.
\newblock \bibinfo{journal}{Proceedings of the American Control Conference} ,
  \bibinfo{pages}{2309--2314}\DOIprefix\doi{10.23919/ACC.2017.7963297}.
\bibitem[{{United States Environmental Protection Agency
  (EPA)}(2020)}]{epacycle}
\bibinfo{author}{{United States Environmental Protection Agency (EPA)}},
  \bibinfo{year}{2020}.
\newblock \bibinfo{title}{Dynamometer drive schedules}.
\newblock \URLprefix
  \url{https://www.epa.gov/vehicle-and-fuel-emissions-testing/dynamometer-drive-schedules}.
\bibitem[{Vahidi and Sciarretta(2018)}]{vahidi2018energy}
\bibinfo{author}{Vahidi, A.}, \bibinfo{author}{Sciarretta, A.},
  \bibinfo{year}{2018}.
\newblock \bibinfo{title}{Energy saving potentials of connected and automated
  vehicles}.
\newblock \bibinfo{journal}{Transportation Research Part C: Emerging
  Technologies} \bibinfo{volume}{95}, \bibinfo{pages}{822--843}.
\bibitem[{Vortisch(2020)}]{vortischpdf}
\bibinfo{author}{Vortisch, P.}, \bibinfo{year}{2020}.
\newblock \bibinfo{title}{Where {I} can find the mathematical formulation of
  {Wiedemann} 99 car following model?}
\newblock \URLprefix
  \url{https://www.researchgate.net/post/Where_I_can_find_the_mathematical_formulation_of_Wiedemann_99_car_following_model}.
\bibitem[{Wan et~al.(2017)Wan, Zhang and Vahidi}]{wan2017probabilistic}
\bibinfo{author}{Wan, N.}, \bibinfo{author}{Zhang, C.},
  \bibinfo{author}{Vahidi, A.}, \bibinfo{year}{2017}.
\newblock \bibinfo{title}{Probabilistic anticipation and control in autonomous
  car following}.
\newblock \bibinfo{journal}{IEEE Transactions on Control Systems Technology}
  \bibinfo{volume}{27}, \bibinfo{pages}{30--38}.
\bibitem[{{Washington State Department of Transportation
  (WSDOT)}(2014)}]{WashingtonStateDepartmentofTransportationWSDOT2014ProtocolSimulation}
\bibinfo{author}{{Washington State Department of Transportation (WSDOT)}},
  \bibinfo{year}{2014}.
\newblock \bibinfo{title}{{Protocol for VISSIM simulation}}.
\newblock \URLprefix \url{http://www.oregon.gov/ODOT/TD/TP/APM/AddC.pdf}.
\bibitem[{Wiedemann(1974)}]{Wiedemann1974}
\bibinfo{author}{Wiedemann, R.}, \bibinfo{year}{1974}.
\newblock \bibinfo{title}{Simulation des strassenverkehrsflusses},
  \bibinfo{publisher}{Institut fur Verkehrswesen der Universitat Karlsruhe}.
\bibitem[{Woody(2006)}]{Woody2006CALIBRATINGVISSIM}
\bibinfo{author}{Woody, T.}, \bibinfo{year}{2006}.
\newblock \bibinfo{title}{{Calibrating freeway simulation models in VISSIM}}.
\newblock \bibinfo{type}{Technical Report}. University of Washington.
  \bibinfo{address}{Seattle, WA}.
\bibitem[{Zhang and Vahidi(2011)}]{Zhang2011}
\bibinfo{author}{Zhang, C.}, \bibinfo{author}{Vahidi, A.},
  \bibinfo{year}{2011}.
\newblock \bibinfo{title}{{Predictive cruise control with probabilistic
  constraints for eco driving}}.
\newblock \bibinfo{journal}{ASME 2011 Dynamic Systems and Control Conference
  and Bath/ASME Symposium on Fluid Power and Motion Control, DSCC 2011}
  \bibinfo{volume}{2}, \bibinfo{pages}{233--238}.
\newblock \DOIprefix\doi{10.1115/DSCC2011-5982}.
\bibitem[{Zheng et~al.(2017)Zheng, Li, Li, Borrelli and Hedrick}]{Zheng2017}
\bibinfo{author}{Zheng, Y.}, \bibinfo{author}{Li, S.E.}, \bibinfo{author}{Li,
  K.}, \bibinfo{author}{Borrelli, F.}, \bibinfo{author}{Hedrick, J.K.},
  \bibinfo{year}{2017}.
\newblock \bibinfo{title}{Distributed model predictive control for
  heterogeneous vehicle platoons under unidirectional topologies}.
\newblock \bibinfo{journal}{IEEE} \bibinfo{volume}{25},
  \bibinfo{pages}{899--910}.
\newblock \DOIprefix\doi{10.1109/TCST.2016.2594588}.

\end{thebibliography}
